\documentclass[useAMS,usenatbib]{mnras}

\usepackage{hyperref}

\usepackage{epsfig}
\usepackage{myaasmacros}
\usepackage{amsmath}
\usepackage{amsfonts}
\usepackage{amssymb}
\usepackage{subfig}
\usepackage{color}
\usepackage[dvipsnames]{xcolor}

\usepackage{hyperref}
\usepackage{changes}

\newcommand{\Msun}{\hbox{M$_\sun$}}
\newcommand{\Zsun}{\hbox{Z$_\sun$}}
\newcommand{\hii}{\hbox{H\,{\sc ii}}}
\newcommand{\ha}{\hbox{H$\alpha$}}
\newcommand{\hb}{\hbox{H$\beta$}}
\newcommand{\oiii}{\hbox{[O\,{\sc iii}]$\lambda5007$}}
\newcommand{\oiiiaur}{\hbox{[O\,{\sc iii}]$\lambda4346$}}
\newcommand{\heiiopt}{\hbox{He\,\textsc{ii}\,$\lambda4686$}}
\newcommand{\oi}{\hbox{[O\,{\sc i}]$\lambda6300$}} 
\newcommand{\nii}{\hbox{[N\,{\sc ii}]$\lambda6584$}}

\newcommand{\sii}{\hbox{[S\,{\sc ii}]$\lambda6724$}}

\newcommand{\oiiihb}{\hbox{[O\,{\sc iii}]/H$\beta$}}
\newcommand{\niiha}{\hbox{[N\,{\sc ii}]/H$\alpha$}}
\newcommand{\siiha}{\hbox{[S\,{\sc ii}]/H$\alpha$}}
\newcommand{\oiha}{\hbox{[O\,{\sc i}]/H$\alpha$}}
\newcommand{\oiiioii}{\hbox{[O\,{\sc iii}]/[O\,{\sc ii}]}}

\newcommand{\nv}{\hbox{N\,\textsc{v}\,$\lambda1240$}}

\newcommand{\niv}{\hbox{N\,\textsc{iv}]\,$\lambda1485$}}
\newcommand{\civd}{\hbox{C\,\textsc{iv}\,$\lambda\lambda1548,1551$}}
\newcommand{\civ}{\hbox{C\,\textsc{iv}\,$\lambda1550$}}
\newcommand{\heii}{\hbox{He\,\textsc{ii}\,$\lambda1640$}}
\newcommand{\oiiiuvd}{\hbox{O\,\textsc{iii}]$\lambda\lambda1661,1666$}}
\newcommand{\oiiiuv}{\hbox{O\,\textsc{iii}]\,$\lambda1663$}}
\newcommand{\niii}{\hbox{N\,\textsc{iii}]\,$\lambda1750$}}

\newcommand{\siliiid}{\hbox{[Si\,\textsc{iii}]$\lambda1883$+Si\,\textsc{iii}]$\lambda1892$}}
\newcommand{\siliii}{\hbox{Si\,\textsc{iii}]\,$\lambda1888$}}
\newcommand{\ciiid}{\hbox{[C\,\textsc{iii}]$\lambda1907$+C\,\textsc{iii}]$\lambda1909$}}
\newcommand{\ciii}{\hbox{C\,\textsc{iii}]\,$\lambda1908$}}
\newcommand{\cii}{\hbox{C\,\textsc{ii}]\,$\lambda2326$}}
\newcommand{\oiid}{\hbox{[O\,{\sc ii}]$\lambda\lambda3726,3729$}}
\newcommand{\oii}{\hbox{[O\,{\sc ii}]\,$\lambda3727$}}
\newcommand{\neiii}{\hbox{[Ne\,{\sc iii}]$\lambda\lambda3869,3967$}}

\newcommand{\lha}{\hbox{$L_\mathrm{H\alpha}$}}
\newcommand{\loii}{\hbox{$L_\mathrm{[O\,\textsc{iii}]}$}}

\newcommand{\loiiihb}{\hbox{$L_\mathrm{[O\,\textsc{iii}]+H\beta}$}}
\newcommand{\oiiinl}{\hbox{[O\,\textsc{iii}]}}
\newcommand{\oiinl}{\hbox{[O\,\textsc{ii}]}}
\newcommand{\heiinl}{\hbox{He\,\textsc{ii}}}

\title[Emission-line properties of IllustrisTNG galaxies] 
{Emission-line properties of IllustrisTNG galaxies: from local
  diagnostic diagrams to high-redshift predictions for JWST}     
\author[Hirschmann et al.]{Michaela Hirschmann$^{1,2}$\thanks{E-mail:
michaela.hirschmann@epfl.ch}, Stephane Charlot$^{3}$, Anna Feltre$^{4}$,
Emma Curtis-Lake$^{5}$, \newauthor Rachel S. Somerville$^{6}$,  Jacopo Chevallard$^{7}$, Ena
Choi$^{8,9}$, Dylan Nelson$^{10}$, \newauthor  Christophe
Morisset$^{11}$,  Adele Plat$^{12}$,  Alba Vidal-Garcia$^{13}$  
\\
$^{1}$Institute for Physics, Laboratory for Galaxy Evolution and
Spectral modelling, Ecole Polytechnique Federale de Lausanne,\\
Observatoire de Sauverny, Chemin Pegasi 51, 1290 Versoix,
Switzerland\\
$^{2}$ INAF, Osservatorio Astronomico di Trieste, Via Tiepolo 11,
34131 Trieste, Italy\\
$^{3}$Sorbonne Universit\'e, UPMC-CNRS, UMR7095, Institut
d'Astrophysique de Paris, F-75014 Paris, France\\
$^{4}$INAF  -  Osservatorio  di  Astrofisica  e  Scienza  dello
Spazio  di  Bologna,  Via  P.  Gobetti  93/3,  40129  Bologna,
Italy\\ 
$^{5}$Centre for Astrophysics Research, Department of Physics,
Astronomy and Mathematics, University of Hertfordshire, Hatfield AL10\\
9AB, UK\\
$^{6}$ Center for Computational Astrophysics, Flatiron Institute, 162
5th Avenue, New York, NY 10010, USA\\ 
$^{7}$Sub-department of Astrophysics, Department of Physics,
University of Oxford, Denys Wilkinson Building, Keble Road, 
Oxford OX1,\\ 3RH, UK\\
$^8$Department of Physics, University of Seoul, 163 Seoulsiripdaero,
Dongdaemun-gu, Seoul 02504, Republic of Korea\\ 
$^9$Korea Institute for Advanced Study, 85 Hoegiro, Dongdaemun-gu,
Seoul 02455, Republic of Korea\\
$^{10}$Universit\"at Heidelberg, Zentrum f\"ur Astronomie, Institut
f\"ur theoretische Astrophysik, Albert-Ueberle-Str. 2, D-69120
Heidelberg,\\ Germany\\ 
$^{11}$ Universidad Nacional Aut\'onoma de M\'exico, Instituto de
Astronomía, AP 106,  Ensenada 22800, BC, Mexico\\
$^{12}$ Steward Observatory, University of Arizona, 933 N Cherry
Ave,Tucson, AZ 85721, USA\\ 
$^{13}$Observatorio Astron\'omico Nacional, C/ Alfonso XII 3, 28014
Madrid, Spain\\ 
}

\begin{document}

%\date{Accepted ???. Received ??? in original form ???}

\pagerange{\pageref{firstpage}--\pageref{lastpage}} \pubyear{2002}

\maketitle

\label{firstpage}

\begin{abstract}We compute synthetic, rest-frame optical and ultraviolet (UV) 
emission-line properties of galaxy populations at redshifts 
from $z\approx0$ to $z=8$ in a full cosmological framework. We 
achieve this by coupling, in  post-processing, the state-of-the-art 
cosmological IllustrisTNG simulations with new-generation
nebular-emission models, accounting for line emission from young
stars, post-asymptotic-giant-branch (PAGB) stars, accreting black 
holes (BHs) and, for the first time, fast radiative shocks. The optical 
emission-line properties of simulated galaxies dominated by different 
ionizing sources in our models are largely consistent with those expected 
from classical diagnostic diagrams and reflect the observed 
increase in \oiiihb\ at fixed \niiha\ and the evolution of the \ha, \oiii\ 
and \oii\ luminosity functions from $z\approx0$ to $z\sim2$. At higher 
redshift, we find that the emission-line galaxy population is dominated by 
star-forming and active galaxies, with negligible fractions of shock- and 
PAGB-dominated galaxies. We highlight 10 UV-diagnostic diagrams 
able to robustly identify the dominant ionizing sources in high-redshift
galaxies. We also compute the evolution of several optical- and UV-line 
luminosity functions from $z=4$ to $z=7$, and
the number of galaxies expected to be detectable per field of view in deep, 
medium-resolution spectroscopic observations with the NIRSpec instrument 
on board the {\it James Webb Space Telescope}. We find that 2-hour-long 
exposures are sufficient to achieve unbiased censuses of \ha\ and \oiii\ 
emitters, while at least 5 hours are required  for \hb, and even 10 hours
will detect only progressively smaller fractions of \oii, \oiiiuv,
\ciii, \civ, \nii, \siliii\ and \heii\ emitters, especially in the presence of dust.
\end{abstract}


%*****************************************************************************************************
%*****************************************************************************************************
\section{Introduction}\label{intro}
%*****************************************************************************************************
%*****************************************************************************************************

Optical and ultraviolet (UV) emission from interstellar gas provides important
information about the ionizing sources and the properties of the
interstellar medium (ISM) in galaxies. Extensive studies of this kind have been
enabled for galaxies in the local Universe by large spectroscopic surveys,
such as the Sloan Digital Sky Survey \citep[SDSS, e.g.,][]{York00,
  Blanton17}.
%%1. Start with observations of optical diagnostic diagrams in the
%%local Universe
For example, optical nebular-emission lines provide useful diagnostics to 
distinguish between ionization by young massive
stars (tracing the star formation rate, hereafter SFR), an active galactic
nucleus (hereafter AGN), evolved, post-asymptotic giant branch
(hereafter post-AGB) stars and fast radiative shocks\footnote{
    Fast radiative shocks refer to shocks with velocities large
    enough (typically $\ga$ 200 km s$^{-1}$) for the photoionization 
    front to detach from the shock front and lead to the formation of 
    a precursor \hii\ region ahead of the shock \citep[see][]{Allen08}.}
\citep[e.g.,][]{Izotov99, Kobulnicky99, Kauffmann03, Nagao06,
  Kewley08, Morisset16}. In fact, the intensity ratios of strong
emission lines, such as \oii, \oi,\hb, \oiii, \ha, \nii\ and \sii,
exhibit well-defined correlations sensitive to the nature of
ionizing radiation. Widely used line-ratio diagnostic diagrams
  are those defined by \oiii/\hb, \nii/\ha, \sii/\ha\ and \oi/\ha\
  (hereafter simply \oiiihb, \niiha, \siiha\ and \oiha), 
  originally proposed by \citet[][hereafter BPT]{Baldwin81} and
  \citet{Veilleux87}. These diagrams have proven efficient to
  distinguish active from inactive star-forming galaxies in large samples of
  local galaxies \citep[e.g.][]{Kewley01,Kauffmann03}. Instead,
  identifying unambiguously galaxies for which the ISM is primarily ionized
  by post-AGB stellar populations or fast radiative shocks appears
more difficult. In the former case, recent studies suggest that
diagnostic diagrams involving the \ha\ equivalent width (EW), such as 
EW(\ha) versus \niiha\  \citep[hereafter the WHAN
  diagram,][]{CidFernandes10, CidFernandes11}, appear more promising.
  In the latter case, \citet{Baldwin81, Kewley19} proposed a
  diagnostic diagram defined by \oiii/\oii\ (hereafter simply \oiiioii)
  versus \oiha.

%%2. Evolution of optical diagostic diagram, in particular evolution
%%of OIII/Hb from z=2 to z=0
Over the past two decades, rest-frame optical spectra have become available 
for increasingly large samples of {\it more distant} galaxies, at redshifts
$z \sim 0.5$--3, through near-infrared (NIR)  spectroscopy
\citep[e.g.,][]{Pettini04, Hainline09, Steidel14,  Shapley15}, in
particular with the NIR multi-object spectrographs MOSFIRE
\citep{McLean10} and FMOS \citep{Kimura10}. Interestingly,
these observations indicate uniformly that star-forming galaxies at $z>1$ 
have systematically larger \oiiihb\ ratio at fixed \niiha\ ratio than their
local SDSS counterparts \citep[see, e.g.,][]{Shapley05, Lehnert09,
  Yabe12,  Steidel14,  Shapley15, Strom17}. The physical origin of
this observation is still being debated,  and several
explanations have been proposed, such as higher ionization
parameter, enhanced N/O abundances, harder ionizing-radiation field
(e.g. due to binary star models) 
or weak, unresolved AGN emission in high-redshift galaxies
compared to low-redshift ones \citep[for a detailed discussion,
see][and references therein]{Hirschmann17}. 

%%3. Evolution of emission line luminosity functions from z=2-0
The above spectroscopic surveys, together with deep narrow-band 
surveys over wide sky areas (e.g. HiZELS, \citealp{Geach08}), have
also allowed studies of the
statistics of rest-frame optical emission-line galaxies, quantified
through luminosity functions of the \ha, \oiii+\hb\ and
\oii\ emission lines at redshifts $z \sim 0$--3 \citep[e.g.,][and
references therein]{Tresse02,
  Fujita03, Ly07, Villar08,   Shim09, Hayes10, Sobral12, Sobral13,
  Sobral15, Colbert13,   Khostovan15, Khostovan18, Khostovan20,
  Hayashi18}. 
These emission lines, which are sensitive to the ionizing radiation from
hot, massive O- and B-type stars, are often used as tracers of star 
formation on timescales of $\sim10$ million years, providing insight
into the evolution of the cosmic star-formation-rate density and the
associated build-up of stellar mass in galaxies (as a complement to 
indicators based on the UV and far-infrared luminosities). Yet,
the luminosity functions derived in the above studies can exhibit large 
scatter, because of uncertain corrections for dust attenuation, AGN 
contamination, selection biases, sample completeness, sample variance, 
filter profiles and various other factors.

In general, the observed line luminosity functions have been
  found to be well described by a \citet{Schechter76} 
  functional form. In this context, the most
  recent studies appear to agree on a negligible evolution of the
  faint-end slope of line-luminosity functions with redshift and an
  increase by one  or two orders of magnitude of the cutoff luminosity
  at the bright end  ($L_*$) from $z=0$ to $z=2$--3 for \ha,
  \oiii+\hb\ and \oii, reflecting the similar increase in cosmic SFR density
\citep[e.g.,][]{Sobral15, Khostovan15}.

%%4. Prospects of future observational studies to look for EL galaxies
%%at z>3, UV lines become accessible
At redshift $z > 3$, emission-line observations are generally
still limited, except for a few pioneering studies \citep[see][for a
review]{Robertson21}. This situation is being revolutionized by the
recently launched {\it James Webb Space Telescope}  ({\it JWST}),
on board of which the sensitive NIRSpec instrument 
\citep{Jakobsen22, Ferruit22} is already detecting rest-frame 
optical emission lines in $z > 3$ galaxies \citep{Curti22, Katz22}.  
Large such samples will be built up over the first year of operations by
multiple teams, with rest-frame UV emission lines being observable 
out to extremely high redshifts. UV lines are
particularly prominent in metal-poor, actively star-forming dwarf
galaxies in the nearby Universe and have been detected in small
samples of high-redshift galaxies with NIR spectrographs
\citep[e.g.,][]{Pettini04, Hainline09, Erb10, Stark14, Vanzella17,
  Senchyna17, Talia17}. In addition to JWST, large surveys of
  high-redshift emission-line galaxies are (expected to be) observed
  with other current and future observational facilities, such as the 
  Dark Energy Spectroscopic Instrument (DESI), {\it Euclid}, and the 
  {\it Nancy Grace Roman Space Telescope}.

%%5. Current unknowns/challenges for these future high-z surveys
Despite this observational progress, interpreting observations of rest-frame 
optical and UV emission lines of high-redshift galaxies remains challenging,
as illustrated by the persisting debate about the nature of the  ionizing radiation
in galaxies at $z\sim3$ (see above). Also, at the very low
metallicities expected in the youngest galaxies at high redshifts
\citep[e.g.,][]{Maiolino08}, the spectral signatures of AGN-dominated 
and star-formation (SF)-dominated galaxies overlap in standard optical (BPT) 
diagnostic diagrams \citep{Groves06, Feltre16, Hirschmann19}.
This led \citet[][see also \citealt{Nakajima18}]{Feltre16} to explore the
location of a wide range of photoionization models of SF- and AGN-dominated
galaxies in alternative, UV emission-line dianostic diagrams. 
While highly instructive, these pioneering studies also present some
limitations: (i) the exploration of purely SF-dominated and AGN-dominated 
models did not account for mixed contributions by these ionizing sources, 
nor for contributions by
post-AGB stars nor radiative shocks; (ii) the results may be biased by the
inclusion of parameter combinations not found in nature; (iii) the conclusions
drawn about line-ratio diagnostic diagrams do not incorporate potential effects
linked to the evolution of galaxy properties with cosmic time; and (iv) these 
studies, based on grids of photoionization models, are not designed to
explore the evolution of emission-line properties of galaxy populations (such
as luminosity functions) with redshift.

%%6.a Start with theoretical modelling, first photo-ionsiation models:\\

%%6.b previous theoretical modelling of simulations plus
%%photo-ionsiation models
A way to gain insight into the connection between emission-line properties 
and ionizing sources for the evolving galaxy population is to model galaxy 
nebular emission in a full cosmological framework. Fully
self-consistent models of this kind are currently limited, mainly
because of insufficient resolution of the ISM and the neglect of 
radiation-gas coupling in cosmological simulations. As an
alternative, some studies proposed the post-processing of cosmological
hydrodynamic simulations and semi-analytic models with
photoionization models to compute nebular emission of galaxies in a
cosmological context, albeit with the drawbacks of either following
emission lines produced by only young star clusters \citep{Kewley13,
  Orsi14, Shimizu16, Wilkins20, Pellegrini20, Shen20, Garg22,
  Baugh22}, or  
being limited to a relatively small sample of massive galaxies
\citep{Hirschmann17, Hirschmann19}. In the latter two studies, we 
included nebular emission from young star clusters, AGN and post-AGB 
stellar populations, but not the contribution from fast radiative shocks.

In this paper, we appeal to the same methodology as in our previous work 
\citep{Hirschmann17, Hirschmann19} to model in a self-consistent way 
the emission-line properties of galaxy populations, but with two main 
improvements: we consider the evolution over cosmic time of a much larger sample 
of galaxies from the IllustrisTNG simulation, covering a full range of masses; and we 
incorporate the contribution to nebular emission from fast radiative shocks, in addition to 
the emission from young star clusters, AGN, and post-AGB stellar populations.
We achieve this by coupling photoionization models for young stars 
\citep{Gutkin16}, AGN \citep{Feltre16}, post-AGB stars \citep{Hirschmann17} 
and fast, radiative shocks \citep{Alarie19} with the IllustrisTNG cosmological 
hydrodynamic simulations. Our methodology, which by design captures 
SF-dominated, composite, AGN-dominated, post-AGB-dominated and 
shock-dominated galaxy populations in a full cosmological framework, offers
a unique way to address several important questions:  
\begin{itemize}
\item Are the predicted emission-line properties of IllustrisTNG galaxy
  populations {\it statistically } consistent with the observed, optical
  emission-line properties of galaxies out to $z\sim 3$, as was previously
  shown for only a small sample of massive galaxies  \citep{Hirschmann17}?
\item Which optical and UV diagnostic diagrams can help robustly 
distinguish between the main ionizing source(s) in metal-poor galaxy populations 
at redshifts $z>3$, and how do these diagnostics compare with the UV-based criteria
derived from the limited sample of \citet{Hirschmann19}?
\item What is the expected evolution of optical and UV line-luminosity functions of
galaxies at redshifts $z>3$, and how many galaxies of a given type can we expect
to be detected by planned {\it JWST}/NIRSpec surveys?  
\end{itemize}
Answers to these questions will allow a robust interpretation of the
high-quality spectra of very distant galaxies that will be collected in the
near future with {\it JWST}, providing valuable insight into, for example, 
the census of emission-line galaxy populations out to $z \sim 8$, and the
relative contributions by star formation and nuclear activity to reionization.

The paper is structured as follows. We start by describing the
theoretical framework of our study in Section~\ref{theory}, including
the IllustrisTNG simulation set, the photoionization models and the
coupling methodology between simulations and emission-line
models. Section~\ref{ionizingsources} presents our main findings
about tracing the nature of ionizing sources in galaxy populations
over cosmic time, via classical optical and novel UV diagnostic
diagrams. In Section~\ref{luminosityfunctions}, we explore: (i) which UV
line luminosities reliably trace the SFR in high-redshift galaxies; (ii) the
cosmic evolution of optical and UV emission-line luminosity functions;
and (iii) number counts of galaxy populations of different types
expected to be detected with {\it JWST}/NIRSpec at high redshift.
We address possible caveats of our approach and put our results into
the context of previous theoretical studies in Section
\ref{discussion}. Section~\ref{summary} summarizes our main results. 

%*****************************************************************************************************
%*****************************************************************************************************
\section{Theoretical framework}\label{theory} 
%*****************************************************************************************************
%*****************************************************************************************************

In this paper, we model the optical and UV emission-line properties of
galaxy  populations in a wide redshift range using the
IllustrisTNG300, IllustrisTNG100 and IllustrisTNG50 simulation suite
\citep[hereafter TNG300, TNG100 and TNG50;][]{Pillepich18a, Springel18,
  Nelson18, Naiman18, Marinacci18, Nelson19, Pillepich19}. In the following 
paragraphs, we briefly summarise the simulation details (Section~\ref{TNG}), 
the emission-line models (Section~\ref{ELmodels}, \citealp{Gutkin16, Feltre16, 
Hirschmann17}) and the coupling methodology (Section~\ref{coupling},
\citealp{Hirschmann17, Hirschmann19}), referring the reader to the
original studies for more details. Our novel incorporation of emission-line 
models of fast radiative shocks \citep[from][]{Alarie19}, and their coupling with 
shocked regions identified in simulated IllustrisTNG galaxies, are described in     
Sections~\ref{shockmodels} and \ref{couplingshocks}, respectively.      

\subsection{The IllustrisTNG simulation suite}\label{TNG}

IllustrisTNG is a suite of publicly available, large volume, cosmological,
gravo-magnetohydrodynamical simulations, run with the moving-mesh code
Arepo \citep{Springel10}, and composed of three simulations with
different volumes and resolutions: TNG300,
TNG100 and TNG50. Each 
IllustrisTNG run self-consistently solves the coupled evolution of dark matter, %cosmic
gas, luminous stars, and supermassive black holes from a starting
redshift of $z=127$ to $z=0$, assuming the currently favoured Planck
cosmology \citep[$\sigma_8 = 0.8159$, $\Omega_{\mathrm{m}}=0.3089$,
$\Omega_{\Lambda}=0.6911$ and $h= 0.6774$; see e.g.,][]{Planck16}.

In brief, TNG300, TNG100 and TNG50 comprise volumes 
of 302.6$^3$, 106.5$^3$ and 51.7$^3$ cMpc$^3$, respectively, populated by
particles with masses $m_{\mathrm{dm}} = 5.9\times 10^7 \Msun$,
$7.5\times 10^6 \Msun$ and $4.5\times 10^5 \Msun$ for dark matter (DM), 
and $m_{\mathrm{bar}} = 1.1\times10^7\Msun$, $1.4\times 10^6\Msun$, and
$8\times 10^4\Msun$ for baryons. The gravitational softening length for DM 
particles at $z=0$ is $1.48\ \mathrm{kpc}$, $0.74\ \mathrm{kpc}$ and 
$0.29\ \mathrm{kpc}$ for TNG300, TNG100 and TNG50, 
respectively.

All IllustrisTNG simulations include a model for galaxy
formation physics \citep{Weinberger17, Pillepich18b}, which has been
tuned to match observational constraints on the galaxy stellar-mass function
and stellar-to-halo mass relation, the total gas-mass content within the
virial radius of massive galaxy groups, the stellar-mass/stellar-size relation
and the relation between black-hole (BH) mass and galaxy mass at $z=0$. 
Specifically, the simulations follow 
(i) primordial and metal-line radiative cooling in the presence of an ionizing 
background radiation field;
(ii) a pressurised ISM via an effective equation-of-state model;
(iii) stochastic star formation;
(iv) evolution of stellar populations and the associated chemical
enrichment (tracing individual elements) by supernov\ae\ (SN)
Ia/II, AGB stars and neutron-star (NS)-NS  
mergers;
(v) stellar feedback via an kinetic wind scheme;
(vi) BH seeding, growth and feedback (two-mode model for
high-accretion and low-accretion phases);
(vii) amplification of a small, primordial-seed magnetic field and
dynamical impact under the assumption of ideal MHD.
The IllustrisTNG simulations have been shown in various studies to 
provide a fairly realistic representation of the properties of galaxies 
evolving across cosmic time \citep[e.g.,][]{Torrey19}.

In an IllustrisTNG simulation, the properties of galaxies, galaxy groups,
subhaloes and haloes (identified using the FoF and Subfind 
substructure-identification algorithms, see \citealp{Davis85} and
\citealp{Springel01}, respectively), 
are computed `on the fly' and saved 
for each snapshot. In the following, we always consider both central
and satellite Subfind haloes and their galaxies.  In addition, an
on-the-fly cosmic shock finder coupled to the code \citep{Schaal15}
uses a ray-tracing method to identify shock surfaces and measure their
properties.

\citet{Schaal16} describe the functioning
and performances of this algorithm on Illustris-simulation data. In a
first step, cells in shock-candidate regions are flagged according to
certain criteria (e.g., negative velocity divergence; temperature and
density gradients pointing in a same direction). Secondly, the shock
surface is defined by the ensemble of cells exhibiting the maximum
compression along the shock normal. Thirdly, the Mach number is
estimated from the Ranking-Hugoniot temperature jump condition, and a 
shock is defined only if the Mach-number jump exceeds 1.3. We
note that, by default, shock finding is disabled for star-forming
cells, which are those lying close to  the effective equation of state
of the ISM in density-temperature ($\rho$-$T$)  phase space. That is,
shocks are not followed when either the pre-shock cell, the post-shock
cell, or the cell in the shock fulfil $\rho > \rho_{\mathrm{sfr}}$
and $T < 2\times10^4 (\rho/\rho_{\mathrm{sfr}})\,$K, where
$\rho_{\mathrm{sfr}}$ is the density threshold for star formation. By
definition, therefore, the IllustrisTNG  simulations are not capturing
shocks in the cold/warm/star-forming ISM. This also implies that shocks 
from supernovae explosions, which primarily occur in star-forming regions, are
not captured. For more details on the shock-finder algorithm, we refer
the reader to \citet{Schaal15} and \citet{Schaal16}.

\subsection{Emission-line models}\label{ELmodels}

We use the same methodology as in \citet{Hirschmann17, Hirschmann19}
to compute nebular-line emission of galaxies from the post-processing
of the three tiers of IllustrisTNG simulations. At $z=0$, TNG50
has $\sim8,800$ galaxies with stellar masses greater than $10^8  
\Msun$, TNG100 $\sim18,000$ galaxies with stellar masses greater than 
$3\times 10^{9} \Msun$ and TNG300 $\sim47,000$ galaxies with stellar 
masses greater than $10^{10} \Msun$. For all these galaxies and their progenitors
at $z>0$, nebular emission from young star clusters \citep{Gutkin16}, narrow-line 
regions (NLR) of AGN \citep{Feltre16} and post-AGB stars \citep{Hirschmann17} is
computed using version c13.03 of the photoionization  code \textsc{Cloudy} 
(\citealp{Ferland13}), while the emission from fast radiative shocks \citep{Alarie19} 
is computed using \textsc{Mappings V} \citep{Sutherland17}. All photoionization 
calculations considered in this work were performed adopting a common set of 
element abundances down to metallicities of a few per cent of Solar \citep[from][]{Gutkin16}.

\subsubsection{Emission-line models of young star clusters, AGN and post-AGB
  stellar populations}\label{cloudymodels}
  
To model nebular emission from \hii\ regions around young stars, AGN 
narrow-line regions and post-AGB stellar populations, we adopt the
same photoionization models as described in section~2.2.1 of 
\citet[][to which we refer for details]{Hirschmann19}. In brief, the emission 
from young star clusters is described with an updated version of the
\citet{Gutkin16} photoionization calculations combining the latest version 
of the \citet{Bruzual03} stellar population
synthesis model with \textsc{Cloudy}, for 10\,Myr-old stellar populations
with constant SFR and a standard \citet{Chabrier03} initial mass function
(IMF) truncated at 0.1 and 300\,$\Msun$. The emission from AGN narrow-line
regions is described with an updated version of the \citet{Feltre16}
photoionization models. Finally, we adopt the `PAGB' models of
\citet{Hirschmann17} to describe the emission from interstellar gas
photoionized by evolved stellar populations of various ages. 

The \hii-region, AGN-NLR and PAGB models are available for full ranges 
of interstellar metallicity, ionization parameter, dust-to-metal mass ratio, 
ionized-gas density, and C/O abundance  ratio [except for the PAGB models 
with fixed (C/O)$_\odot$=0.44], as listed in table~1 of  \citet{Hirschmann17}.

\subsubsection{Emission-line models of fast radiative
  shocks}\label{shockmodels}

To describe the contribution by shock-ionized gas to nebular
emission in IllustrisTNG galaxies, we appeal to the 3MdBs 
database3 of fully-radiative shock models computed by \citet{Alarie19}
using \textsc{Mappings V}. The models, computed in plane-parallel geometry,
are available over a similarly wide range of interstellar metallicities
($Z_{\mathrm{shock}}$) to those adopted in the \hii-region, AGN-NLR and 
PAGB photoionization models described in the previous section [albeit 
with only two values of the C/O ratio:  (C/O)$_{\mathrm{shock}}=
0.25\times$(C/O)$_\odot$ and (C/O)$_\odot$]. Metal depletion onto dust
grains is not included in this case, as in fast shocks, dust can be
efficiently destroyed by grain-grain collisions, through both
shattering and spallation, and by thermal sputtering \citep{Allen08}.

The other main adjustable parameters defining the shock-model grid are 
the shock velocity ($100\leq v_{\mathrm{shock}}\leq1000\,$km\,s$^{-1}$), 
pre-shock density ($1\leq n_{\mathrm{H,shock}}\leq10^4\,$cm$^{-2}$) and
magnetic-field strength ($10^{-4}\leq B_{\mathrm{shock}}\leq10\,\mu$G). 
We focus here on the predictions of models including nebular
emission from both shocked and shock-precursor (i.e., pre-shock) gas
photoionized by the shock (see \citealp{Alarie19} for more details). 

%The pre-shock density (nH) and transverse magnetic field (noted B) have a
%much weaker influence than shock velocity on most emission lines of
%interest to us (see Section~4).
%Thus, in the following, to probe
%global trends in the influence of radiative shocks on the nebular
%emission from star-forming galaxies, we consider for simplicity models
%with fixed nH = 102 cm−3 and B = 1 μG in the full available range of
%shock velocities. 

%*****************************************************************************************************
%*****************************************************************************************************
\subsection{Coupling nebular-emission models with IllustrisTNG simulations}\label{coupling} 
%*****************************************************************************************************
%*****************************************************************************************************

We combine the IllustrisTNG simulations of galaxy populations described in 
Section~\ref{TNG} with the photoionization models of Section~\ref{ELmodels} 
by associating, at each time step, each simulated galaxy single \hii-region, 
AGN-NLR, PAGB and radiative-shock models, which, when combined,
constitute the total nebular emission of the galaxy. We achieve this using
the procedure described in \citet{Hirschmann17,Hirschmann19}, 
by self-consistently matching the model parameters available from the 
simulations with those of the emission-line models, a novelty here being
the incorporation of the radiative-shock models (see below). The ISM
and stellar parameters of simulated galaxies are evaluated by considering
all `bound' gas cells and star particles (as identified by the Subfind algorithm;
Section~\ref{TNG}) for the coupling with the \hii-region and PAGB models, 
and within 1~kpc around the BH for the coupling with the AGN-NLR
models. We note that while the coupling methodology described in the 
next paragraphs pertains to galaxy-wide nebular emission (as in
\citealp{Hirschmann17, Hirschmann19}), the versatile nature of our approach
allows its application to spatially-resolved studies of galaxies (Hirschmann et al., 
in prep.).

A few photoionization-model parameters cannot be defined from the 
simulation, such as the slope of the AGN ionizing spectrum ($\alpha$),
the hydrogen gas density in individual ionized regions ($n_{\mathrm{H}}$),
the dust-to-metal mass ratio ($\xi_{\rm d}$) and the pre-shock density
($n_{\mathrm{H,shock}}$), since individual ionised regions are
  unresolved and dust formation is not followed. For simplicity, we
fix the slope of the AGN ionizing spectrum to $\alpha = -1.7$ and
adopt standard hydrogen densities  $n_{\mathrm{H}, \star}
=10^2\,$cm$^{-3}$ in the \hii-region models,  
$n_{\mathrm{H},\bullet} = 10^3\,$cm$^{-3}$ in the AGN-NLR models, and 
$n_{\mathrm{H}, \diamond} = 10\,$cm$^{-3}$ in the PAGB models. We 
mitigate the impact of the dust-to-metal mass ratio on the predicted nebular 
emission of the \hii-region, AGN and PAGB models by sampling two values, 
$\xi_{\rm d}=0.3$ and 0.5, around the Solar-neighbourhood value of 
$\xi_{\rm d,\odot}=0.36$ \citep{Gutkin16}. \citet{Hirschmann17, 
Hirschmann19} discuss in detail the influence of the above parameters 
on predicted optical and UV emission-line fluxes. For the pre-shock 
density, which has a comparably weak influence on most emission lines of 
interest to us, we adopt for simplicity $n_{\mathrm{H,shock}}$ = 1 cm$^{-3}$, 
the lowest value available in the shock models, closest to the typical 
galaxy-wide densities sampled by the simulations.

In the next paragraphs, we provide slightly more details on the coupling 
of the IllustrisTNG simulations with \hii-region, AGN-NLR and PAGB 
photoionization models (Section~\ref{couplingSFAGN}) and radiative-shock 
models (Section~\ref{couplingshocks}).

\subsubsection{Coupling simulated galaxies with \hii-region, AGN-NLR 
and PAGB photoionization models}\label{couplingSFAGN}

At each simulation time step, we associate each simulated galaxy 
with the \hii-region model of closest metallicity, C/O ratio and
ionization parameter. The ionization parameter of gas ionized
by young star clusters in a simulated galaxy, $U_{\mathrm{sim}, \star}$,
depends on the simulated SFR (via the rate of ionizing photons) and 
global average gas density (via the filling factor 
$\epsilon=\rho_{\mathrm{gas, glob}}/n_{\mathrm{H}, \star}$; see 
equations~1--2 of \citealp{Hirschmann17}). As a further 
fine-tuning, in the present study, we calibrate the filling factor in such 
a way that present-day galaxies in the simulation follow the local relation
between ionization parameter and interstellar metallicity, $\log U\approx
-0.8\log(Z_{\mathrm{ISM}}/\Zsun) - 3.58$ 
\citep[see figure~2 of][]{Carton17}.\footnote{This calibration ensures that
the  models reproduce the observed trends between gas-phase oxygen 
abundance and strong-line ratios in present-day galaxy populations.
We investigate this in a separate study.} 
At higher redshift, we take the evolution of the filling factor to follow
that of the global average gas density. 

The procedure to associate each simulated galaxy with an AGN-NLR
and a PAGB photoionization models is identical to that described in
\citet{Hirschmann19}. At each time step, we select the AGN-NLR 
model of metallicity, C/O ratio and ionization parameter closest to 
the average values computed in a co-moving sphere of 1-kpc radius 
around the central BH (roughly appropriate to probe the NLR for AGN
with luminosities in the range found in our simulations). The PAGB 
model is selected to be that with age and metallicity closest to the average 
ones of stars older than 3\,Gyr
in the simulated galaxy, and with same warm-gas properties as 
the \hii-region model (since the simulation does not resolve gas in 
star-forming clouds versus the diffuse ISM).

\subsubsection{Coupling simulated galaxies with radiative-shock
  models}\label{couplingshocks}

To compute the contribution from radiative shocks to nebular-line 
emission in a simulated galaxy at a given time step, we consider all cells 
flagged by the shock-finder algorithm (Section~\ref{TNG}) and 
compute their mean magnetic-field strength ($B_{\mathrm{shock,sim}}$),
metallicity ($Z_{\mathrm{shock,sim}}$), C/O abundance ratio 
[(C/O)$_{\mathrm{shock,sim}}$] and velocity ($v_{\mathrm{shock,sim}}$). 
The individual shock-cell velocities are estimated from the Mach numbers
($M$, available for each shock cell) as $v_{\mathrm{shock,sim}} = M \times 
v_{\mathrm{sound}}$, where $v_{\mathrm{sound}}$ is the sound speed for
an ideal gas at the temperature of the cell. Then, we associate the galaxy 
with the radiative-shock model of closest magnetic-field strength 
($B_{\mathrm{shock}}$), metallicity ($Z_{\mathrm{shock}}$),  C/O 
abundance ratio [(C/O)$_{\mathrm{shock}}$] and velocity 
($v_{\mathrm{shock}}$).

%*****************************************************************************************************
%*****************************************************************************************************
\subsection{Total emission-line luminosities, line ratios and
  equivalent widths of IllustrisTNG galaxies}\label{ELlums} 
%*****************************************************************************************************
%*****************************************************************************************************

The procedure described in the previous paragraphs allows us to
compute the contributions of young stars, an AGN, post-AGB stars and
fast radiative shocks to the luminosities of various emission lines (such as
$L_{\mathrm{H}\alpha}$, $L_{\mathrm{H}\beta}$,  $L_{\mathrm{[OIII]}}$,
etc.) in a simulated galaxy. The {\it total} emission-line
luminosities of the galaxy can then be calculated by summing 
these four contributions. For line luminosity ratios, we adopt for
simplicity the notation
$L_{\mathrm{[OIII]}}/L_{\mathrm{H}\beta}=\oiiihb$. In this study, we
focus on exploring line ratios built from six optical lines, 
%%optical lines
\hb, \oiii, \oi, \ha, \nii\ and \sii\,
%%UV lines, ordered according to the wavelength
and nine UV lines, \nv\ (multiplet), \civd\ (hereafter simply \civ),
\heii, \oiiiuvd\ (hereafter simply \oiiiuv), \niii\ (multiplet),
\siliiid\ (hereafter simply \siliii), \ciiid\ (hereafter simply
\ciii), \cii\ and \oiid\ (hereafter simply \oii). 

We note that, throughout this paper (except in Section~\ref{numbercounts}), we do not
consider attenuation by dust outside \hii\ regions and compare our
predictions with observed emission-line ratios corrected for this
effect
%(when not provided by the original studies, we apply a
%correction based on the \ha/\hb\ ratio and the \citealt{Calzetti00}
%attenuation curve).
By design, 
the above optical-line ratios are anyway rather insensitive to dust, as
they are defined by lines close in wavelength (for reference, the 
corrections are less than $\sim0.015$~dex for a $V$-band attenuation 
of $A_V\sim0.5$\,mag). 

In addition to line luminosities and line ratios, we also compute the
total equivalent widths of some nebular emission lines. We obtain
the EW of an emission line by dividing the total line luminosity by
the total continuum flux $C$ at the line wavelength, e.g., EW(\ciii) $=
L_{\rm CIII]}/C_{\rm CIII]}$ (expressed in \AA). The total continuum $C$
is the sum of the contributions by the stellar (SF and PAGB) and AGN
components.\footnote{Note that the SF models provide the
  stellar continuum of stellar populations younger than 1 Gyr, and
  the PAGB models that of stellar populations older than 1 Gyr.
  The radiative-shock models do not include any continuum
  component.} For
the stellar components, we account for both attenuated stellar
radiation and nebular recombination continuum. For the AGN 
component, we consider only the nebular recombination continuum, and
do not account for any attenuated radiation from the accreting
BH. This assumption should be reasonable for type-2 AGN, where direct
AGN radiation is obscured by the surrounding torus, and only emission
from the narrow-line region is observed. Instead, for type-1 AGN, the
EWs computed here should be interpreted as upper limits. In the
remainder of this study, we will focus on exploring the EWs of one
optical and five UV lines: EW(\ha), EW(\ciii), EW(\civ), EW(\oiiiuv),
EW(\siliii) and EW(\niii).

%*****************************************************************************************************
%*****************************************************************************************************
\section{Identification of the main ionizing source(s) in galaxy populations
over cosmic time}\label{ionizingsources} 
%*****************************************************************************************************
%*****************************************************************************************************

In this section, we investigate optical and UV diagnostic
diagrams to identify the main ionizing sources in full galaxy
populations of the TNG50 and TNG100 simulations at $z=0-7$.
We focus on galaxies containing at least $\sim1000$ star particles, corresponding
to stellar masses greater than $10^8 \Msun$ and $3 \times 10^9 \Msun$ in the
TNG50 and TNG100 simulations, respectively. Note that in this section, 
we do not explicitly show results for TNG300 galaxies, whose emission-line
properties hardly differ from those of TNG100 galaxies.

In the simulations, we distinguish between five different
galaxy types, based on the predicted ratio of BH accretion rate
(BHAR) to star formation rate (SFR) and the H$\beta$-line
luminosity. Specifically, SF-dominated, composite, AGN-dominated,
PAGB-dominated and shock-dominated galaxies are defined as follows
\citep[see also][]{Hirschmann17, Hirschmann19}: 
\begin{itemize}
 \item SF-dominated galaxies: BHAR/SFR $< 10^{-4}$ and H$\beta_{\rm
     SF+AGN} > $ H$\beta_{\rm PAGB}$ + H$\beta_{\rm shock}$;
\item Composite galaxies: $10^{-4}<$ BHAR/SFR  $< 10^{-2}$ and
  H$\beta_{\rm SF+AGN} > $ H$\beta_{\rm PAGB} $ + H$\beta_{\rm shock}$;
\item AGN-dominated galaxies: BHAR/SFR $> 10^{-2}$ and H$\beta_{\rm
    SF+AGN} > $ H$\beta_{\rm PAGB}$ + H$\beta_{\rm shock}$;
\item PAGB-dominated galaxies: H$\beta_{\rm SF+AGN}$ + H$\beta_{\rm
    shock} < $ H$\beta_{\rm PAGB}$;
  \item Shock-dominated galaxies: H$\beta_{\rm SF+AGN}$ + H$\beta_{\rm
    PAGB} < $ H$\beta_{\rm shock}$.
\end{itemize}

In Section~\ref{opticaldiagrams_z0} below, we start by comparing the 
predicted locations of these different galaxy types at $z=0$ in the classical 
BPT and Mass-Excitation \citep[MEx,][]{Juneau11} diagnostic diagrams 
with those of SDSS galaxies. We
also propose alternative diagnostics to reliably identify shock-dominated
and PAGB-dominated galaxies. In Section~\ref{physprop}, we examine how
the star formation rates and cold-gas fractions of simulated galaxies 
compare with those of SDSS galaxies. Then, we move on to the redshift
range $z=0.5-7$ and explore, in Section~\ref{opticaldiagrams_evol}, to 
what extent classical BPT diagrams still allow accurate separation of 
SF-dominated, composite and AGN galaxy populations, and  whether 
our simulation predictions can statistically reproduce the observed evolution 
of \oiiihb\ at fixed \niiha. In Section~\ref{galtypefractions}, we investigate
changes in the relative fractions of galaxies of different types with cosmic 
time. Finally, in Section~\ref{uvdiagrams}, we confront our 
simulations of galaxy populations with the UV diagnostic diagrams proposed 
by \citet[][based on a set of only 20 zoom-in simulations of massive
galaxies]{Hirschmann19} to assess the robustness of these diagrams to
identify different galaxy types during the epoch or reionization, as a guidance 
for interpreting planned spectroscopic, high-redshift galaxy surveys
with, e.g., {\it JWST}/NIRSpec.

%*****************************************************************************************************
%*****************************************************************************************************
\subsection{Optical diagnostic diagrams at $z=0$}\label{opticaldiagrams_z0} 
%*****************************************************************************************************
%*****************************************************************************************************

\begin{figure*}
\center
%\textbf{SF galaxies}\par\medskip\hspace{-1.5cm}
\epsfig{file=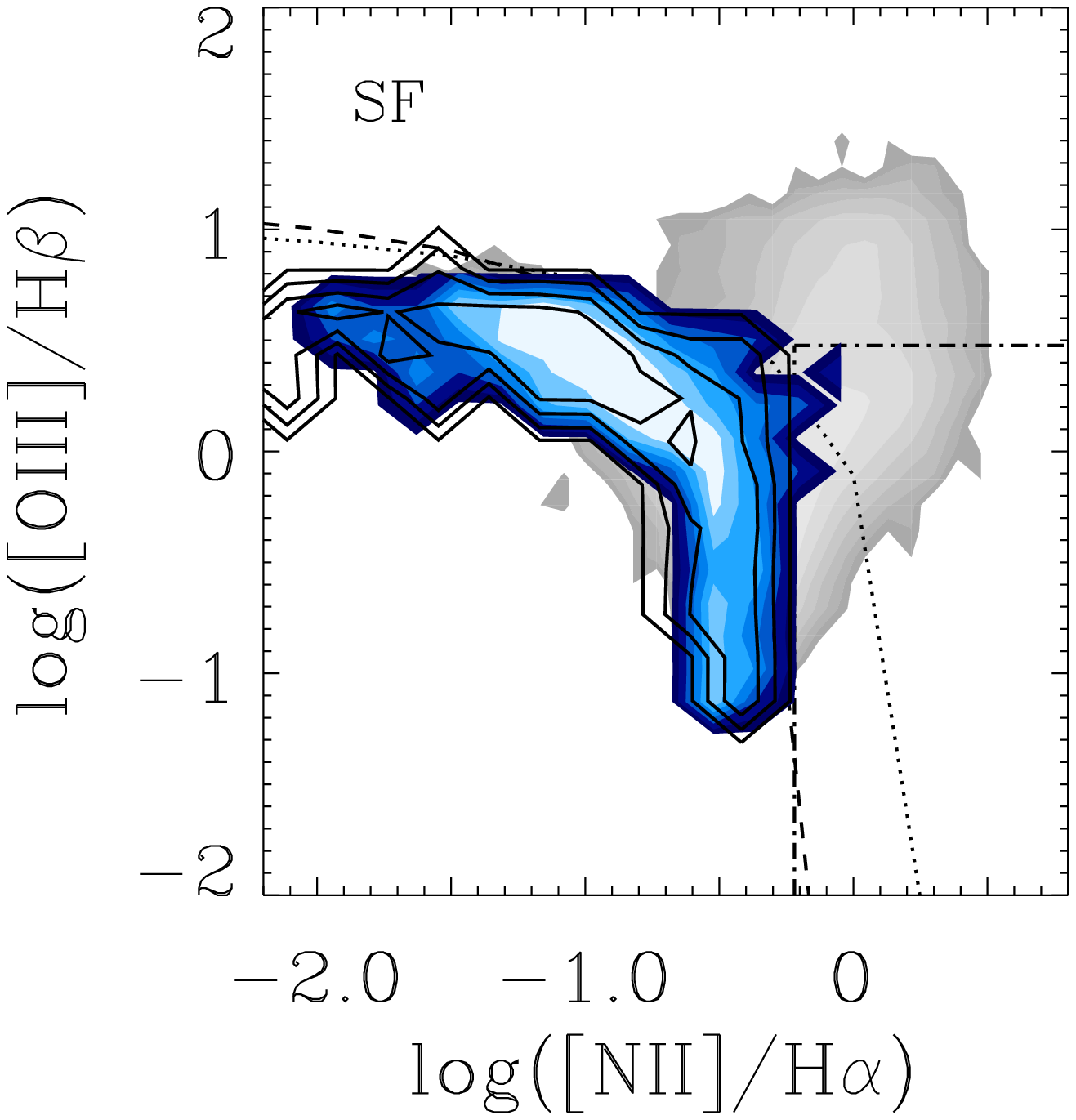,
  width=0.25\textwidth}\hspace{-1.5cm}
\epsfig{file=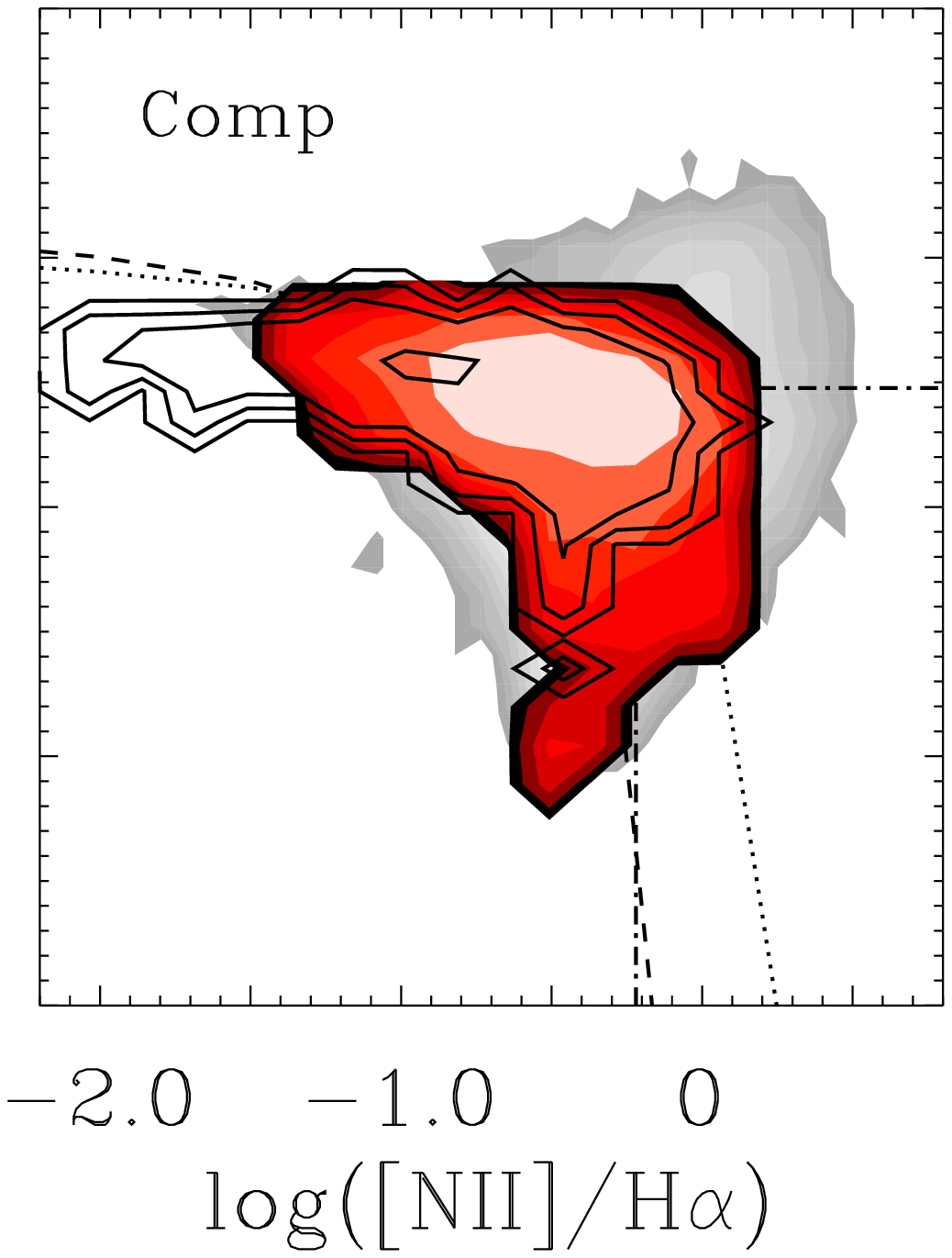,
  width=0.25\textwidth}\hspace{-1.5cm}
\epsfig{file=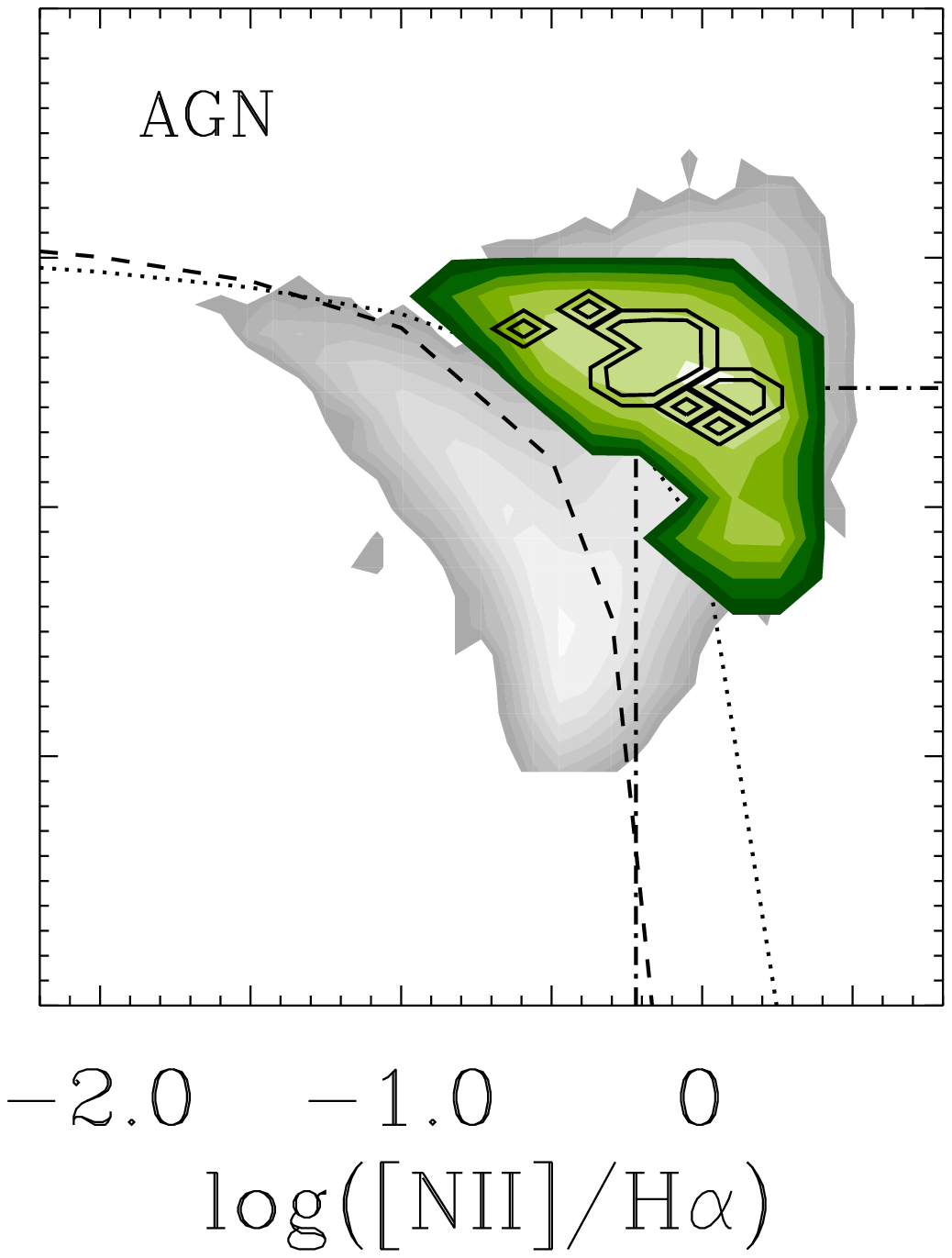,
  width=0.25\textwidth}\hspace{-1.5cm} 
\epsfig{file=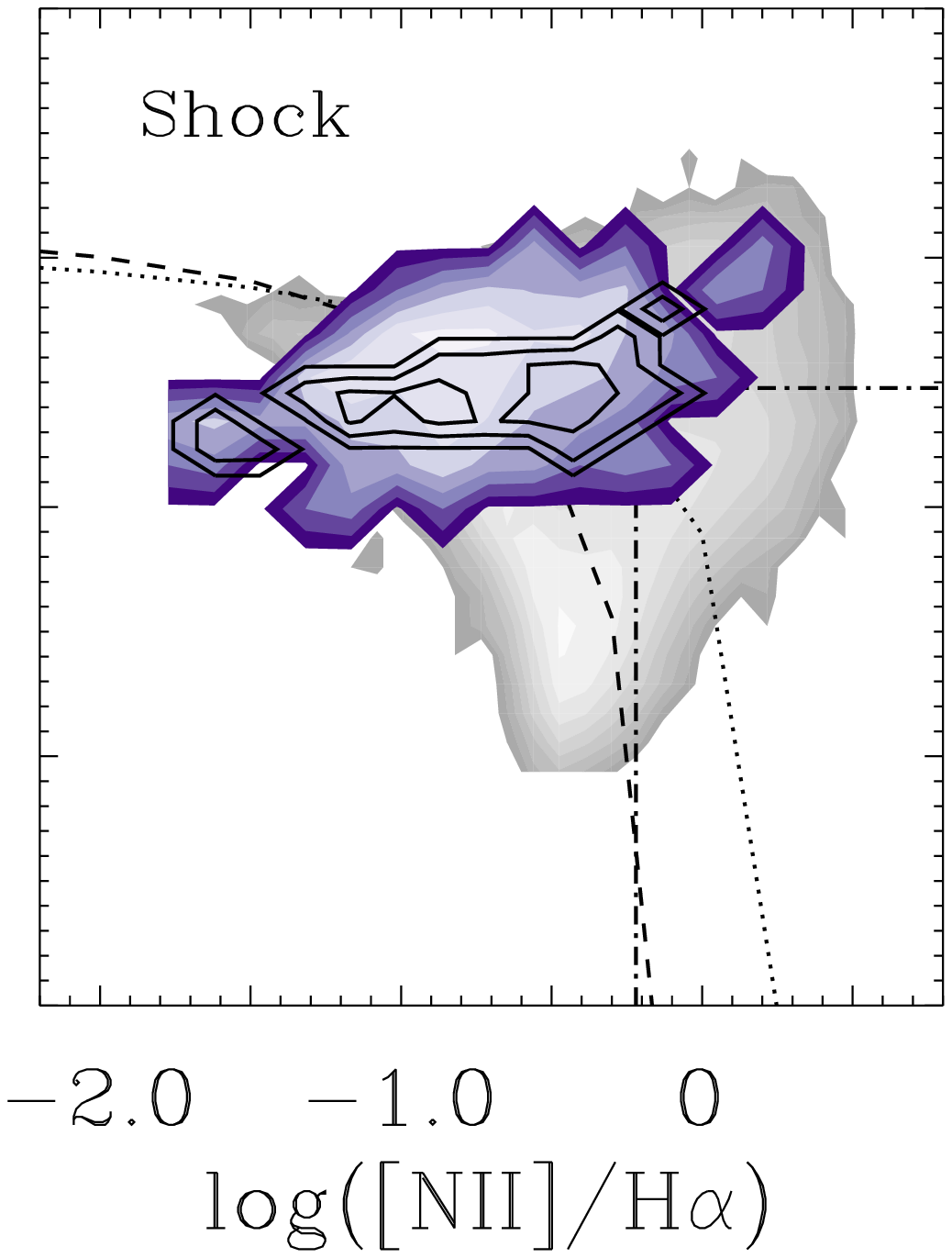,
  width=0.25\textwidth}\hspace{-1.5cm}
\epsfig{file=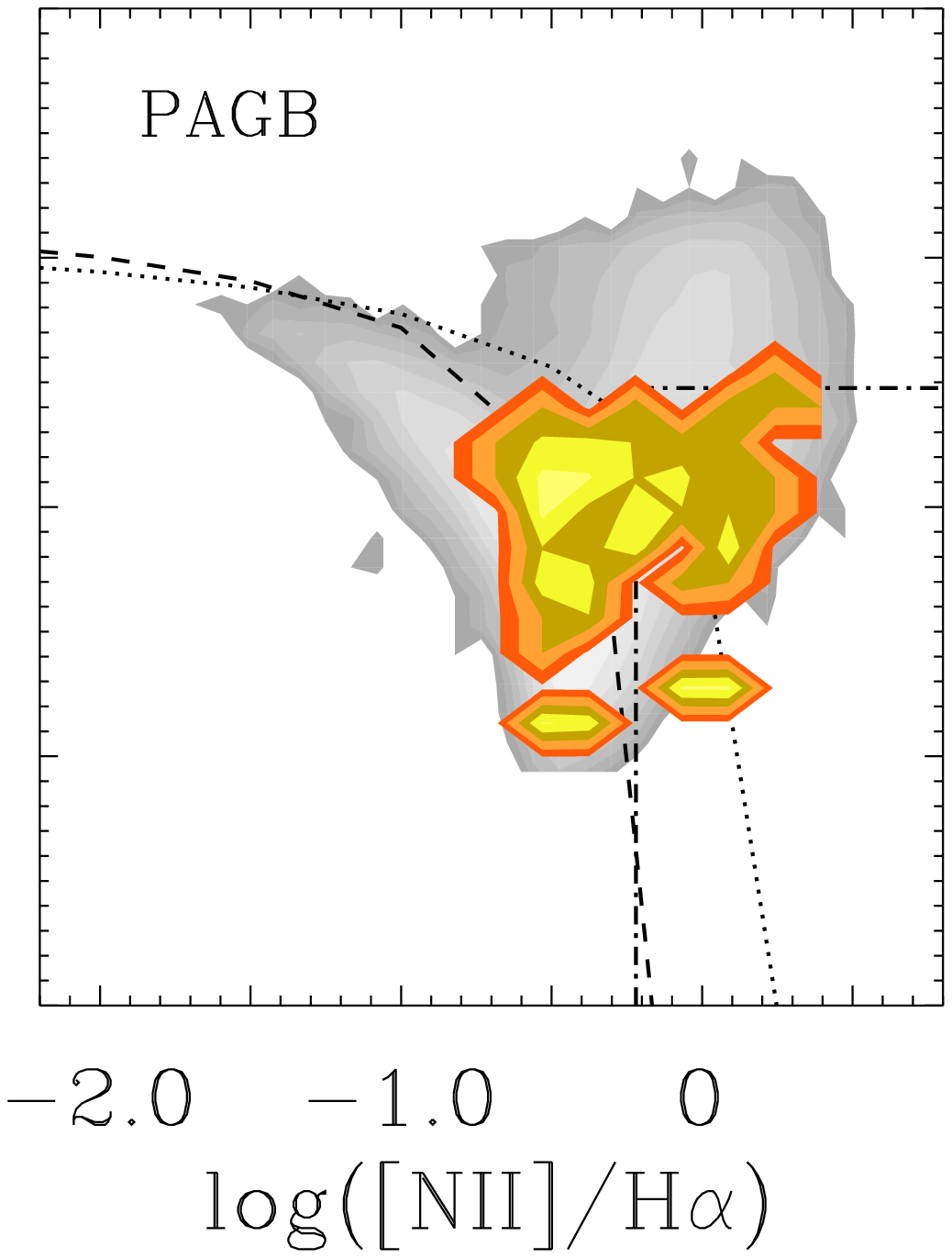,
  width=0.25\textwidth}
\epsfig{file=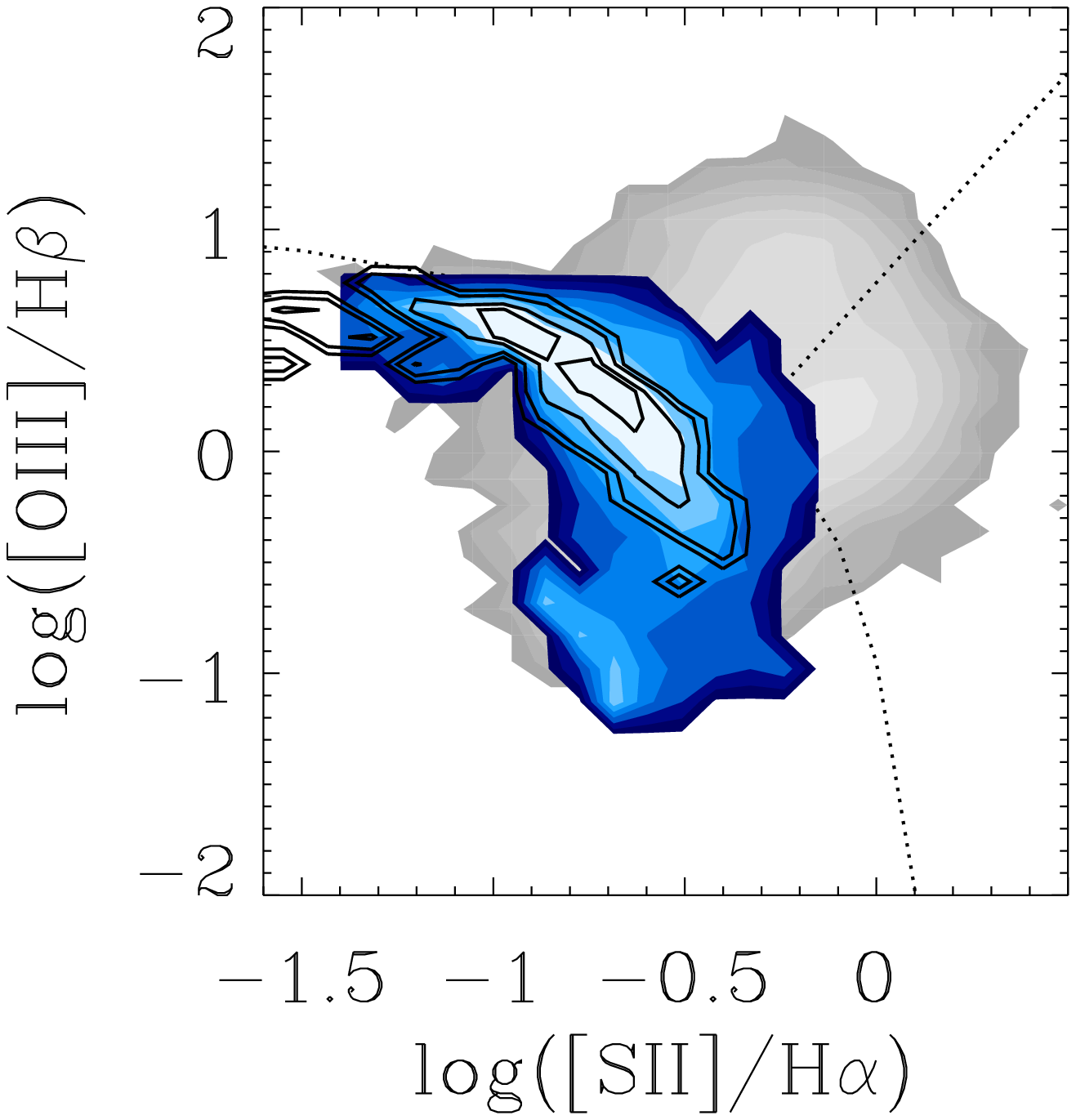,
  width=0.25\textwidth}\hspace{-1.5cm}
\epsfig{file=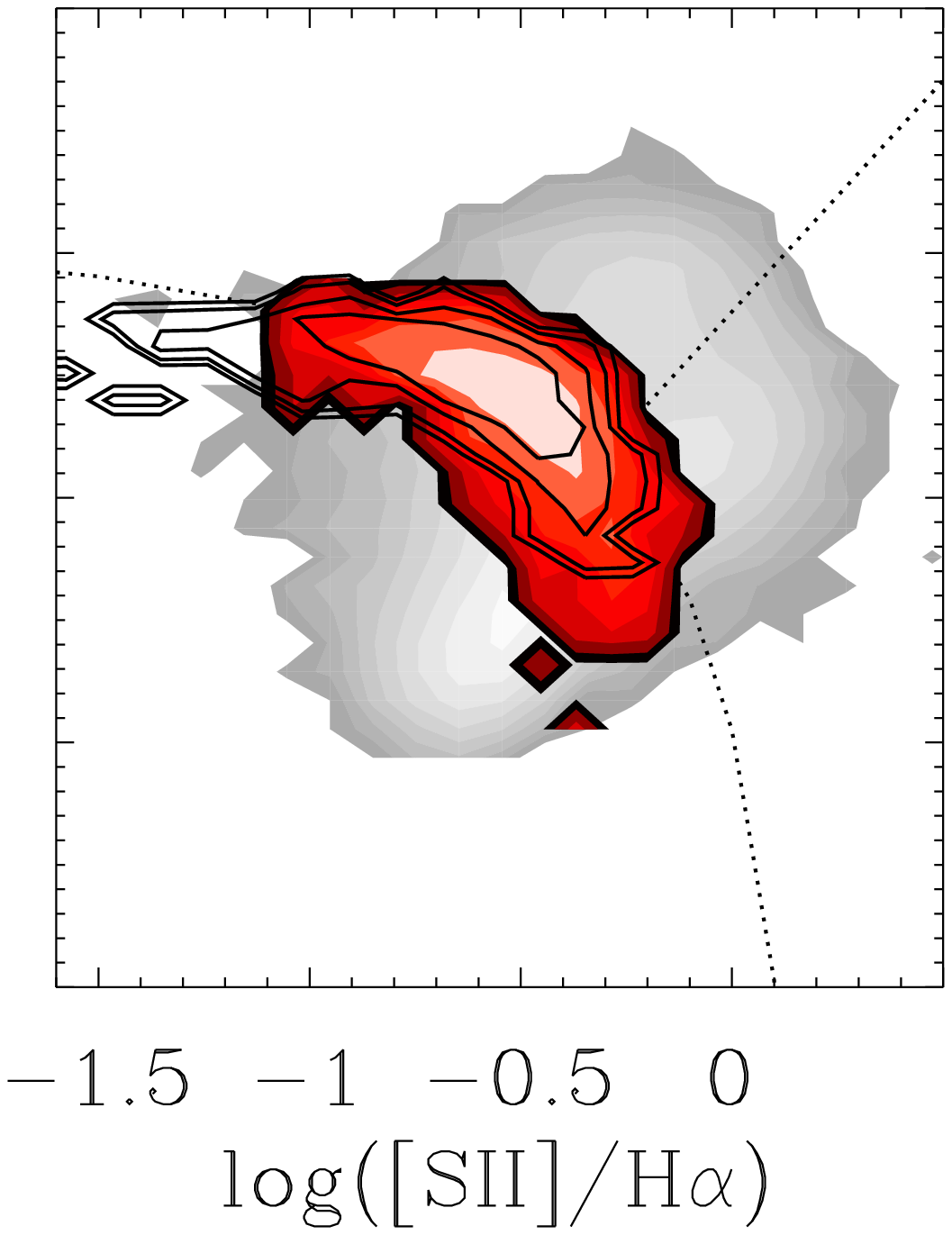,
  width=0.25\textwidth}\hspace{-1.5cm}
\epsfig{file=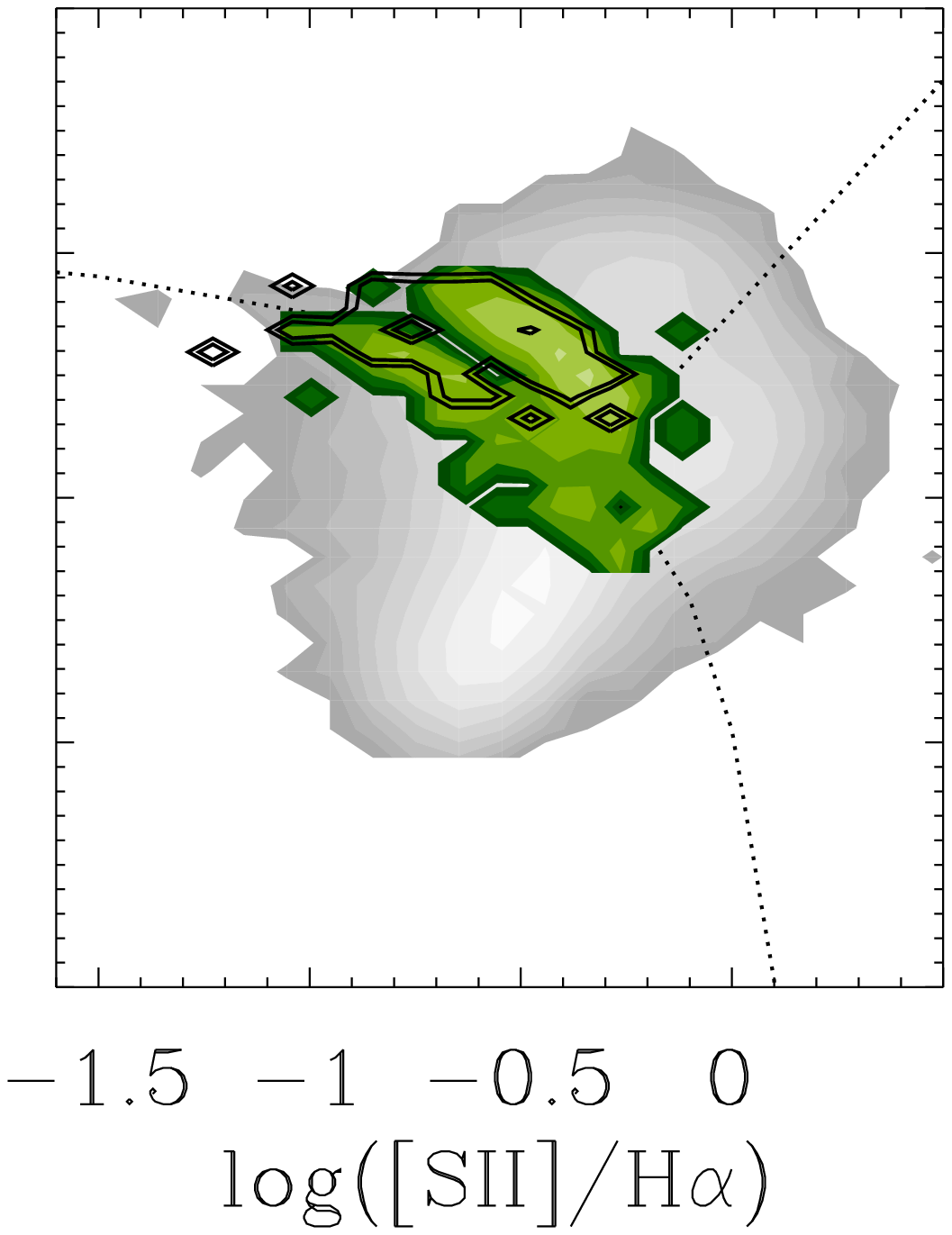,
  width=0.25\textwidth}\hspace{-1.5cm} 
\epsfig{file=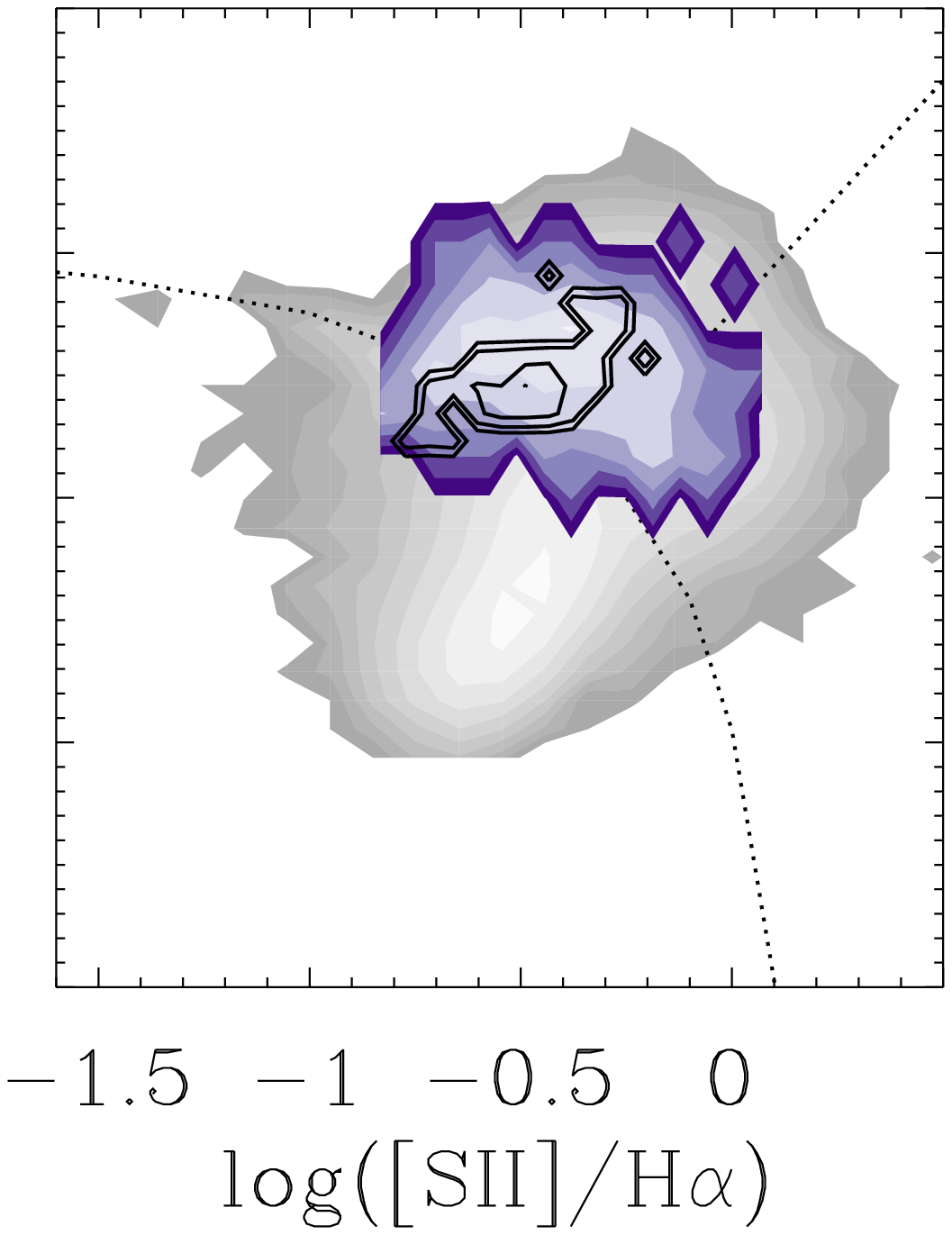,
  width=0.25\textwidth}\hspace{-1.5cm} 
\epsfig{file=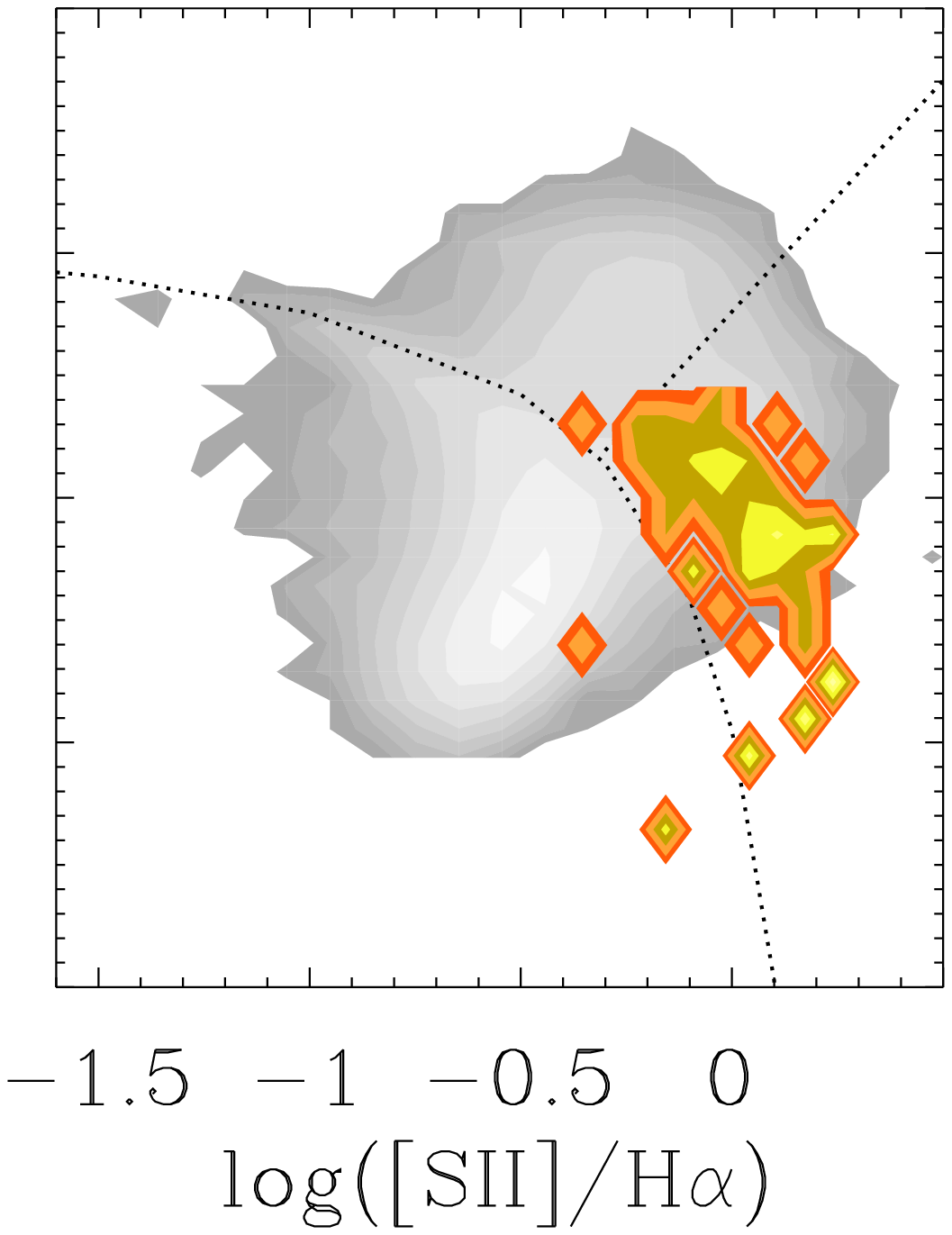,
  width=0.25\textwidth}
\epsfig{file=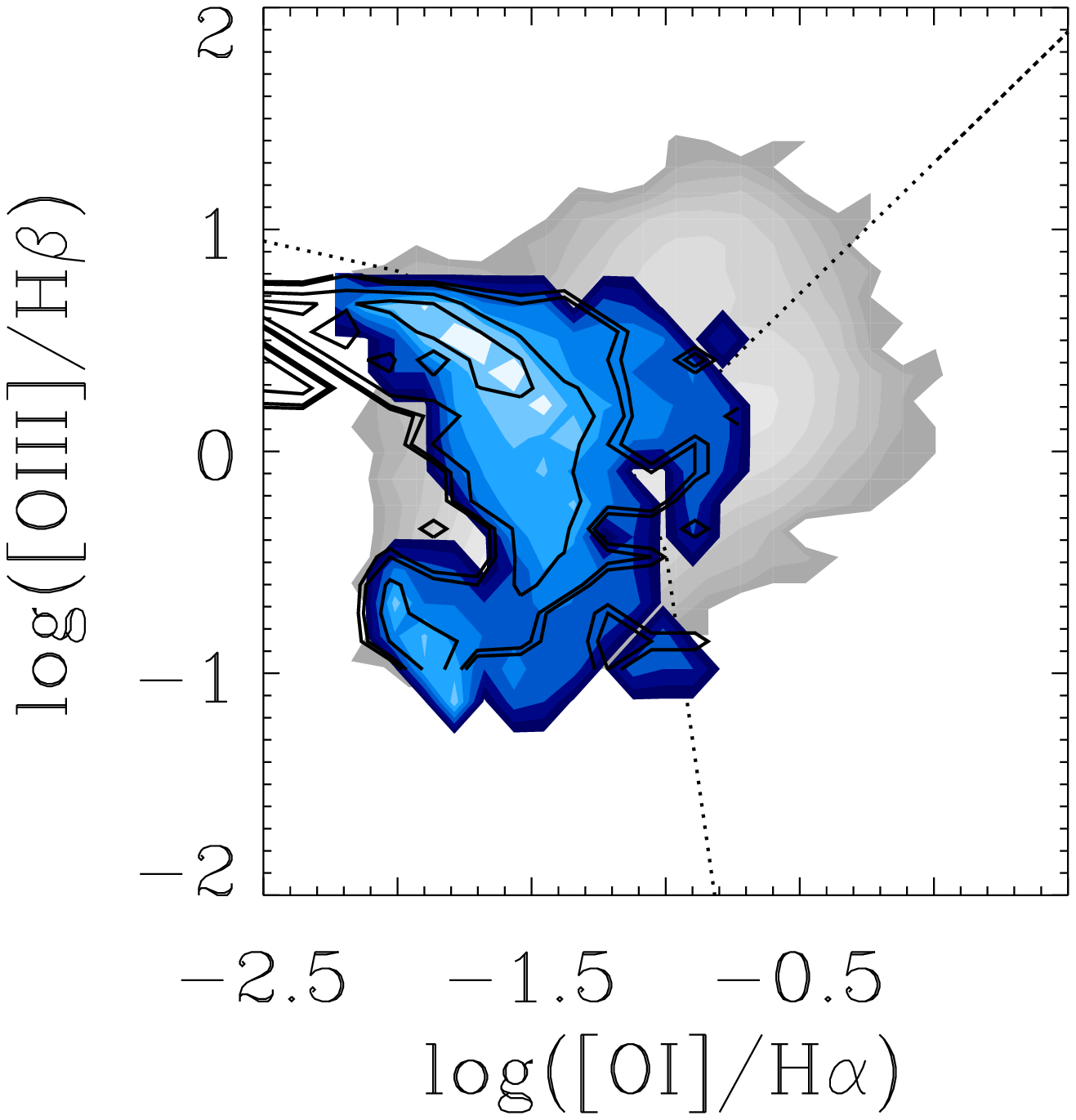,
  width=0.25\textwidth}\hspace{-1.5cm}
\epsfig{file=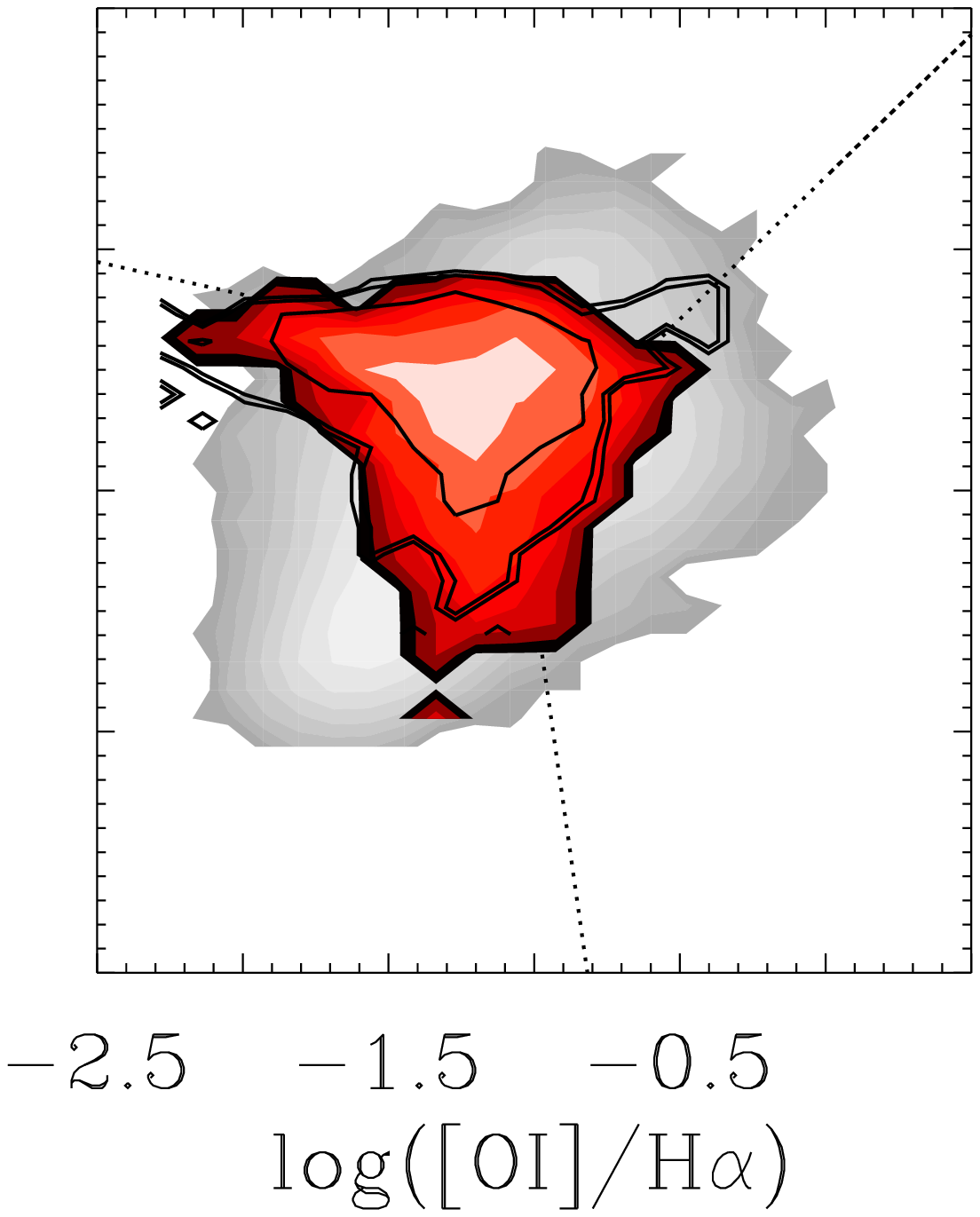,
  width=0.25\textwidth}\hspace{-1.5cm}
\epsfig{file=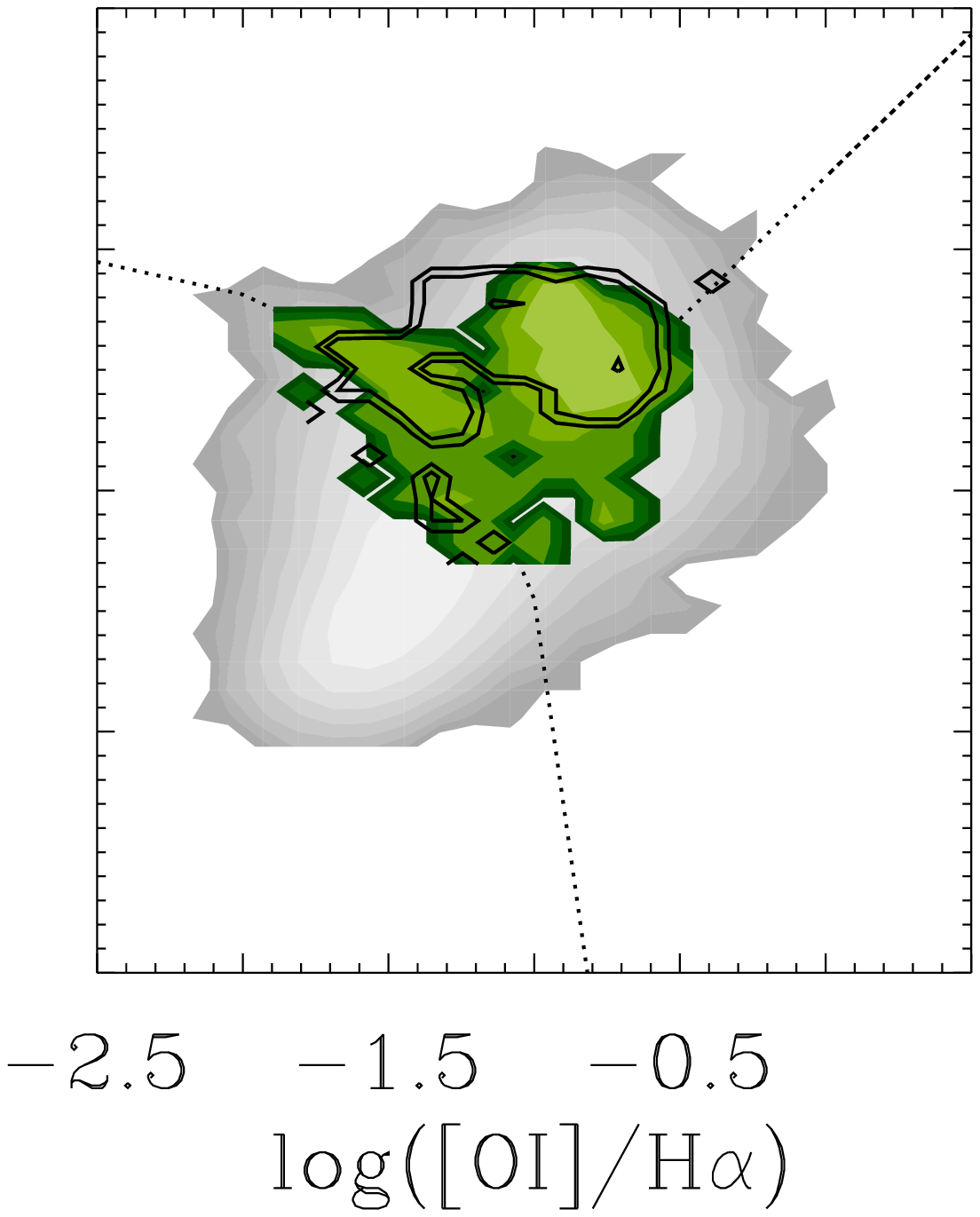,
  width=0.25\textwidth}\hspace{-1.5cm} 
\epsfig{file=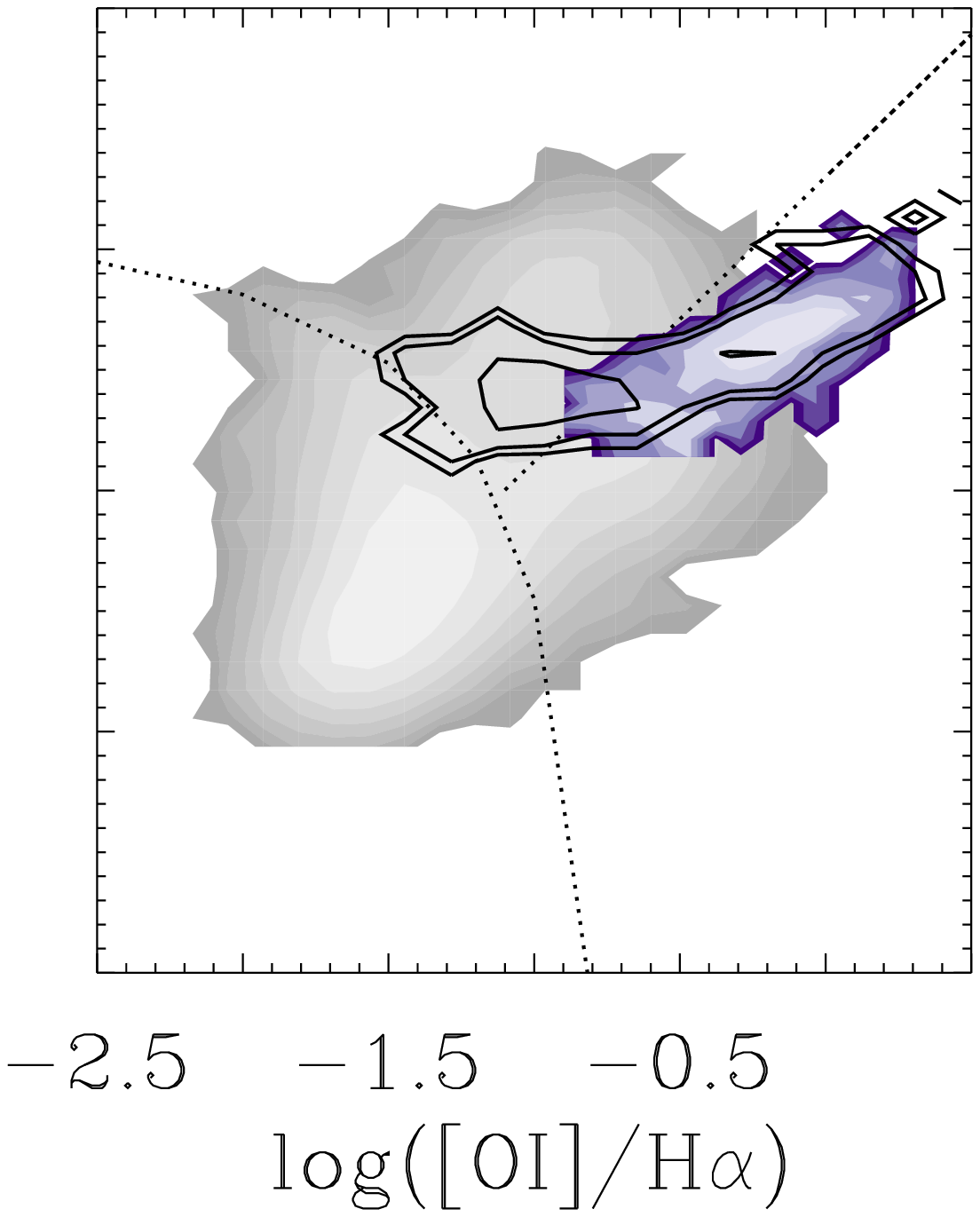,
  width=0.25\textwidth}\hspace{-1.5cm} 
\epsfig{file=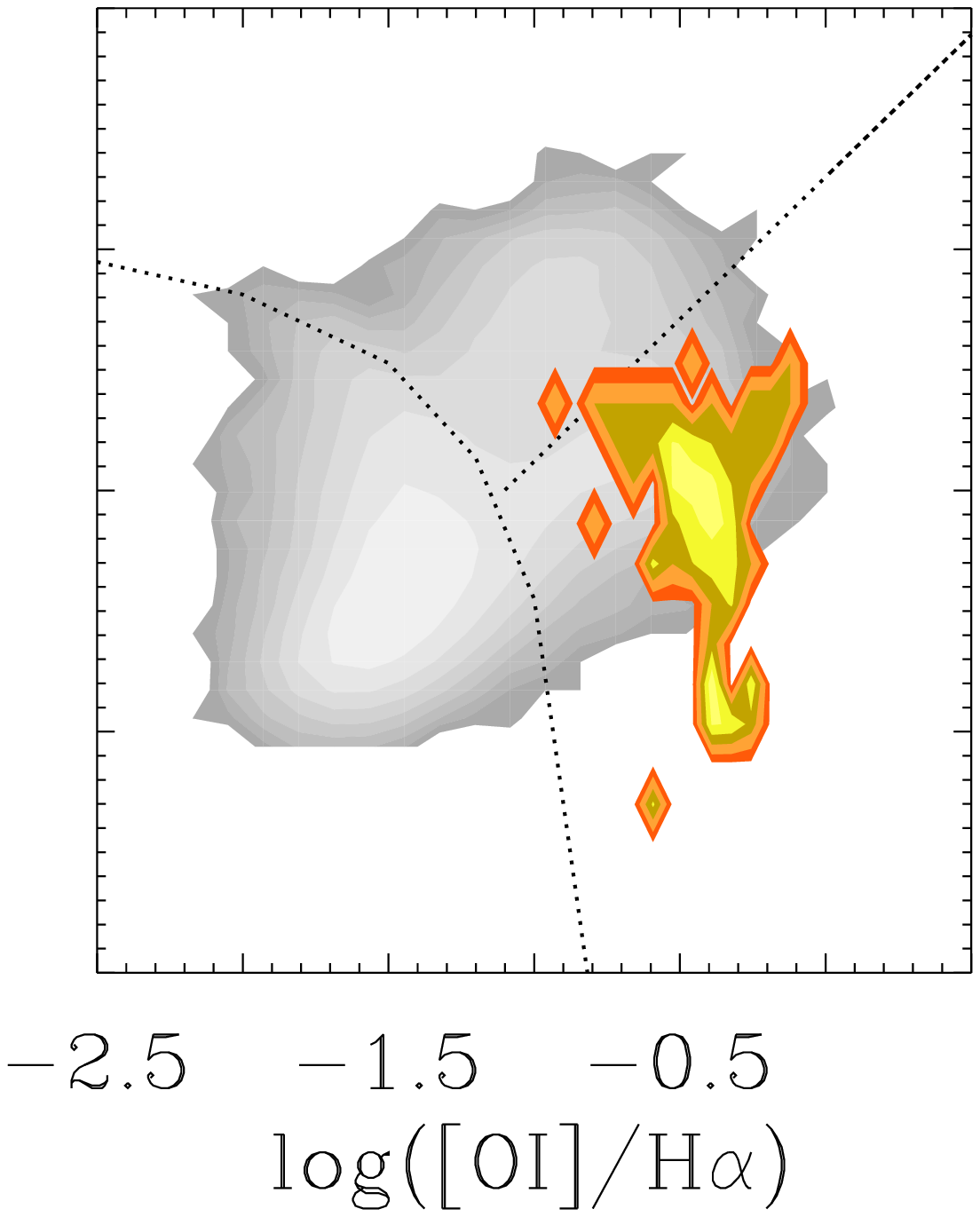,
  width=0.25\textwidth}
\caption{Location of the TNG100 galaxy population at $z=0$ in
  three BPT diagrams (coloured 2D histograms): \oiiihb\ versus \niiha\ (top row), \oiiihb\ versus \siiha\
  (middle row) and \oiiihb\ versus \oiha\ (bottom row). The locations of different galaxy types
  (blue: SF-dominated; red: composite; green: AGN-dominated; lilac: shock-dominated; yellow: 
  Post-AGB dominated) are shown in different columns. In each panel, the location of TNG50 
  galaxies is overplotted as black contours, while the grey shaded 2D histogram shows the observed
  distribution of SDSS galaxies. Also shown are classical empirical selection criteria. 
  In the \oiiihb\ vs. \niiha\ diagram, these are supposed to distinguish SF galaxies 
  (below the dashed line) from composites (between the dashed and dotted lines), 
  AGN (above the dotted line) and LI(N)ER (in the bottom-right quadrant defined 
  by dot-dashed lines), according to \citet[][dotted line]{Kewley01} and 
  \citet[][dashed and dot-dashed lines]{Kauffmann03}. In the
  \oiiihb\ vs. \siiha\ and \oiiihb\ vs. \oiha\ diagrams, SF-dominated galaxies are
  supposed to reside in the bottom left corner (below the curved dotted line),
while AGN-dominated galaxies should occupy the top part (above the
curved and straight dotted lines), and LI(N)ER the right part 
(right next to the curved and below the straight dotted
lines), according to \citet[][curved dotted line]{Kewley01} and
\citet[][straight dotted line]{Kewley06}.}\label{bpts}      
\end{figure*}

Fig.~\ref{bpts} shows the predicted distributions of emission-line ratios
of SF-dominated (left column, blue 2D histogram),  composite (second
column, red 2D histogram), AGN-dominated (third column, green 2D
histogram), PAGB-dominated (fourth column, yellow 2D histograms) and
shock-dominated (right column, lilac 2D histogram TNG100 galaxies 
at $z=0$ in three BPT diagnostic diagrams (top row:  \oiiihb\ versus \niiha; 
middle row: \oiiihb\ versus \siiha; bottom row: \oiiihb\ versus \oiha). In each
panel, the predicted distribution for TNG50 galaxies is overplotted 
as black contours, while the grey shaded 2D histogram shows the observed
distribution of SDSS galaxies. For a fair comparison between models and
observations, we have imposed a typical flux-detection limit of $5 \times 
10^{-17}$\,erg\,s$^{-1}$cm$^{-2}$ for simulated galaxies \citep[motivated 
by table 1 of][as in figure~1 of \citealp{Hirschmann17}]{Juneau14}. 
%\SC{Note
%that table 1 of Juneau+14 does not have an entry for SDSS galaxies?}
%{\bf \MH{True, but I think when discussing these flux limits with Stefanie
%  and Anna (?), they pointed me to this table. Not sure there is any
%  better justification?!}}

\begin{figure*}
\epsfig{file=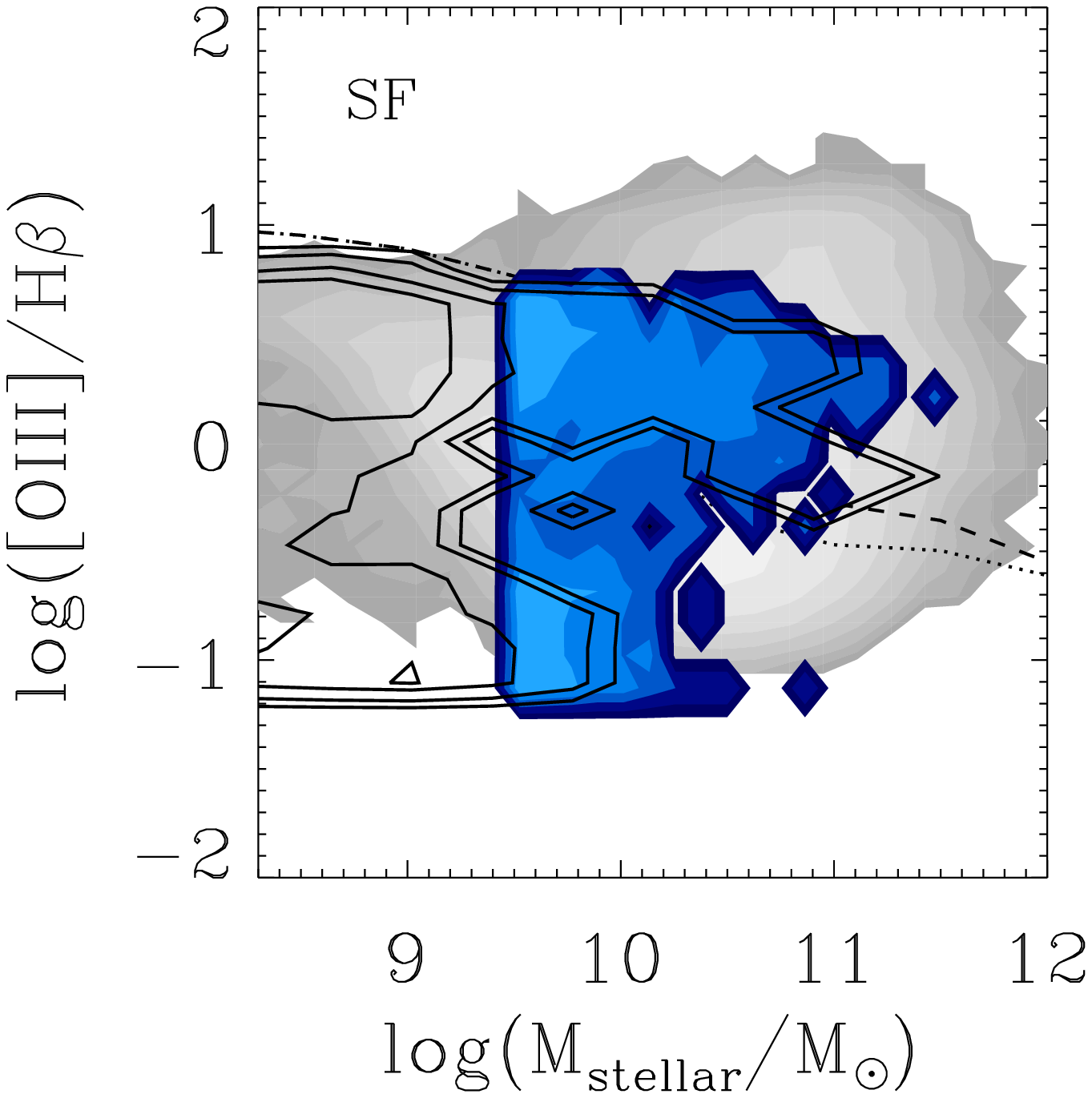,
  width=0.25\textwidth}\hspace{-1.5cm}
\epsfig{file=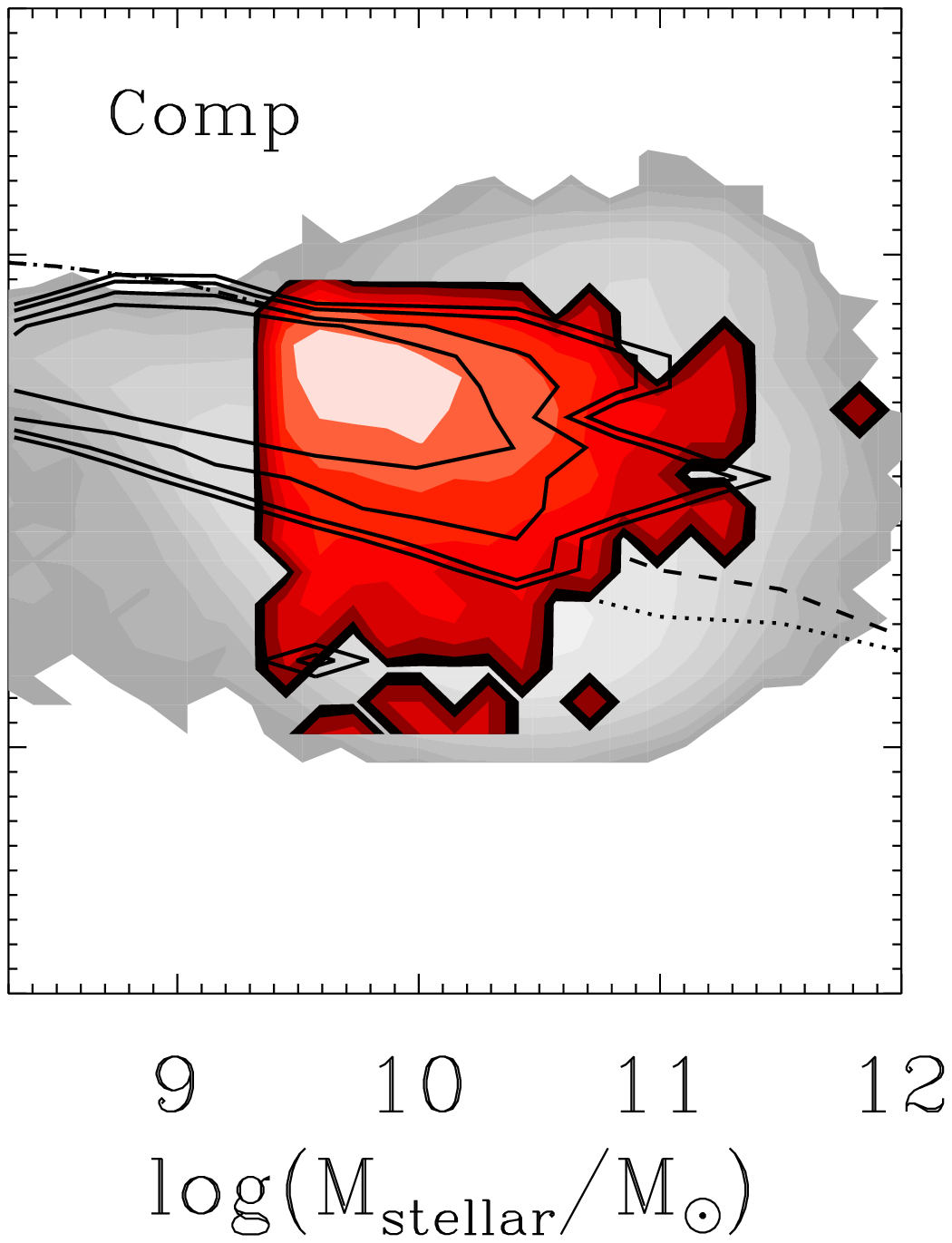,
  width=0.25\textwidth}\hspace{-1.5cm}
\epsfig{file=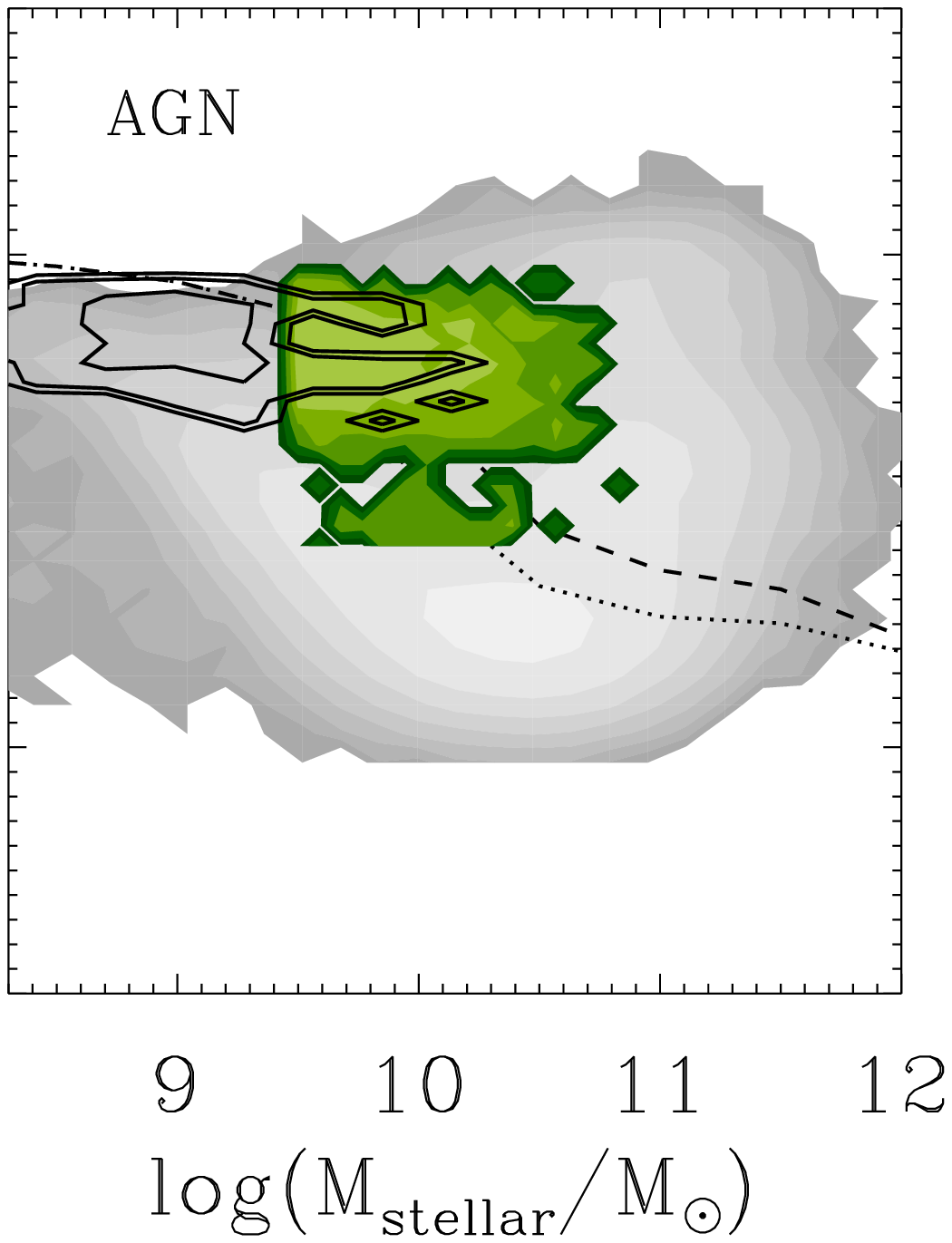,
  width=0.25\textwidth}\hspace{-1.5cm} 
\epsfig{file=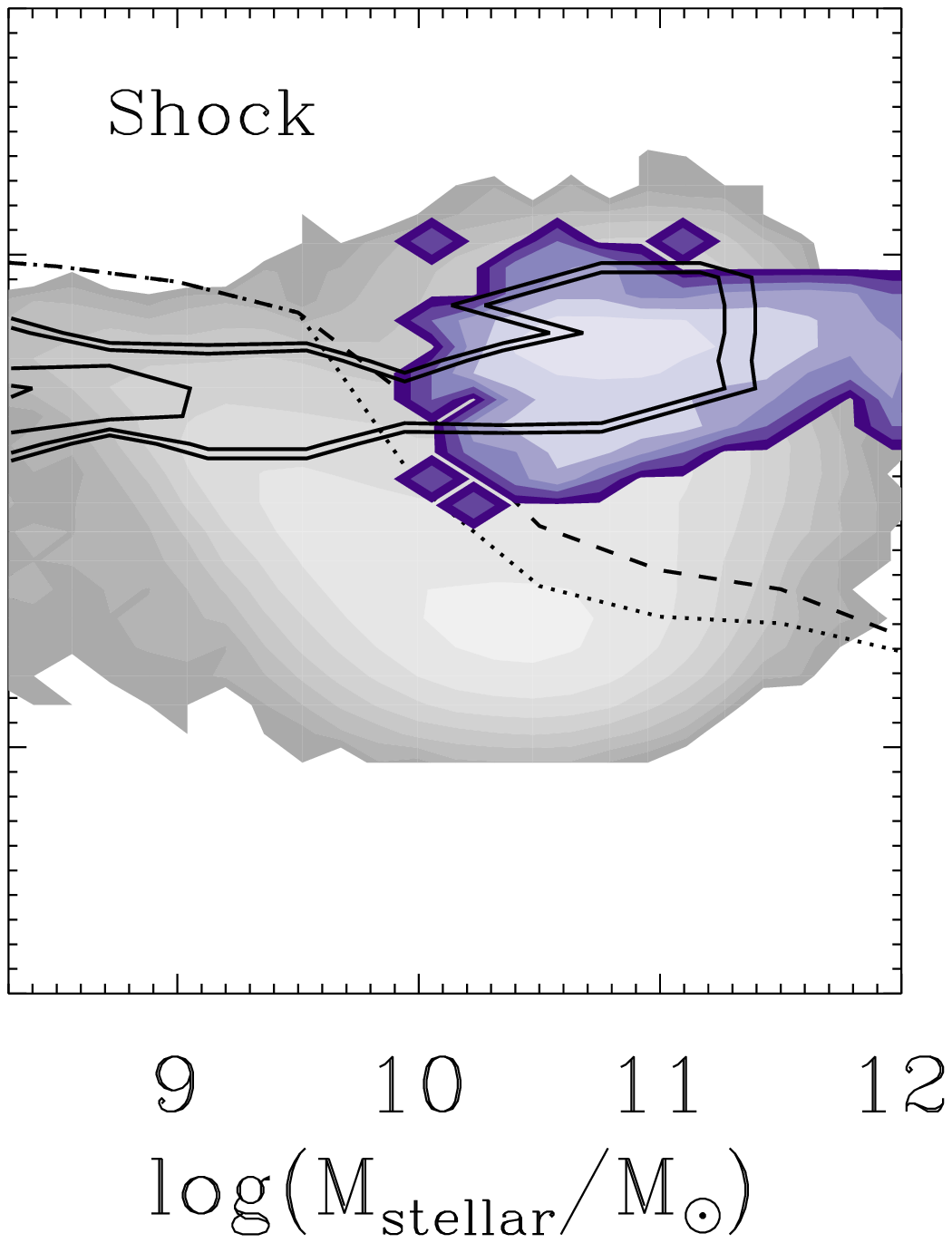,
  width=0.25\textwidth}\hspace{-1.5cm} 
\epsfig{file=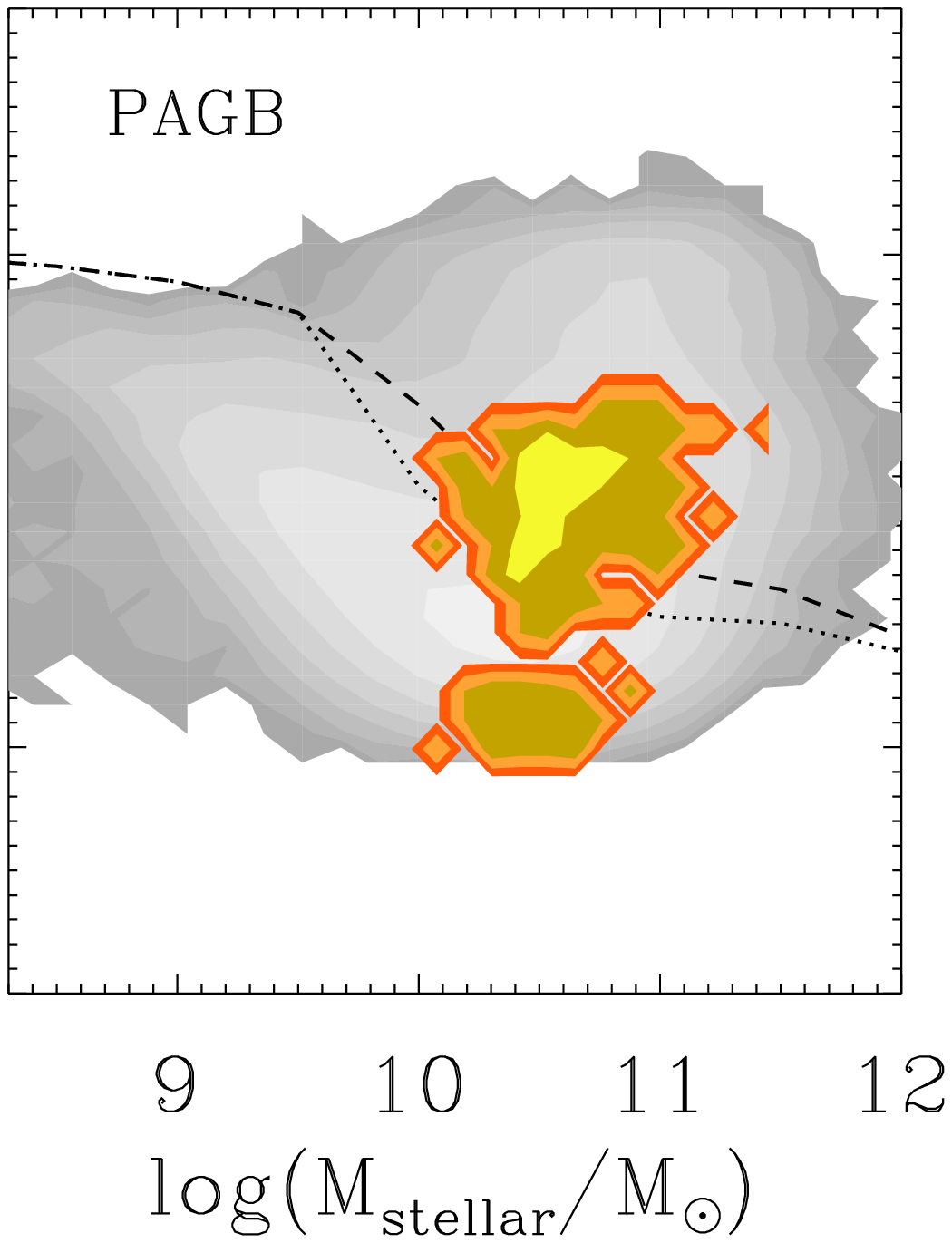,
  width=0.25\textwidth}
\caption{Location of the TNG100 and TNG50 galaxy 
populations at $z=0$ in the Mass-Excitation \citep[MEx,][]{Juneau11} diagram. 
The layout and colour coding are the same as in the top row of Fig.~\ref{bpts}. Overplotted
as black dotted and dashed lines are the empirical criteria of \citet{Juneau14}
to distinguish between SF-dominated (below the 
   dotted line), composite (between dashed and dotted lines) and
   AGN-dominated (above the dashed line) galaxies.}\label{mex}
\end{figure*}

In the top-row diagrams, the black dashed and dotted lines indicate the 
classical empirical criteria of \citet[]{Kauffmann03} and \citet[]{Kewley01}, 
respectively,  to differentiate SF galaxies from  composites, AGN and 
LI(N)ER [low-ionization (nuclear) emission]  galaxies. Observationally,
galaxies with line ratios below the dashed  line are classified as
SF-dominated,  those with line ratios between  the dashed and dotted
lines as composites, and  those with line ratios  above the dotted
line as AGN-dominated. In addition, galaxies with  line ratios in the
bottom-right quadrant defined by dot-dashed lines  are classified as
LI(N)ER, whose main ionizing sources are still  debated \citep[e.g.,
faint AGN, post-AGB stellar populations, shocks  or a mix of these
sources; e.g.,][]{Belfiore16}. The lines in the middle- and bottom-row 
diagrams of Fig.~\ref{bpts} show other criteria by \citet{Kewley01,Kewley06}
to separate SF-dominated (below the curved dotted line) from 
AGN-dominated (above both dotted lines) and LI(N)ER (in the bottom 
right area) galaxies.
 
As expected from the results obtained by \citet{Hirschmann17, Hirschmann19}
for a small sample of massive, re-simulated galaxies, we find that 
the TNG50 and TNG100 galaxy populations at $z=0$
overlap with SDSS galaxies in all BPT diagrams in Fig.~\ref{bpts}. 
Moreover, the predicted locations of SF-, AGN- and PAGB-dominated 
galaxy populations largely coincide with the SF, AGN and LI(N)ER areas 
defined by standard criteria.
Composite galaxies extend over a larger area than expected from 
the standard criteria toward the AGN region, while shock-dominated 
galaxies lie preferentially at high $\log(\oiiihb) \ga 0$ and
$\log(\oiha) \ga -1$, with \niiha\ and \siiha\ similar to those of 
other galaxy types.

We note that SF-dominated and composite galaxies reach lower \niiha, 
\siiha\ and \oiha\ ratios in the TNG50 population, which samples 
lower stellar masses and metallicities than the TNG100  simulation.
Likewise, there are hardly any PAGB-dominated TNG50 galaxies. 
Such galaxies are typically more massive than $3 \times 10^{10} \Msun$ 
and therefore significantly less numerous than in the eight-fold larger 
TNG100 volume. 

It is interesting to note that, while PAGB dominated galaxies account for
only 3 percent of all emission-line galaxies in the TNG100 simulation at $z=0$, they
make up $\sim20$ per cent of galaxies residing in the LI(N)ER area of the \oiiihb\ vs. \niiha\ 
diagram. In the TNG300 simulation, which has a higher stellar 
mass cut and better statistics for massive galaxies, PAGB-dominated galaxies 
make up $\sim40$ per cent of all LI(N)ER. Thus, the assumption often made 
in observational studies that LI(N)ER are typically weak AGN can
potentially strongly bias statistics, such as, for example, 
the optical AGN luminosity function and the AGN luminosity-SFR relation. 

Similarly to PAGB-dominated galaxies, shock-dominated galaxies are relatively 
rare, representing at most 10~per cent of all emission-line galaxies
at $z=0$ in TNG100 (the fraction is lower in TNG50,
because of the lower percentage of massive galaxies).
%\SC{(In which simulation? Does this depend on the Illustris
%  simulation?)} {\bf \MH{Yes, it does. I have added a few words for
%  clarification. Do you think this is suffcient?}}.
Nevertheless, shocks contribute 5 to 20 per cent of the \hb-line emission
from 81~per cent of SF-dominated galaxies, 98~per cent of composites, 
97~per cent of AGN-dominated galaxies and 51~per cent of PAGB-dominated
galaxies.

In Fig.~\ref{mex}, we show the analogue of Fig.~\ref{bpts} for the MEx diagram
\citep{Juneau11}, defined by the \oiiihb\ ratio and galaxy stellar mass 
$M_{\mathrm{stellar}}$. The dotted and dashed lines indicate empirical
criteria from \citet{Juneau14} to distinguish between SF-dominated
(below the dotted line), composite (between dotted and dashed line)
and AGN-dominated (above the dashed line) galaxies. We find that AGN-dominated 
galaxies in the TNG100 simulation mostly overlap with the 
observationally-defined AGN region. However, the \oiiihb\ ratios of 
AGN-dominated galaxies with lower stellar masses sampled by TNG50
fall below the expectation. Shock-dominated galaxies lie 
in the same area and are therefore hardly distinguishable from AGN. Instead, 
SF-dominated and composite galaxies extend beyond the area identified by 
\citet{Juneau14} in this diagram. In the stellar mass range $M_{\mathrm{stellar}} 
= 3 \times 10^{10}$--$10^{11}\,\Msun$, both populations can reach
$\oiiihb\sim1$, i.e., an order of magnitude above expectation.

\begin{figure*}
\epsfig{file=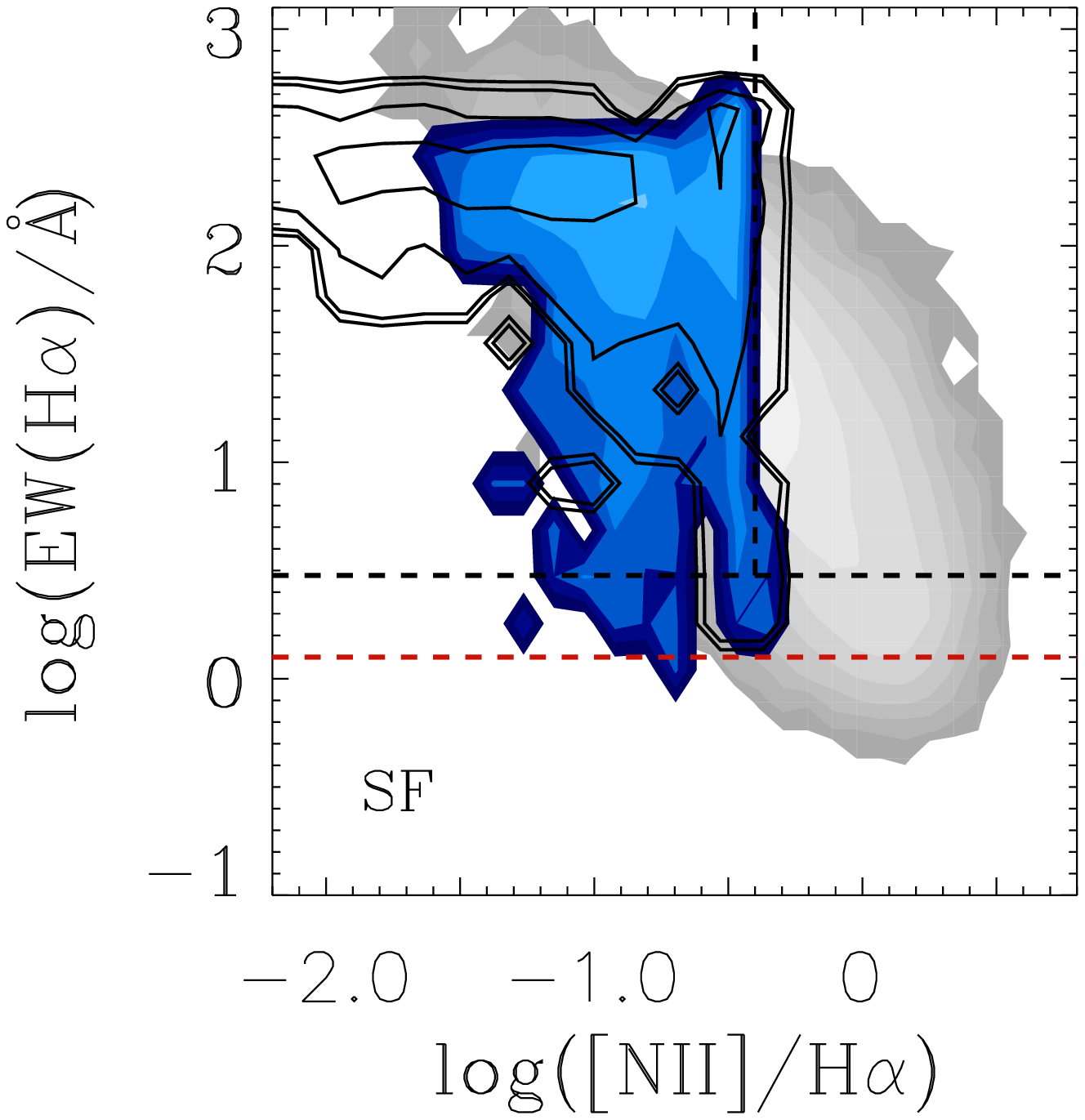,
  width=0.25\textwidth}\hspace{-1.5cm}
\epsfig{file=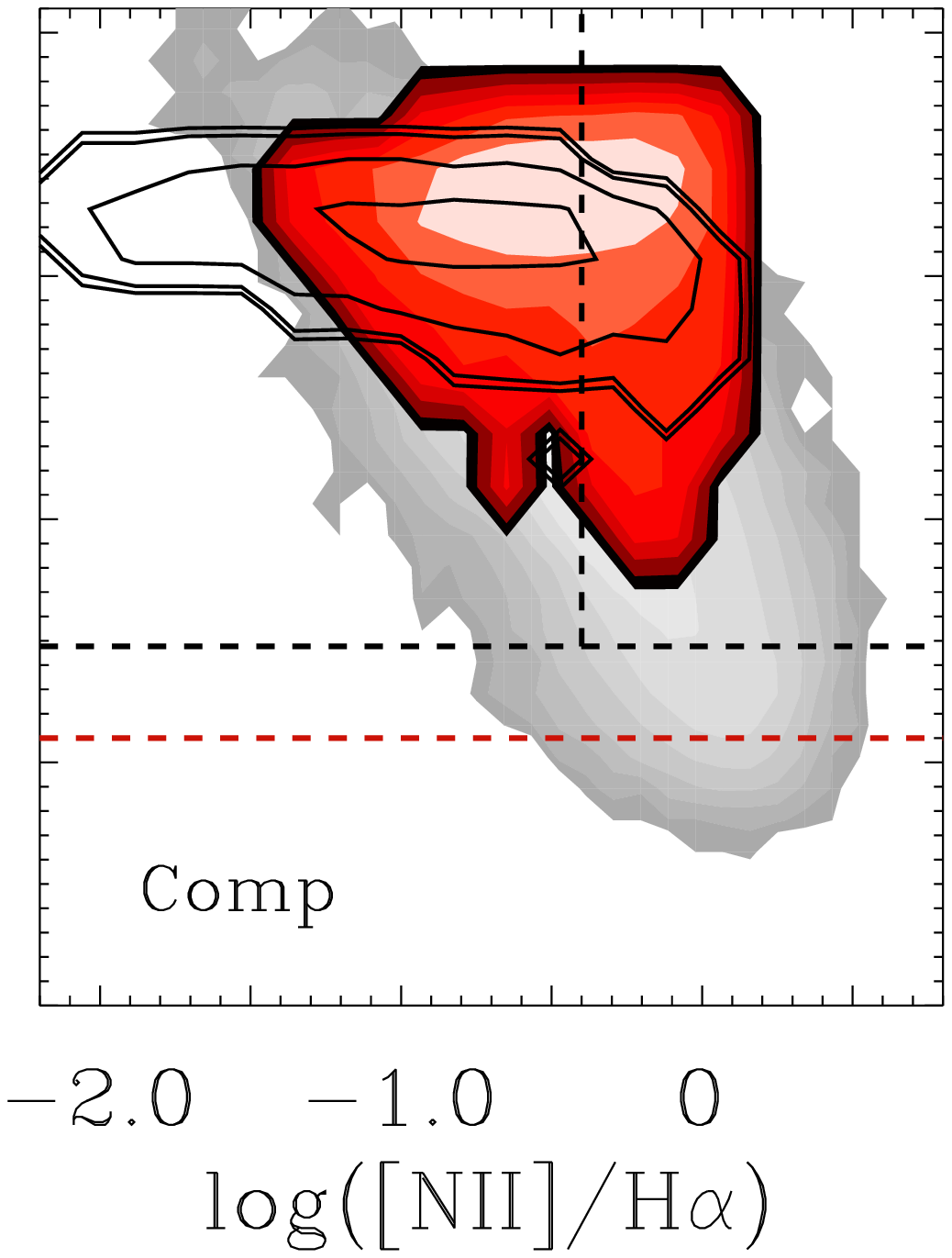,
  width=0.25\textwidth}\hspace{-1.5cm}
\epsfig{file=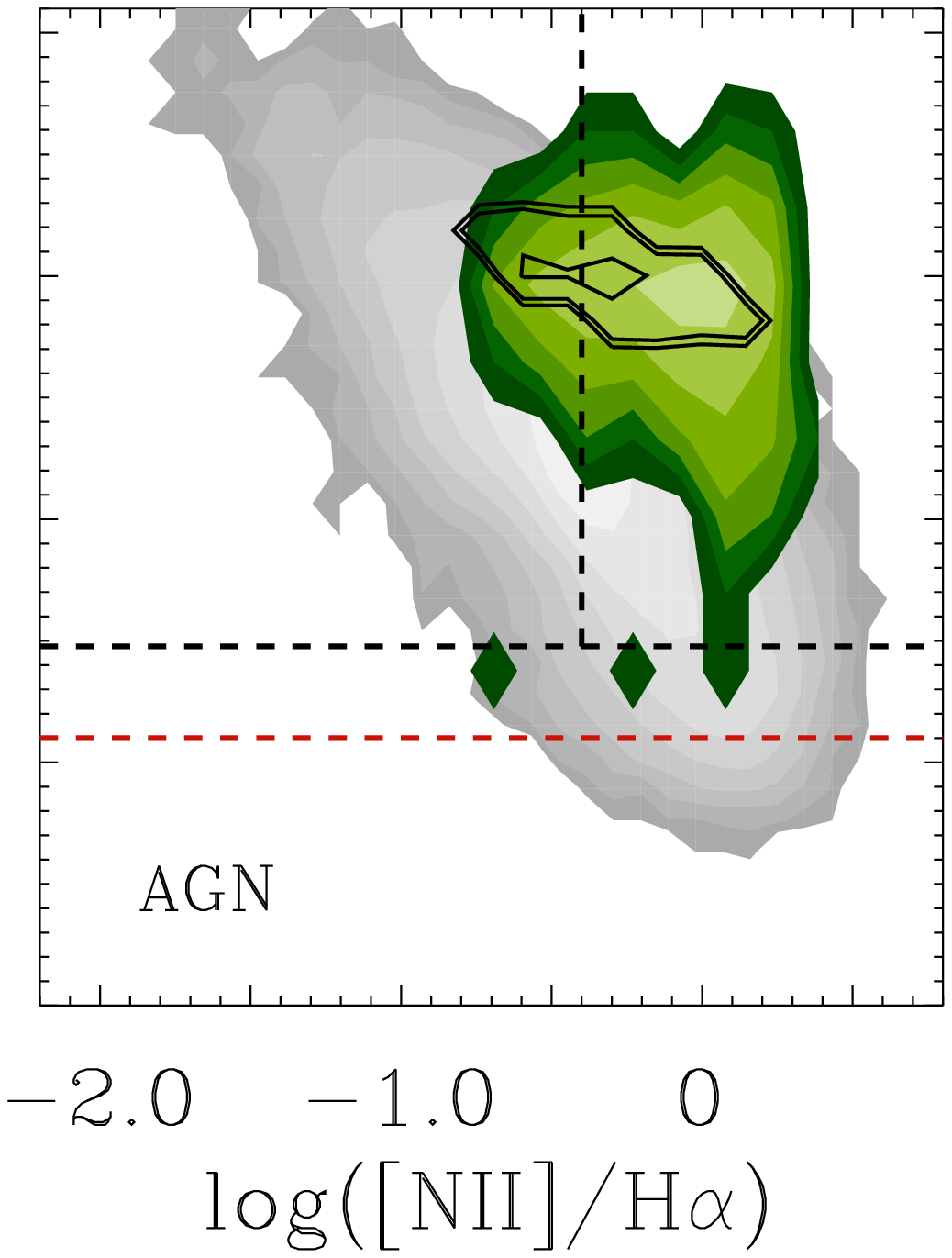,
  width=0.25\textwidth}\hspace{-1.5cm} 
\epsfig{file=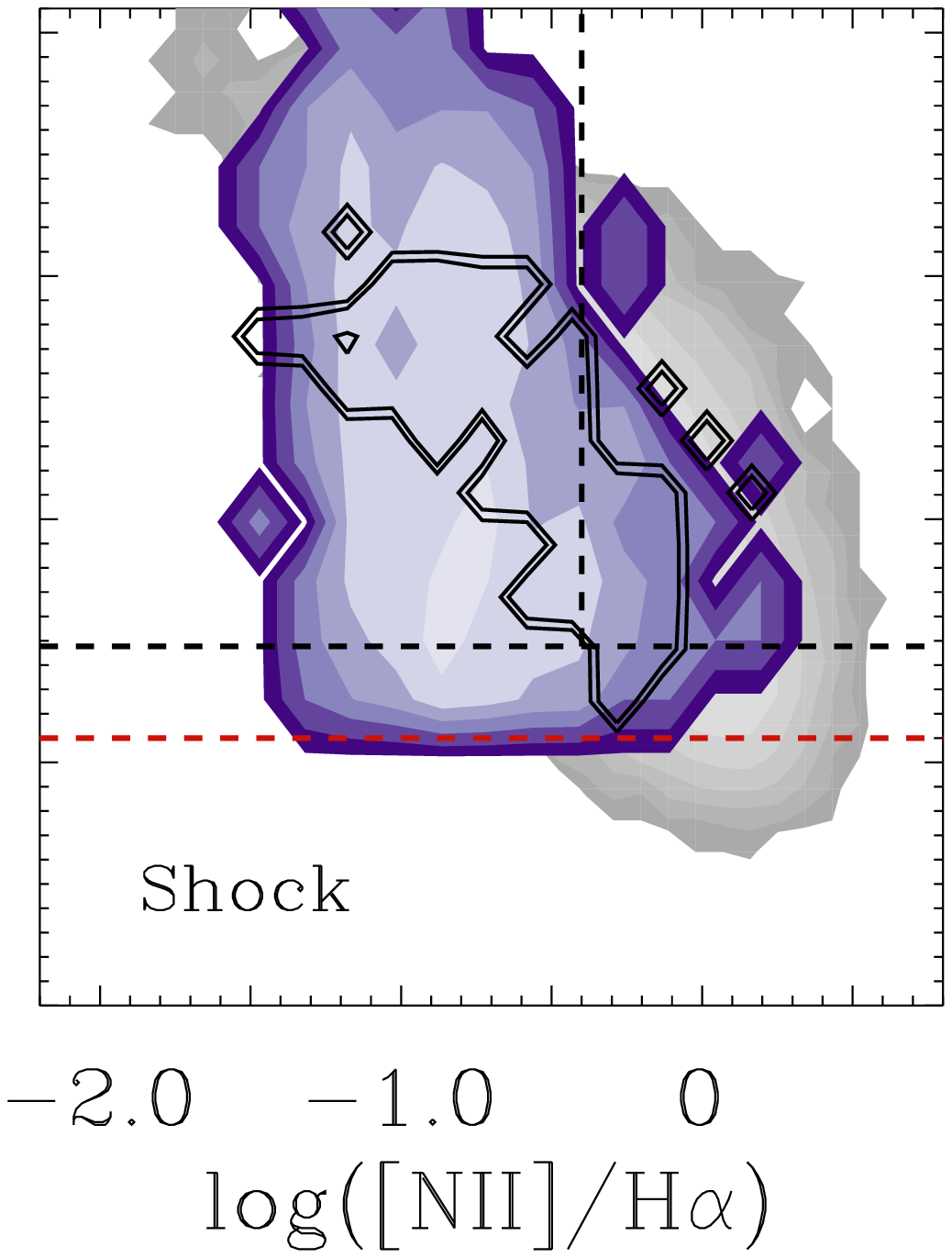,
  width=0.25\textwidth}\hspace{-1.5cm} 
\epsfig{file=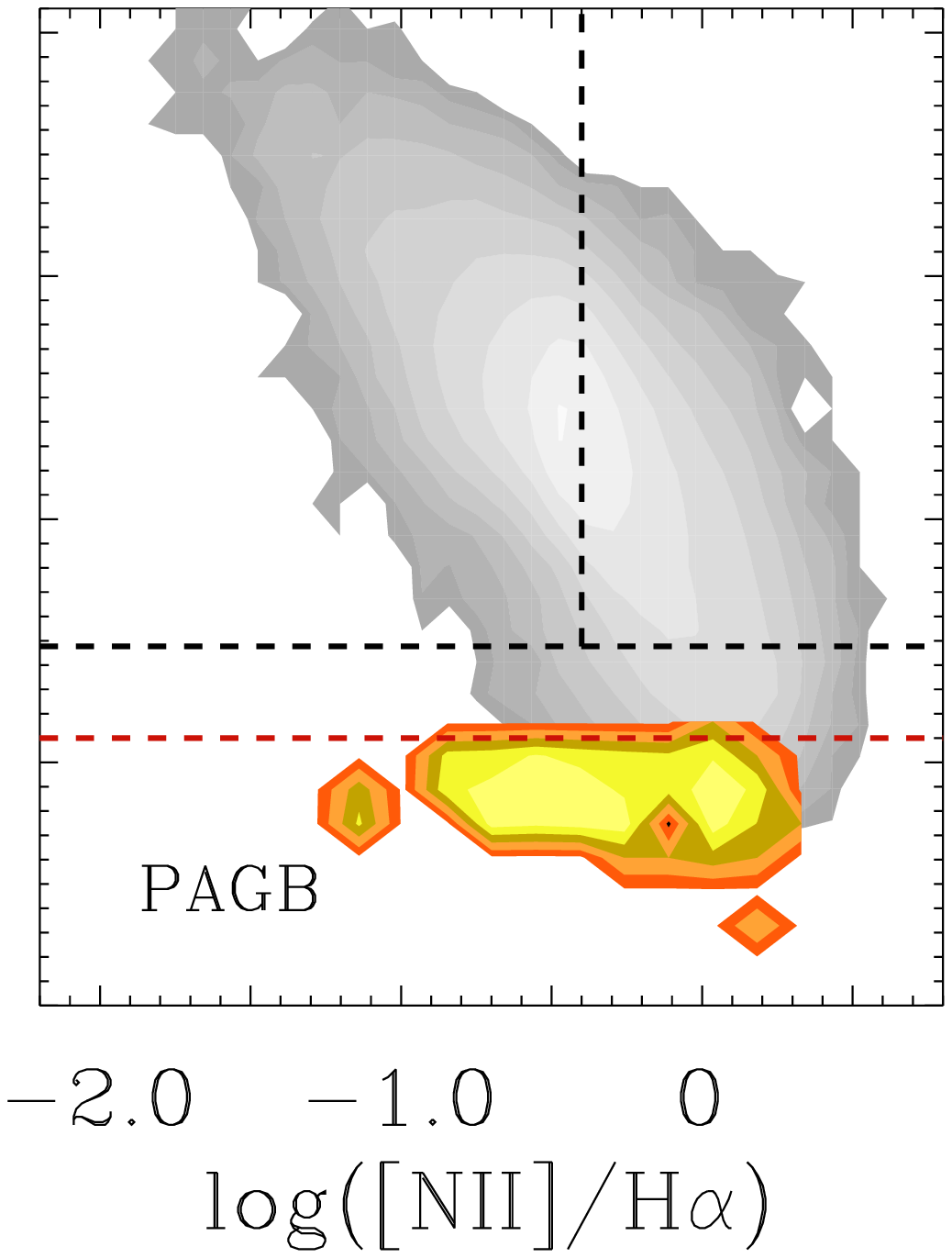,
  width=0.25\textwidth}
\caption{Location of the TNG100 and TNG50 galaxy 
populations at $z=0$ in the EW(\ha)-versus-\niiha\ (WHAN) diagnostic diagram \citep{CidFernandes10}. 
The layout and colour coding are the same as in the top row of Fig.~\ref{bpts}. 
Overplotted as black dashed lines in each panel are the empirical criteria of \citet{CidFernandes11} to 
distinguish between SF-dominated (top-left quadrant), AGN-dominated (top-right quadrant), and 
retired, post-AGB dominated (below horizontal dashed black line) galaxies. The red dashed line
shows the criterion suggested by our simulated emission-line galaxies to robustly identify 
post-AGB-dominated galaxies.}\label{whan}     
\end{figure*}
\begin{figure*}
  \center
\epsfig{file=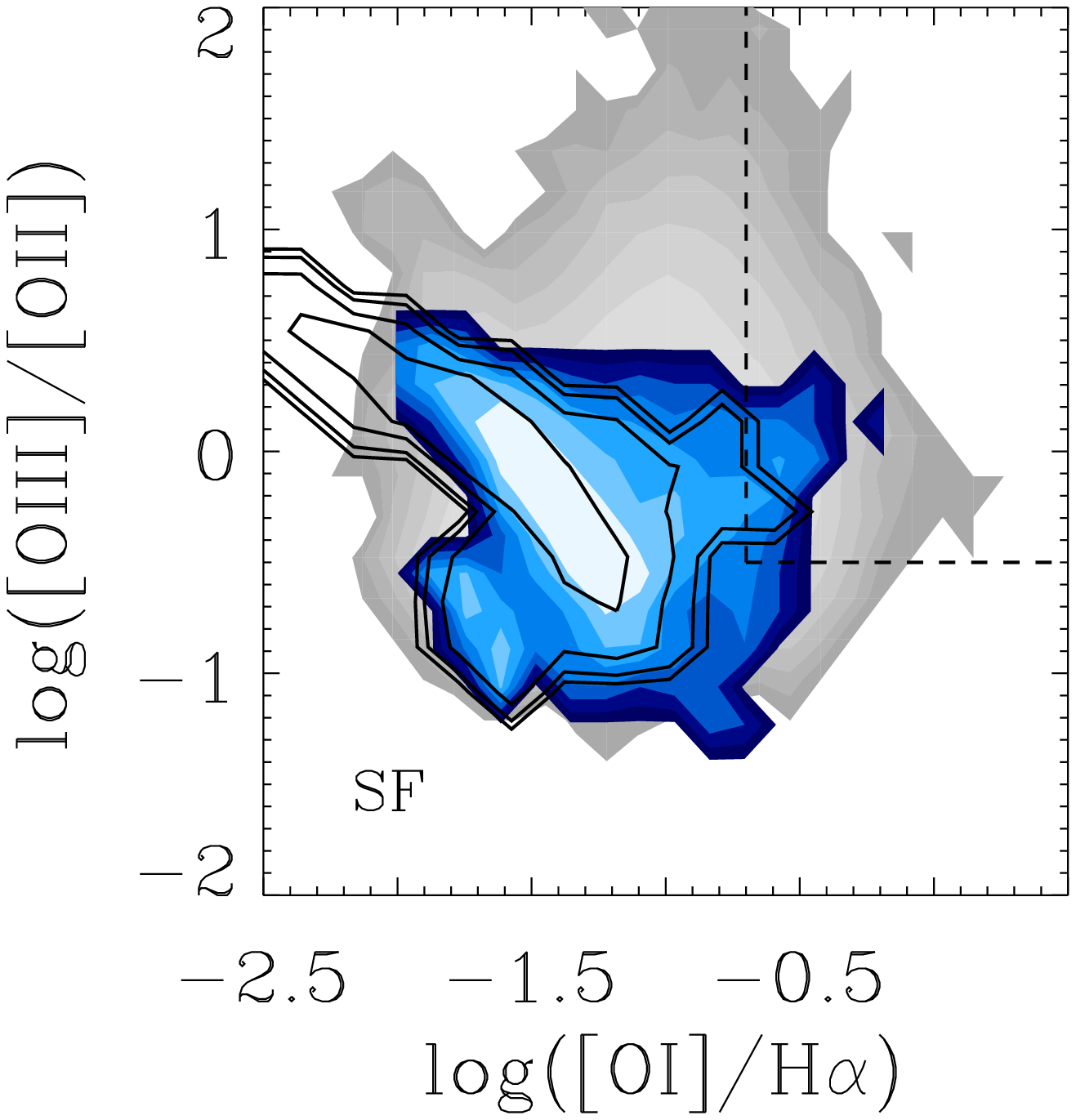,
  width=0.25\textwidth}\hspace{-1.5cm}
\epsfig{file=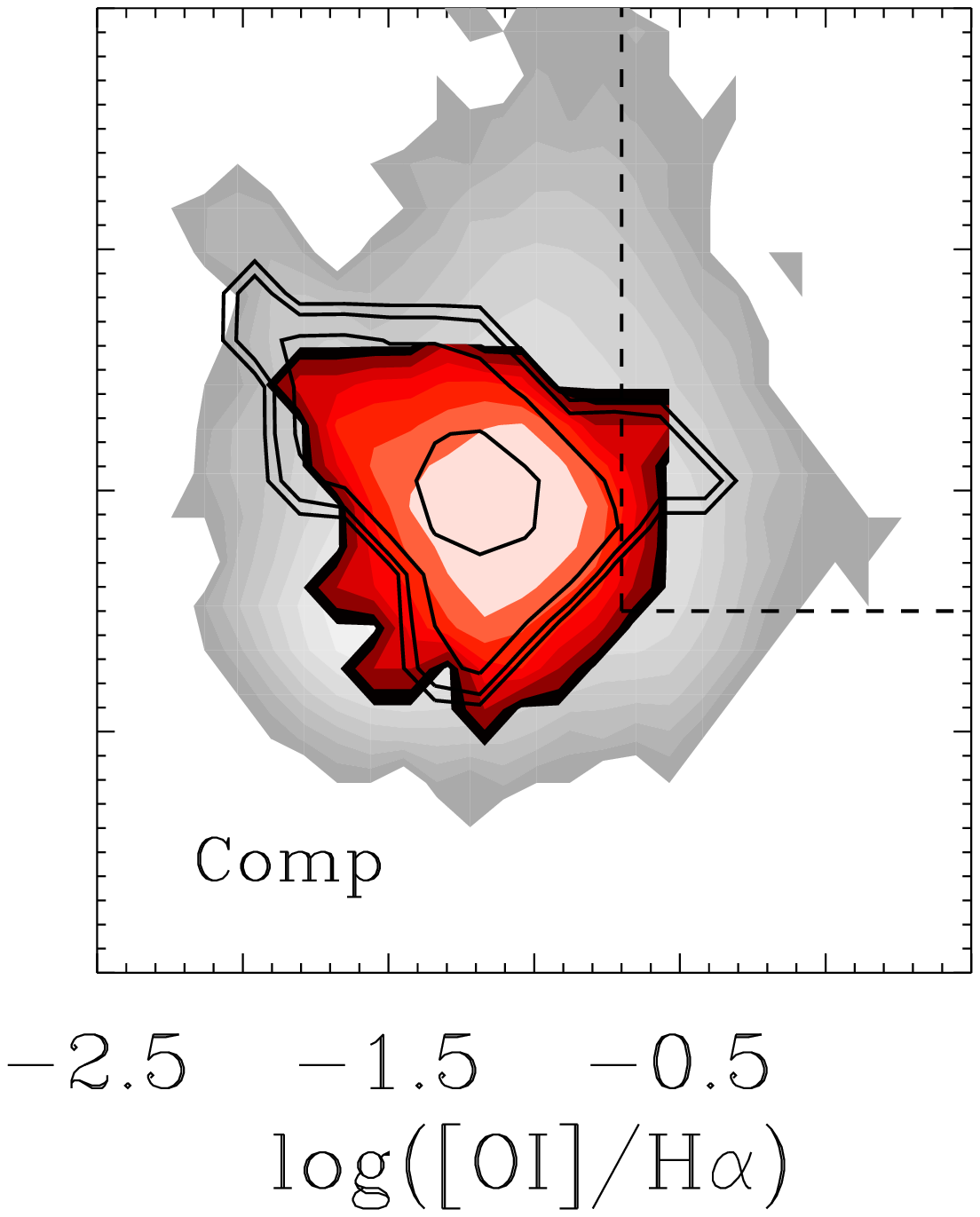,
  width=0.25\textwidth}\hspace{-1.5cm}
\epsfig{file=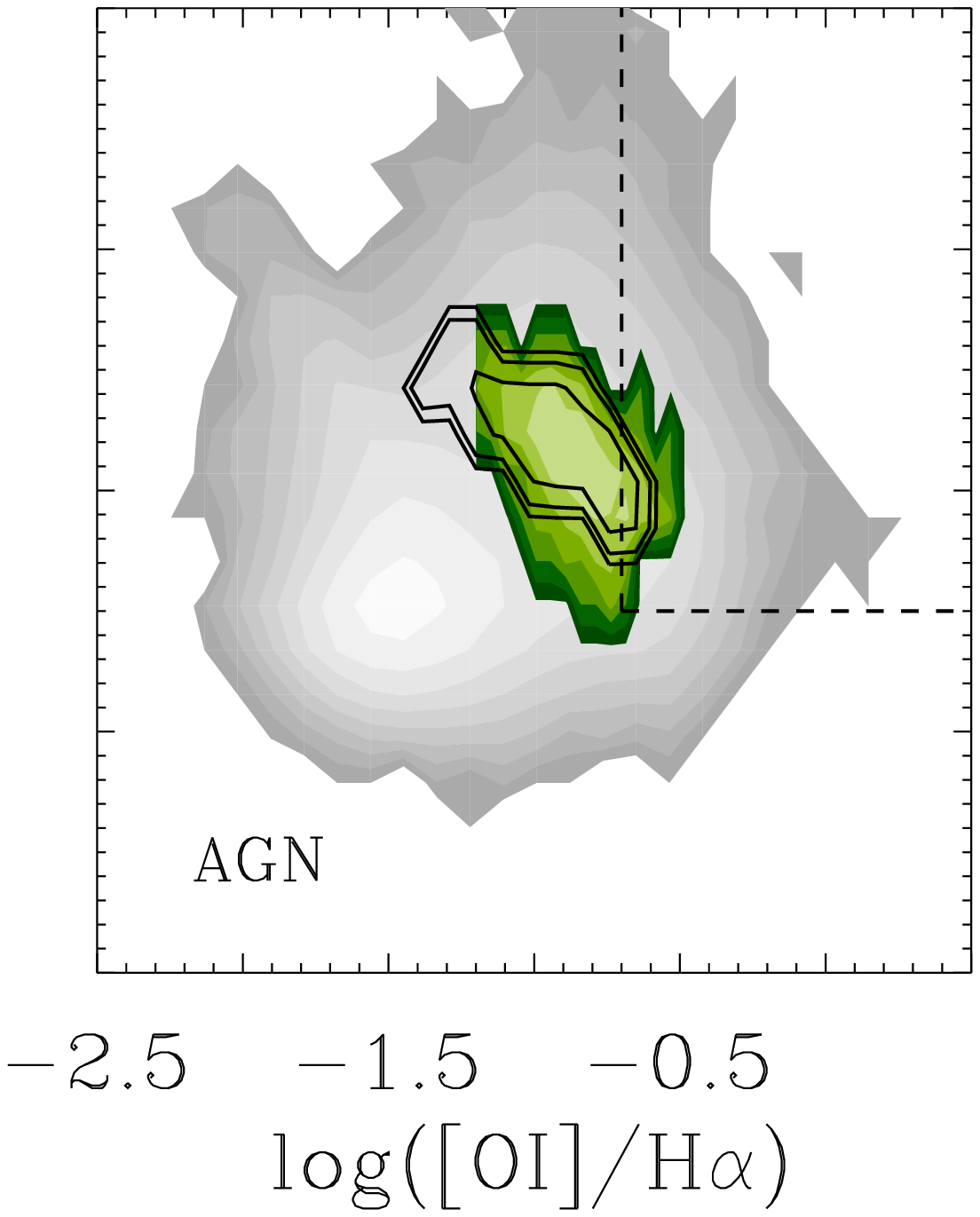,
  width=0.25\textwidth}\hspace{-1.5cm} 
\epsfig{file=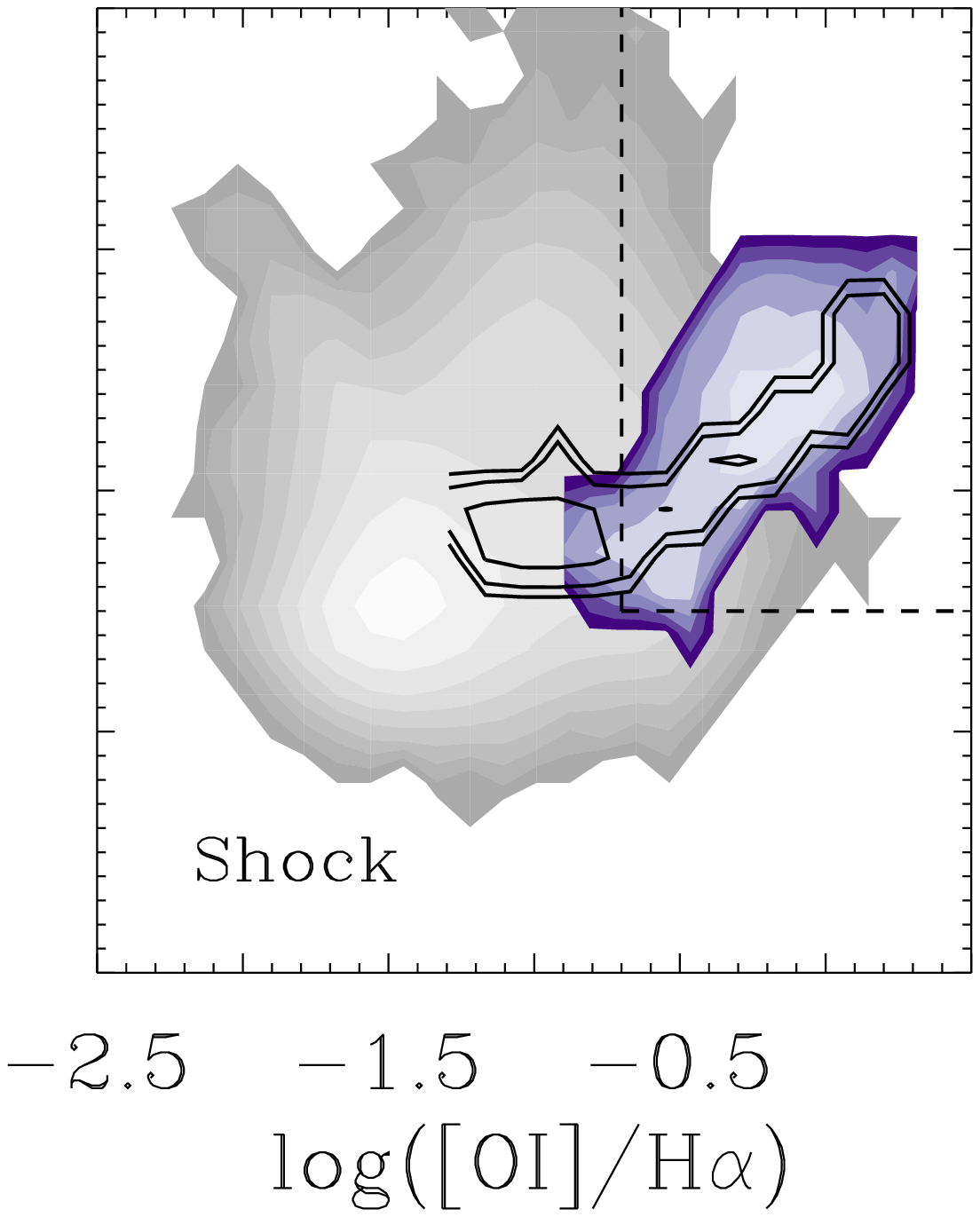,
  width=0.25\textwidth}\hspace{-1.5cm} 
\epsfig{file=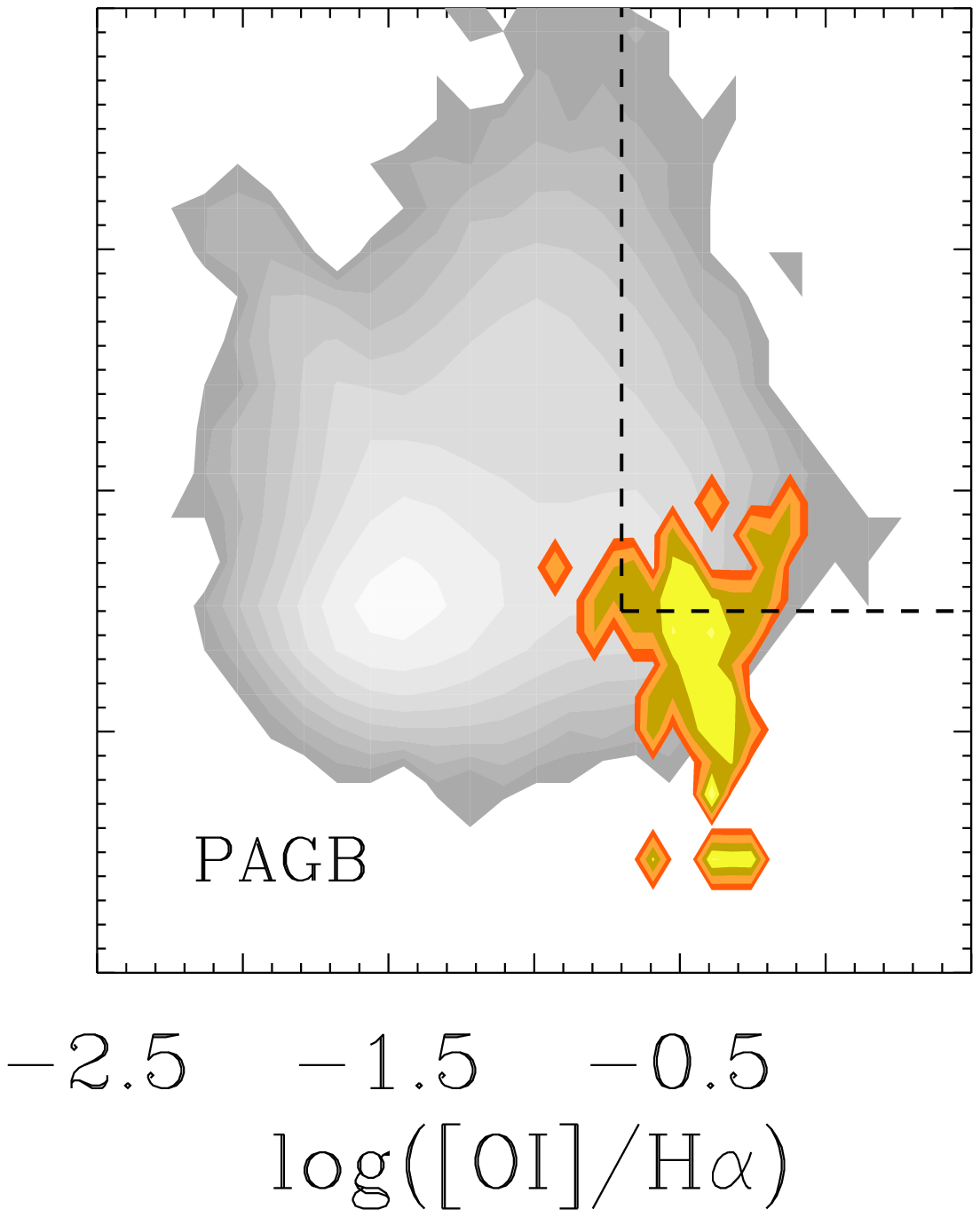,
  width=0.25\textwidth}
\caption{Location of the TNG100 and TNG50 galaxy 
populations at $z=0$ in the  \oiiioii-versus-\oiha\ diagnostic diagram proposed by \citet{Kewley19}.
The layout and colour coding are the same as in the top row of Fig.~\ref{bpts}. 
In each panel, the top-right quadrant delimited by dashed black lines is the region 
expected to be populated only by galaxies dominated by fast radiative shocks 
\citep[see figure~11 of][]{Kewley19}.}\label{shockdiagram}     
\end{figure*}

\begin{figure*}
\center
\epsfig{file=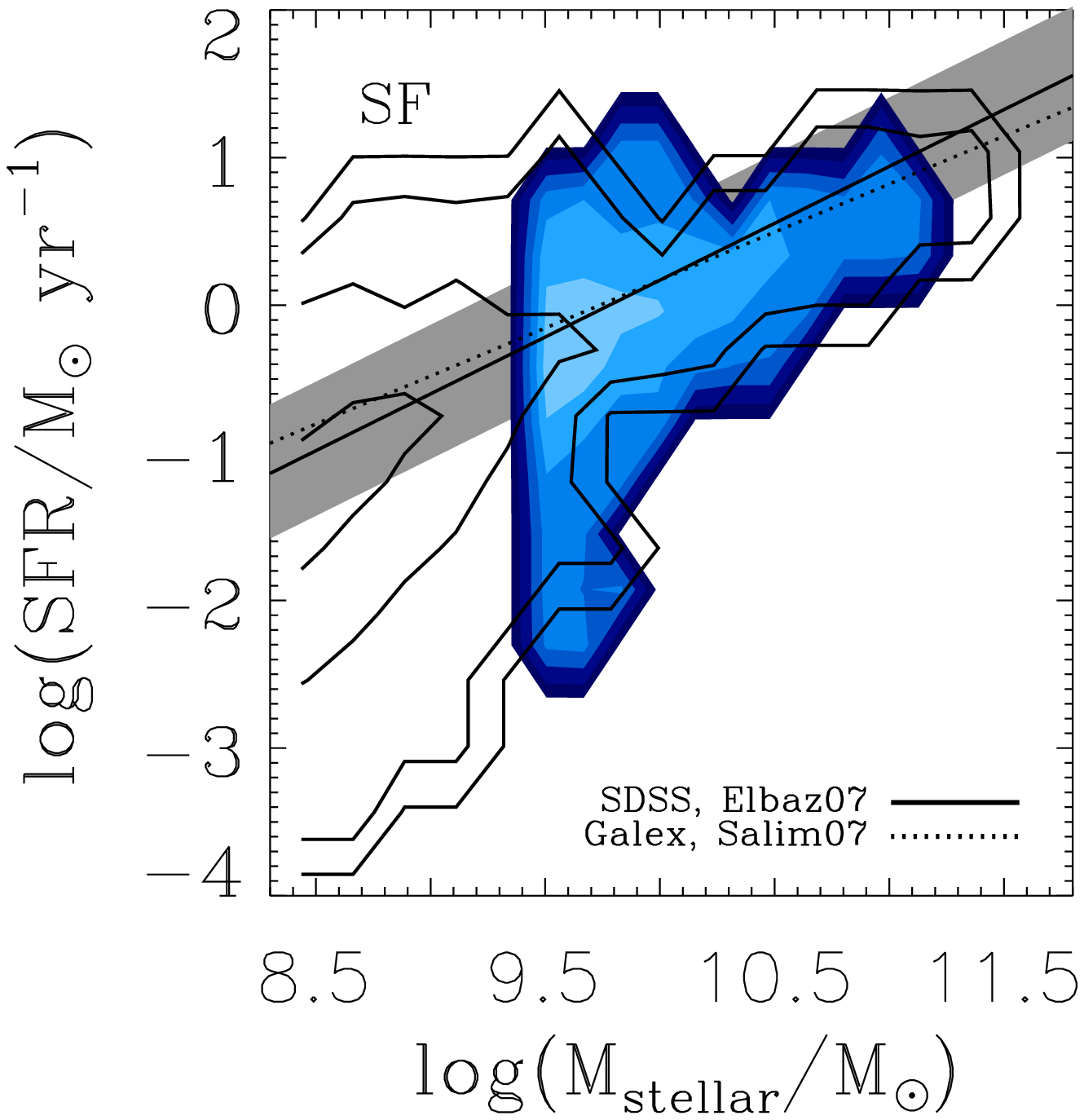, width=0.25\textwidth}\hspace{-1.5cm}
\epsfig{file=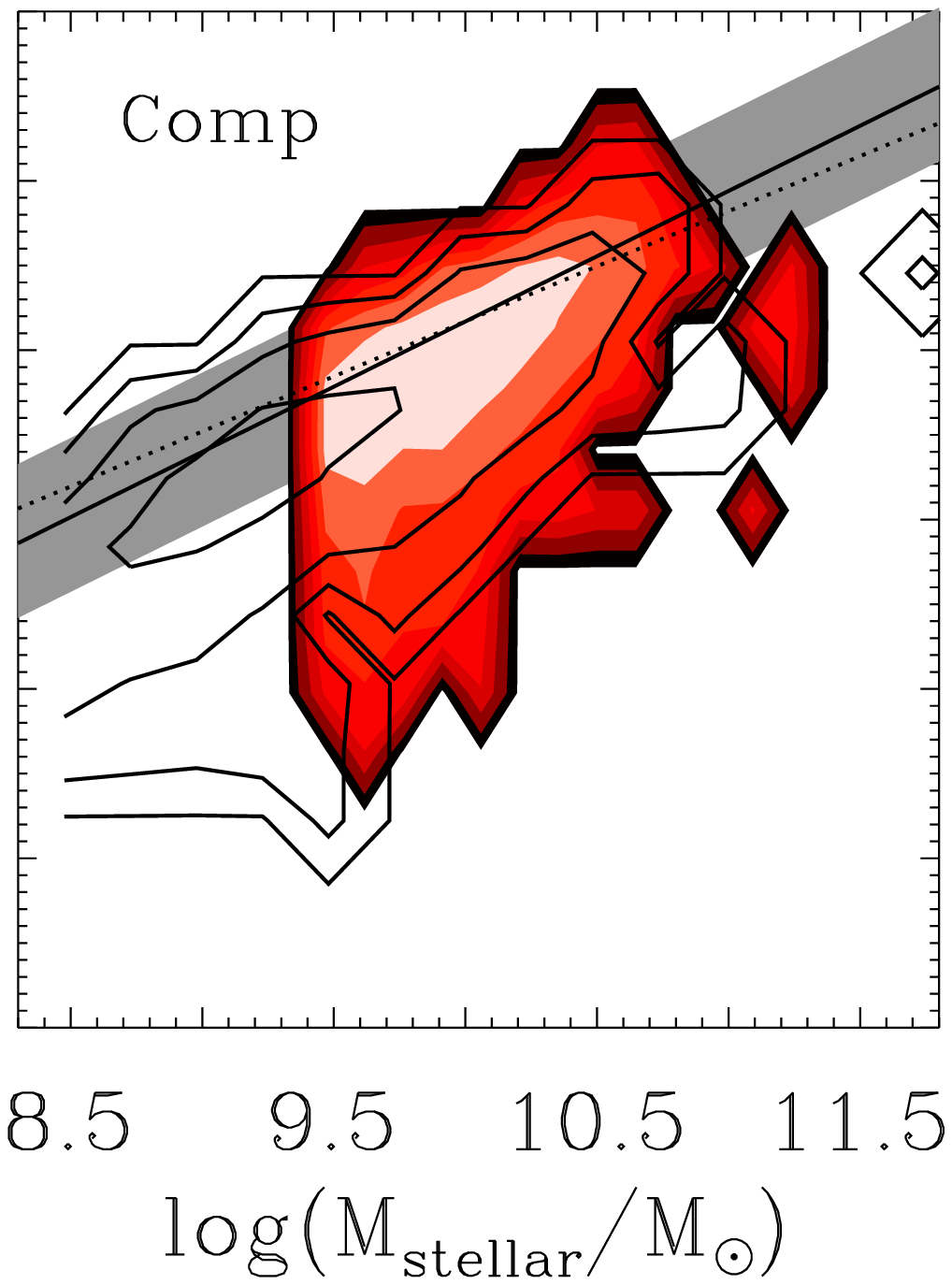, width=0.25\textwidth}\hspace{-1.5cm}
\epsfig{file=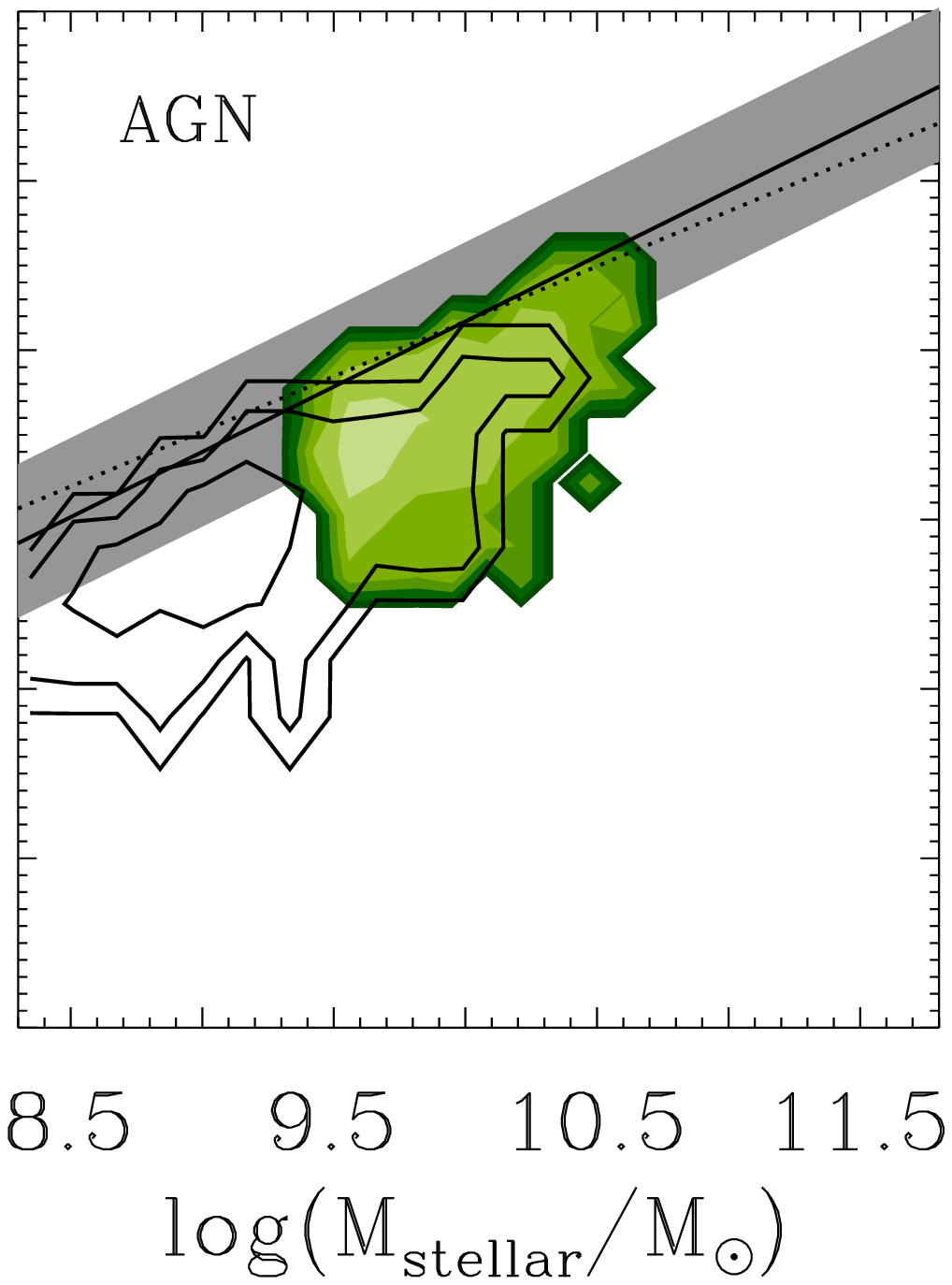, width=0.25\textwidth}\hspace{-1.5cm} 
\epsfig{file=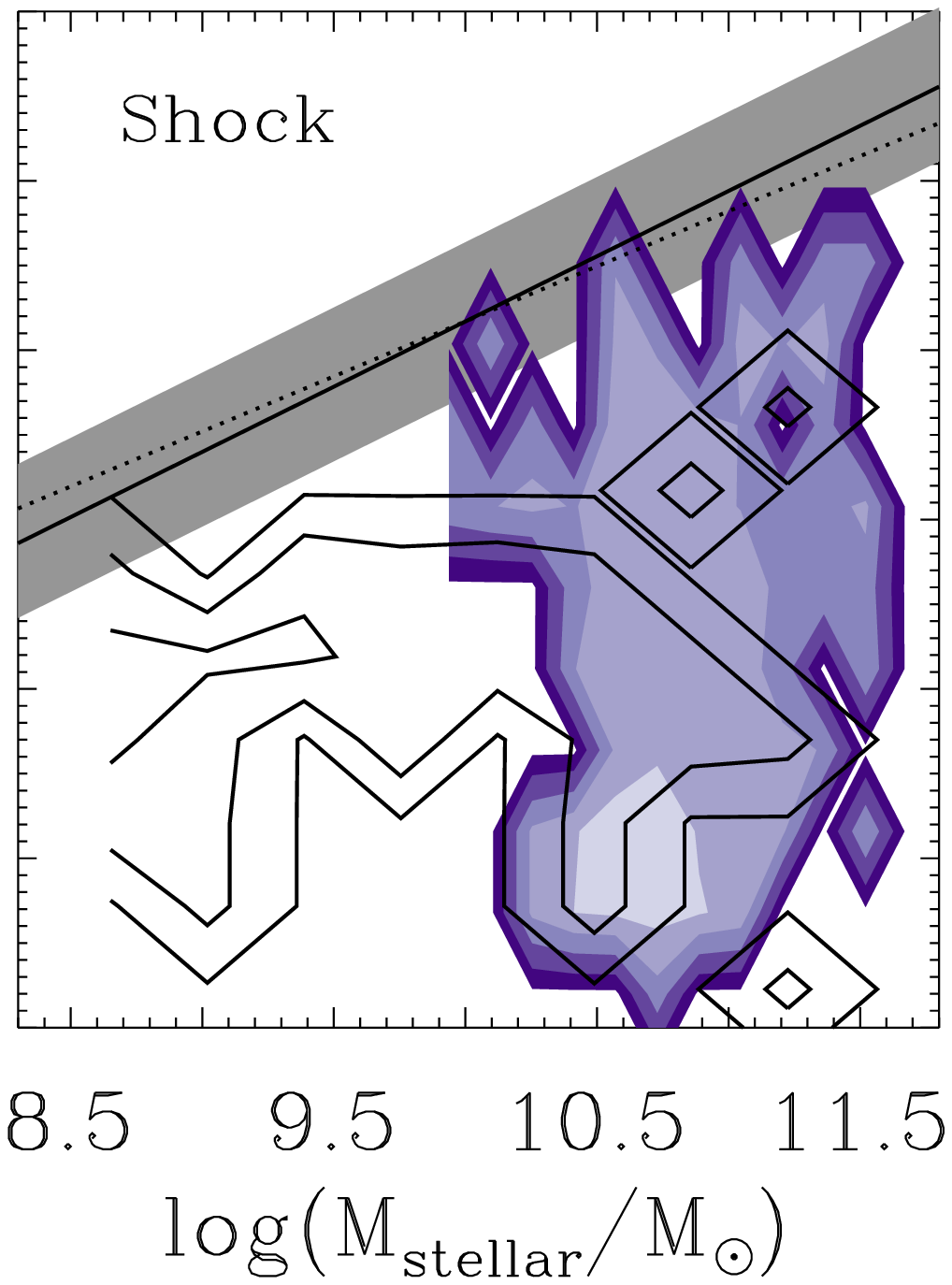, width=0.25\textwidth}\hspace{-1.5cm} 
\epsfig{file=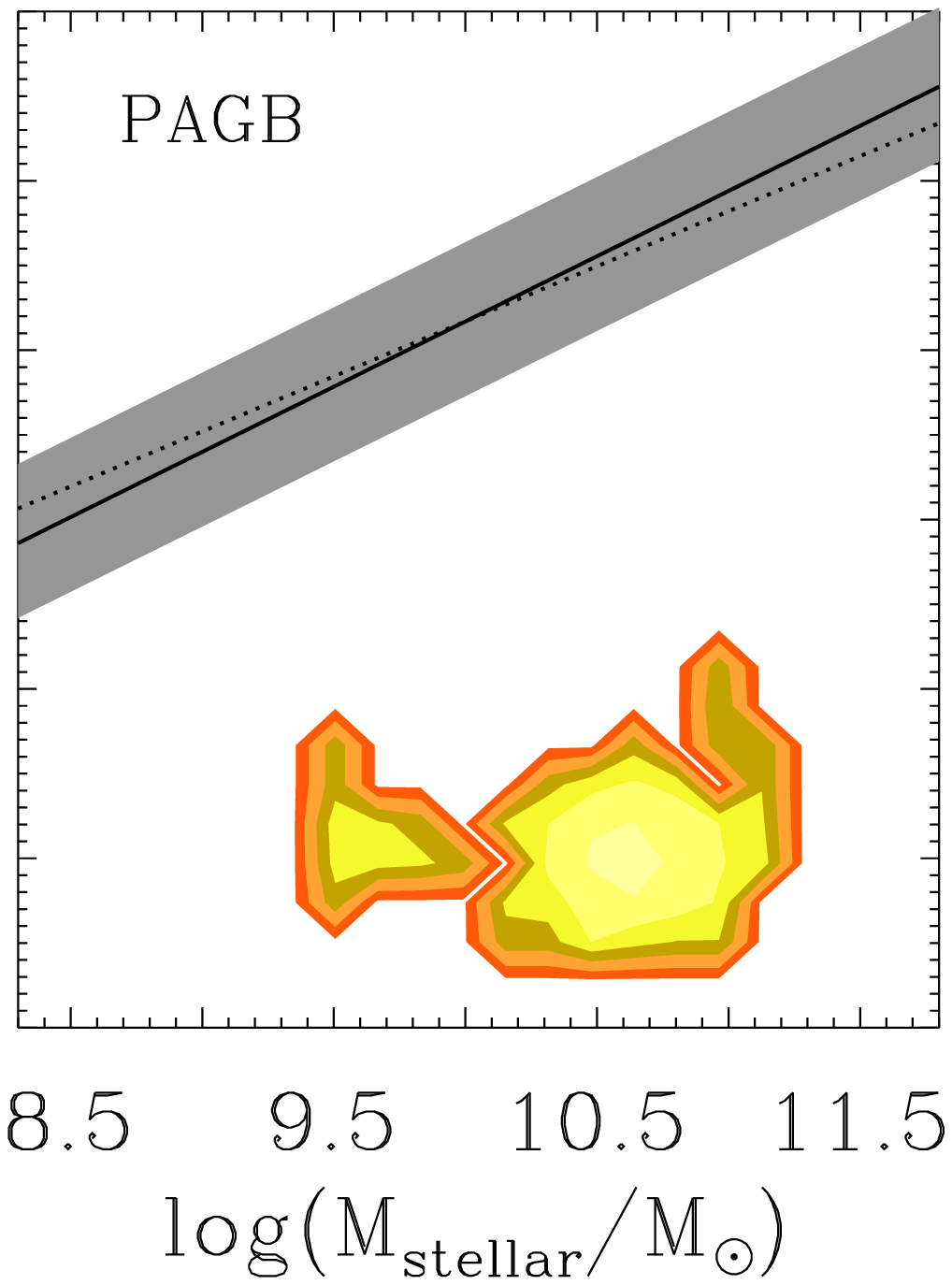,
  width=0.25\textwidth}
\epsfig{file=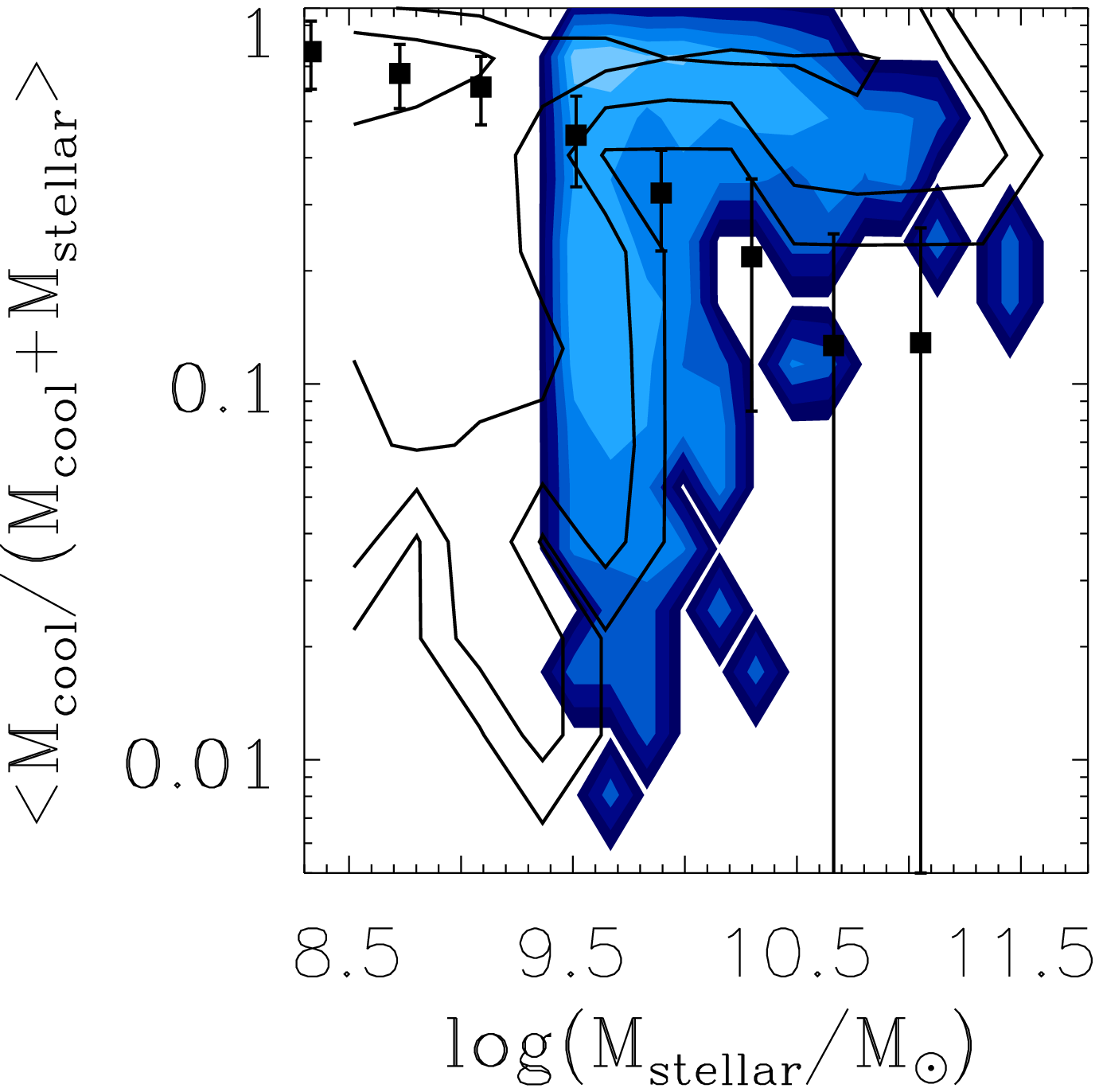, width=0.25\textwidth}\hspace{-1.5cm}
\epsfig{file=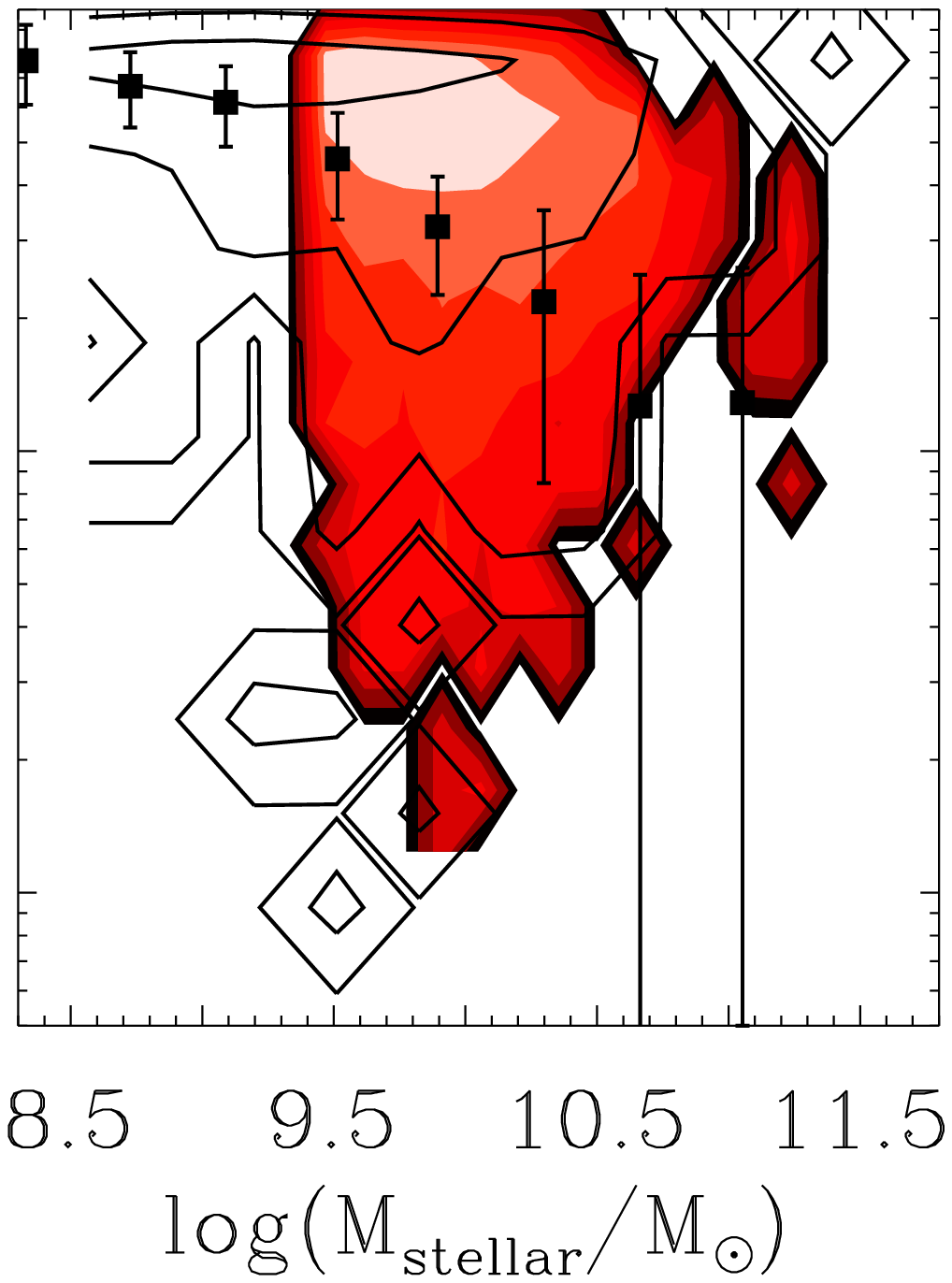, width=0.25\textwidth}\hspace{-1.5cm}
\epsfig{file=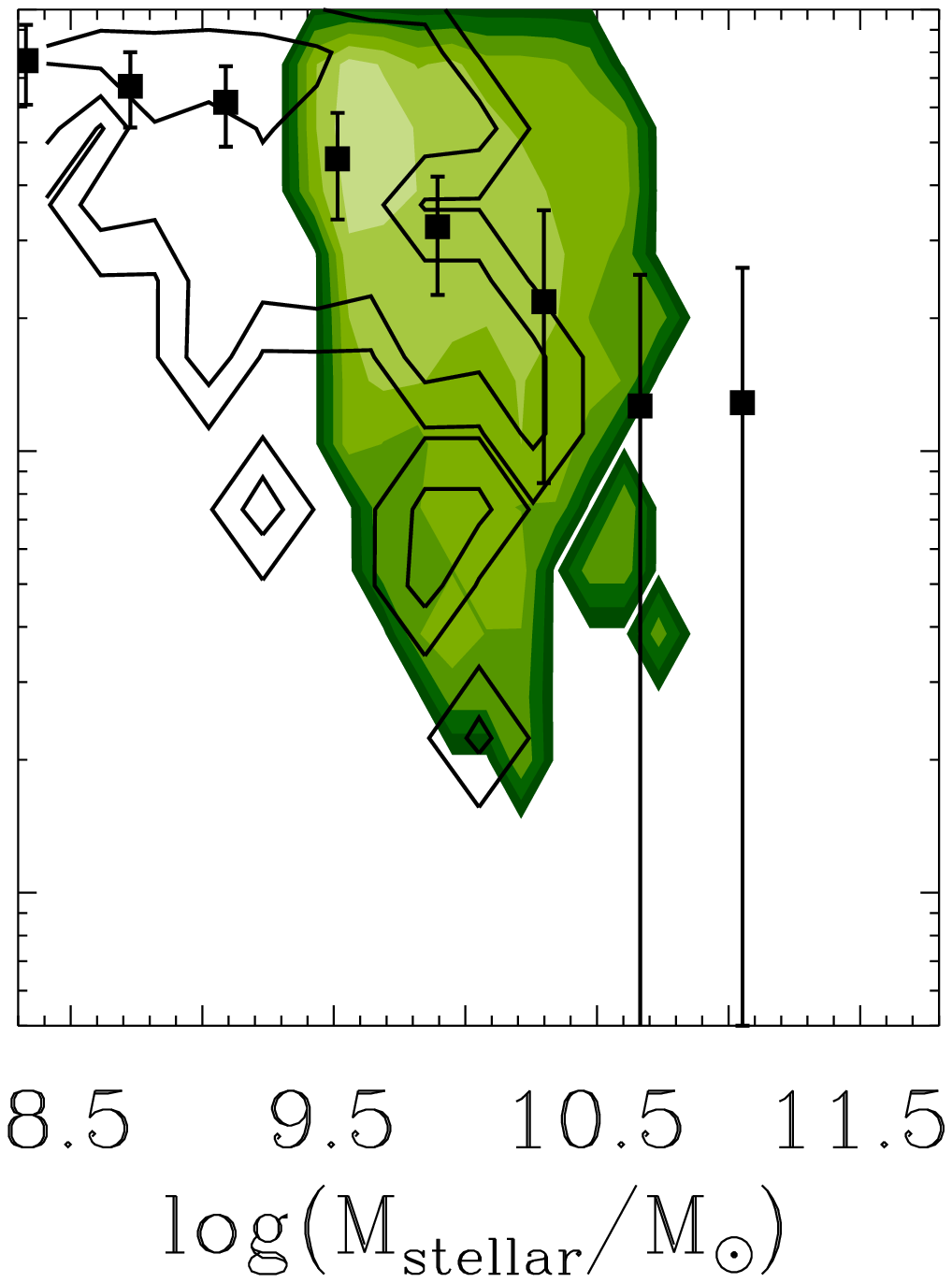, width=0.25\textwidth}\hspace{-1.5cm} 
\epsfig{file=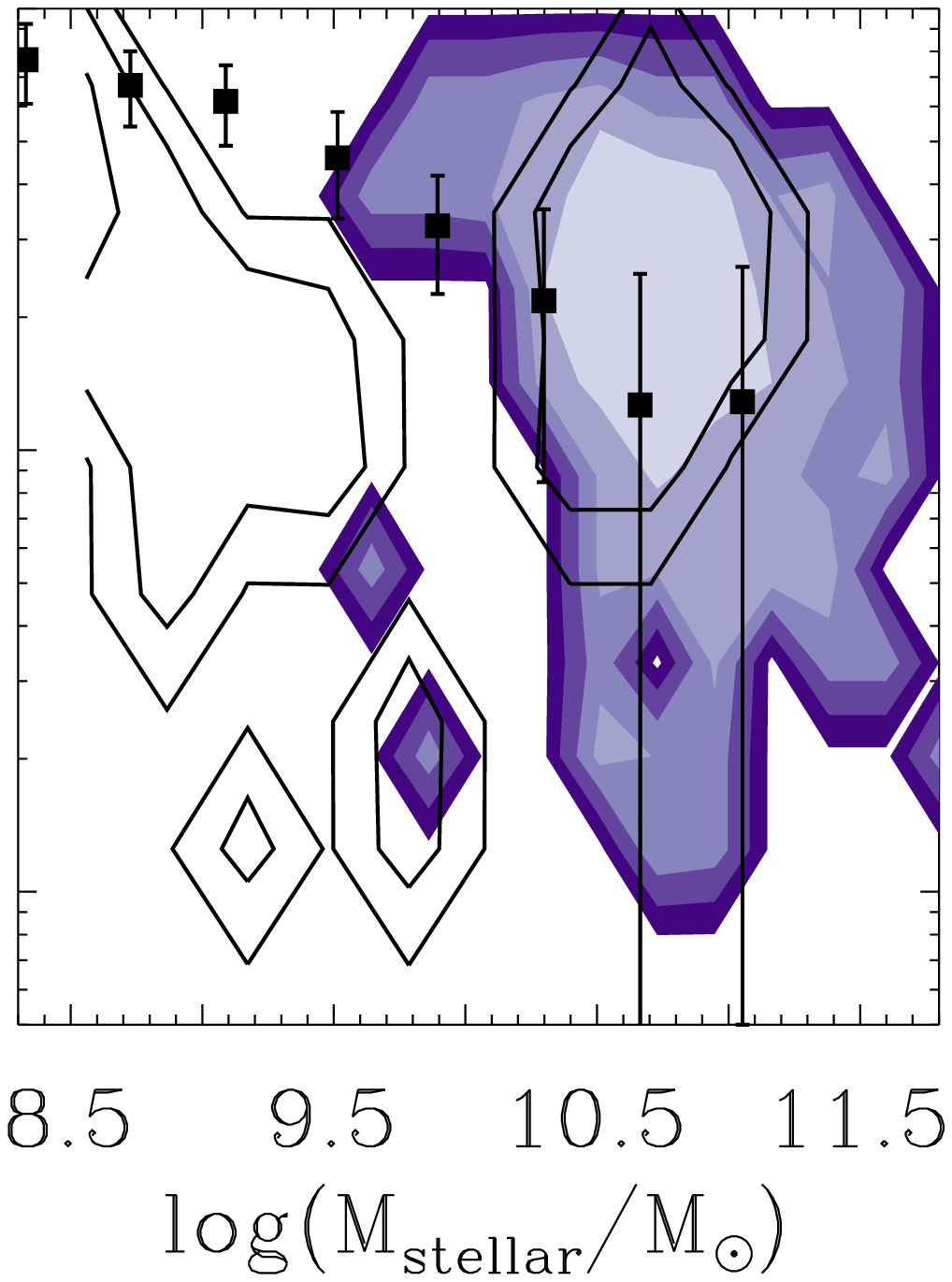,
  width=0.25\textwidth}\hspace{-1.5cm} 
\epsfig{file=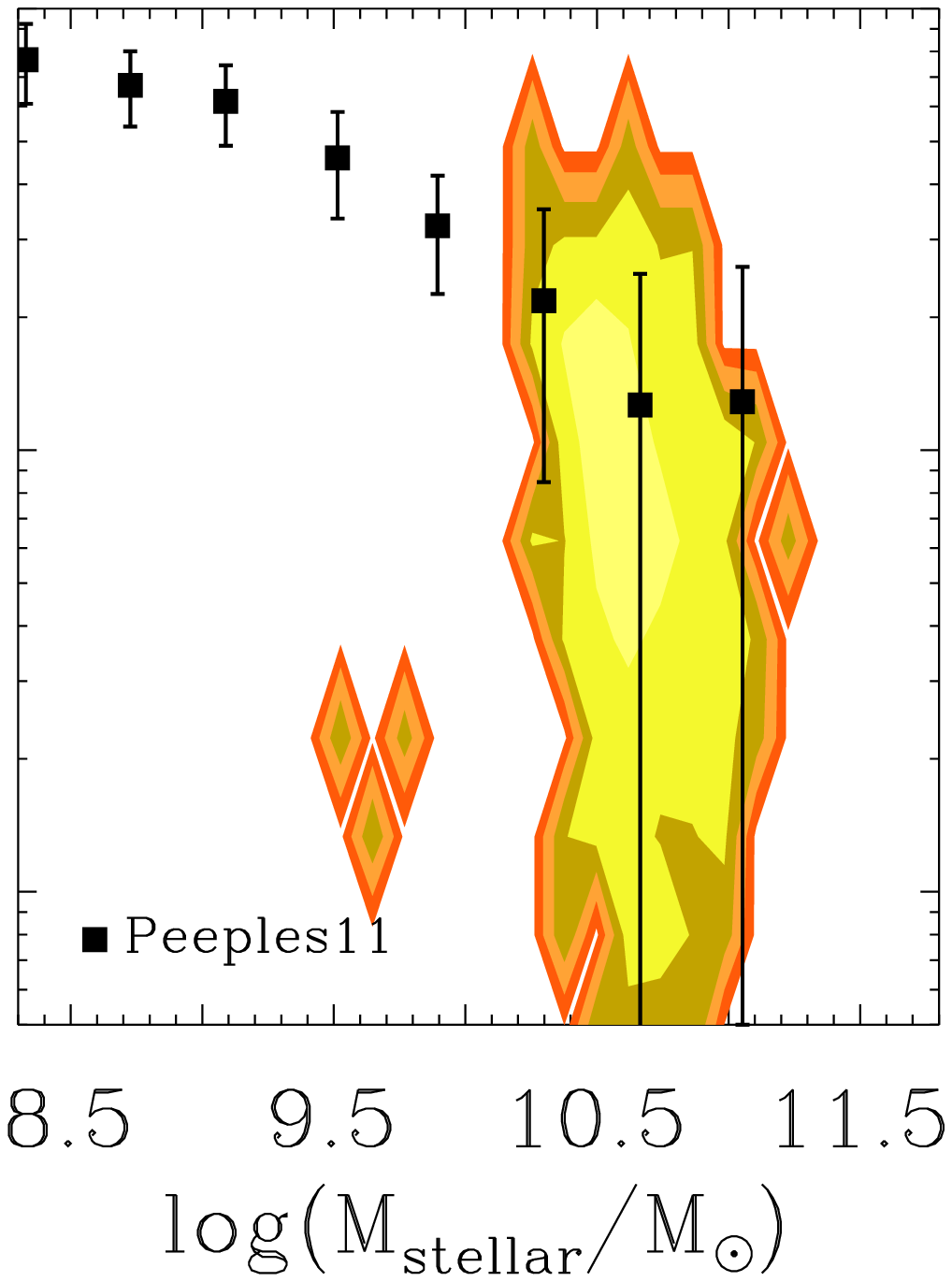, width=0.25\textwidth}
\caption{Location of the TNG100 and TNG50 galaxy populations at $z=0$ in the 
SFR-versus-stellar mass (top row) and cool-gas fraction-versus-stellar mass (bottom row) 
diagrams. The layout and colour coding are the same as in the top row of Fig.~\ref{bpts}. 
Also shown by the black solid and dotted lines in the top diagrams are the observed main 
sequences of local star-forming galaxies from SDSS and {\it GALEX} data, respectively 
\citep[][the grey shaded area indicates the 1$\sigma$ scatter about the solid line]{Elbaz07, Salim07}. 
In the bottom panels, the black squares with 1$\sigma$ error bars are observed 
mean gas fractions from various samples of $z\sim0$ galaxies compiled by 
\citet{Peeples11}.}\label{ms_fcold}       
\end{figure*}

The diagnostic diagrams discussed so far in Figs~\ref{bpts} and
\ref{mex} do not allow us to uniquely identify shock-dominated and
PAGB-dominated galaxies. We may appeal for this to other diagnostic 
diagrams, such as the EW(\ha)-versus-\niiha\ (WHAN) diagram
\citep{Stasinska08, CidFernandes10, 
CidFernandes11,Stasinska15}, expected to help distinguish PAGB-dominated 
galaxies among LI(N)ER, and the \oiiioii-versus-\oiha\ diagram 
\citep{Kewley19}, expected to help distinguish between shock-dominated 
and other types of galaxies. We show the analogues of Fig.~\ref{bpts} for these
two additional diagnostic diagrams in Figs~\ref{whan} and \ref{shockdiagram},
respecively. %\MH{Change the y-axis label in the figure to EW(\ha)!!!}

The black dashed lines in Fig.~\ref{whan} show the empirical selection 
criteria of \citet{CidFernandes11}, which divide the WHAN diagram into 
three rectangles. SF-dominated  galaxies are expected to reside in the
top-left rectangle, AGN-dominated galaxies in the top-right one, and
PAGB-dominated galaxies \citep[referred to as `retired' galaxies 
by][]{CidFernandes11} in the bottom one, corresponding to $\rm EW(\ha)<3\,$\AA.

Fig.~\ref{whan} shows that the TNG100 and TNG50 simulated
galaxies roughly overlap with SDSS galaxies in the WHAN diagram. 
As expected, the predicted \ha\ equivalent widths of PAGB-dominated 
galaxies lie below those of any other considered galaxy type, with
$\rm  EW(\ha) \la 1.25$\,\AA. This is less than half the value of 
3\,\AA\  proposed by \citet{CidFernandes11}, although this 
difference may arise in part from our definition of PAGB-dominated
galaxies. Had we considered as PAGB-dominated all galaxies
where PAGB stars contribute at least 40 (instead of 50) per cent of the total \hb\
luminosity, the upper limit (red line) in Fig.~\ref{whan} would lie 
around $\rm  EW(\ha)\sim3$--4\,\AA. 
Also, in our models, the predicted \ha\ luminosities of PAGB-dominated 
galaxies rely on PAGB evolutionary tracks from \citet[][incorporated in 
the latest version of the \citealt{Bruzual03} code]{MBertolami16},
with shorter lifetimes (but higher luminosities) than in previous 
widely used prescriptions. Adopting older PAGB models would 
enhance the predicted EW(\ha) of  PAGB-dominated galaxies 
(Section~\ref{cavphotoionization}). In any case, neither changing
the definition of PAGB-dominated galaxies nor using older PAGB models
would introduce any confusion between PAGB-dominated and SF- or
shock-dominated galaxies in Fig.~\ref{whan}. Only the separating 
EW(\ha) value would change.

In Fig.~\ref{shockdiagram}, the black dashed lines delimit the top-right 
region of the \oiiioii-versus-\oiha\ diagram expected to be populated by 
galaxies whose line emission is dominated by fast radiative shocks, 
according to \citet[][their figure~11]{Kewley19}. The different types of
TNG50 and TNG100 galaxy populations in the different 
diagrams of Fig.~\ref{shockdiagram} all fall on the footprint of SDSS galaxies,
indicating the ability of our models to account for these observations.
Furthermore, the simulated shock-dominated galaxies are the only ones to populate
the top-right region noted by \citet[][]{Kewley19}, confirming the robustness 
of this diagnostic diagram to separate shock-dominated galaxies
from other galaxy types. 

Overall, we consider the agreement between simulated and observed 
galaxy populations of different types at $z=0$ in the different diagnostic 
diagrams of Figs.~\ref{bpts}--\ref{shockdiagram} as remarkable, given that,
in our  approach, different galaxy types are assigned on the basis of physical 
parameters,  such as fractional contribution to the total H$\beta$ luminosity and 
BHAR/SFR  ratio, rather than of observables. Our results confirm previous
suggestions that the \oiiihb-versus-\niiha, EW(\ha)-versus-\niiha\ and
\oiiioii-versus-\oiha\  diagrams, a 5D line-ratio diagnostics,
provide powerful means of distinguishing between SF-, AGN-, PAGB-  
and shock-dominated galaxies in the nearby Universe.

% \noindent **Main conclusions:\\
% 1. consistent with SDSS galaxies $\rightarrow$ reliable basis for further high-z 
% predictions\\
% 2. Post-AGB-dominated galaxies best selected via the
% WHAN diagram. \\
% 3. Shock-dominated galaxies best identified via [OIII]/[OII]-[OI]/Ha
% diagram.\\ 

%*****************************************************************************************************
%*****************************************************************************************************
\subsection{SFR and cool-gas fraction}\label{physprop} 
%*****************************************************************************************************
%*****************************************************************************************************

After having divided the simulated galaxy populations into different 
types based on their dominant ionizing sources, and explored their 
location in various diagnostic diagrams, we continue in this Section 
to study some of the predicted properties of the different 
galaxy types at $z=0$, such as SFR and baryonic
fraction in the `cool-gas' phase (i.e.,  gas with $T<10^5\,$K).
We define the cool-gas fraction as
$M_{\mathrm{cool}}/(M_{\mathrm{cool}}+M_{\mathrm{stellar}})$, 
where $M_\mathrm{cool}$ and $M_{\mathrm{stellar}}$ are the cool-gas
and stellar masses.

In the top row of Fig.~\ref{ms_fcold}, we show the distributions of SFR versus 
$M_{\mathrm{stellar}}$ for the different types of TNG100 and TNG50 
galaxy populations, using the same layout and colour coding as in Fig.~\ref{bpts}. 
The observed `main sequence' (MS) of local star-forming galaxies is 
depicted by the black solid and dotted lines for SDSS and
{\it GALEX} data, respectively \citep{Elbaz07, Salim07}. SF-dominated and
composite galaxies in our simulations largely follow this observed MS (left two
panels), while AGN-dominated galaxies have SFRs typically 1--1.5~dex lower 
at fixed $M_{\mathrm{stellar}}$ (central panel). The shock- and PAGB-dominated
galaxies (right two panels) are typically more massive, with $M_{\mathrm{stellar}}\ga3 
\times 10^{10}\Msun$, and even less star-forming that the other galaxy types. 
Shock-dominated galaxies span a wide range of SFRs, between 1 and 10$^{-4}$\,\Msun\,yr$^{-1}$,
filling the so-called `green valley' of the SFR-versus-stellar mass plane \citep[e.g.,][]{Salim14}.
PAGB-dominated galaxies are fully quiescent (retired) massive galaxies, with SFRs below
$0.01\,\Msun\,$yr$^{-1}$. 

The sequence of decreasing SFRs from AGN-, to shock-, 
to PAGB-dominated galaxies at fixed stellar mass in Fig.~\ref{ms_fcold} 
reflects the process of SF quenching by AGN feedback in
IllustrisTNG galaxies \citep{Weinberger17, Nelson18, Terrazas20}: in
an AGN-dominated galaxy, 
feedback from the heavily accreting central black hole reduces star
formation. This also regulates BH growth, via a drop in gas-accretion rate. 
The kinetic-feedback mode associated with the low BH-accretion phase 
(Section~\ref{TNG}) causes shocks to emerge in the ISM, heating and
expelling gas from the galaxy. This further lowers the SFR and produces
shock-dominated line emission. After this phase, the galaxy becomes
quiescent/retired, with very low levels of on-going star formation and BH
accretion, so that PAGB stars can dominate line emission.

In the bottom row of Fig.~\ref{ms_fcold}, we show the distributions of
cool-gas fraction versus stellar mass for the different TNG100 
and TNG50 galaxy populations. Also plotted for reference are observed 
mean  gas fractions from various samples of $z\sim0$ galaxies compiled by 
\citet[][black squares]{Peeples11}. 
%\SC{(Note that figure~2 of Peeples+11
%has Mgas/Mstars on the y axis; I assumed you made the conversion to plot 
%Mgas/(Mstars+Mstars) on Fig.~\ref{ms_fcold}?)}\MH{Yes, I did it.} 
Consistently with our findings
above regarding star formation, we find that SF-dominated, 
composite, AGN-dominated, shock-dominated and PAGB-dominated galaxies 
represent a sequence of decreasing cool-gas fractions. We note that, at
the end of the sequence, over
95~per cent of the stars in PAGB-dominated galaxies are older than 1~Gyr,
with mean stellar ages greater than 6~Gyr and mean stellar metallicities between 
0.6 and 1.2 times solar, i.e., typically more metal-rich than other galaxy types.

%*****************************************************************************************************
%*****************************************************************************************************
\subsection{Evolution of galaxy populations in the [OIII]/H$\beta$-versus-[NII]/H$\alpha$
diagram}\label{opticaldiagrams_evol} 
%*****************************************************************************************************
%*****************************************************************************************************

Based on the overall good agreement between predicted and observed distributions
of different types of galaxy populations at $z=0$ (Section~\ref{opticaldiagrams_z0}), 
we now investigate the evolution of TNG100 and TNG50 galaxy 
populations in the \oiiihb-versus-\niiha\ diagnostic diagram at higher redshift. In 
particular, we are interested in the extent to which the larger
statistics together with different physical models offered by the
IllustrisTNG simulations alter the conclusions drawn by \citet{Hirschmann17},
using a small set of 20 cosmological zoom-in simulations of massive galaxies and
their progenitors, regarding the apparent rise in \oiiihb\ at fixed \niiha\ from
$z=0$ to $z=2$ in observational samples.

\begin{figure*}
%\center
%\textbf{SF galaxies}\par\medskip\hspace{-1.5cm}
\epsfig{file=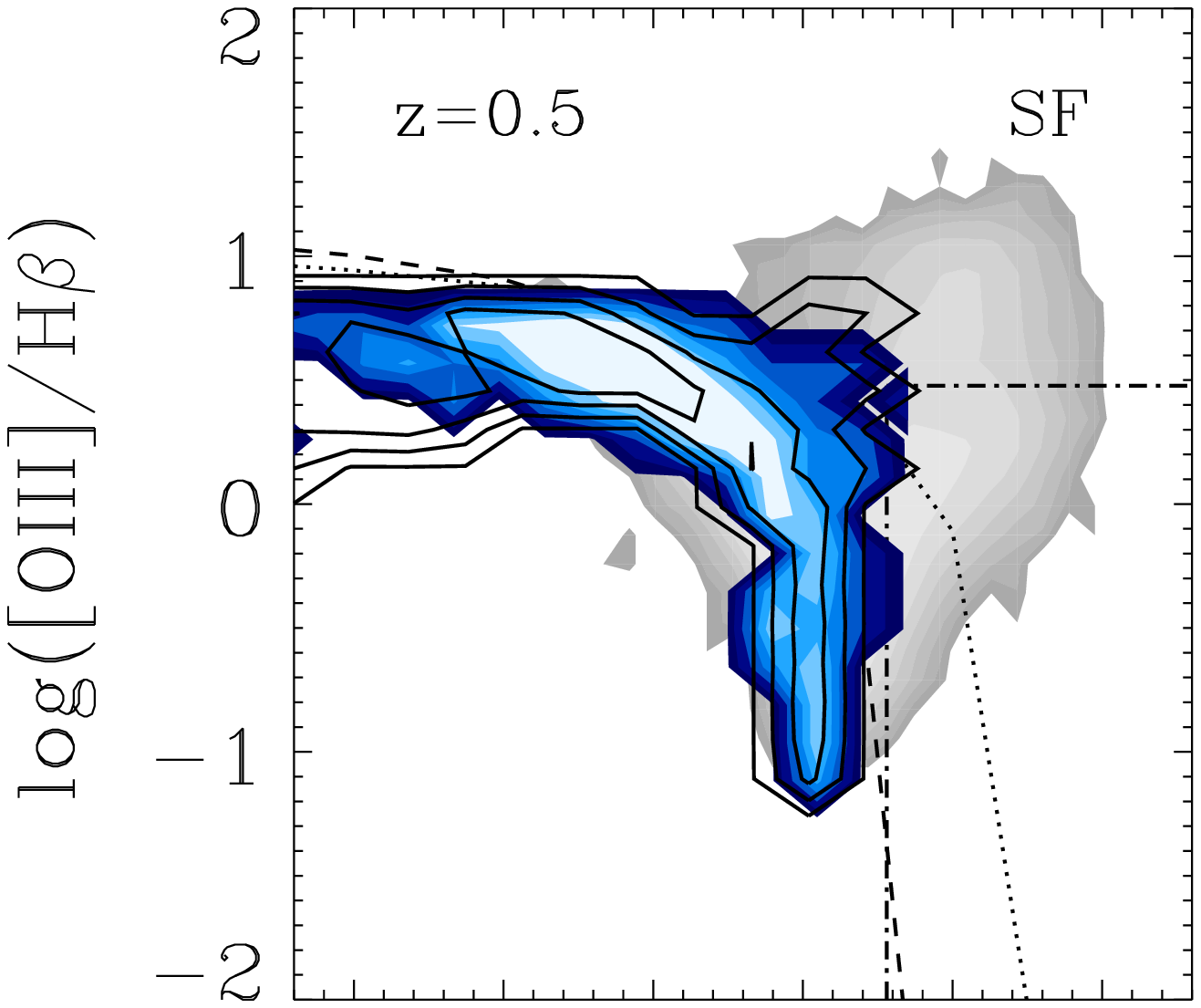,
  width=0.25\textwidth}\hspace{-1.5cm}
\epsfig{file=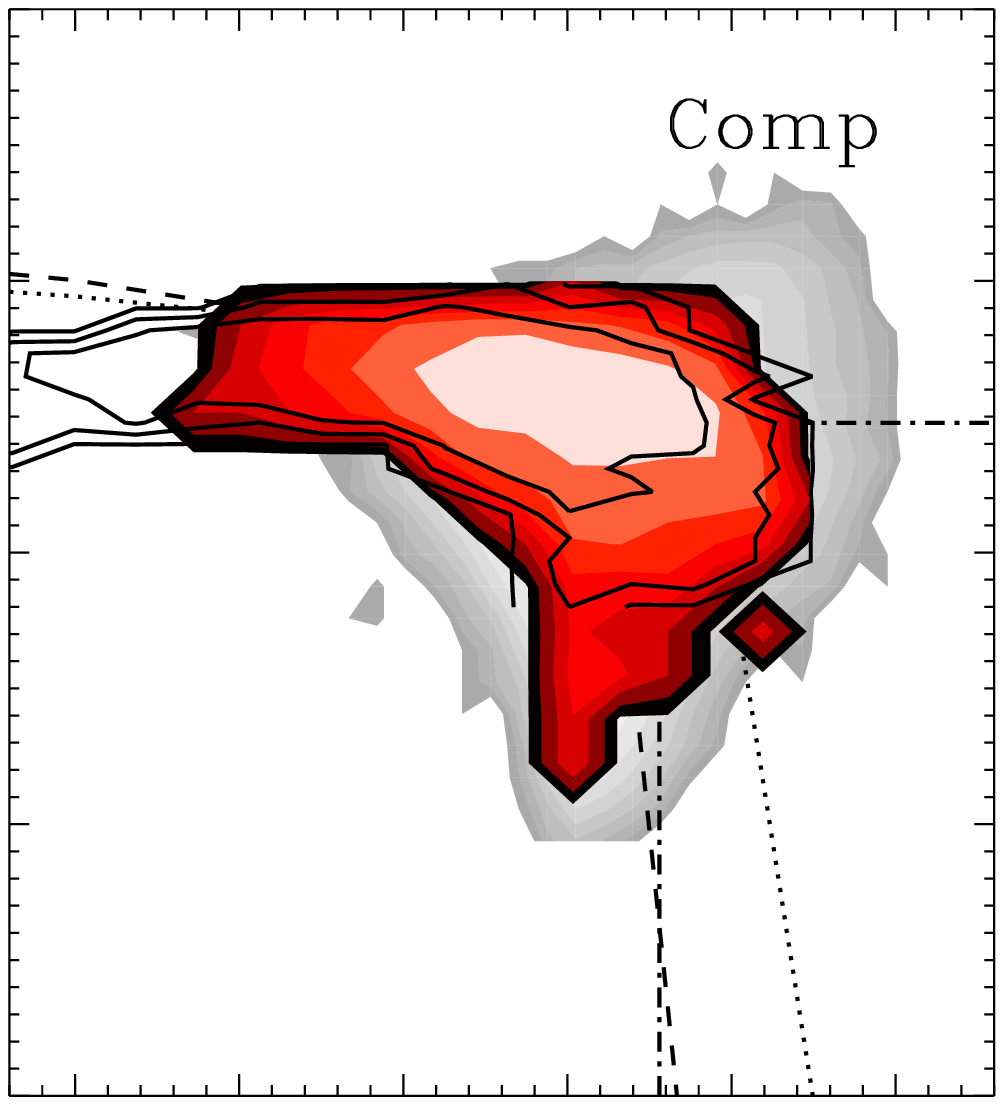,
  width=0.25\textwidth}\hspace{-1.5cm}
\epsfig{file=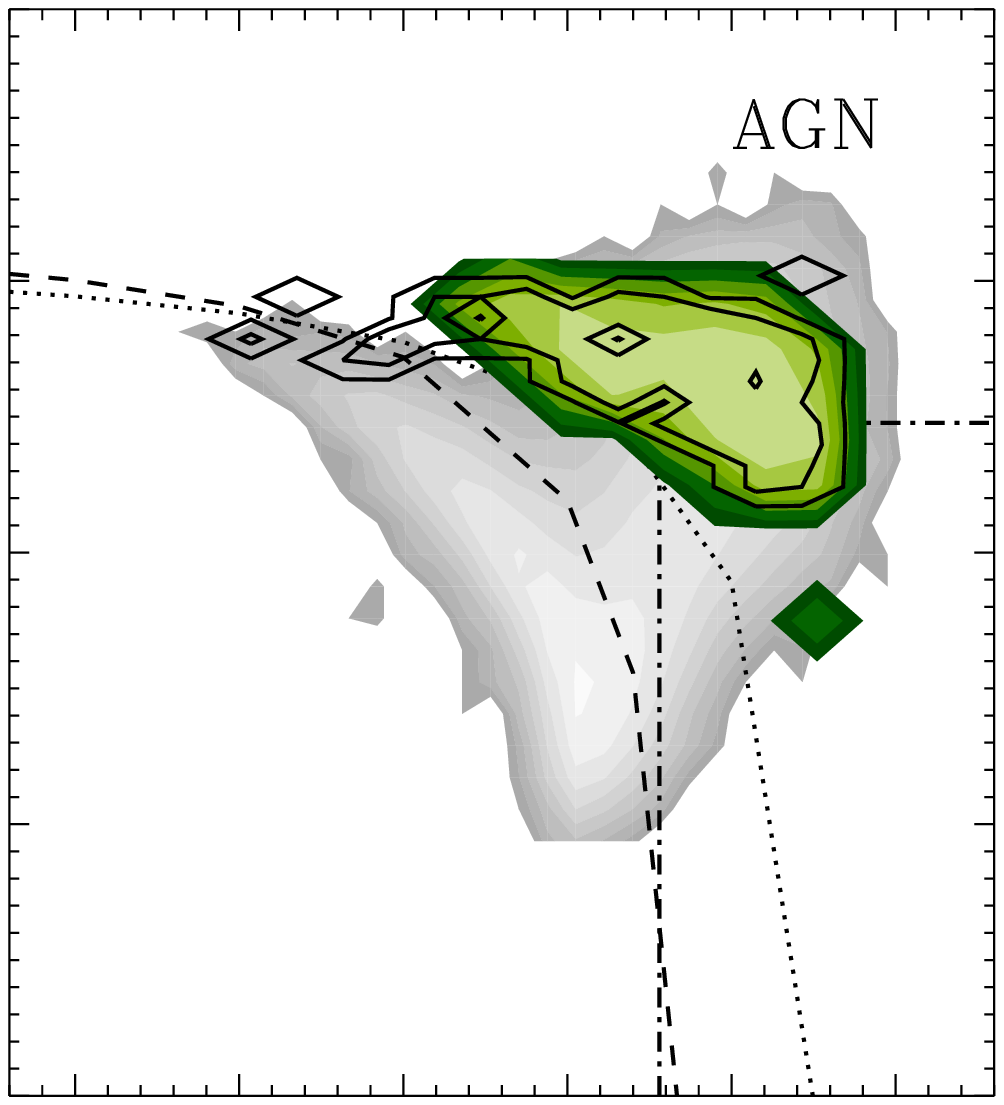,
  width=0.25\textwidth}\hspace{-1.5cm} 
\epsfig{file=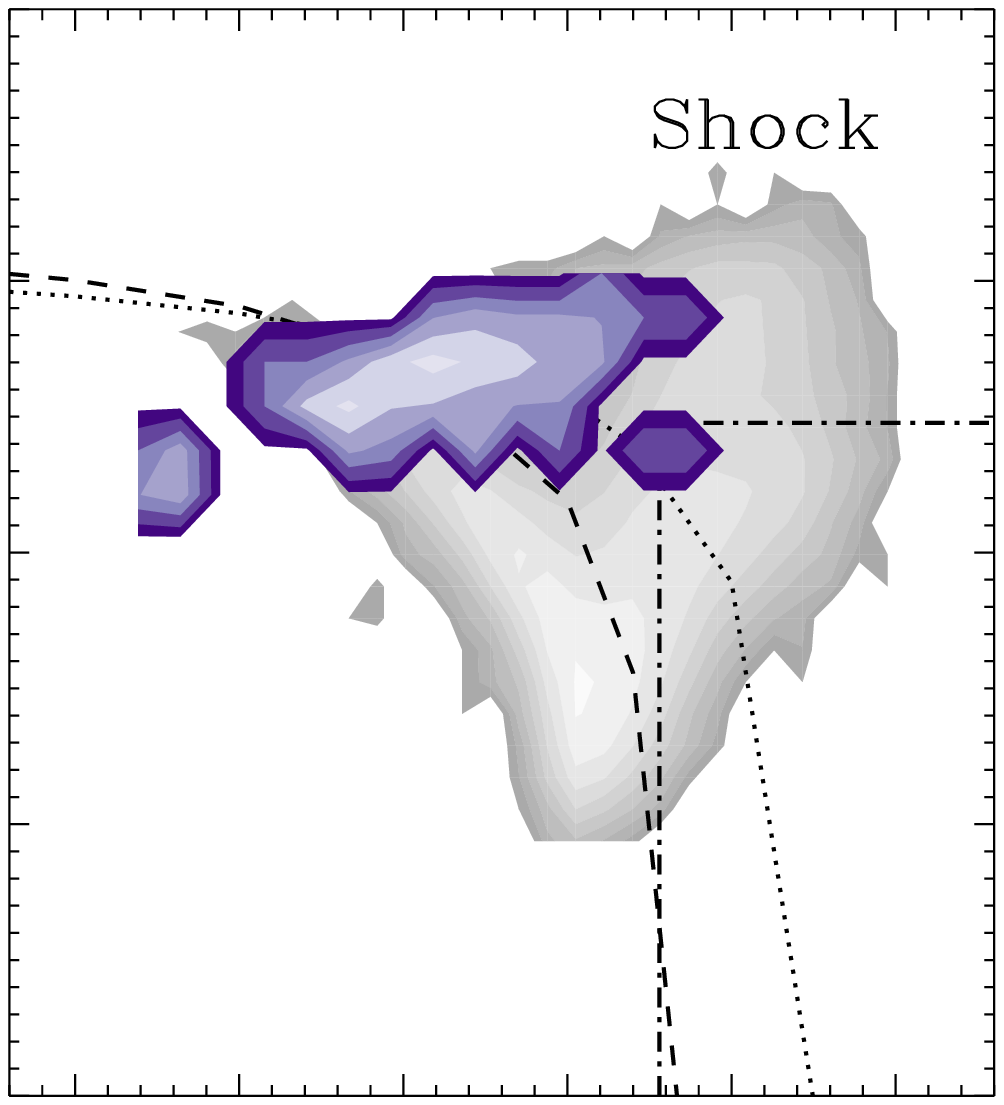,
  width=0.25\textwidth}\hspace{-1.5cm} 
\epsfig{file=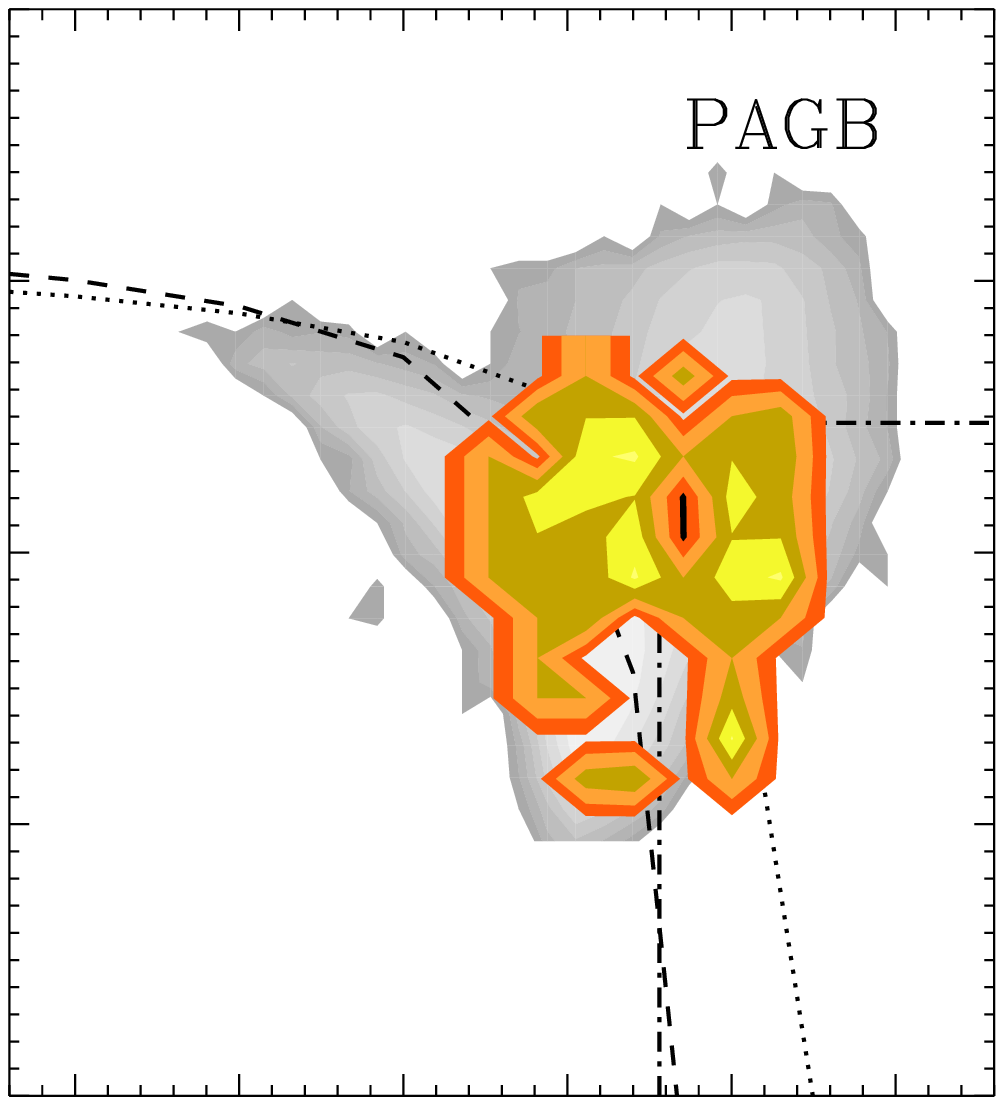,
  width=0.25\textwidth}\vspace{-1.1cm}
\epsfig{file=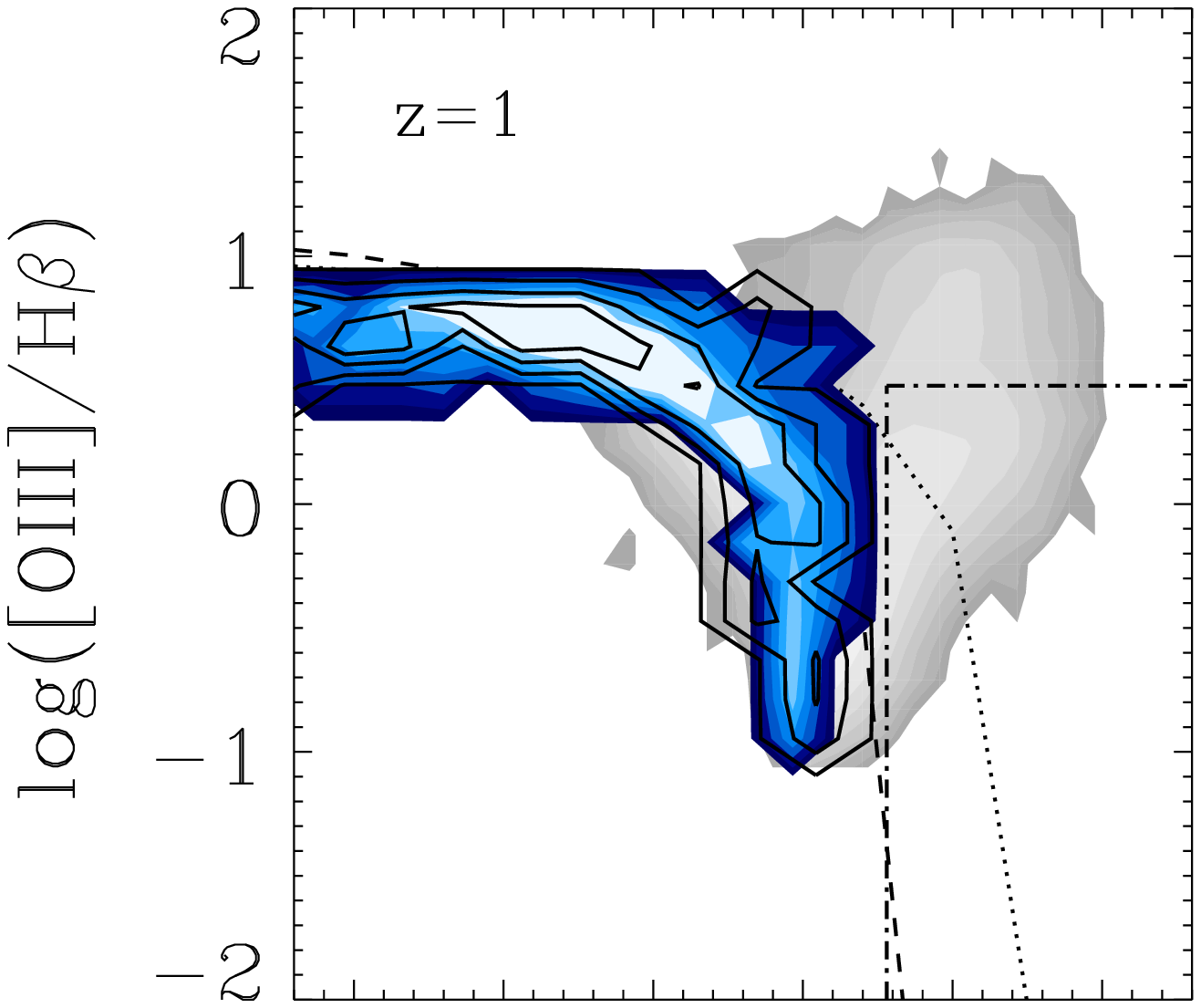,
  width=0.25\textwidth}\hspace{-1.5cm}
\epsfig{file=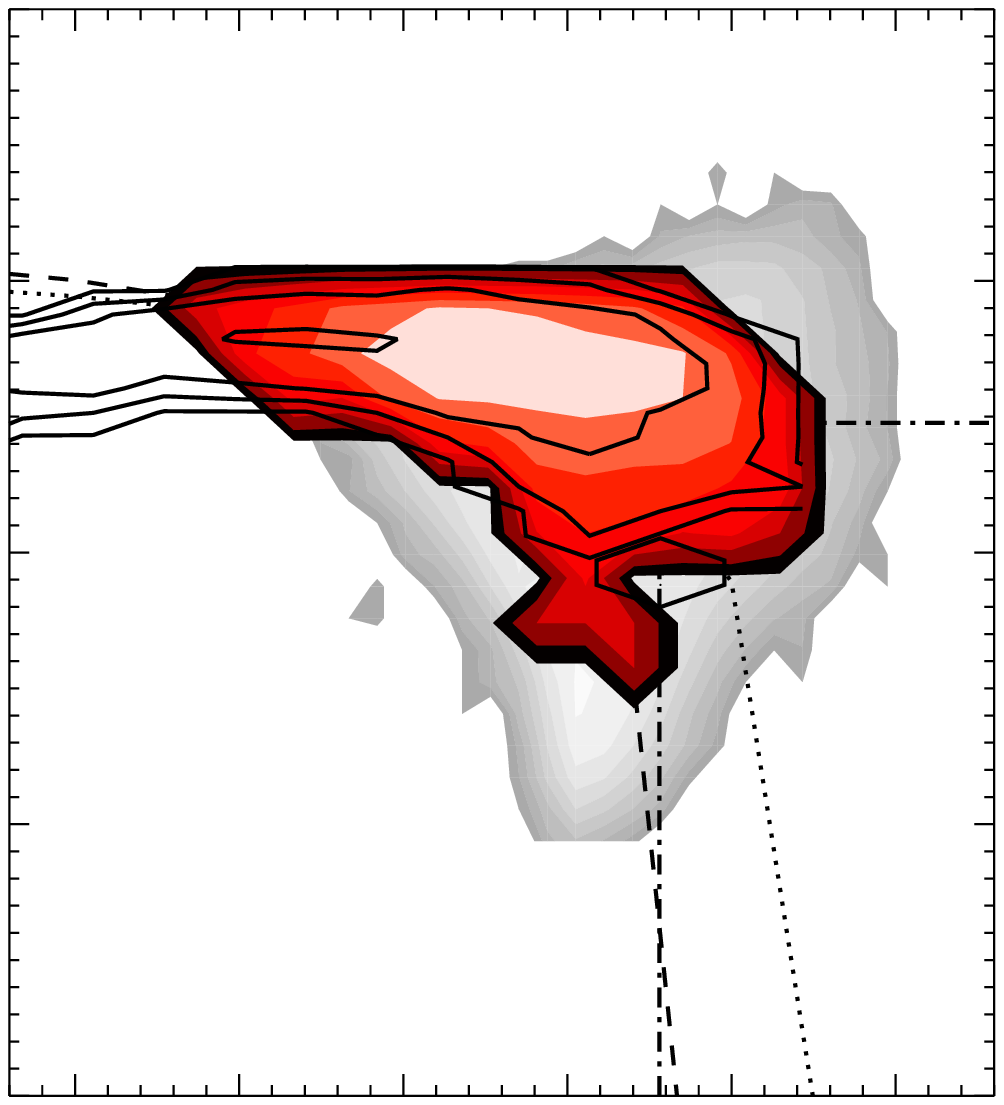,
  width=0.25\textwidth}\hspace{-1.5cm}
\epsfig{file=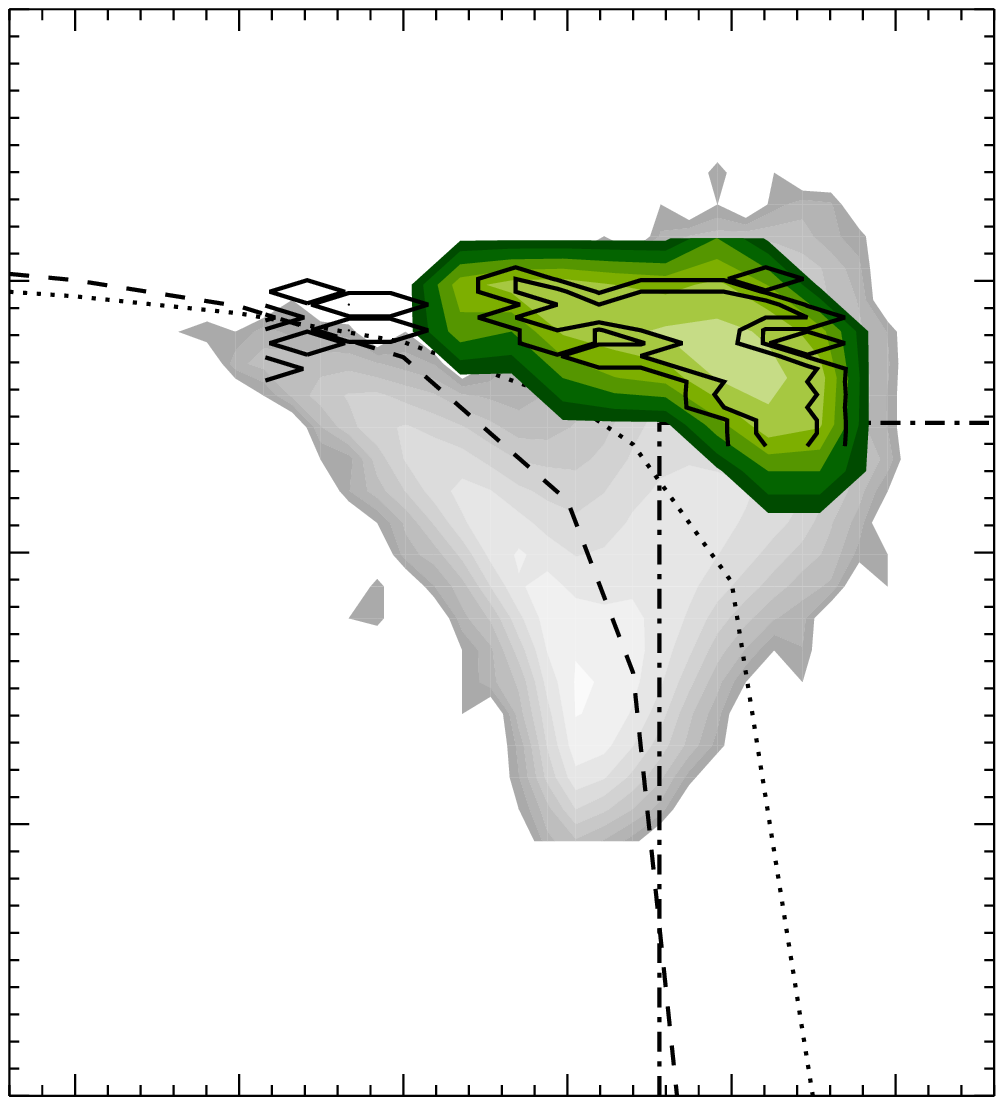,
  width=0.25\textwidth}\hspace{-1.5cm} 
\epsfig{file=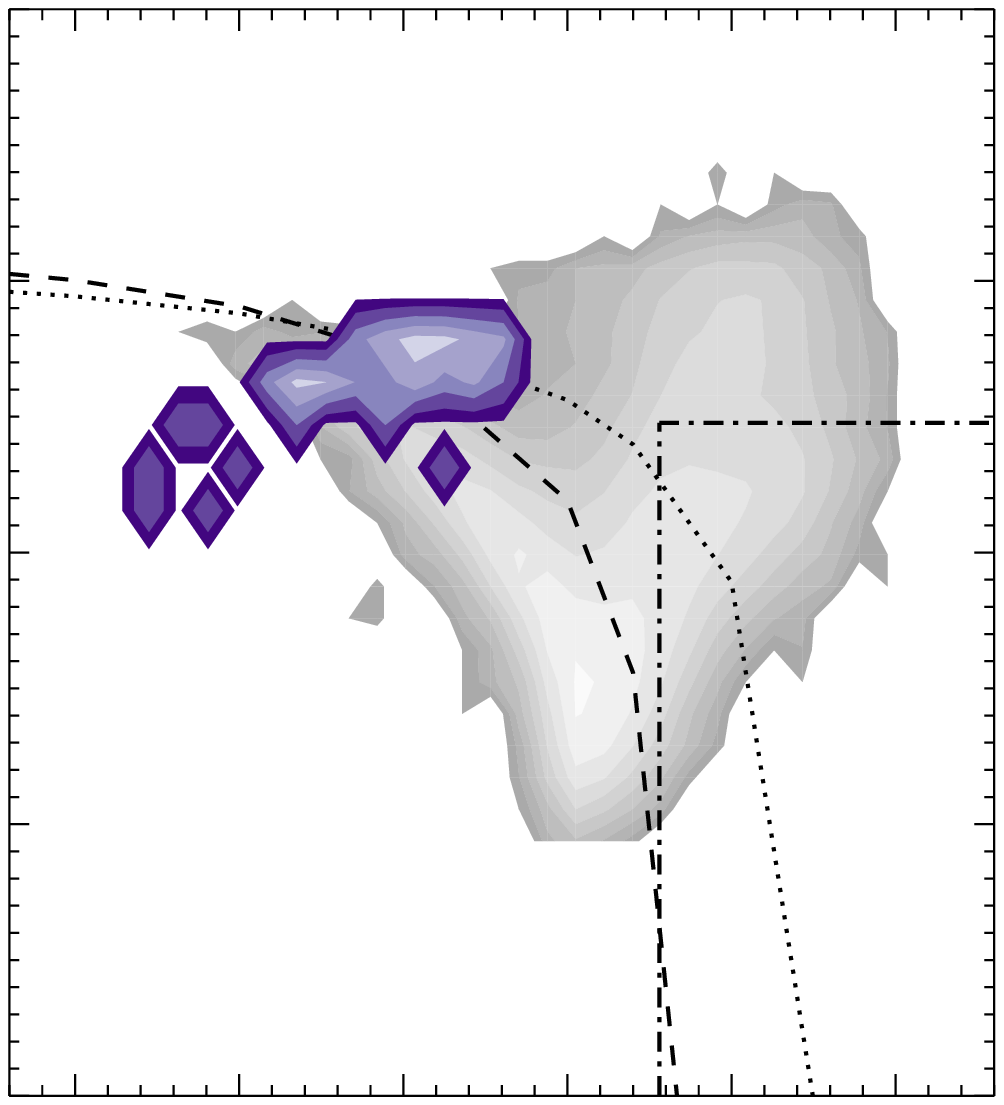,
  width=0.25\textwidth}\hspace{-1.5cm} 
\epsfig{file=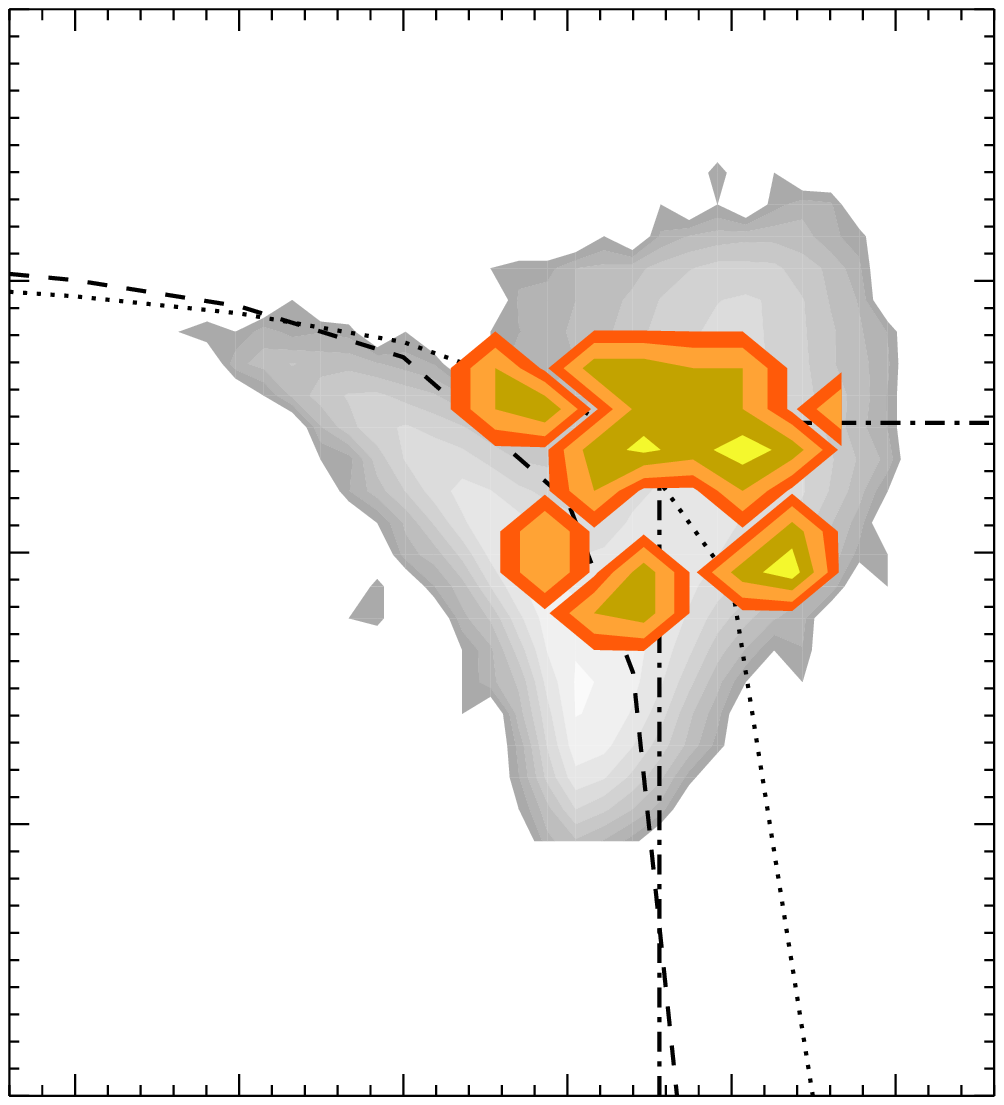,
  width=0.25\textwidth}\vspace{-1.1cm}
\epsfig{file=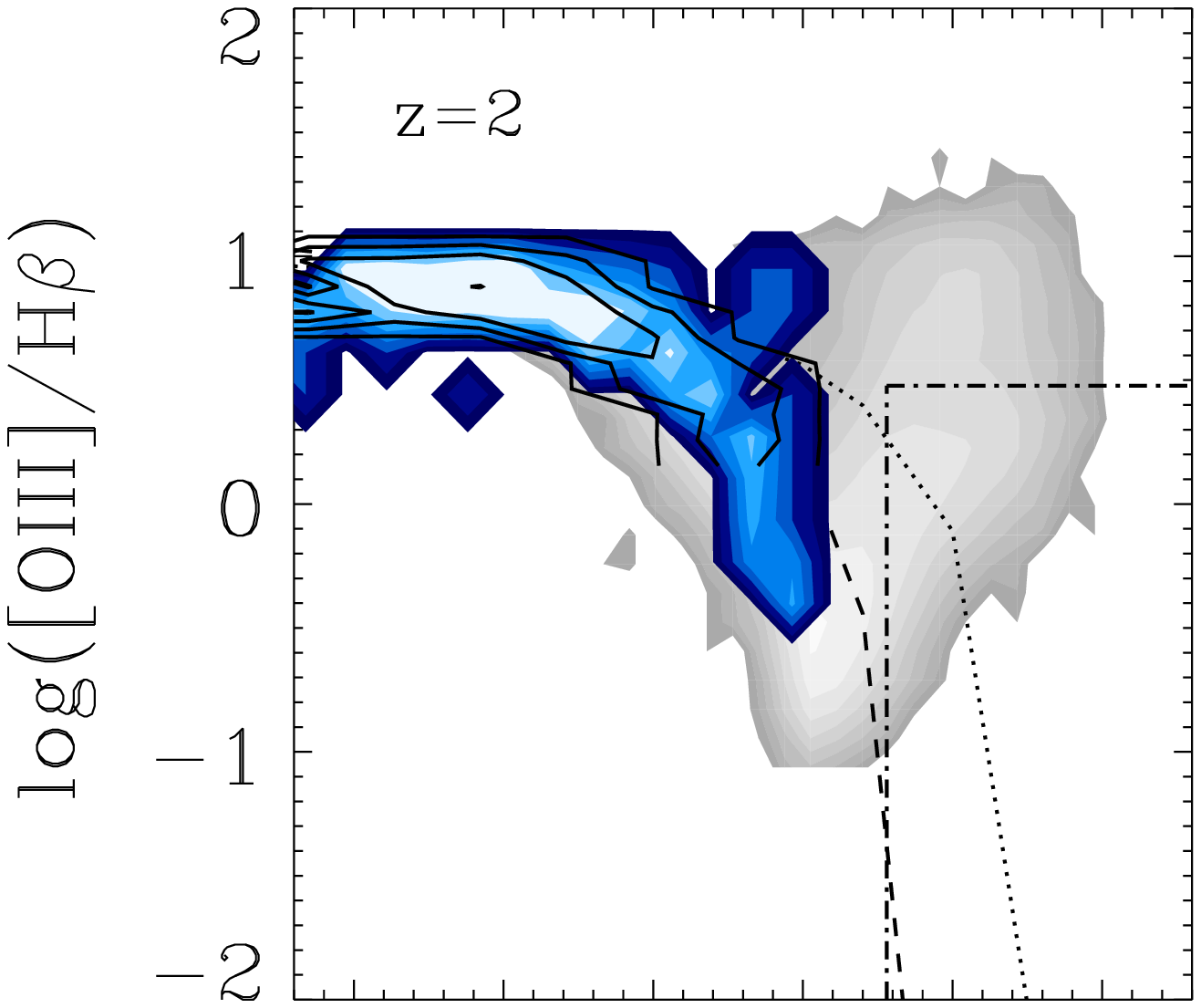,
  width=0.25\textwidth}\hspace{-1.5cm}
\epsfig{file=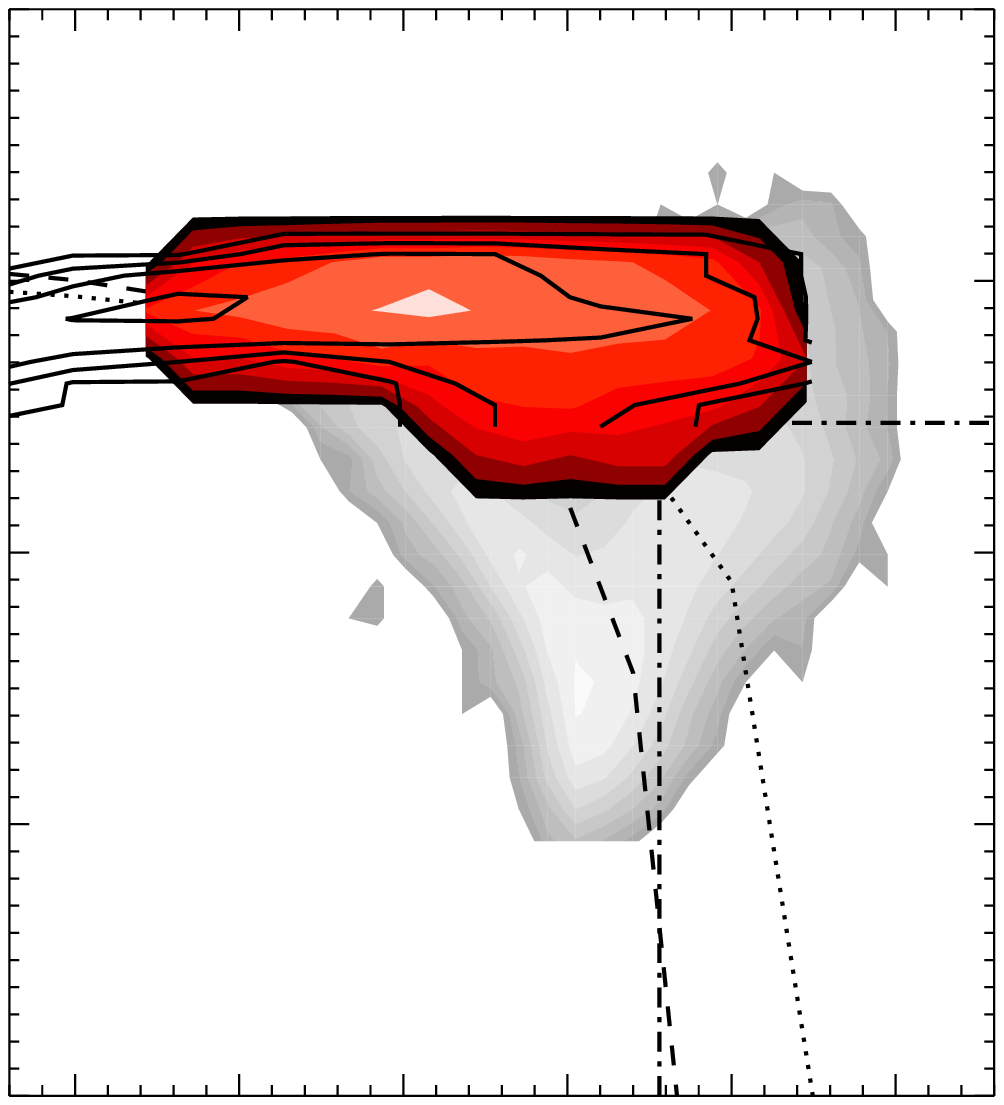,
  width=0.25\textwidth}\hspace{-1.5cm}
\epsfig{file=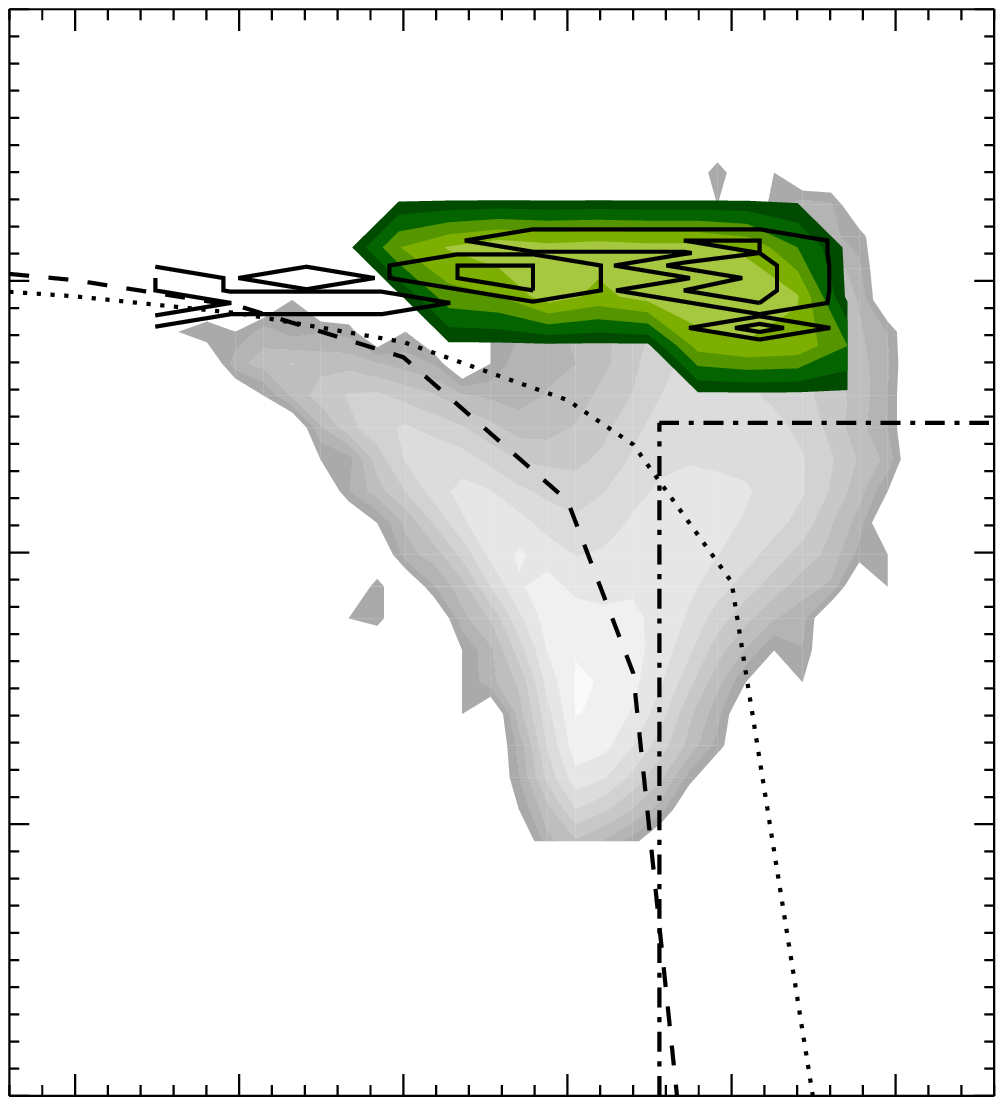,
  width=0.25\textwidth}\hspace{-1.5cm} 
\epsfig{file=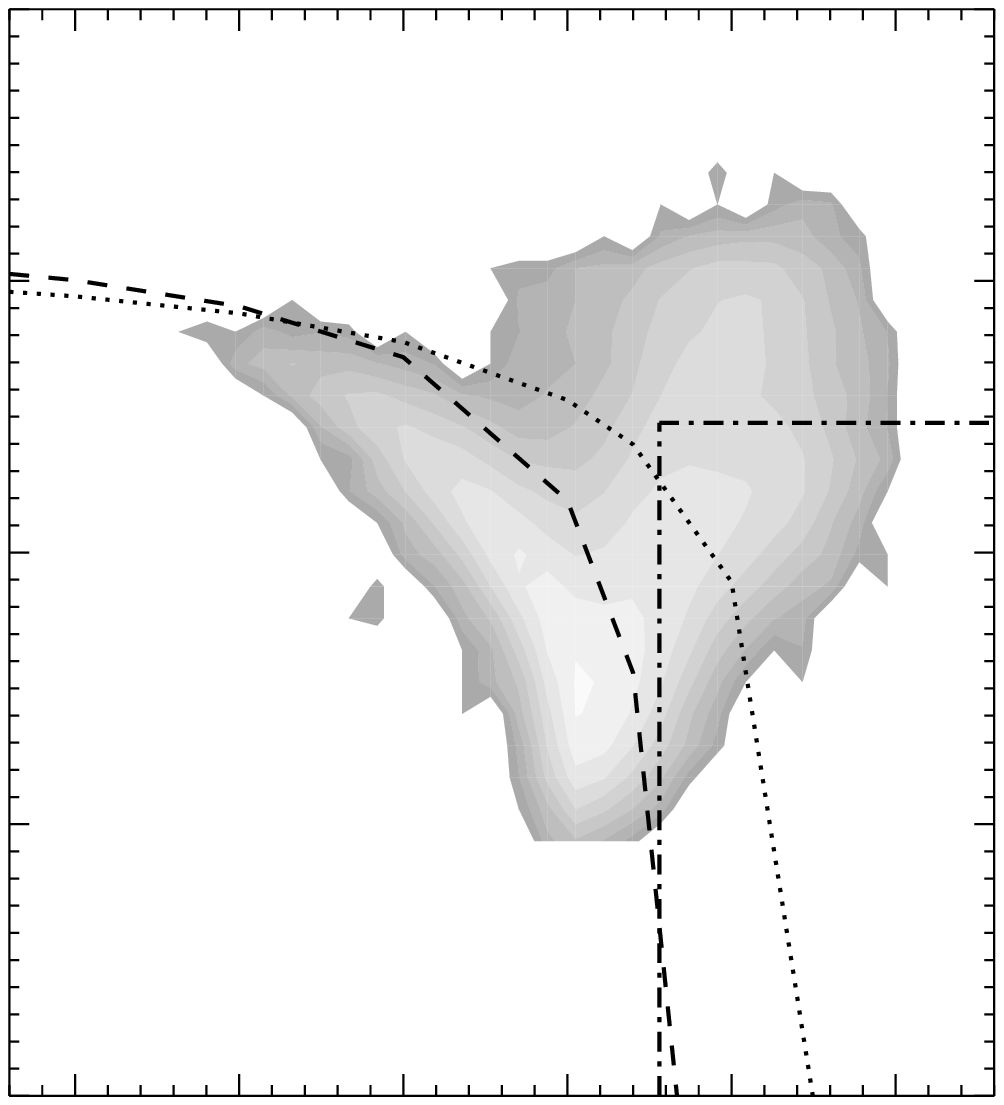,
  width=0.25\textwidth}\hspace{-1.5cm} 
\epsfig{file=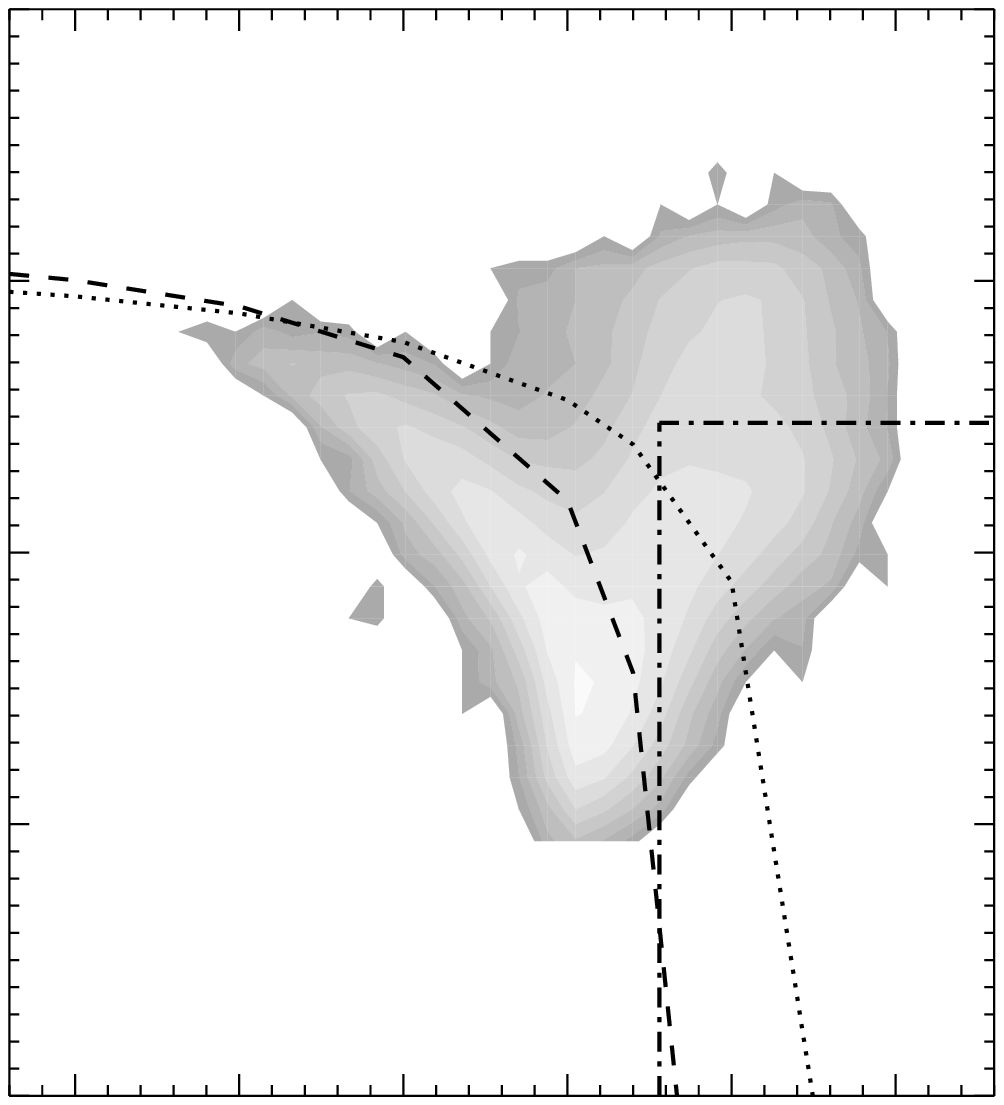,
  width=0.25\textwidth}\vspace{-1.1cm}
\epsfig{file=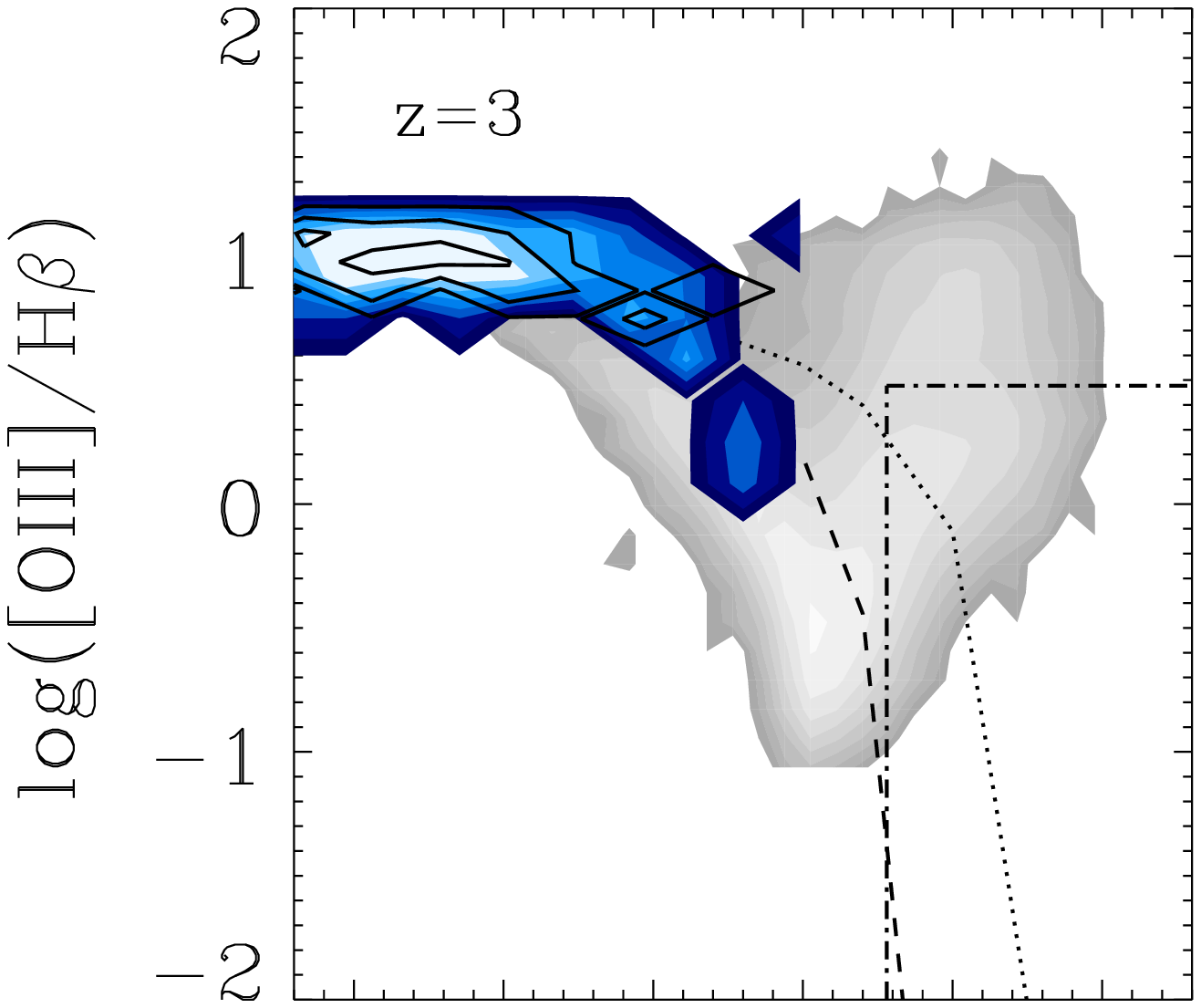,
  width=0.25\textwidth}\hspace{-1.5cm}
\epsfig{file=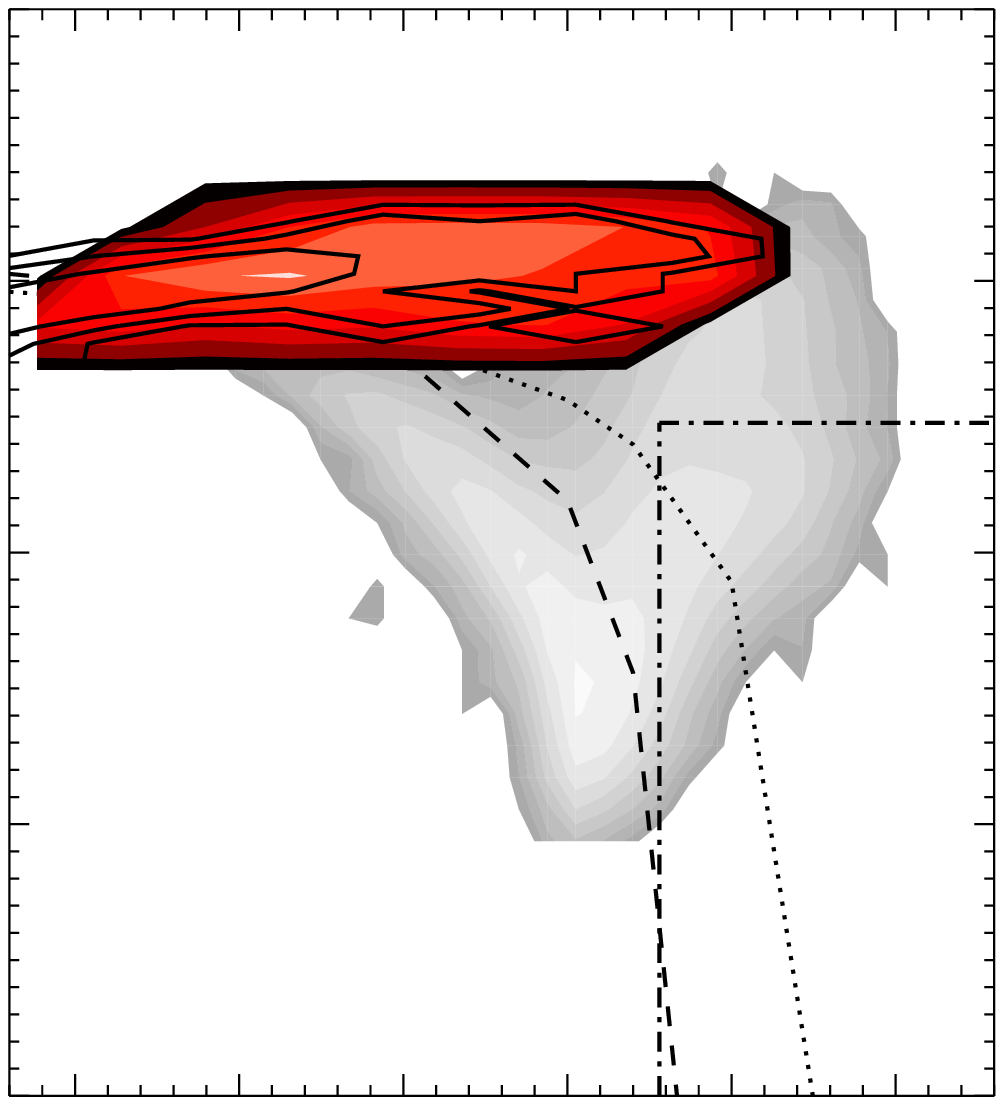,
  width=0.25\textwidth}\hspace{-1.5cm}
\epsfig{file=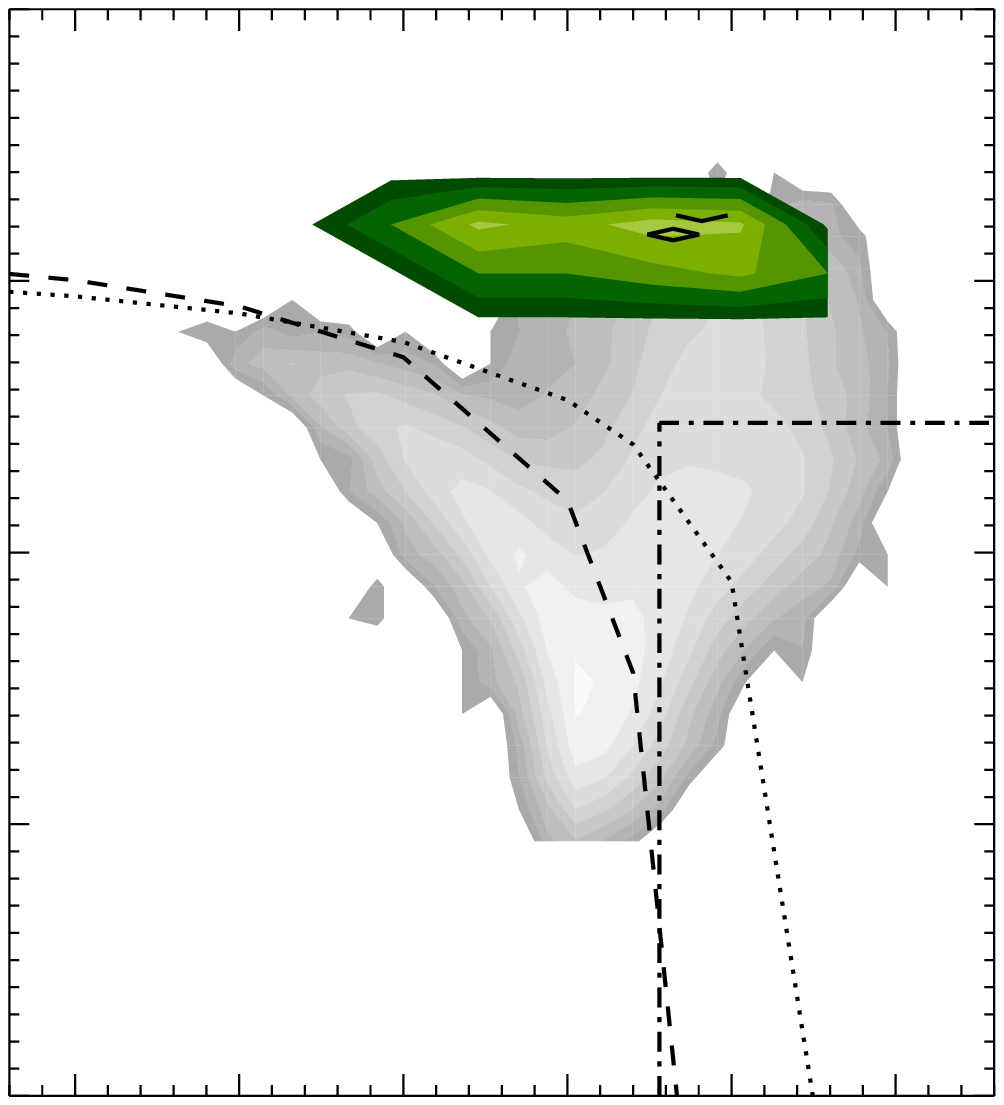,
  width=0.25\textwidth}\hspace{-1.5cm} 
\epsfig{file=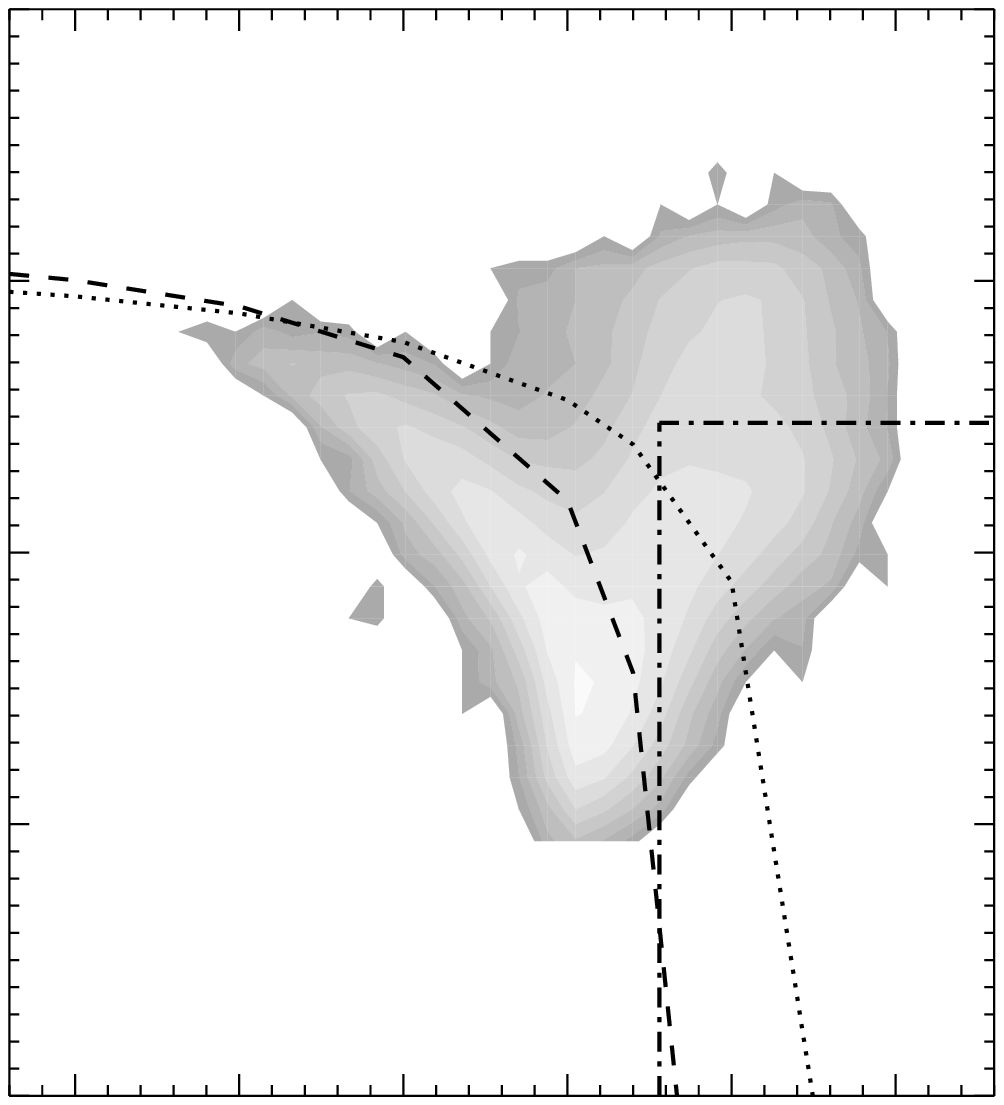,
  width=0.25\textwidth}\hspace{-1.5cm} 
\epsfig{file=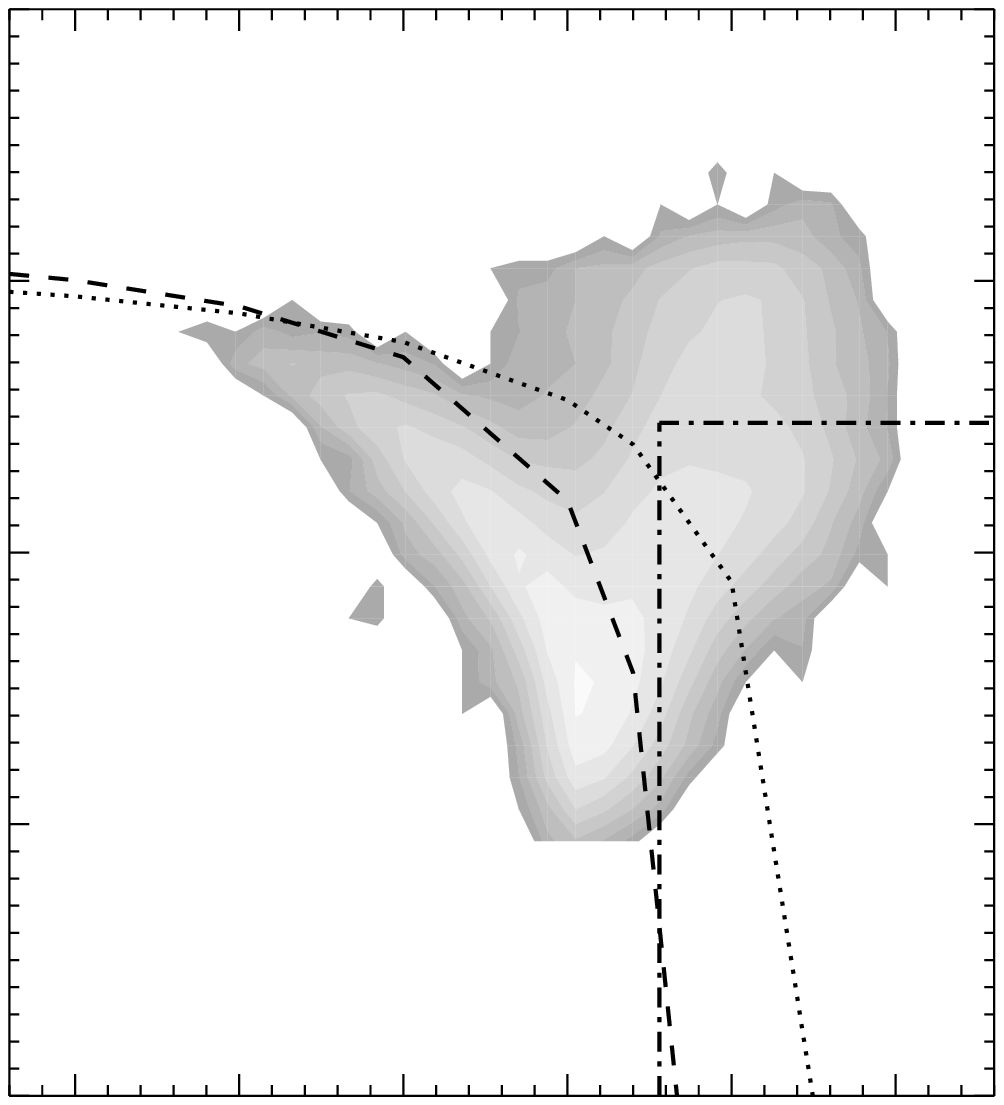,
  width=0.25\textwidth}\vspace{-1.1cm}
\epsfig{file=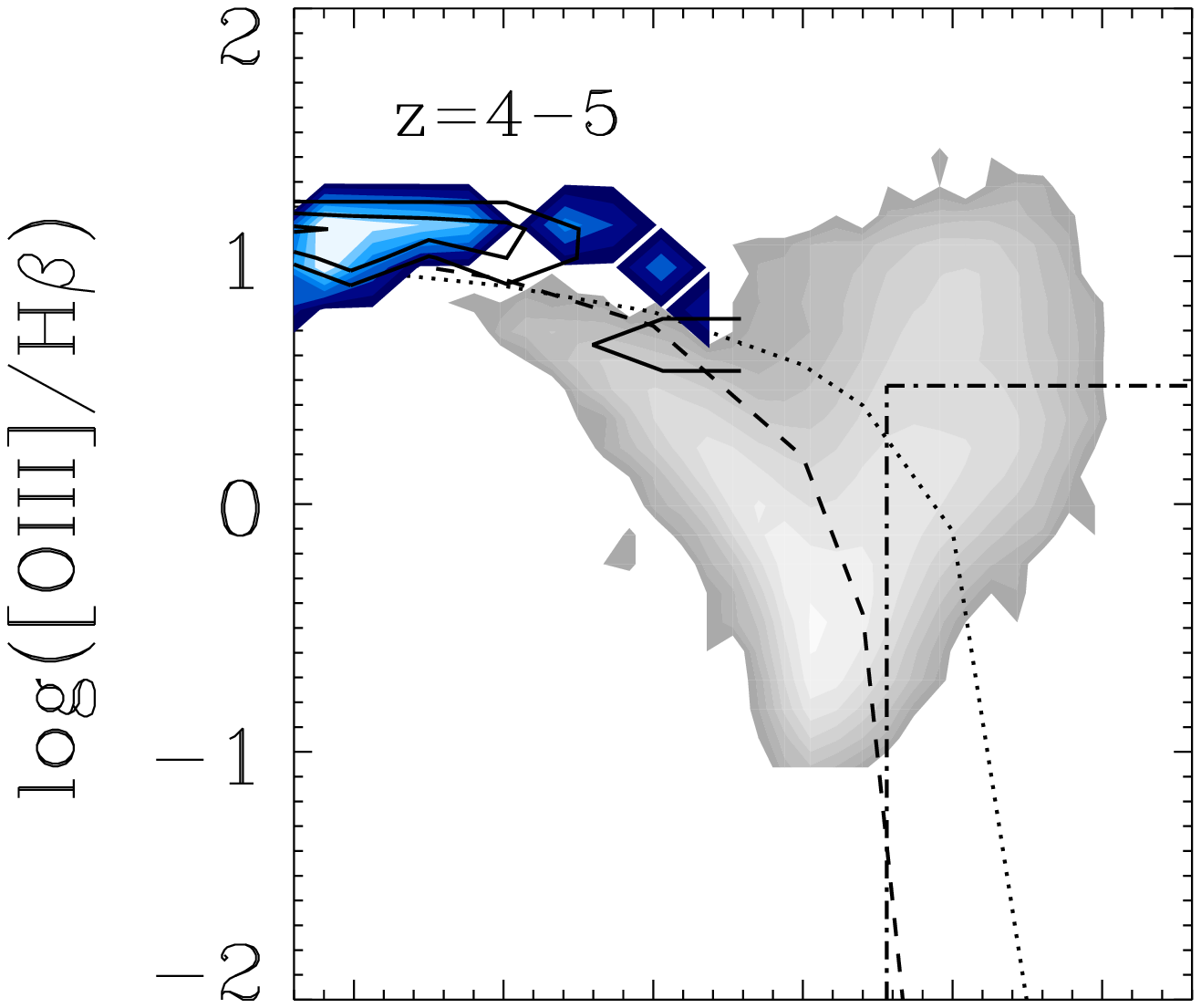,
  width=0.25\textwidth}\hspace{-1.5cm}
\epsfig{file=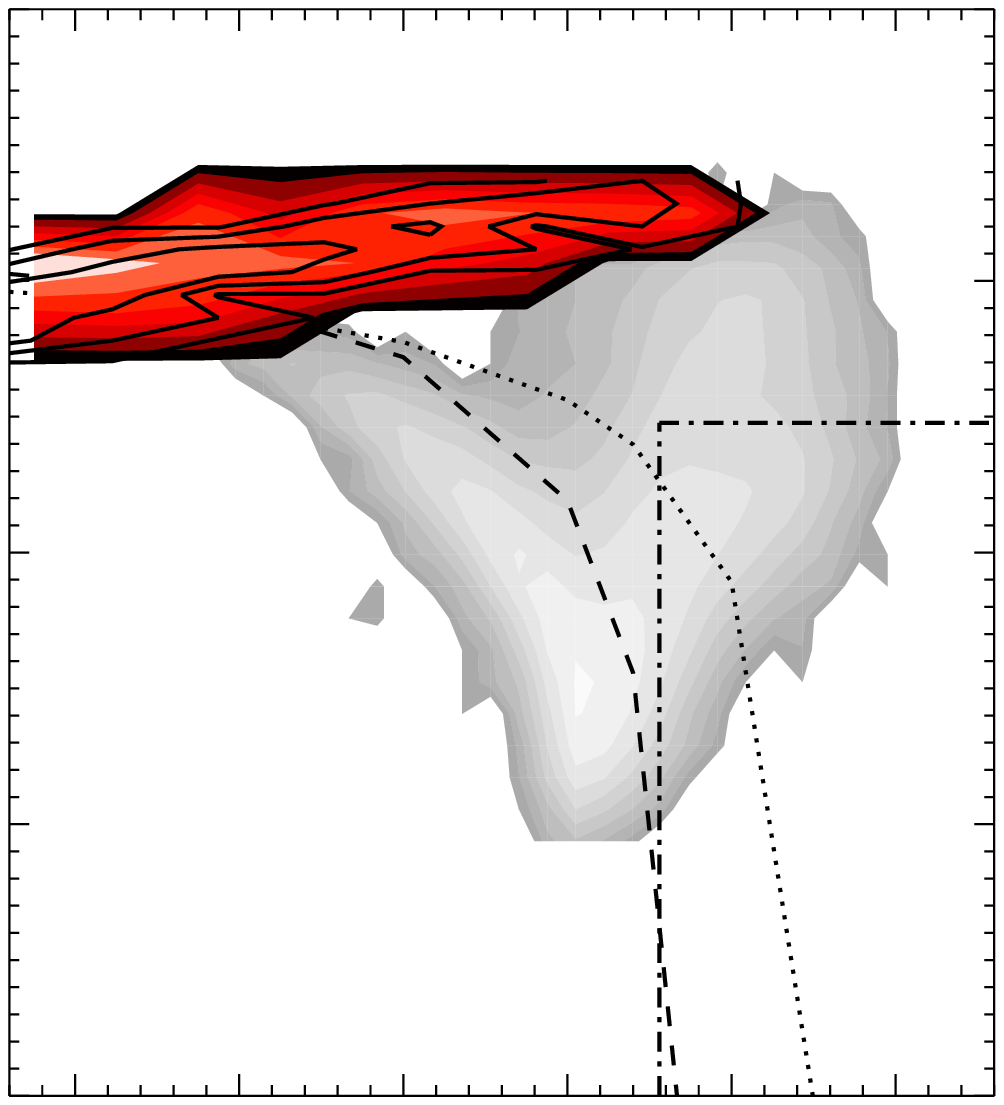,
  width=0.25\textwidth}\hspace{-1.5cm}
\epsfig{file=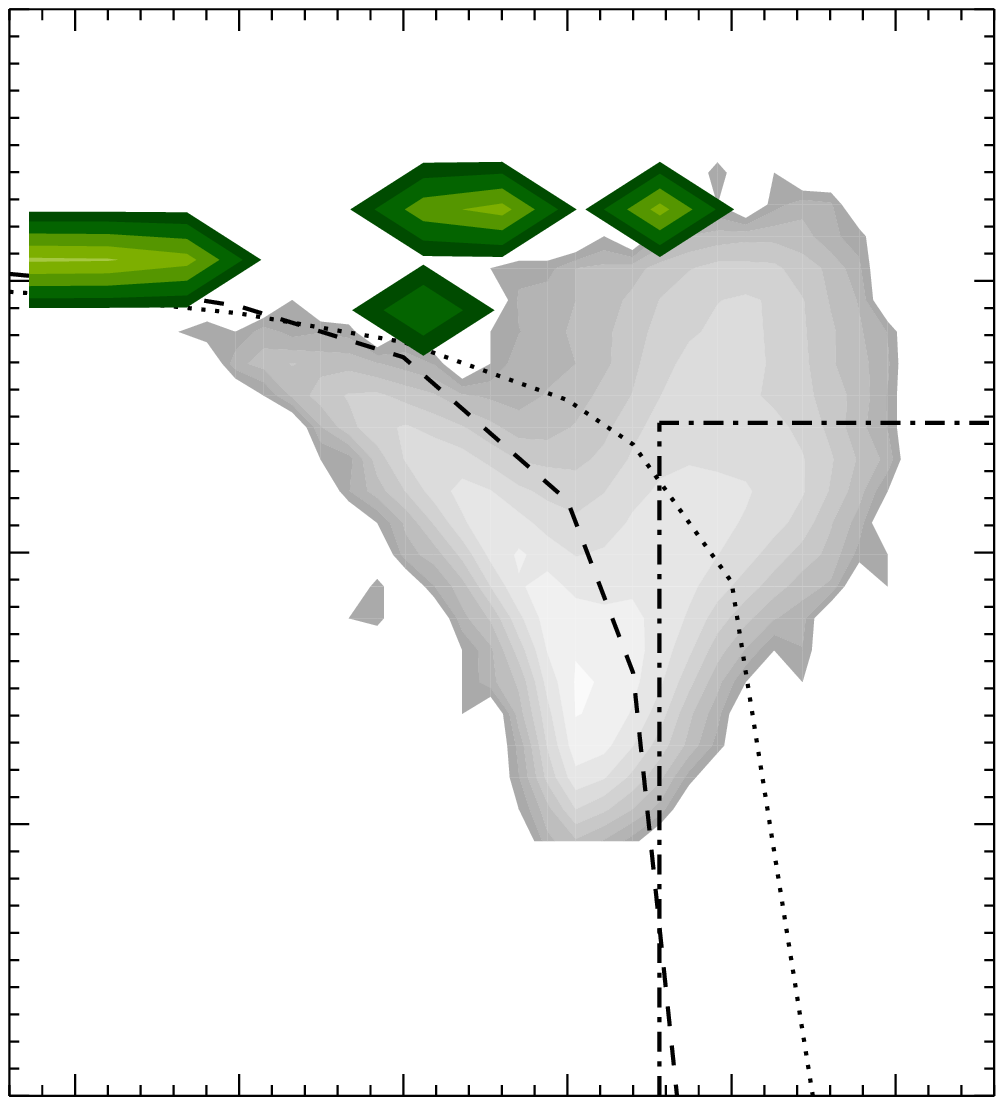,
  width=0.25\textwidth}\hspace{-1.5cm} 
\epsfig{file=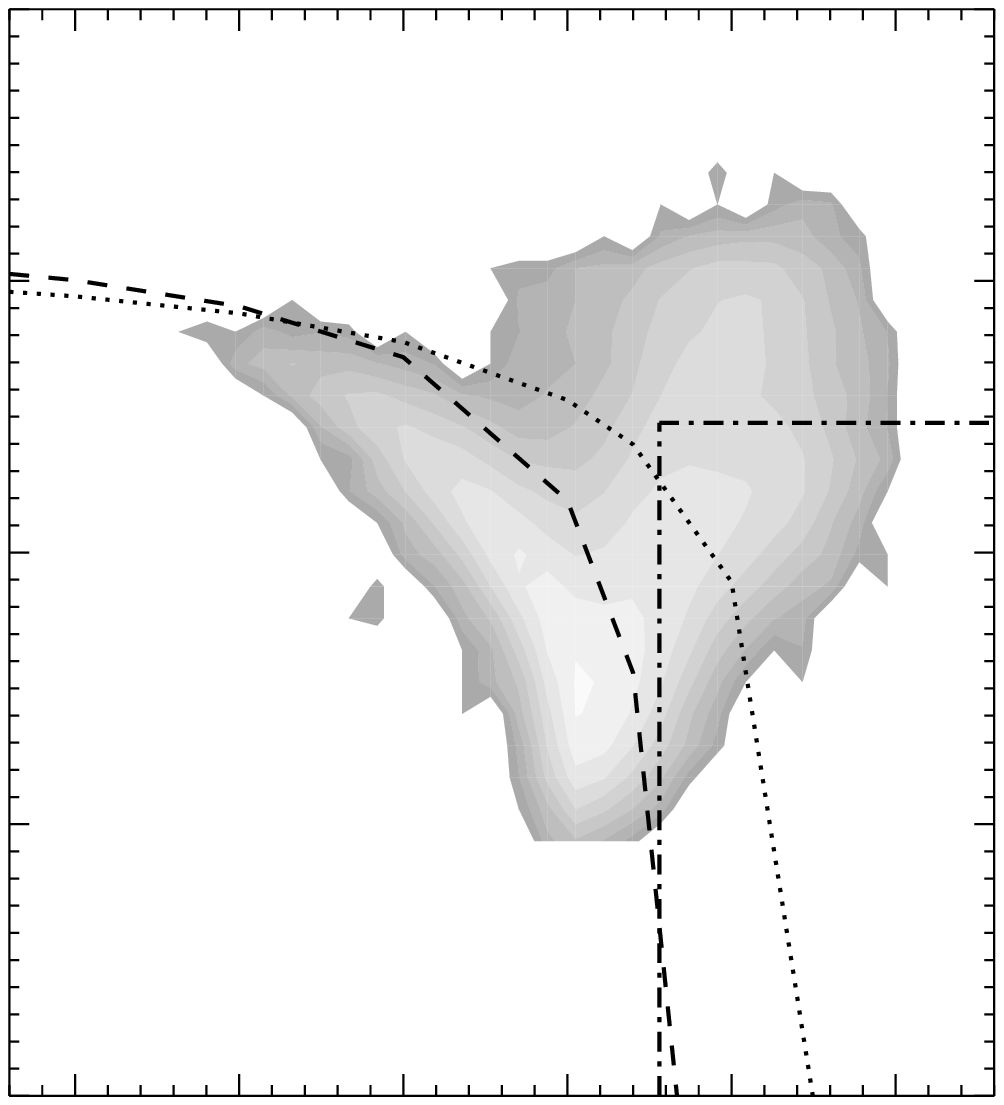,
  width=0.25\textwidth}\hspace{-1.5cm} 
\epsfig{file=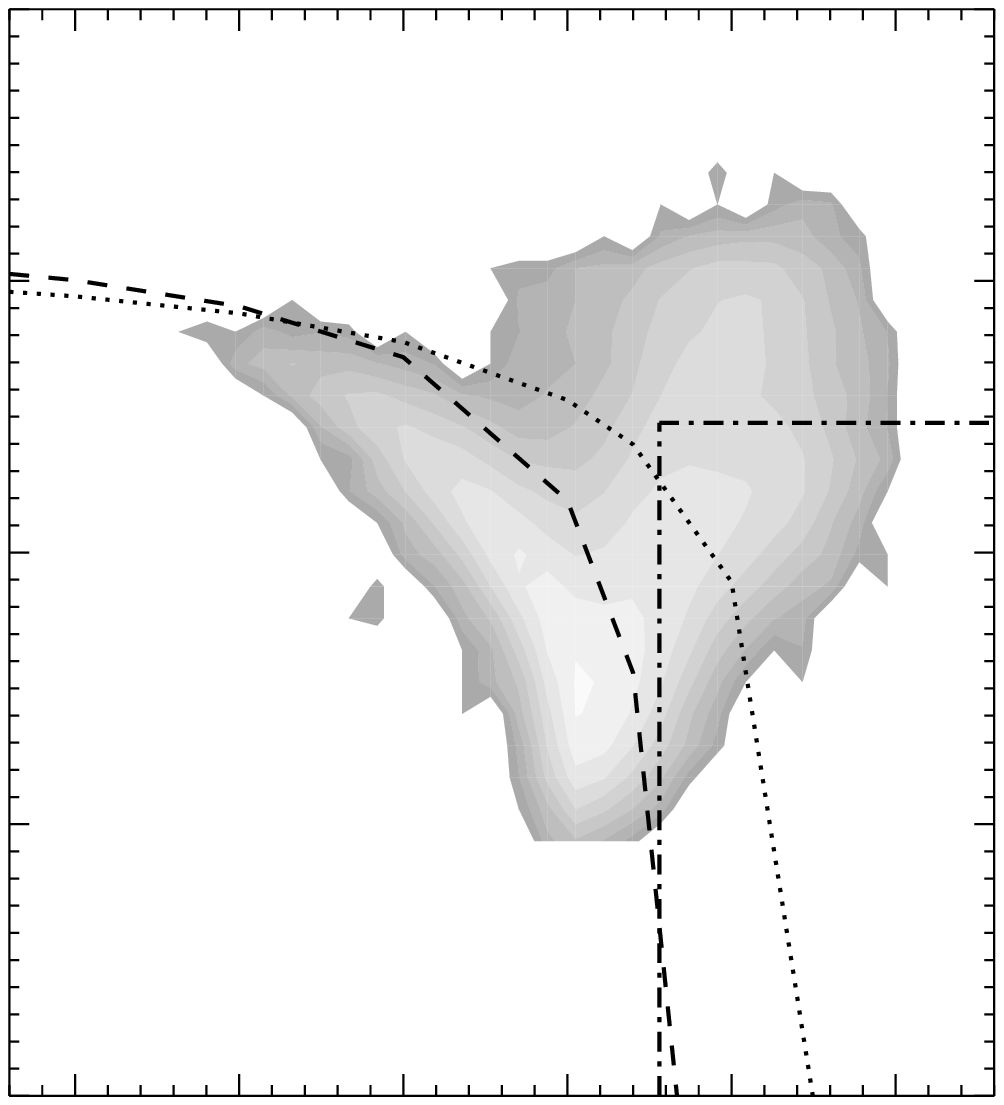,
  width=0.25\textwidth}\vspace{-1.1cm}\\
\epsfig{file=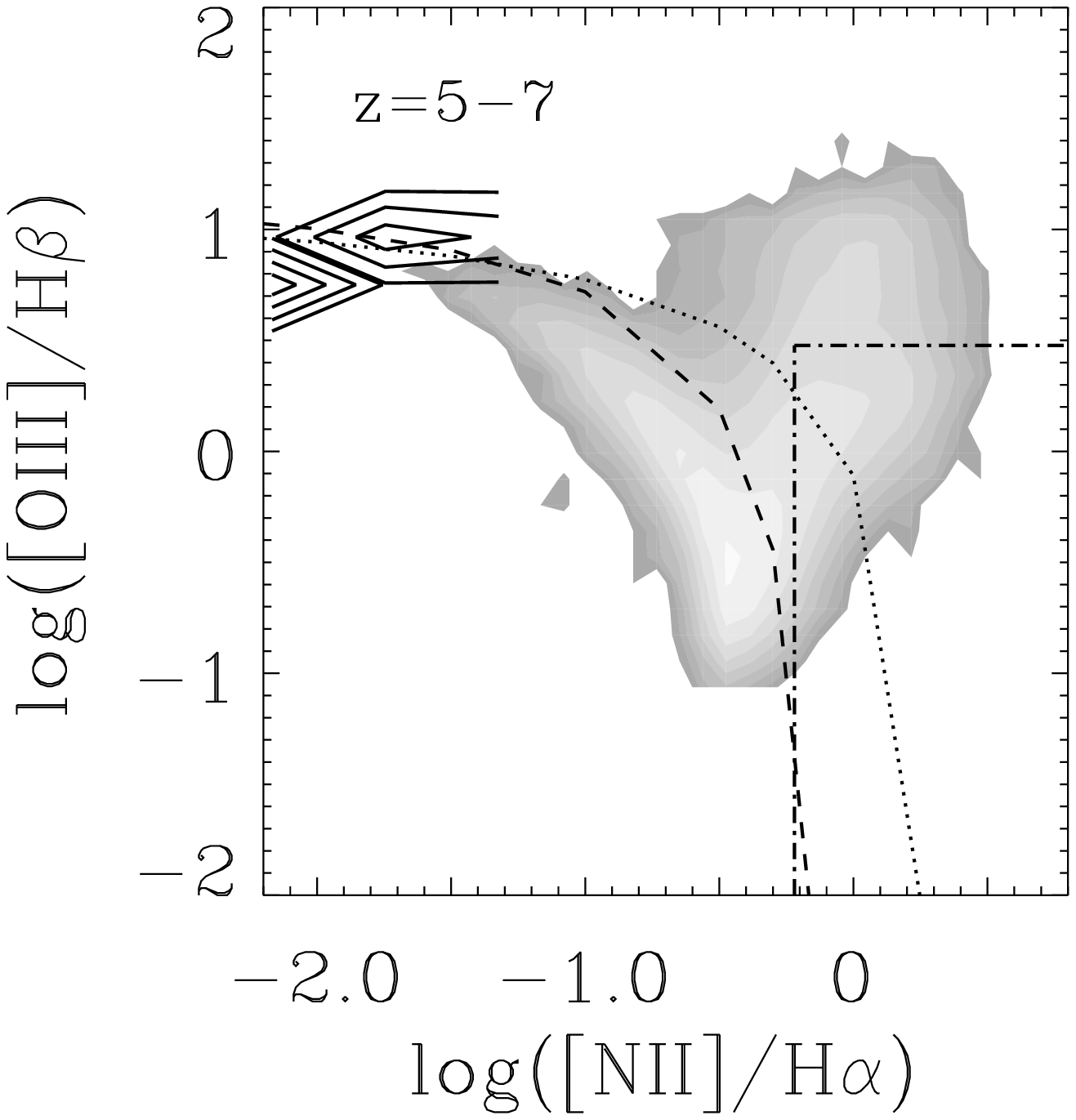,
  width=0.25\textwidth}\hspace{-1.5cm}
\epsfig{file=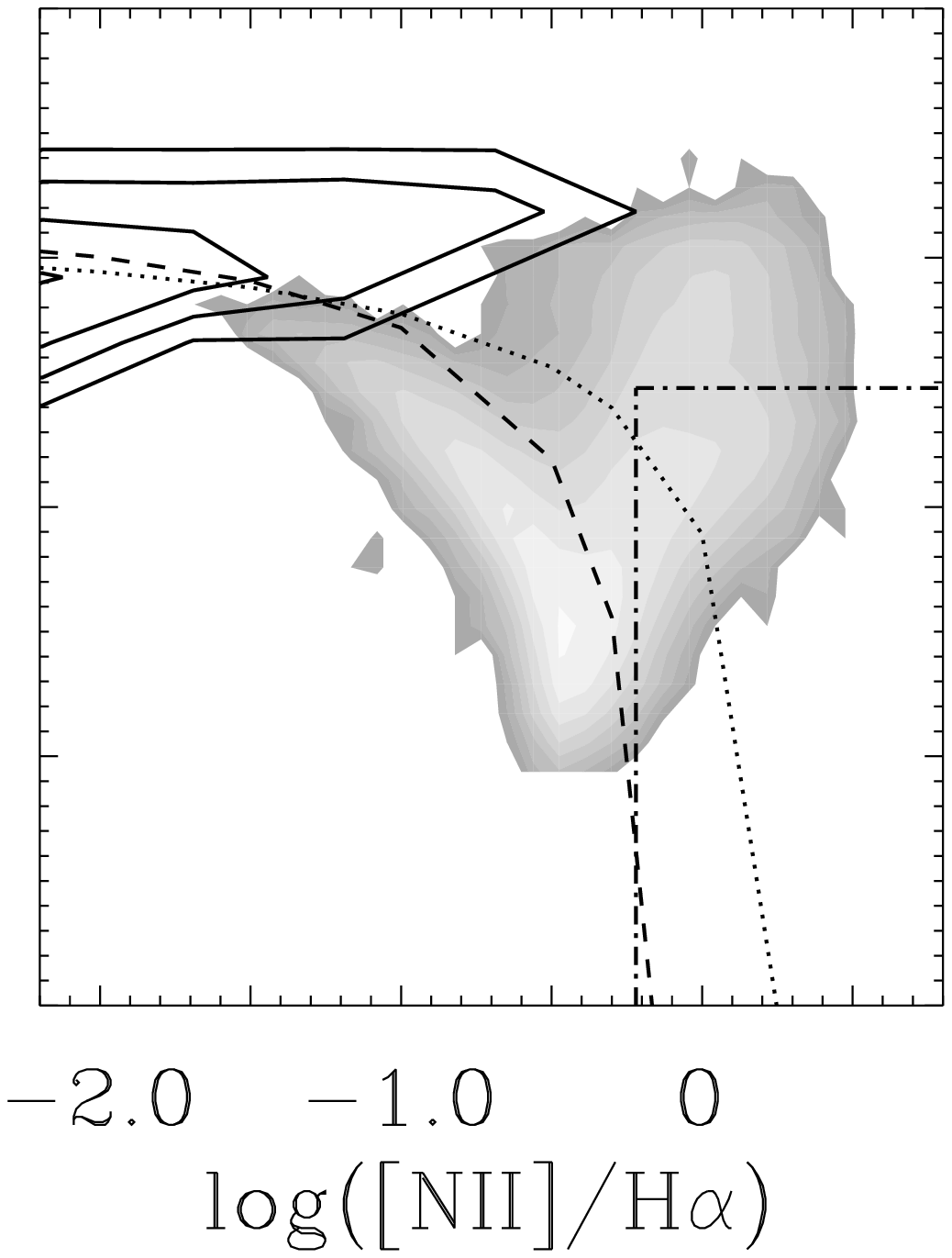,
  width=0.25\textwidth}\hspace{-1.5cm}
\epsfig{file=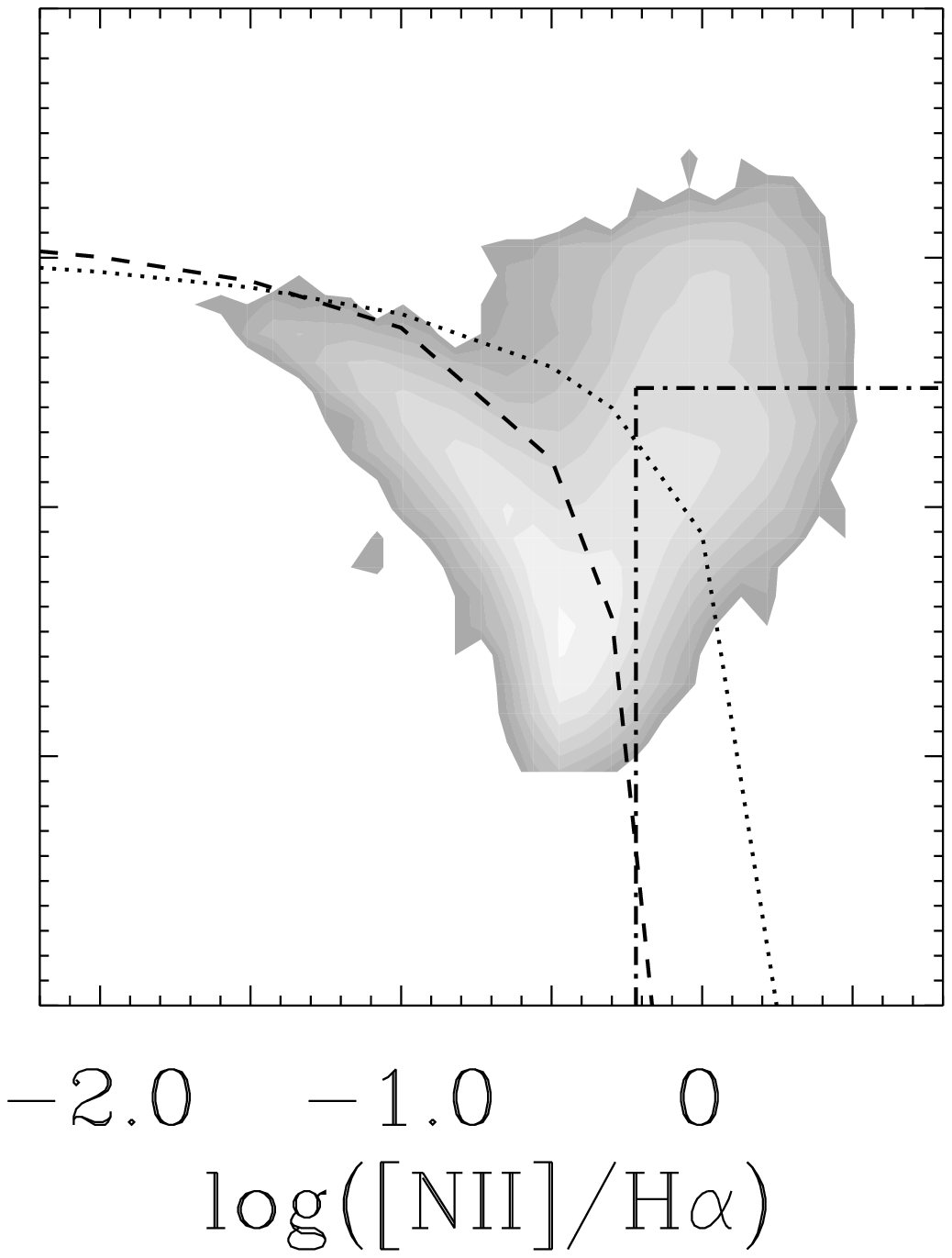,
  width=0.25\textwidth}\hspace{-1.5cm} 
\epsfig{file=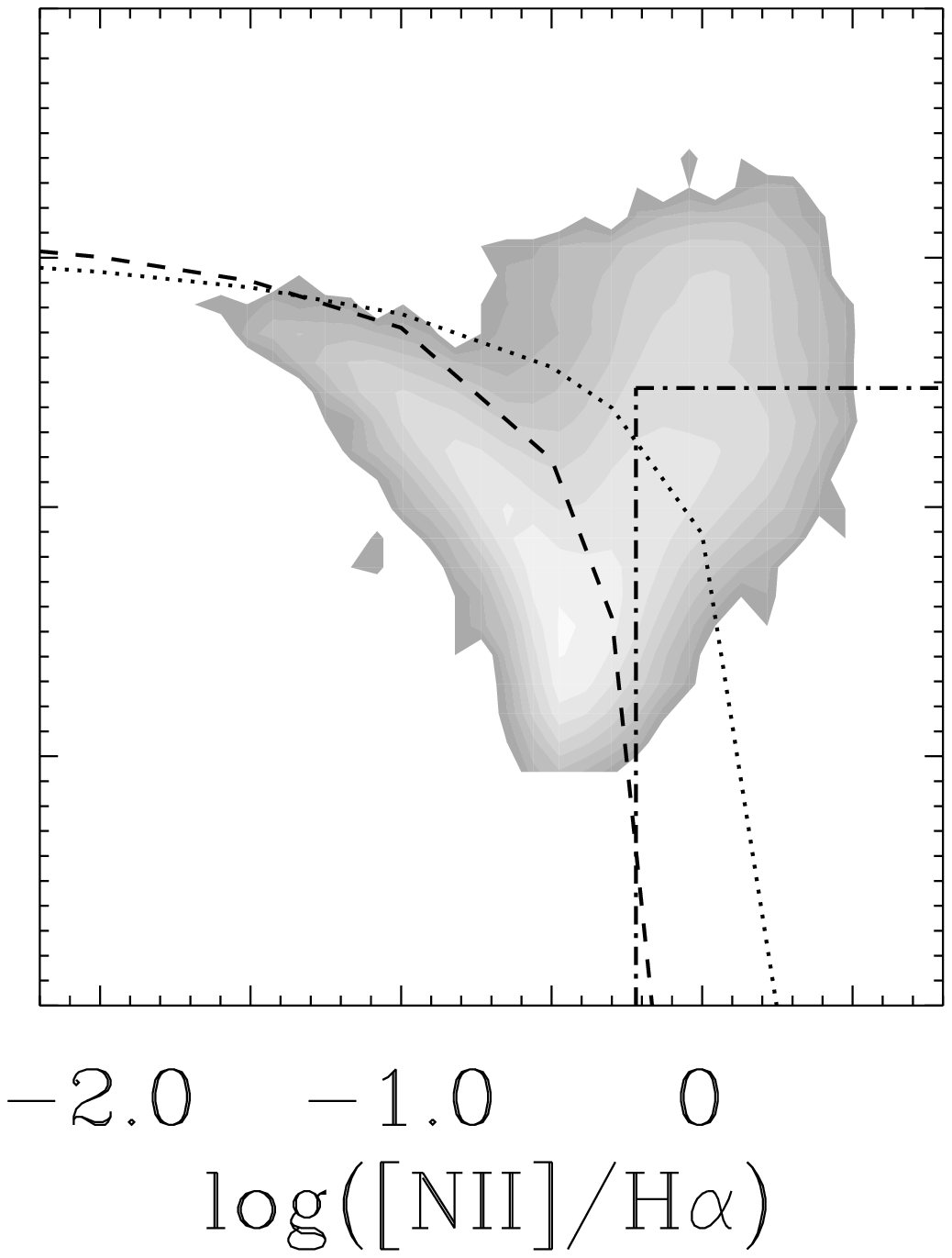,
  width=0.25\textwidth}\hspace{-1.5cm} 
\epsfig{file=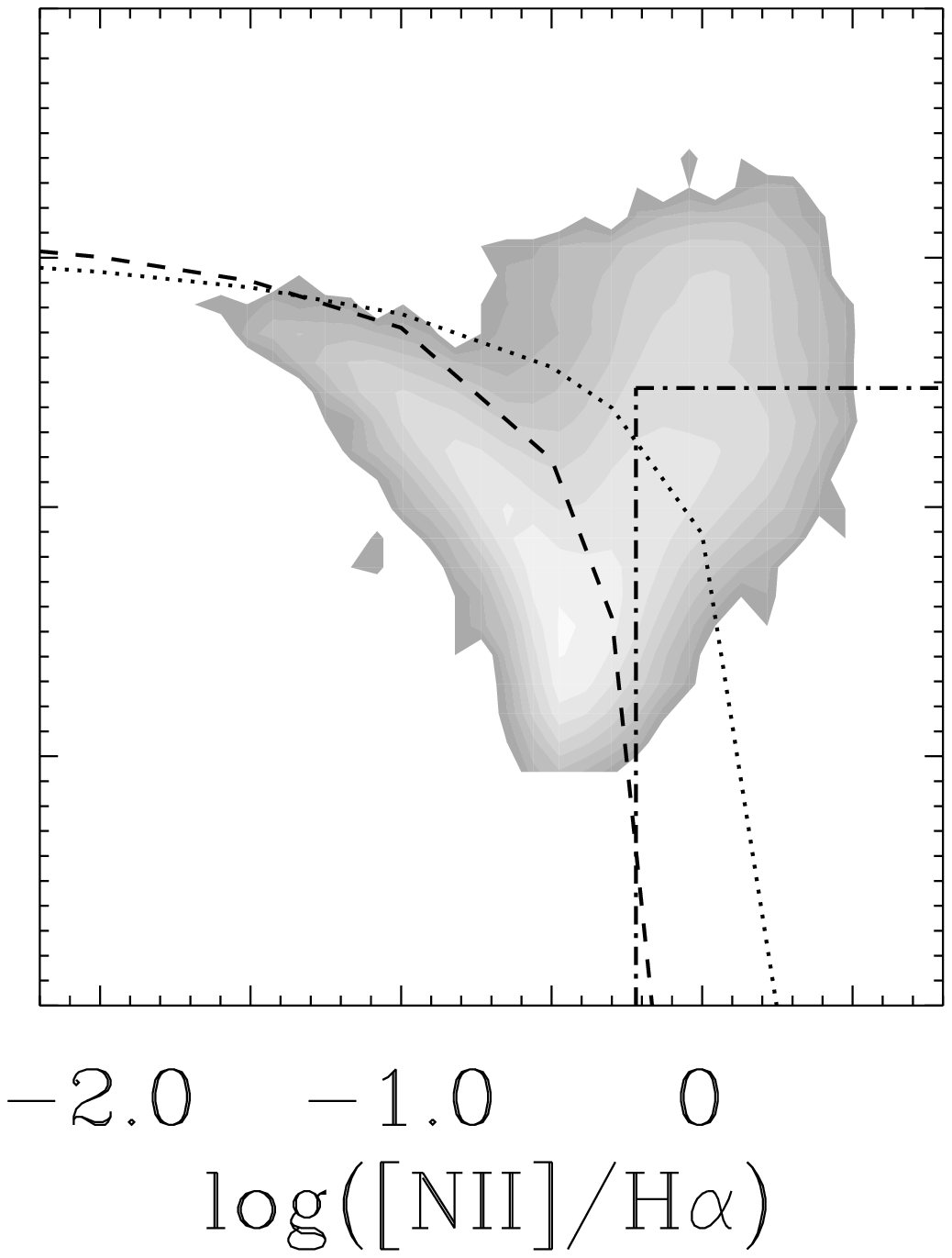,
  width=0.25\textwidth}
%\raggedright
%\epsfig{file=BPTdiagrams_IllustrisTNG50Mpc_2DHisto_shocks_013_OIIINII_SFR_logU-Zlinrel_zevol_scat_interpol.eps,
%  width=0.25\textwidth}\hspace{-1.5cm}
%\epsfig{file=BPTdiagrams_IllustrisTNG50Mpc_2DHisto_shocks_013_OIIINII_Comp_logU-Zlinrel_zevol_scat_interpol.eps,
%  width=0.25\textwidth}\hspace{-1.5cm}
\caption{Redshift evolution of the location of simulated TNG100 
and TNG50 galaxy populations in the \oiiihb-versus-\niiha\ diagnostic
diagram. Rows from top to bottom show the distributions at redshifts $z=0.5$, 1, 2, 3, 
4--5 and 5--7, as indicated. The layout and colour coding are the same as 
in the top row of Fig.~\ref{bpts}.
%except for the highest-redshift bin, where 
%the few assembled TNG100 galaxies are shown as individual symbols.
}\label{bpt_evol}      
\end{figure*}

\begin{figure*}
\center
\epsfig{file=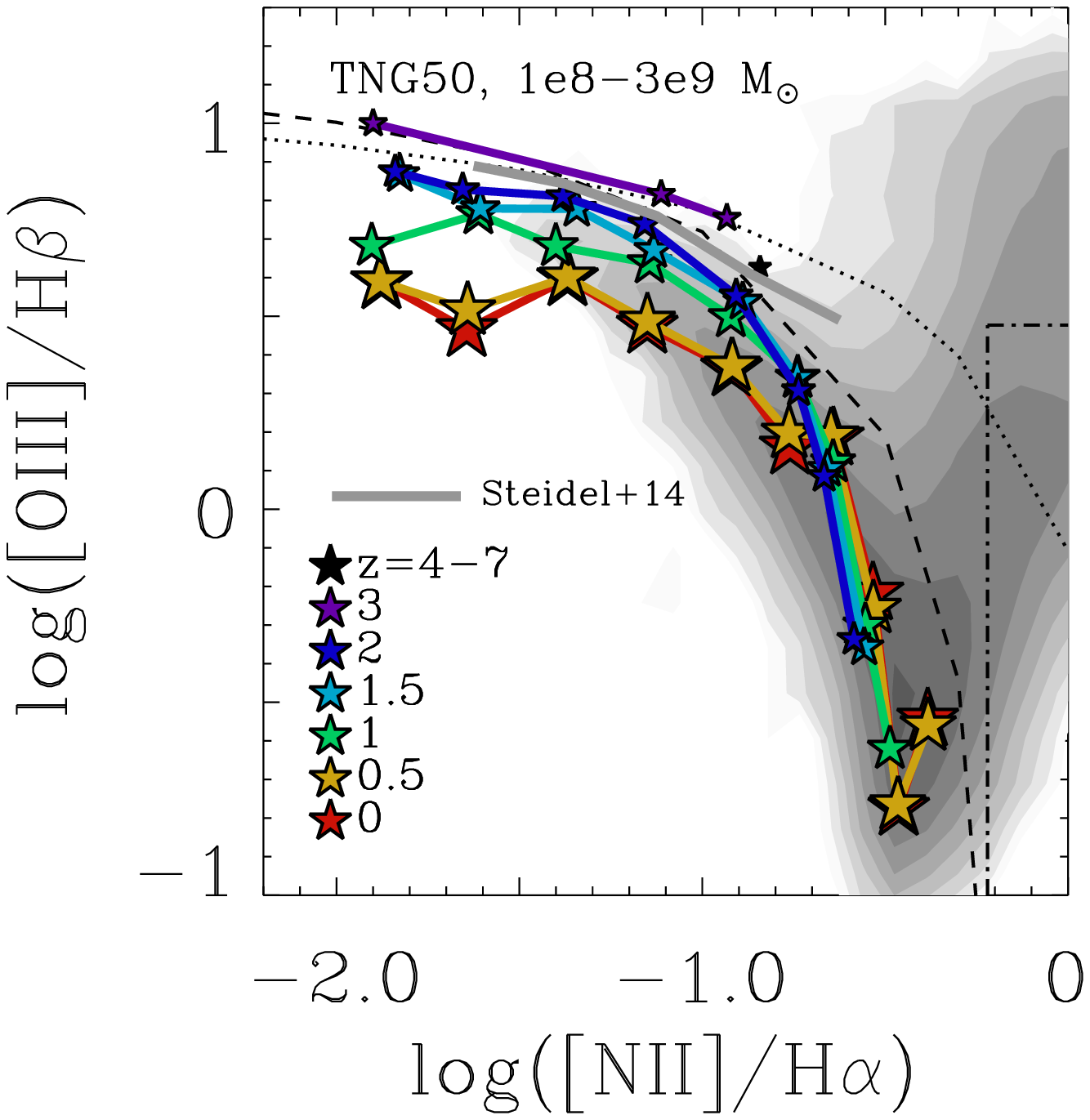,
  width=0.4\textwidth}\hspace{-2cm} 
\epsfig{file=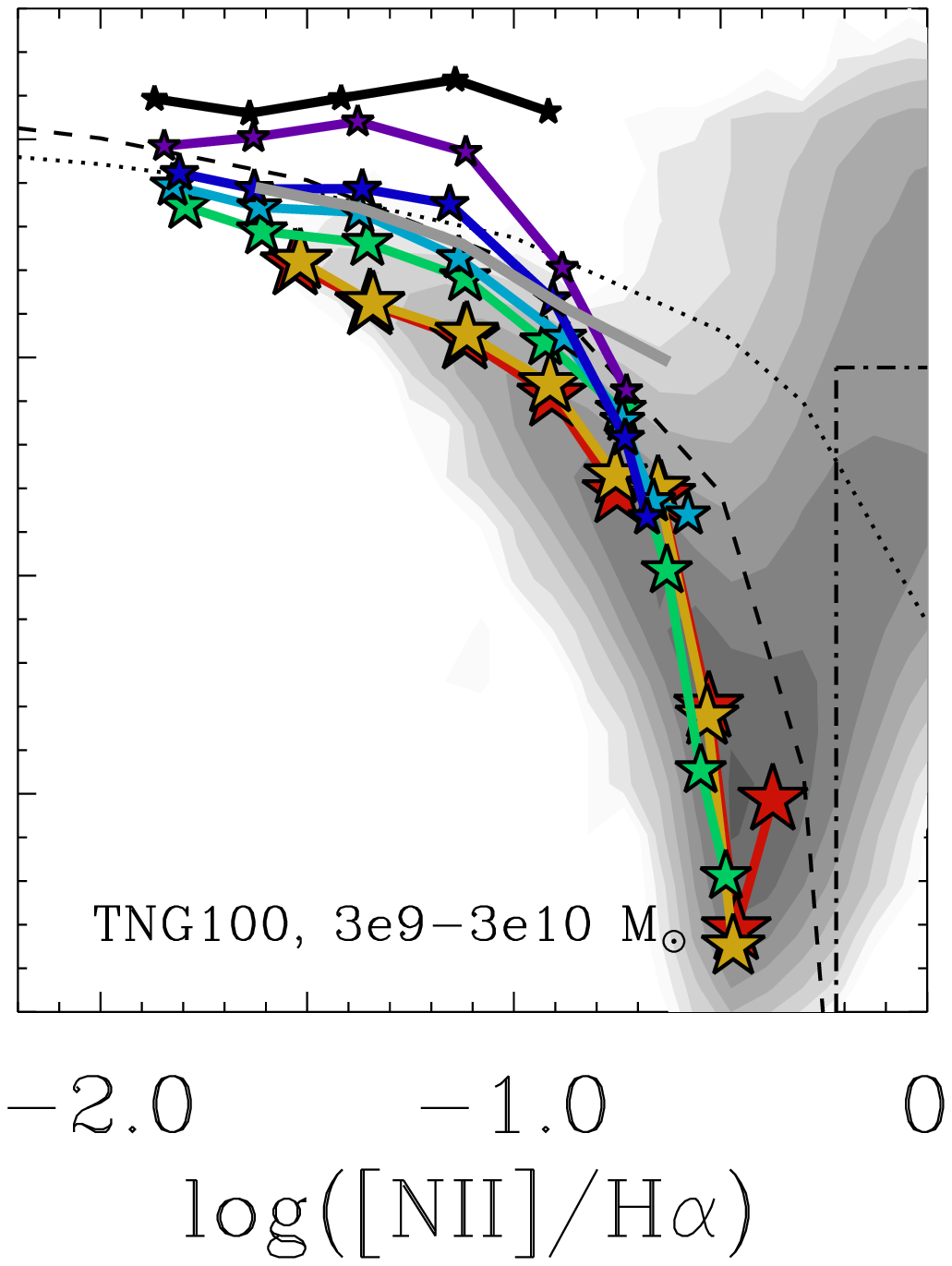,
  width=0.4\textwidth}\hspace{-2cm} 
\epsfig{file=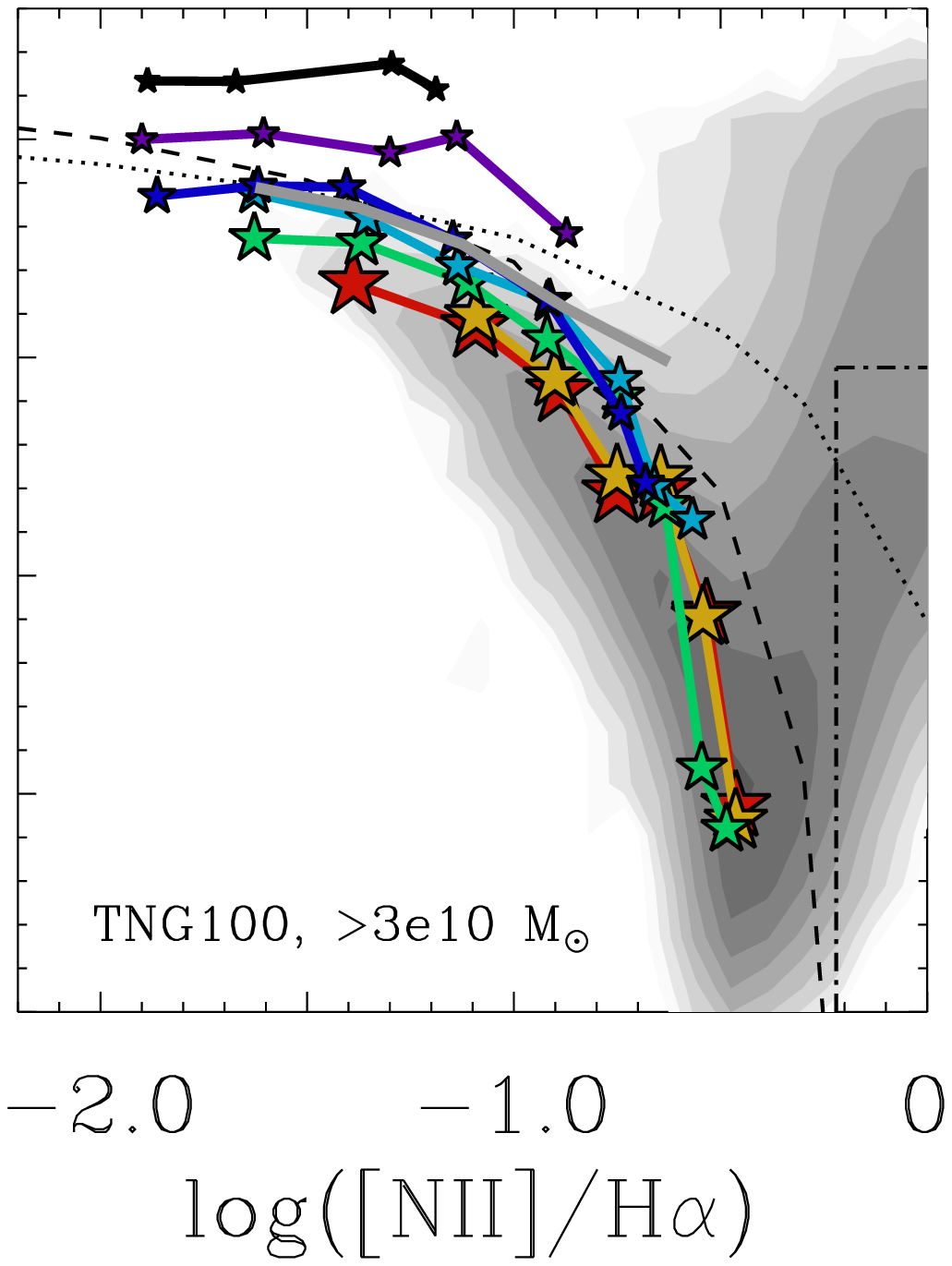,
  width=0.4\textwidth} 
\caption{Average \oiiihb\ ratio in bins of \niiha\ ratio for low-mass TNG50 galaxies ($M_{\mathrm{stellar}} 
 =0.1$--$3 \times 10^9\, \Msun$, left panel), intermediate-mass TNG100 galaxies
($M_{\mathrm{stellar}} = 0.3$--$3\times 10^{10}\, \Msun$, middle panel) and massive 
TNG100 galaxies ($M_{\mathrm{stellar}}> 3 \times 10^{10}\, \Msun$, right panel)
at different redshifts (different colours, as indicated).
Also shown for reference are the same SDSS data and empirical selection criteria 
as in the top row of Fig.~\ref{bpts}, along with the mean relation observed in
a sample of 251 star-forming galaxies 
at $z\sim2.3$ by \citet[][thick grey solid line]{Steidel14}.
}\label{SFbranch}    
\end{figure*}

The different rows of Fig.~\ref{bpt_evol} show the distributions of
TNG100 and TNG50 galaxies in the \oiiihb-versus-\niiha\ 
diagram at redshifts $z=0.5$, 1, 2, 3, 4--5 and 5--7 (from top to bottom), 
using the same layout and colours as for the $z=0$ distributions in the 
top row of Fig.~\ref{bpts} (except for the highest-redshift bin in the 
bottom row of Fig.~\ref{bpt_evol}, where we
show the few assembled TNG100 galaxies as individual symbols). The
adopted flux-detection limit cut is the same as in Fig.~\ref{bpts}.

The lack of shock-dominated and PAGB-dominated galaxies at redshifts
$z>1$ in Fig.~\ref{bpt_evol} reflects the generally higher star-formation 
and BH-accretion rates at these early epochs. Moreover, for all the simulated galaxy 
types, \oiiihb\ appears to globally increase and \niiha\ decrease from low to 
high redshift. This can be traced back to a drop in interstellar metallicity 
(the models include secondary Nitrogen production) combined with a rise in SFR 
and global gas density (controlling the ionization parameter) in the models, 
as discussed in detail in \citet[][]{Hirschmann17, Hirschmann19}.
The global drop in \niiha\ toward higher redshifts implies that the SF-dominated, 
composite and AGN-dominated galaxies become less distinguishable in this
classical BPT diagram \citep[see also figure~14 of][]{Feltre16}. This validates,
on the basis of a much larger sample of simulated galaxies, the finding by  
\citet{Hirschmann19} that standard optical selection criteria to identify 
the nature of the dominant ionizing source break down for metal-poor 
galaxies at $z>1$.

Motivated by a number of observational studies \citep[e.g.,][]{Steidel14,
Shapley15,Kashino17,Strom17}, we focus on the redshift evolution of 
\oiiihb\ for SF-dominated galaxies. Fig.~\ref{SFbranch} shows the average
 \oiiihb\ ratio in \niiha\ bins for low-mass TNG50 galaxies ($M_{\mathrm{stellar}} 
 =0.1$--$3 \times 10^9\, \Msun$, left panel), intermediate-mass TNG100 galaxies
($M_{\mathrm{stellar}} = 0.3$--$3\times 10^{10}\, \Msun$, middle panel) and massive 
TNG100 galaxies ($M_{\mathrm{stellar}}> 3 \times 10^{10}\, \Msun$, right panel)
at different redshifts (different colours, as indicated).
%(red line: $z=0$, yellow line: $z=0.5$, green
%line: $z=1$, light blue line: $z=1.5$, dark blue line: $z=2$, lilac
%line: $z=3$, black line: $z=4-7$).
In all stellar-mass ranges, \oiiihb\ exhibits a significant rise at a fixed \niiha\ 
from low to high redshifts, most pronounced with an
increase of up to 1~dex for intermediate-mass galaxies (middle panel). 

A remarkable result from Fig.~\ref{SFbranch} is the agreement between
the predicted emission-line properties of simulated galaxies at $z\sim2$ 
(blue symbols) and the observations of a sample of 251 star-forming
galaxies with stellar masses in the range $M_{\mathrm{stellar}} 
 =4\times 10^8$--$2.5 \times 10^{11}\, \Msun$
at $z\sim2.3$ by \citet[][thick grey solid line]{Steidel14}. To gain insight
into the physical origin of the rise in \oiiihb\ at high redshift, we have conducted 
an analysis similar to that discussed in section~4 of \citet{Hirschmann17} by
exploring separately the relative influence of different physical parameters 
on observables, but using now the much larger sample of simulated IllustrisTNG 
galaxies. Our analysis confirms our previous finding that the increase
in \oiiihb\ at a fixed galaxy stellar mass and \niiha\ from low to high redshifts
is largely driven by an increase in the ionization parameter resulting from
the rise in SFR and global gas density.
%(see Fig.~\ref{galproperties} of 
%Appendix~\ref{galpropevol}). {\bf \MH{Do we need the appendix?}}

%*****************************************************************************************************
%*****************************************************************************************************
\subsection{Galaxy-type fractions over cosmic time}\label{galtypefractions} 
%*****************************************************************************************************
%*****************************************************************************************************

\begin{figure}
   \center
   \epsfig{file=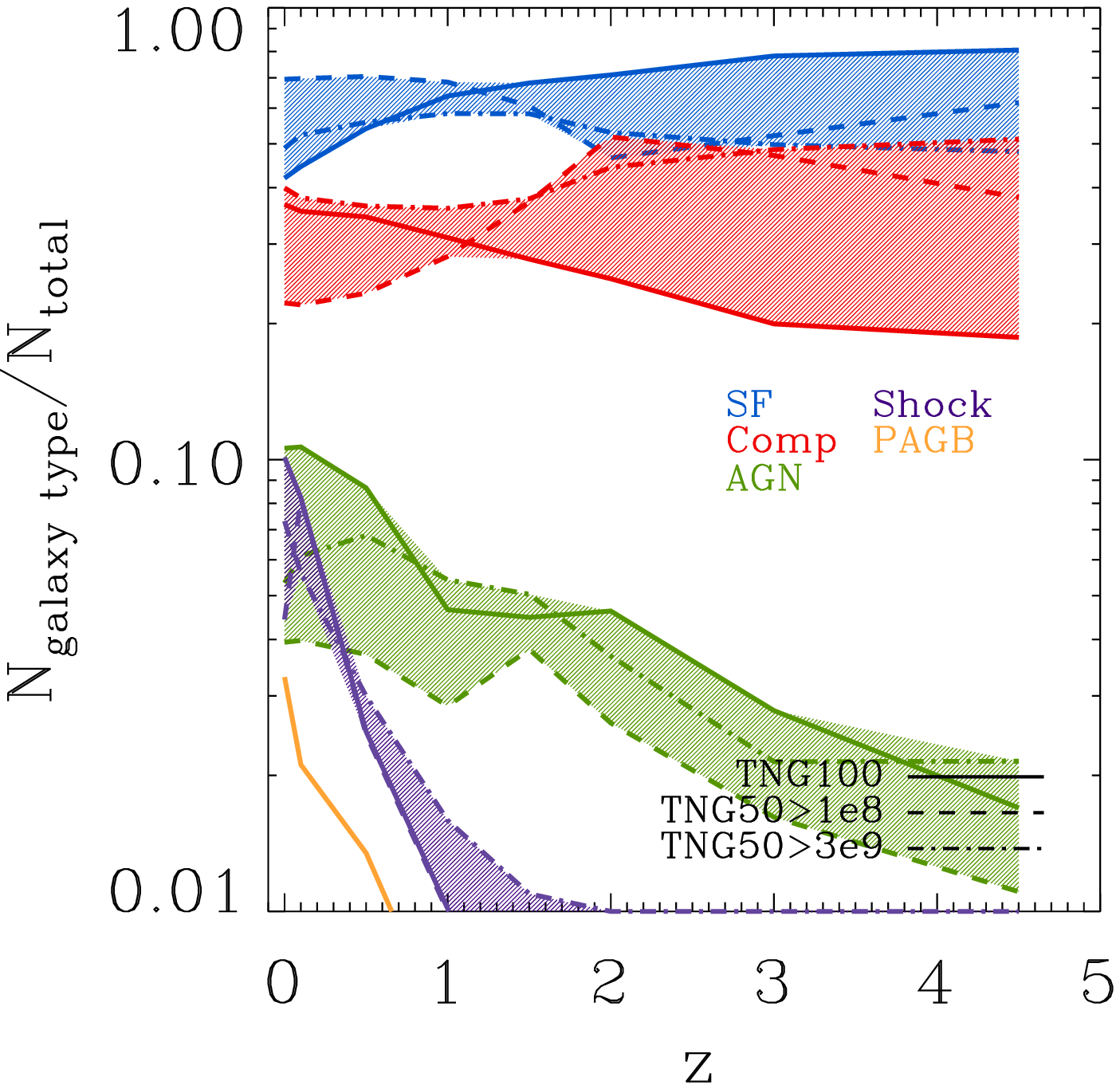,
     width=0.45\textwidth}
   \caption{Fractions of galaxies of different types in the
IllustrisTNG simulations plotted against redshift. The colour coding 
is the same as in Figs~\ref{bpts}--\ref{bpt_evol} (SF-dominated:
blue; composites: red; AGN-dominated: green; shock-dominated: lilac;
PAGB-dominated: yellow). Solid lines refer to TNG100 galaxies 
more massive than $M_{\mathrm{stellar}}=3\times 10^9\Msun$ and
dashed and dot-dashed lines to TNG50 galaxies more massive
than $M_{\mathrm{stellar}} = 1\times10^8$ and $3\times 10^9\,\Msun$, 
respectively. For clarity, the areas encompassed by all
curves referring to a given galaxy type are colour-shaded.}\label{galtypefrac}     
 \end{figure}

In Fig.~\ref{bpt_evol} above, we have seen that PAGB-
and shock-dominated galaxies tend to disappear from emission-line
diagrams at redshifts $z\ga1$. Fig.~\ref{galtypefrac} quantifies how the 
fraction of galaxies of different types evolves with redshift in the 
TNG100 (solid lines) and 
TNG50 (dashed lines) simulations (using the same colour 
coding as in Figs~\ref{bpts}--\ref{bpt_evol}).
To assess the possible impact of resolution effects on
inferred galaxy type fractions, we also show the results obtained when
adopting for TNG50 galaxies the same stellar-mass cut as in 
the TNG100 simulation, i.e., $M_{\mathrm{stellar}} > 3 \times 10^9
\,\Msun$ (dot-dashed lines). For clarity, the areas encompassed by all
curves referring to a given galaxy type are colour-shaded. 

Fig.~\ref{galtypefrac} shows that, at low redshift ($z\la1$), the
emission-line populations in the TNG100 and TNG50 simulations
consist primarily of SF-dominated galaxies. Specifically, at $z=0$, the
TNG100 (TNG50) population consists of about
40~(69)~per cent of SF-dominated galaxies, 37~(21)~per
cent of composites, 10~(4)~per cent of
AGN-dominated galaxies, 10~(6)~per cent of
shock-dominated and 3~($\ll1$)~per cent of  
PAGB-dominated galaxies. 
%\SC{The sums should add to unity in both cases
%(presently 101 and 103 instead)}, \MH{Indeed, I double-checked and now
%the numbers should be consistent}. 
The different fractions found
for the TNG100 and TNG50 simulations (solid and dashed
lines, respectively) follow from the different
stellar-mass distributions: at fixed mass cut, these
fractions (solid and dot-dashed lines, respectively) are roughly
consistent with one another, indicating no significant resolution effect at low redshift.    

At higher redshift, the fraction of SF-dominated galaxies increases from 
60 to 80~per cent over the range $1\la z\la5$ in the TNG100 
simulation, while the fraction of composites drops from 25 to 
18~ per cent. In the TNG50 simulation, instead, the
fraction of composites increases slightly at the expenses of
SF-dominated galaxies, irrespective of the stellar-mass cut. This difference 
relative to TNG100 may arise from resolution effects, with higher
BH-accretion rates -- likely due to higher gas densities in the
accretion regions around the BHs -- being captured by the TNG50 
simulation over this redshift interval. In
both simulations, the fractions of shock- and PAGB-dominated galaxies
drop below 1~per cent at $z>1$, implying  
negligible contributions to line emission (Fig.~\ref{bpt_evol}). Also, the 
fraction of AGN-dominated galaxies falls to only 2--3 per cent from
$z\sim1$ to $z\sim5$, because of the higher star-formation activity of 
high-redshift galaxies compared to their local counterparts, which can 
more efficiently outshine AGN emission.

 %*****************************************************************************************************
%*****************************************************************************************************
\subsection{UV emission-line diagnostic diagrams for metal-poor galaxies}\label{uvdiagrams} 
%*****************************************************************************************************
%*****************************************************************************************************

We have seen in Section~\ref{opticaldiagrams_evol} that optical emission-line
diagnostic diagrams can help reliably distinguish active from
inactive galaxies out to $z \sim 1$, but start to break down for
metal-poor galaxies above this redshift. We now investigate the 
ability of 12 UV emission-line diagnostic diagrams highlighted by \citet[][using 
a small set of 20 cosmological zoom-in simulations of massive galaxies and
their progenitors]{Hirschmann19} to help identify the dominant ionizing sources 
in metal-poor TNG100 and TNG50 galaxies at redshifts $z > 1$.
These diagrams are:\\

\noindent(i) EW(\ciii) versus \ciii/\heii;\\
(ii) EW(\civ) versus \civ/\heii;\\
(iii) EW(\oiiiuv) versus \oiiiuv/\heii;\\
(iv) EW(\siliii) versus \siliii/\heii;\\
(v) EW(\niii) versus \niii/\heii; \\
(vi) \ciii/\heii\ versus \ciii/\civ; \\
(vii) \ciii/\heii\ versus \nv/\heii; \\
(viii) \ciii/\heii\ versus \oiiiuv/\heii; \\
(ix) \ciii/\heii\ versus \siliii/\heii; \\
(x) \ciii/\heii\ versus \niii/\heii; \\
(xi) \civ/\ciii\ versus \ciii/\cii; \\
(xii) \civ/\ciii\ versus \oi/\ha. \\

\begin{figure*}
  \center
  %\vspace{-0.2cm}
\epsfig{file=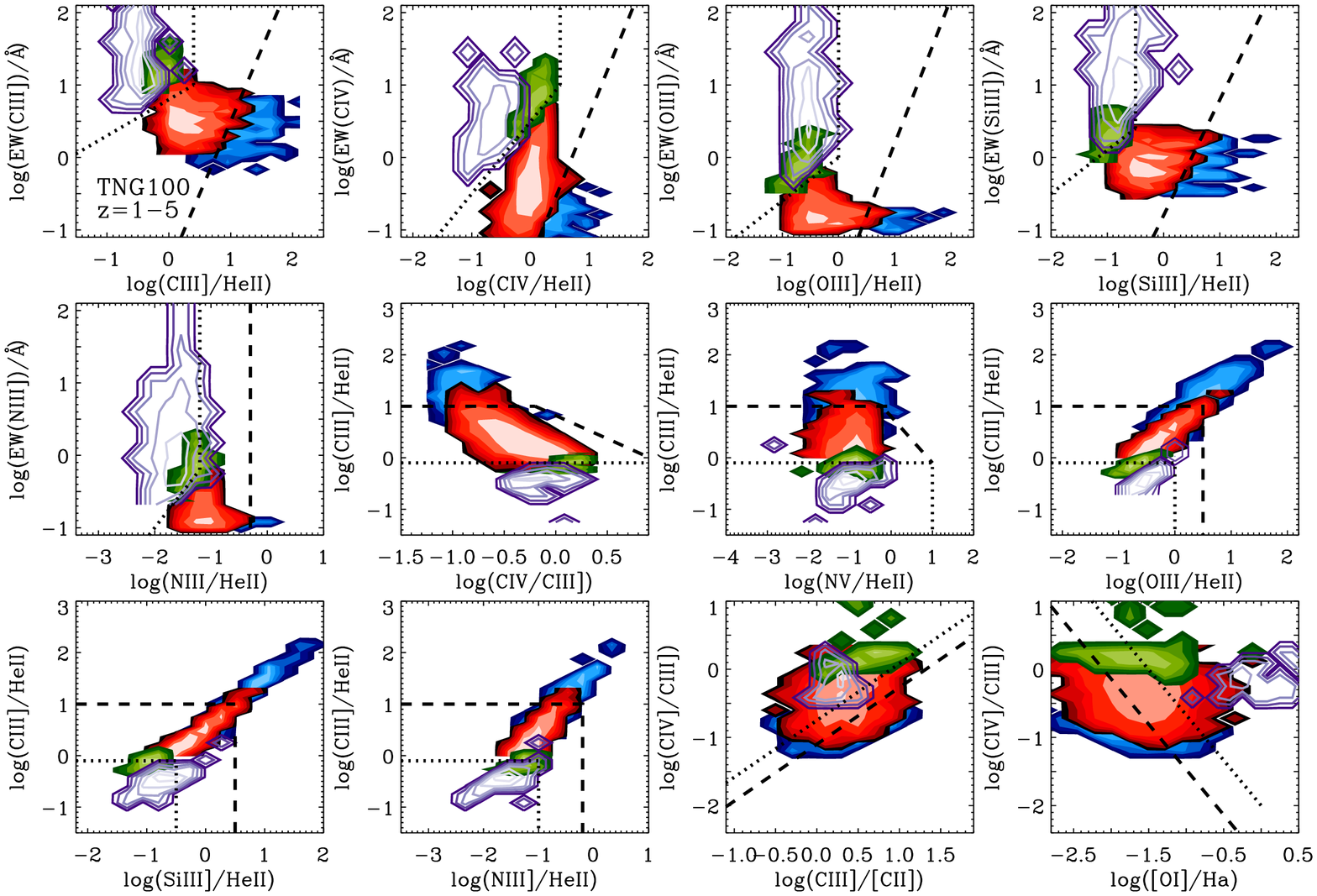,
  width=0.88\textwidth}\vspace{0.4cm}
\epsfig{file=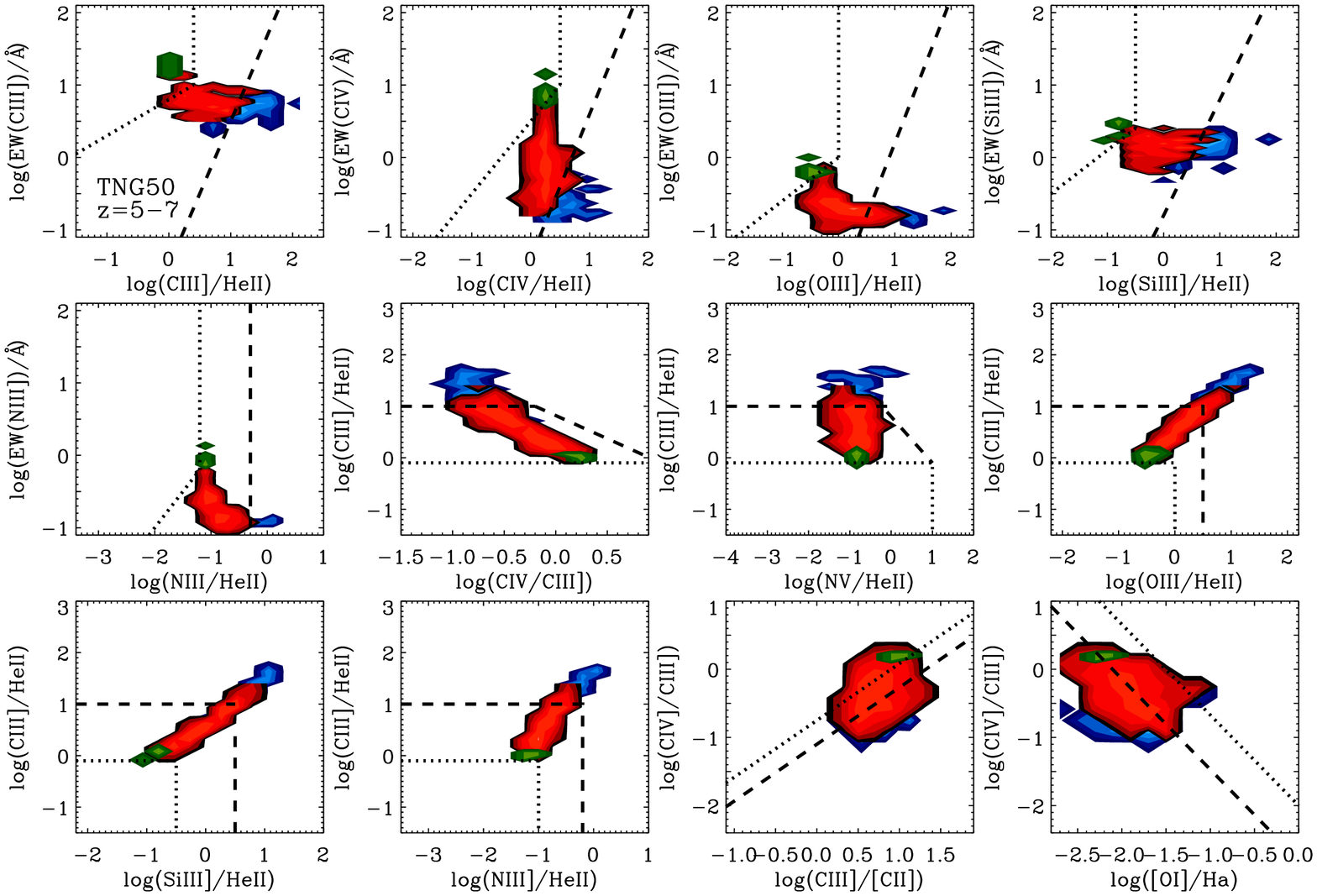,
  width=0.88\textwidth}
\caption{Location of metal-poor (${\rm N2O2}< -0.8$) galaxy populations in the
TNG100 simulation at $z=1$--5 (top 12 panels) and the TNG50
simulation at $z=5$--7 (bottom 12 panels), in 12 different UV-diagnostic diagrams 
highlighted by \citet[][see Section~\ref{uvdiagrams} for details]{Hirschmann19}. A
flux-detection limit of $\rm 10^{-18}\,erg\,s^{-1}cm^{-2}$ was applied to all UV
emission lines. Each
panel shows the 2D distributions of SF-dominated (blue), composite (red) 
and AGN-dominated (green) galaxies; the top 12 panels also show the distribution
of shock-dominated galaxies (lilac contours; no such galaxies are predicted
to be found at $z=5$--7). Overplotted in each panel are the selection criteria of
\citet{Hirschmann19} to discriminate between SF-dominated and composite galaxies 
(black dashed line), and between composite and AGN-dominated galaxies (black 
dotted line).}\label{uvdiagnostics}      
\end{figure*}

In Fig.~\ref{uvdiagnostics}, we show the locations in these UV diagnostic 
diagrams of metal-poor galaxy populations of different types from the 
TNG100 simulation in the redshift interval $z=1$-5 (top 12 panels),
and the TNG50 simulation in the redshift interval $z=5$--7 
(bottom 12 panels). The galaxies were selected to have ${\rm N2O2}< -0.8$, 
corresponding roughly to metallicities below $0.5\Zsun$ \citep{Hirschmann19}. 
This selection makes the contribution to line emission by post-AGB stellar populations 
negligible, and we ignore it here \citep[this contribution is most important in 
evolved, metal-rich galaxies at low redshift; see, e.g.,][]{Hirschmann17}.
We note that, by analogy with \citet{Hirschmann19}, we have applied
a flux-detection limit of $\rm 10^{-18}\,erg\,s^{-1}cm^{-2}$ to all
emission lines when building Fig.~\ref{uvdiagnostics}. This corresponds 
roughly to the point-source emission-line flux sensitivity reached in $10^4$\,s 
using NIRSpec on board {\it JWST}, with a signal-to-noise ratio of 10. 
We also applied a detection limit of 0.1~\AA\ on emission-line EWs. 
This low EW limit is justified by our desire to keep our analysis as general as 
possible for the proposed diagnostic diagrams to be useful also for future
instruments with very high sensitivity (e.g., MOSAIC and HARMONI on
the Extremely Large Telescope). We have checked 
that the results shown in this section do not change qualitatively when applying
more rigorous cuts in UV-line fluxes and EWs. 

Each panel of Fig.~\ref{uvdiagnostics} also indicates the selection
criteria identified by \citet{Hirschmann19} to discriminate between
SF-dominated and composite galaxies (black dashed line), and
between composite and AGN-dominated galaxies (black dotted line).
Since we already validated these UV-selection criteria against observational 
data in \citet{Hirschmann19}, for clarity, we do not report these observations 
in Fig.~\ref{uvdiagnostics}.

The first 12 panels of Fig.~\ref{uvdiagnostics} show how
shock-dominated galaxies largely fall in areas populated by
AGN-dominated (and composite) galaxies in these 
diagrams. Thus, the UV-diagnostic diagrams shown here do not 
allow unique identification of galaxies, whose nebular emission is 
dominated by fast, radiative shocks. Given the scarcity of such 
galaxies in the simulations at $z>1$ (Fig.~\ref{galtypefrac}), 
we do not regard this degeneracy as problematic. 

In contrast, Fig.~\ref{uvdiagnostics} clearly demonstrates that the first
10 UV-diagnostic diagrams [labelled (i) to (x)] in the above list can 
robustly discriminated between SF-dominated, composite and 
AGN-dominated galaxies over the full redshift range $1\la z\la7$,
based on the large statistical sample of galaxies in the TNG100 
and TNG50 simulations. This confirms the results obtained 
using a much smaller 
set of 20 cosmological zoom-in simulations of massive galaxies and
their progenitors by \citet{Hirschmann19}. As described in that study, the
reason for the good separability of different galaxy types is mainly
that the ratio of any of the five considered (collisionally excited) metal
lines to the \heii\ (recombination) line successively
decreases from SF-dominated, to composite, to AGN-dominated
galaxies. This is because of the harder ionizing radiation of
accreting BHs compared to that of stellar populations, which increases the
probability of producing doubly-ionized helium. We note that this is less the
case for \nv/\heii, since \nv\ requires photons of even higher energy
than \heii\ to be produced (77.5 versus 54.4\,eV). Hence, \nv/\heii\
is not a clear indicator of the hardness of the ionizing
radiation. Also the \civ/\ciii\ ratio is   sensitive to the presence of hard 
AGN radiation, which increases the probability of triply ionizing carbon. 
%
%The hard radiation from AGN also makes
%the EW of high-ionization, collisionally excited metal lines stronger,
%and more so for luminous than for faint AGN. We recall that the EW in
%Fig.~\ref{UV2lines}, which account for the emission from narrow-line
%regions but not for direct attenuated radiation from accreting BHs,
%should be appropriate for type-2 AGN and taken as upper limits for
%type-1 AGN (see Section~\ref{eqwidth}). 
%
We note that the completeness and purity fractions (not shown)\footnote{The
completeness (purity) fraction provides a measure of how complete 
(uncontaminated) a population of a given type selected using empirical criteria 
is with respect to our theoretically defined galaxy types 
\citep[see, e.g.,][]{Hirschmann19}.}  of SF-dominated, 
composite and AGN-dominated galaxies in the diagrams of 
Fig.~\ref{uvdiagnostics} corresponding to the diagnostics (i)--(x)  
above all exceed 75~per cent over the full considered redshift range, 
further confirming the reliability of these UV-diagnostic diagrams.

To avoid having to rely on the \heii\ line, whose high strength 
can sometimes be challenging to model in some metal-poor star-forming galaxies
\citep[e.g.,][]{Senchyna17, Steidel16, Berg18, Jaskot16, Wofford21}, 
\citet{Hirschmann19} proposed alternative UV and optical-UV 
diagnostic diagrams to discriminate between different galaxy types.
Among those are the last two diagnostic diagrams [labelled (xi) to 
(xii)] in the above list (we do not include here other diagrams involving 
the very faint \niv\ line). Fig.~\ref{uvdiagnostics} shows that, 
unfortunately, when considering the much larger galaxy 
populations in the TNG50 and TNG100 simulations
(based on different physical models), the 
different galaxy types overlap in these diagrams and cannot be 
separated, with completeness and purity fractions of only
10--25~per cent. 

In this context, it is comforting that if the \heii-line strength were to
increase in the SF-galaxy models \citep[by factors of 2--4, as 
considered in section~5.2 of][]{Hirschmann19}, this would hardly affect the 
purity and completeness fractions of AGN-dominated galaxies, the 
purity fractions of SF-dominated galaxies and the completeness fractions 
of composites in the first 10 diagnostic diagrams of Fig.~\ref{uvdiagnostics}
(all fractions would remain above 75~per cent). This
indicates that UV-selected AGN-dominated galaxies would still be
reliably identified, SF-dominated samples largely uncontaminated by
active galaxies (composites and AGN), and composite samples fairly
complete. In contrast, the completeness fractions of SF-dominated 
galaxies and the purity fractions of composites could drop below 20~per 
cent if the \heii\ luminosity were quadrupled, as SF-dominated galaxies
would move toward the region populated by composites in UV diagnostic diagrams. 
In this case, SF-dominated samples would become incomplete, and composite
samples heavily contaminated by SF-dominated galaxies.

%*****************************************************************************************************
%*****************************************************************************************************
\section{Line-emission census of galaxy populations
  across cosmic time}\label{luminosityfunctions}  
%*****************************************************************************************************
%*****************************************************************************************************

In this section, we explore the statistics of optical- and UV-line emission
from galaxy populations across cosmic time, as enabled by the large
galaxy samples in the IllustrisTNG simulation set. %which covers different
%cosmological volumes for different galaxy stellar-mass cuts (Section~\ref{theory}).
We start by examining how the \ha, \oiii\ and \oii\ luminosities predicted
by these simulations trace galaxy SFR, and compare the \ha, \oiiinl+\hb\ 
and \oiinl\ luminosity functions at various redshifts with available observations 
from the literature (Section~\ref{opticallf}). Then, in Section~\ref{uvlf}, 
we explore the luminosity
functions of different UV lines in the redshift range $2\leq z\leq7$, and 
which UV lines are the most promising tracers of cosmic star-formation activity 
at $z\ga7$. All luminosity
functions and SFR-line luminosity relations shown in that section are
corrected for attenuation by dust.\footnote{Except for the potential
  attenuation of  ionizing photons by dust before they produce line
  emission   \citep[see, e.g.,][]{Charlot02}.} Finally, in Section~\ref{numbercounts},
  we compute predictions of 
number counts of active and inactive galaxies in various optical and UV lines 
down to the sensitivity limits of deep observing programs currently
planned with {\it JWST}, including the potential effect of attenuation by dust.

\begin{figure}
\center
\epsfig{file=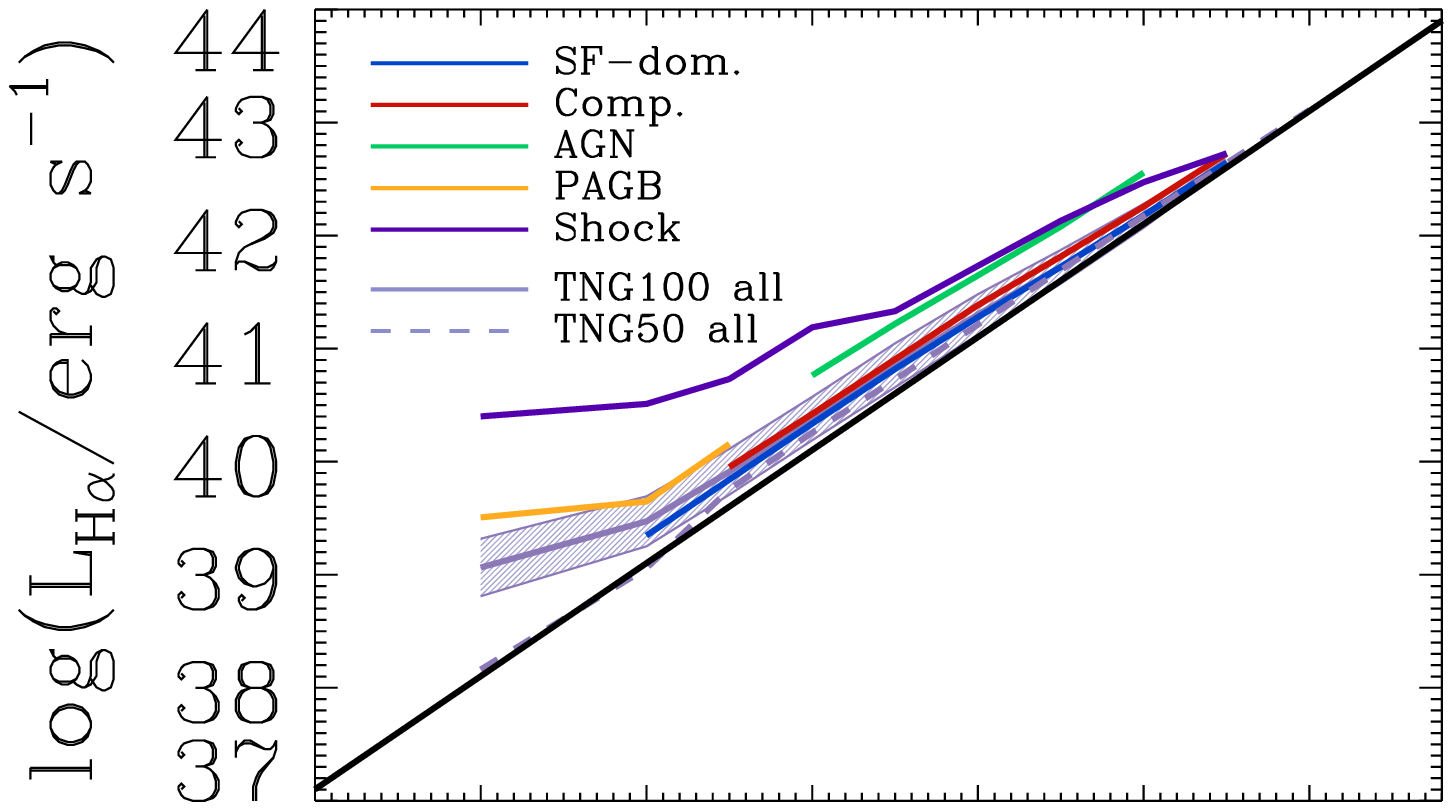, width=0.45\textwidth}\vspace{-1.7cm}
\epsfig{file=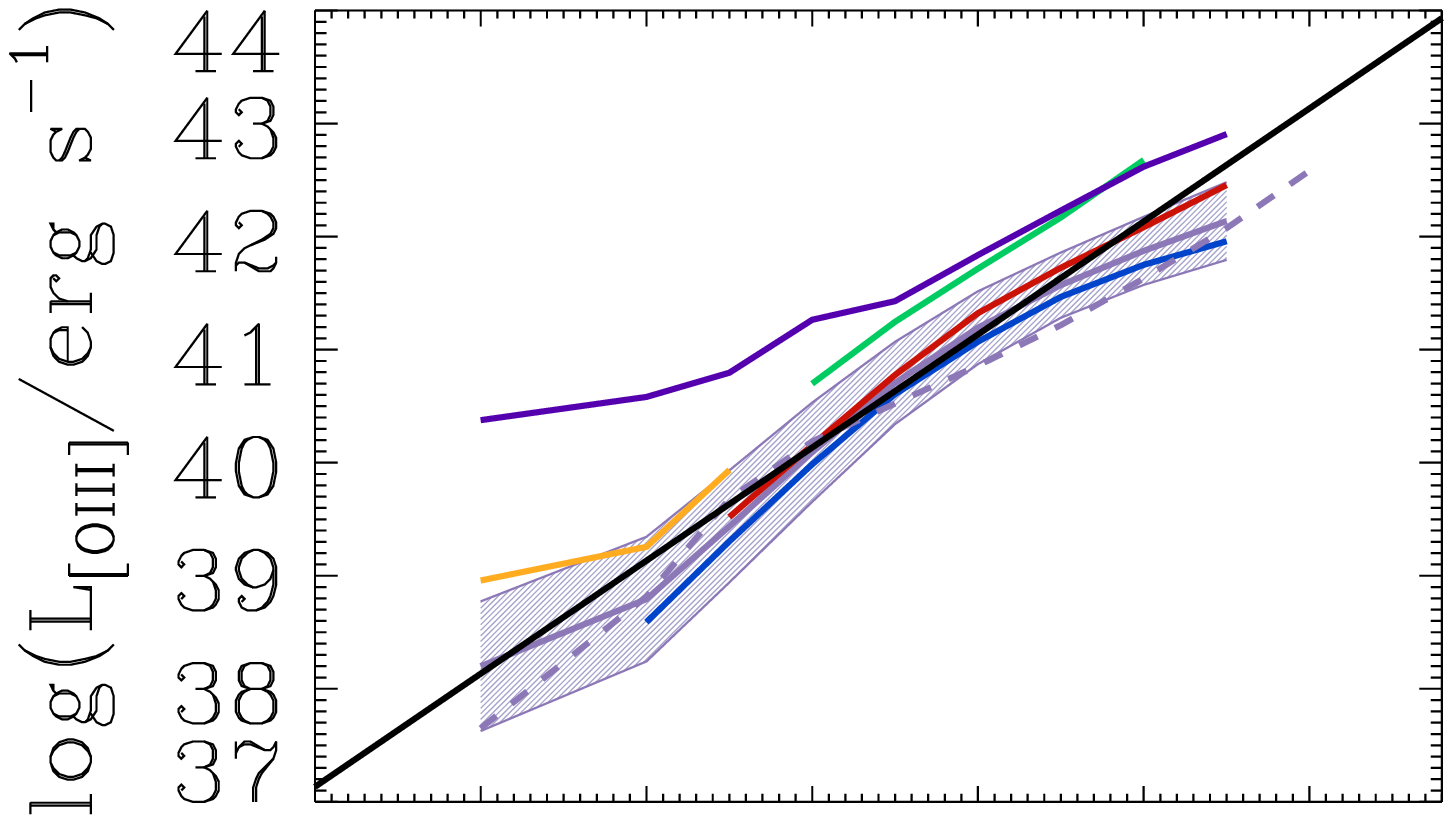, width=0.45\textwidth}\vspace{-1.7cm}
\epsfig{file=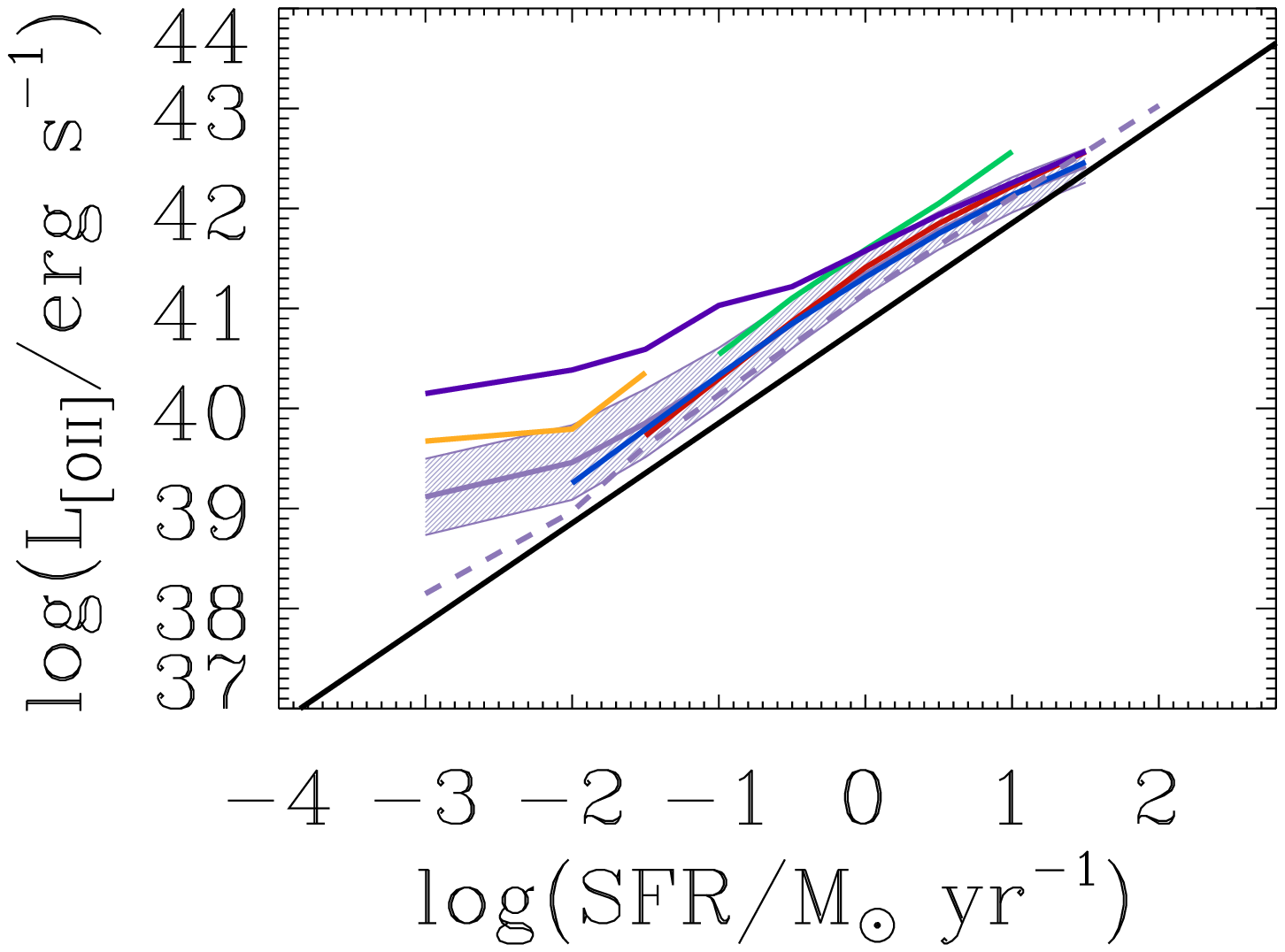, width=0.45\textwidth}
\caption{Average line luminosities of \ha, \oiii\ and \oii\  (from top to bottom) 
in bins of SFR for TNG100 (solid mauve line with shaded
1$\sigma$ scatter) and TNG50 (dashed mauve line) galaxies at
$z=0$. Relations for different galaxy types are also shown
in the case of TNG100 (blue: SF-dominated; red: composite; 
green:  AGN-dominated; orange: PAGB-dominated; lilac: shock-dominated). 
Also shown for completeness are relations often used in the literature
(from \citealt{Kennicutt98} for \ha\ and \oii, and \citealt{Sobral15} for 
\oiii; black solid line in each panel). These should not be compared in 
detail with the IllustrisTNG predictions, as they were derived from purely 
solar-metallicity, SF-galaxy models assuming a \citet{1955ApJ...121..161S} 
IMF truncated at 0.1 and 1\Msun.}\label{sfr_lopt}     
\end{figure}

\begin{figure*}
  \center
  \vspace{-0.2cm}
\epsfig{file=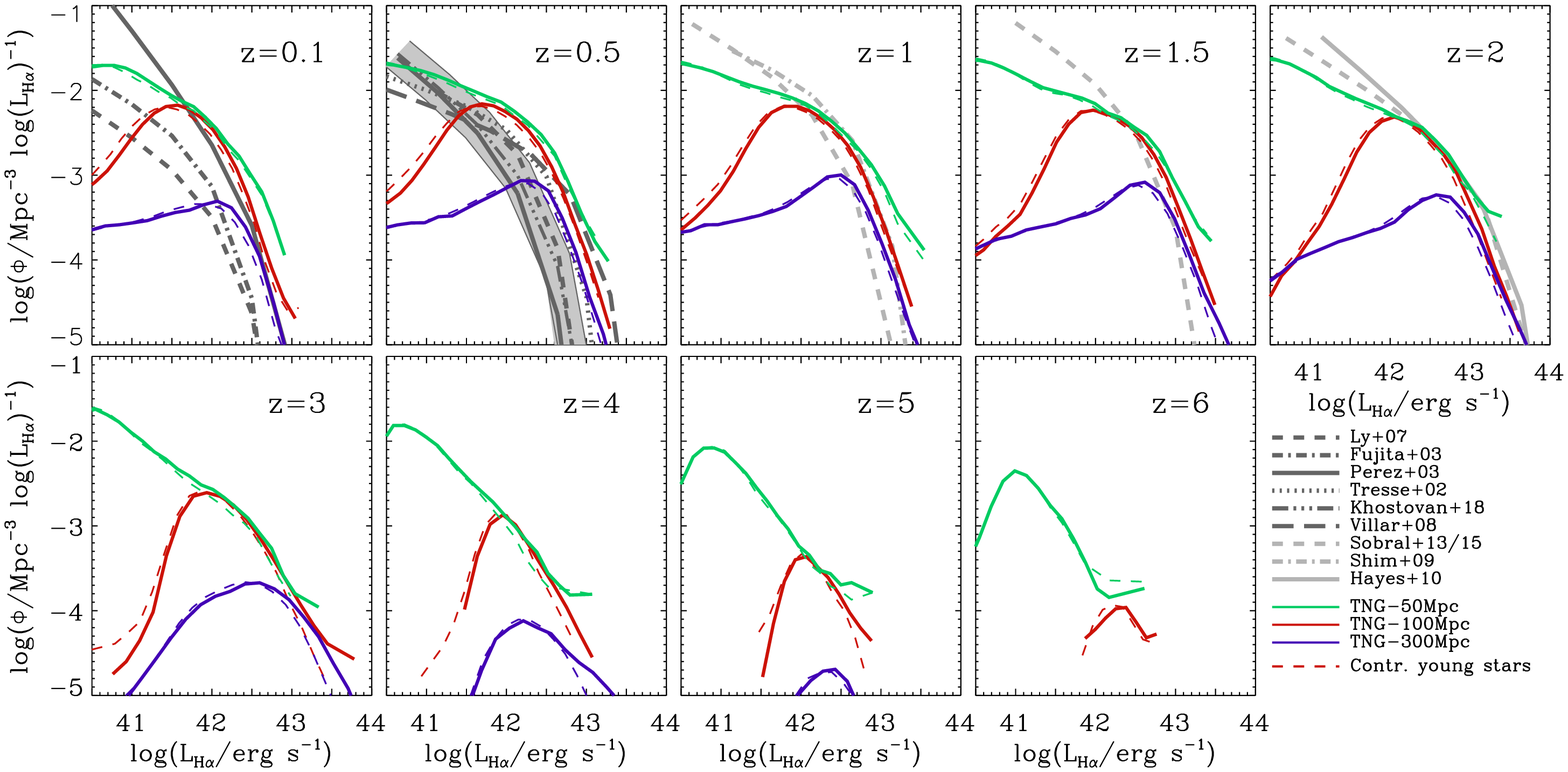,
  width=0.84\textwidth}
\epsfig{file=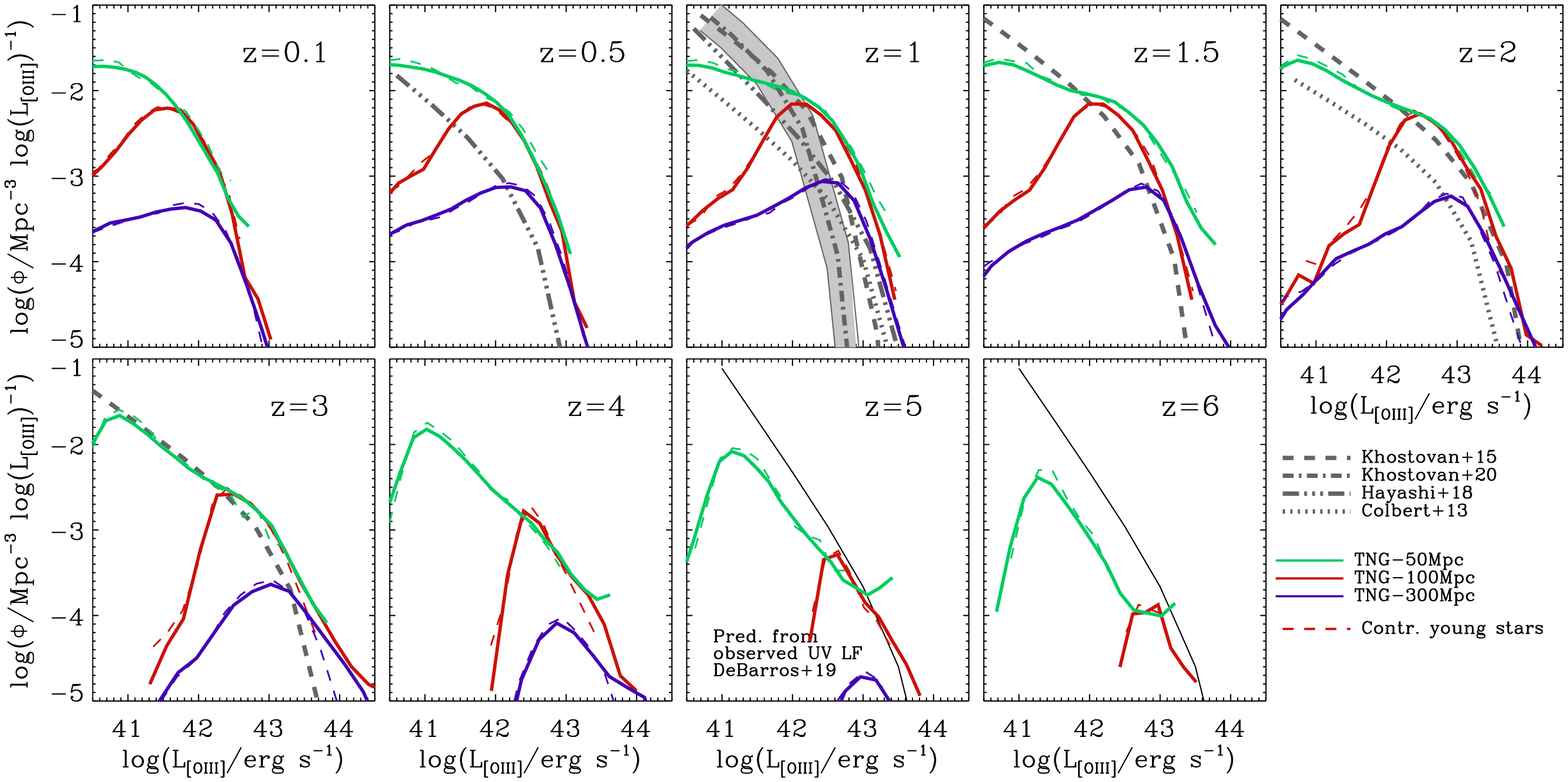,
  width=0.84\textwidth}
\epsfig{file=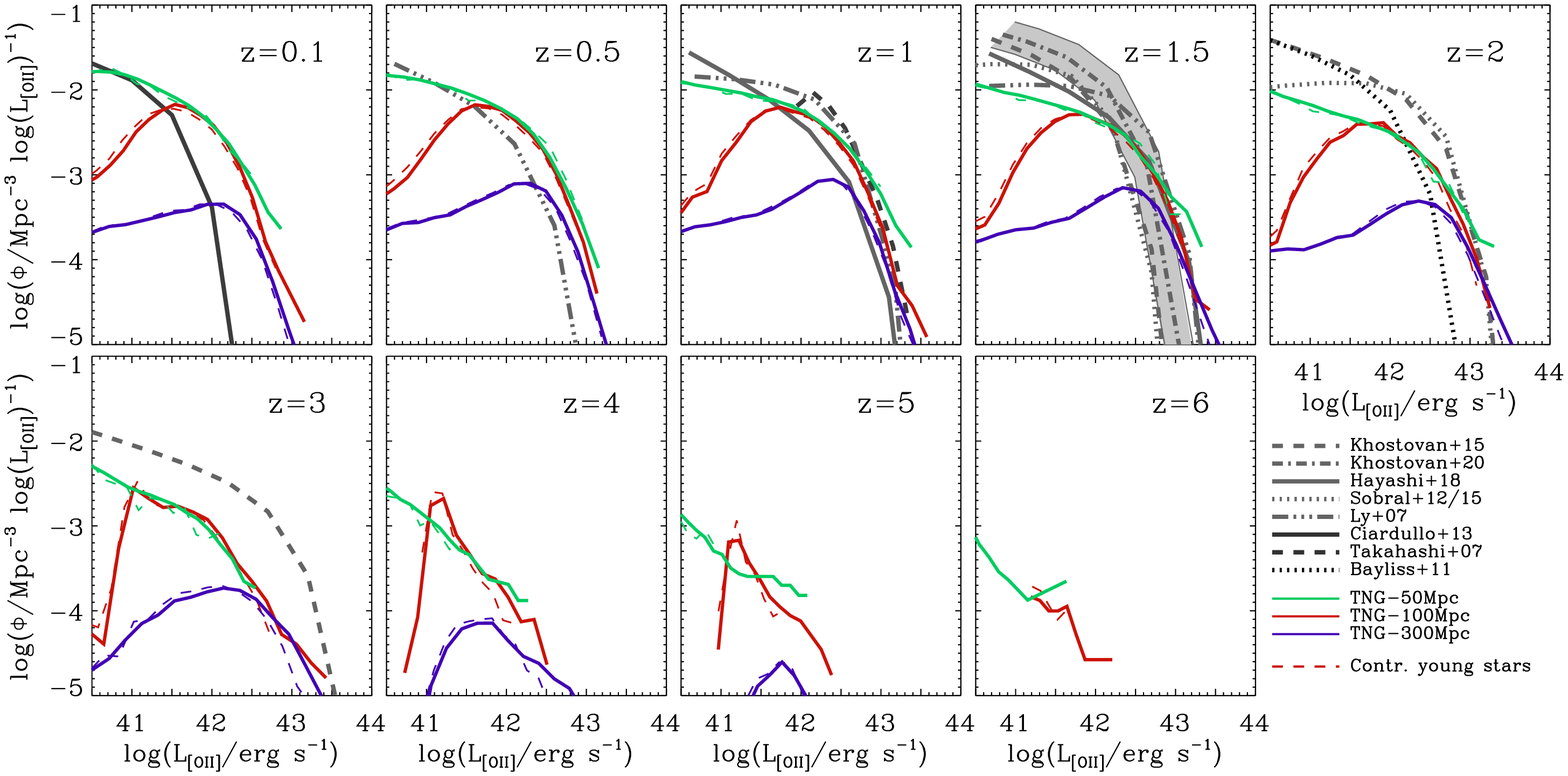,
  width=0.84\textwidth}
\vspace{-0.2cm}
\caption{\ha-\ (top two rows),
 \oiiinl+\hb- (middle two rows) and \oiinl- (bottom two rows) luminosity functions
of TNG50 (green solid lines), TNG100 (red solid
lines) and TNG300 (lilac solid lines) galaxies more massive than
$10^8\, \Msun$, $3 \times 10^9 \, \Msun$ and $10^{10} \,  \Msun$,
respectively, at $z=0.1,$ 0.5, 1, 1.5, 2, 
3, 4, 5 and 6 (different panels, as indicated). Dashed coloured lines 
show the results obtained when considering only the contribution by 
\hii\ regions to the luminosity of each galaxy. 
Also shown are observational
data from various sources listed in the legend (all corrected for
attenuation by dust; different grey- and black-line styles). The 
peaks in the TNG100 and TNG300 luminosity functions arise 
from the resolution limits; galaxy number densities below 
the peak are incomplete and hence not meaningful.} 
\label{Halum}     
\end{figure*}

%*****************************************************************************************************
%*****************************************************************************************************
\subsection{Evolution of optical-line luminosity
  functions}\label{opticallf} 
%*****************************************************************************************************
%*****************************************************************************************************

\subsubsection{Relation between \ha, \oiii and \oii\ luminosities and
  SFR for IllustrisTNG galaxies}

%Before investigating the evolution of the \ha, \oiii and \oii\ 
%luminosity functions,
We start by exploring how optical-line luminosities relate to the 
SFR for IllustrisTNG galaxies.
Fig.~\ref{sfr_lopt} shows the mean relations between \ha, \oiii\ and
\oii\ luminosities (from top to bottom) and SFR for TNG100 and 
TNG50 galaxies at $z=0$ (solid and dashed mauve lines, 
respectively). In the case of TNG100, we distinguish between 
different types of galaxy populations (blue: SF-dominated; red: composite; 
green:  AGN-dominated; orange: PAGB-dominated; lilac: shock-dominated). 
%\SC{Here I have kept the same type-color coding as in 
%all previous figures so far.} \MH{I changed the colors in the figure accordingly.}
Also shown for completeness are relations often used in the literature
(from \citealt{Kennicutt98} for \ha\ and \oii, and \citealt{Sobral15} for 
\oiii; black solid line in each panel). These should not be compared in 
detail with our predictions, as they were derived from purely solar-metallicity, 
SF-galaxy models assuming a \citet{1955ApJ...121..161S} IMF truncated at 0.1 and 1\Msun.
 
The good correlation between line luminosity and SFR for 
SF-dominated galaxies in Fig.~\ref{sfr_lopt} confirms that
\ha, \oiii\ and \oii\ can provide rough estimates of the SFR in such
galaxies. For other galaxy types, contributions to line emission from an AGN, 
shocks and PAGB stars shift the relations to larger luminosities at fixed 
SFR. Hence, the ability to constrain galaxy type (Section~\ref{opticaldiagrams_evol})
is important to refine SFR estimates from optical-line luminosities.

\subsubsection{\ha\ luminosity function}

The top two rows of Fig.~\ref{Halum} show the evolution of the 
\ha\ luminosity function of TNG50 (green lines), TNG100
(red lines) and TNG300 (lilac lines) galaxies over the redshift 
range $1\leq z\leq6$ (different panels, as indicated). It is worth
pointing out that the peaks in the TNG100 and TNG300 
luminosity functions are an artefact of the resolution limit, 
which causes galaxy number densities below the peak to be 
incomplete and hence not meaningful. The highest 
resolution of the TNG50 simulation enables us to follow 
the evolution of the faint end down to luminosities below $\sim10^{41}$\,erg\,s$^{-1}$,
and the larger cosmological volumes of the TNG100 and TNG300
simulations enable quantification of the evolution of the exponential cutoff at luminosities near
$\sim10^{43}$\,erg\,s$^{-1}$.

At any redshift, the bulk of the \ha\ luminosity in Fig.~\ref{Halum} arises
from \hii\ regions ionized by young stars (coloured dashed lines).
Only in the most luminous \ha\ emitters, with $\lha \ga10^{43}$\,erg\,s$^{-1}$,
can the contribution from an AGN become significant. This makes the \ha\
luminosity function a reliable tracer of the cosmic SFR density of galaxy 
populations, as often assumed in observational studies \citep[e.g.,][]{Ly07, 
Sobral13, Sobral15}. In fact, Fig.~\ref{Halum} 
shows that, as a consequence of the rising cosmic SFR density of
IllustrisTNG galaxy populations from $z=0.1$ to $z=3$, the simulated 
\ha\ luminosity function strongly evolves over this redshift range: while 
the faint end hardly changes, the exponential cutoff rises from 
$\lha \ga 10^{42}$\,erg\,s$^{-1}$ to $\lha \ga10^{43}$\,erg\,s$^{-1}$.

The predicted \ha\ luminosity functions in Fig.~\ref{Halum} are in 
reasonable agreement with observational determinations (including 
corrections for dust attenuation) from various
deep and wide, narrow-band and spectroscopic surveys conducted
at $z<2$ \citep[shown as different grey lines]{Tresse02, 
Fujita03, Villar08, Shim09, Hayes10, Sobral13, Sobral15,
Khostovan18}. The large dispersion in these constraints illustrates
the uncertainties inherent in their derivation (e.g., cosmic 
variance, corrections for dust attenuation, filter profile, 
completeness, etc.). In this respect, the constraints from the 
HiZELS survey \citep[][light-grey dashed lines]{Sobral13, Sobral15} 
are of particular interest, as they provide homogeneous estimates
of the \ha\ luminosity function over the redshift range $1\leq z\leq2$.
These are consistent with the evolution predicted by our models,
except at the faint end, where resolution effects and the stellar-mass
cut at $10^8\,\Msun$ in the TNG50 simulation are likely to 
account for the shallower predicted slope.
%% Perez03, ??

At redshifts $z>2$, where future observational constraints will be
gathered by {\it JWST}, {\it Euclid} and the {\it Roman Space Telescope},
the galaxy number density at fixed \ha\ luminosity is predicted to 
strongly decline, e.g., by a factor of $\sim30$ between $z=3$ and $z=6$ 
at $\lha = 10^{42}$\,erg\,s$^{-1}$. While this reflects the decline 
of the cosmic SFR density over this redshift range in the IllustrisTNG 
simulations, these predictions should be taken with caution, because of 
the growing limitation by resolution effects affecting the 
census of small galaxies at high redshift.

\subsubsection{\oiiinl+\hb\ luminosity function}

The middle two rows of Fig.~\ref{Halum} show the evolution 
of the \oiiinl+\hb\ luminosity function of the TNG50,
TNG100 and TNG300 galaxies, using the same layout 
as for the \ha\ luminosity function in the top two rows. Here, the \oiiinl+\hb\ 
luminosity is taken to be the sum of the \oiiiaur, \hb\ and \oiii\ luminosities, which 
are hard to separate in narrow-band surveys. We note that the predicted
\oiiinl+\hb\ luminosity function computed in this way differs noticeably from 
the pure \oiii\ luminosity function only at the faint end.  

As in the case of the \ha\ line, Fig.~\ref{Halum} shows that the 
\oiiinl+\hb\ luminosity function is dominated by \hii\
regions ionized by young stars at all redshifts. Only at
$z=2$--3 do luminosities above $\loiiihb\sim 10^{43.5}$\,erg\,s$^{-1}$ 
start to be dominated by line emission from AGN. The \oiii+\hb\ luminosity 
function also exhibits strong evolution with redshift:
even though the faint end hardly changes from $z=0.1$ to
$z=3$, the exponential cutoff rises from $\loiiihb \sim 10^{41.5}$\,erg\,s$^{-1}$ at
$z=0.1$ to $\loiiihb > 10^{43}$\,erg\,s$^{-1}$ at $z=3$, in fair agreement with
the bulk of observational constraints \citep[][]{Colbert13,
  Hayashi18, Khostovan15, Khostovan20}.

The only constraint we could find on the \oiiinl+\hb\ luminosity function 
at redshift $z>3$ is that derived indirectly from the UV luminosity function 
at $z=8$ by \citet{DeBarros19}.
%[][which {\bf also includes} the minor contribution from 
%\oiiiaur]{DeBarros19}. \SC{Is the statement in parentheses
%  true?}\MH{No, they write in the abstract: '...which we use to derive for
%  the first time the \oiii+\hb', thus, I changed the
%  statement accordingly } 
For reference, we show this as a  thin, 
black solid line in the $z=5$ and $z=6$ panels of the fourth row of 
Fig.~\ref{Halum}. At $z=5$, our predictions are roughly consistent 
with this semi-empirical determination, which however lies $\sim0.5\,$dex above 
the predictions at $z=6$. The difference could arise from uncertainties in the 
derivation of this indirect constraint, as well as from resolution effects in the 
simulations at that redshift.

\subsubsection{\oiinl\ luminosity function}

The evolution of the \oii\ luminosity function, shown in the bottom two
rows of Fig.~\ref{Halum}, has characteristics similar to that of the \oiiinl+\hb\
luminosity function. It is largely dominated by the contribution from \hii\ 
regions ionized by young stars and shows strong evolution of the 
exponential cutoff, from $\loii \sim 10^{42.5}$\,erg\,s$^{-1}$ at $z=0.1$ 
to $\loii \sim 10^{43}$\,erg\,s$^{-1}$ at $z=2$. At redshifts
$z = 1$--1.5, the predicted evolution is in reasonable agreement with
various empirical determinations \citep{Khostovan15, Khostovan20, 
Hayashi18, Sobral15, Ly07, Takahashi07, Bayliss11}. At lower 
redshift, the number density of \oiinl-luminous emitters lies above those
estimated by \citet{Ciardullo13} and \citet{Ly07}, while at higher redshift, 
it lies below that estimated by \citet[][]{Khostovan15}. 

Overall, we conclude that the luminosity functions of the \ha, \oiiinl+\hb\ and 
\oiinl\ emission lines predicted by the IllustrisTNG simulations are in 
reasonable agreement with empirical determination at redshifts from 
$z=0$ to $z=2$--3,  although there are some discrepancies with the few 
existing constraints on the number densities of \oii\ emitters at $z<1$ and $z>1.5$.

%\begin{figure*}
%\center
%\epsfig{file=HbOIII_lumfct_TNG_allz.eps, width=0.95\textwidth}
%\caption{Evolution of the Hbeta+OIII luminosity function.}\label{Hblum}    
%\end{figure*}

\begin{figure}
\center
\epsfig{file=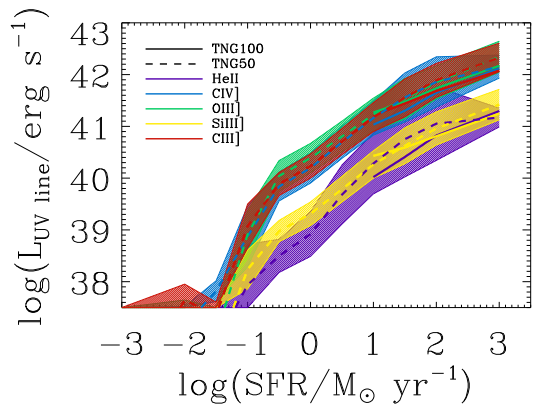, width=0.45\textwidth}
\caption{Average luminosities of different UV emission lines,
  \heii, \civ, \oiiiuv, \siliii\ and \ciii\  (coloured as indicated) 
plotted against SFR for galaxies at redshifts $z=4$--7 in the 
TNG100 (solid lines for $\rm SFR \ga10\,\Msun\,yr^{-1}$) and TNG50 (dashed lines, 
with shaded area indicating the 1$\sigma$ 
scatter) simulations.}\label{sfr_luv}    
\end{figure}
\begin{figure*}
  \centering 
\epsfig{file=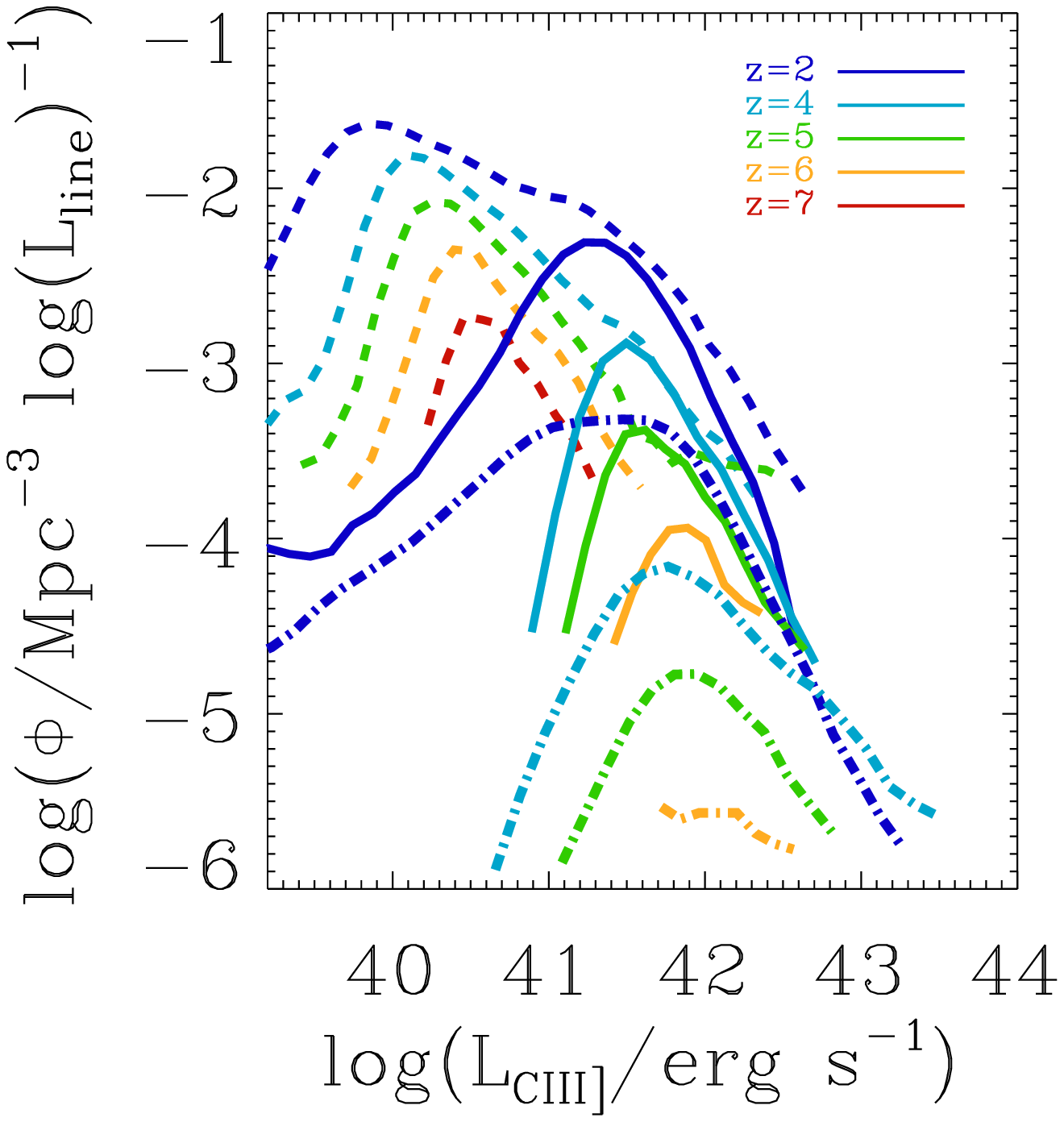,
  width=0.25\textwidth}\hspace{-1.3cm}
\epsfig{file=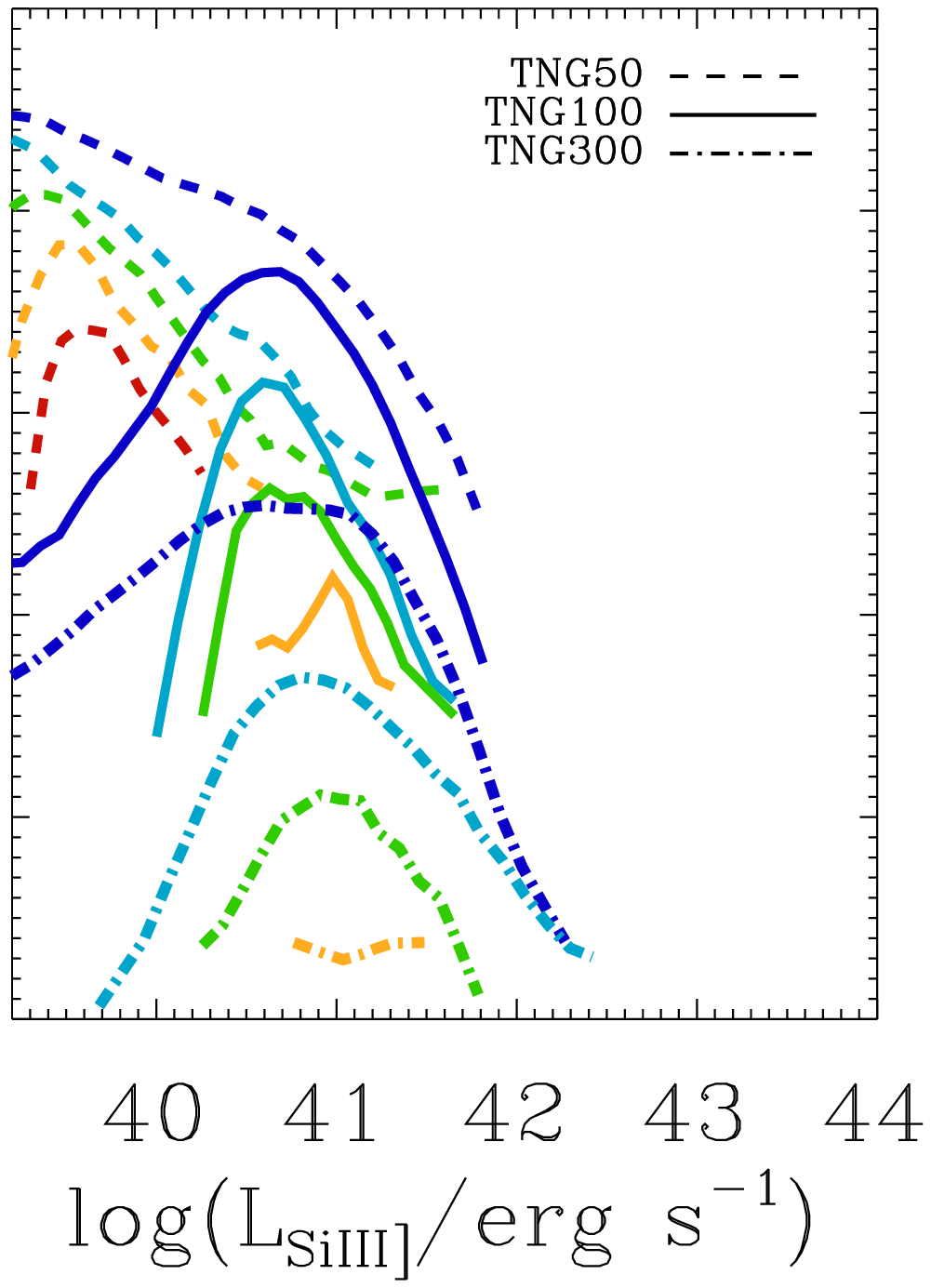,
  width=0.25\textwidth}\hspace{-1.3cm}
\epsfig{file=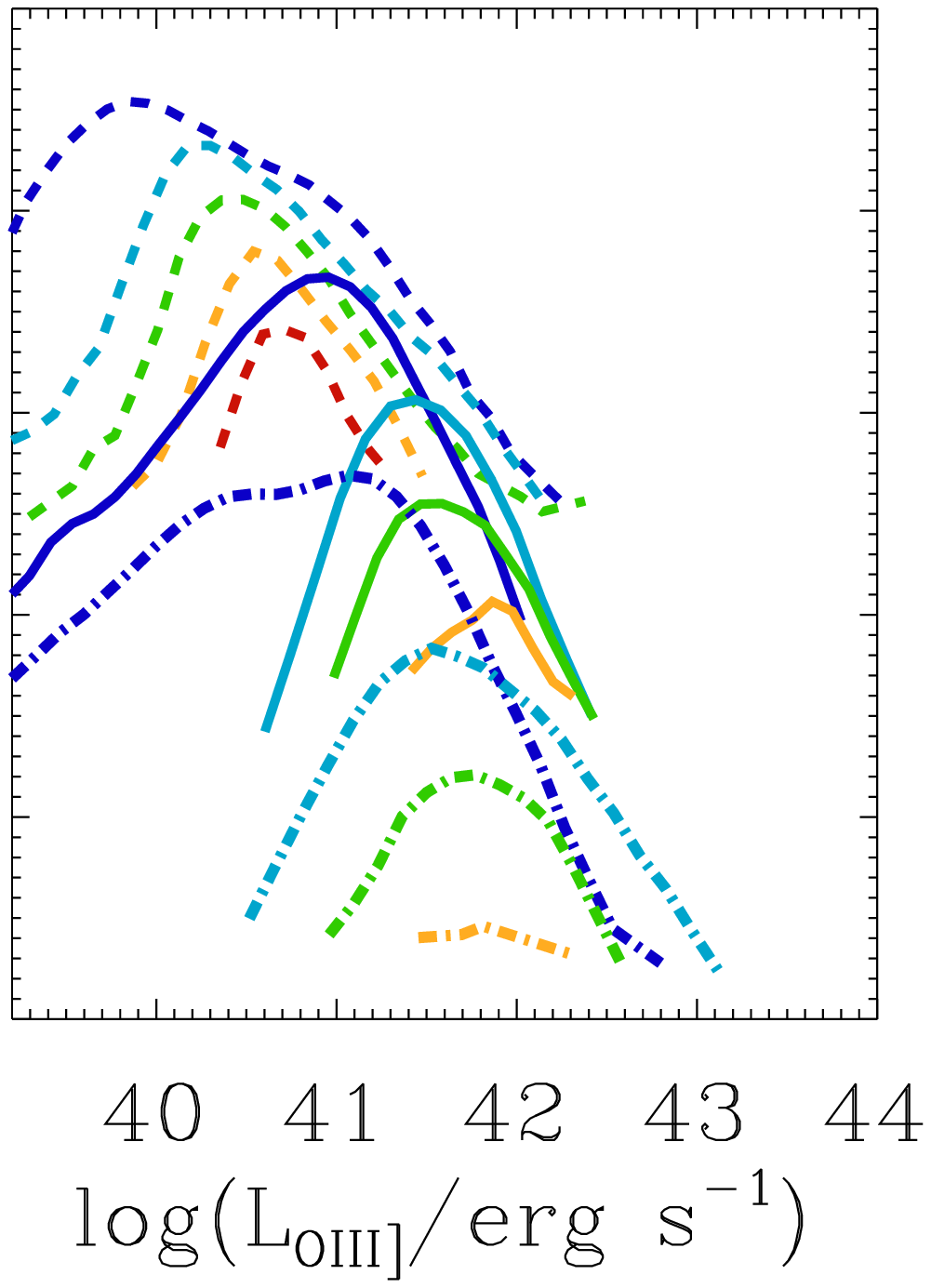,
  width=0.25\textwidth}\hspace{-1.3cm}
\epsfig{file=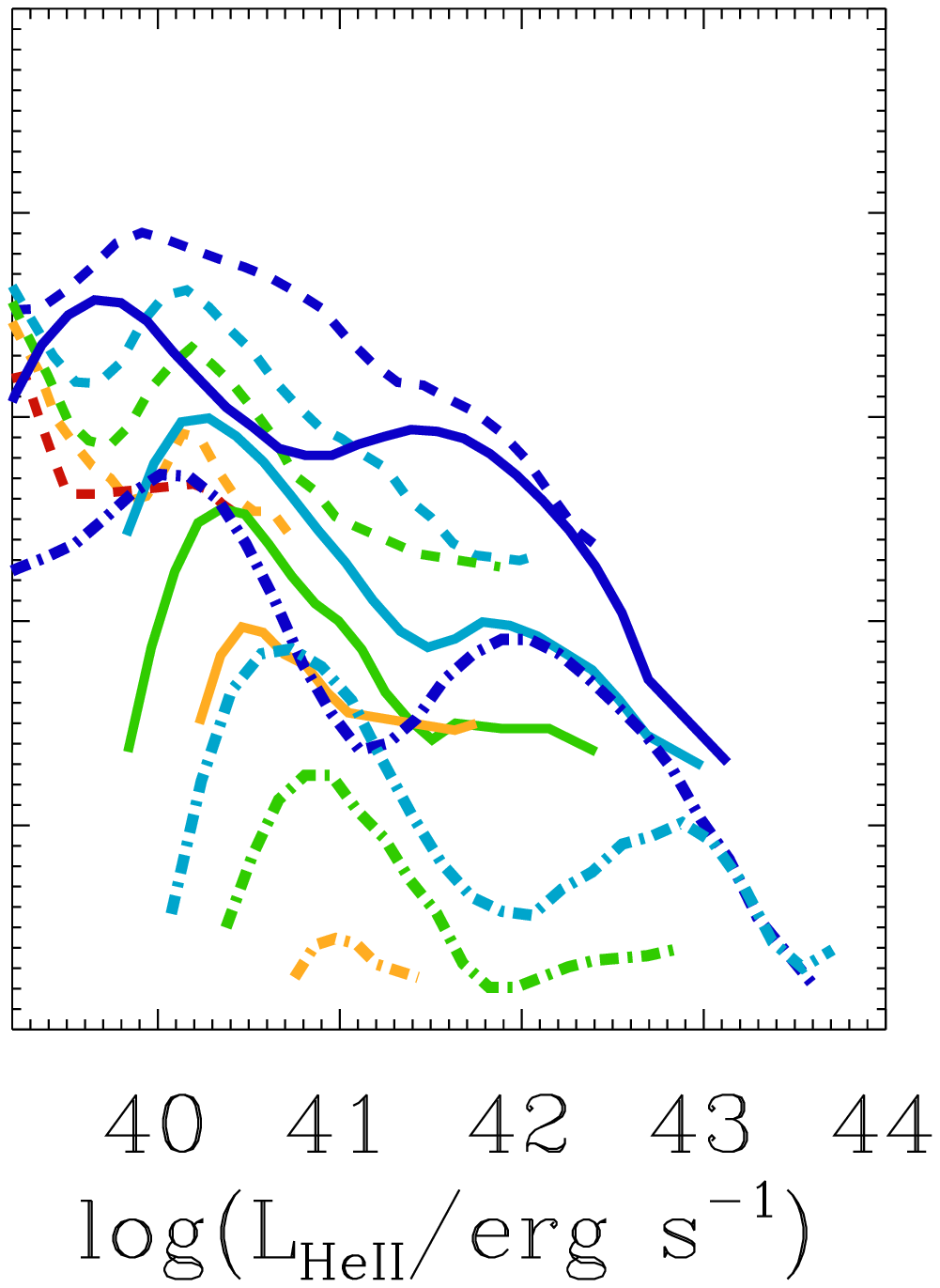,
  width=0.25\textwidth}\hspace{-1.3cm}
\epsfig{file=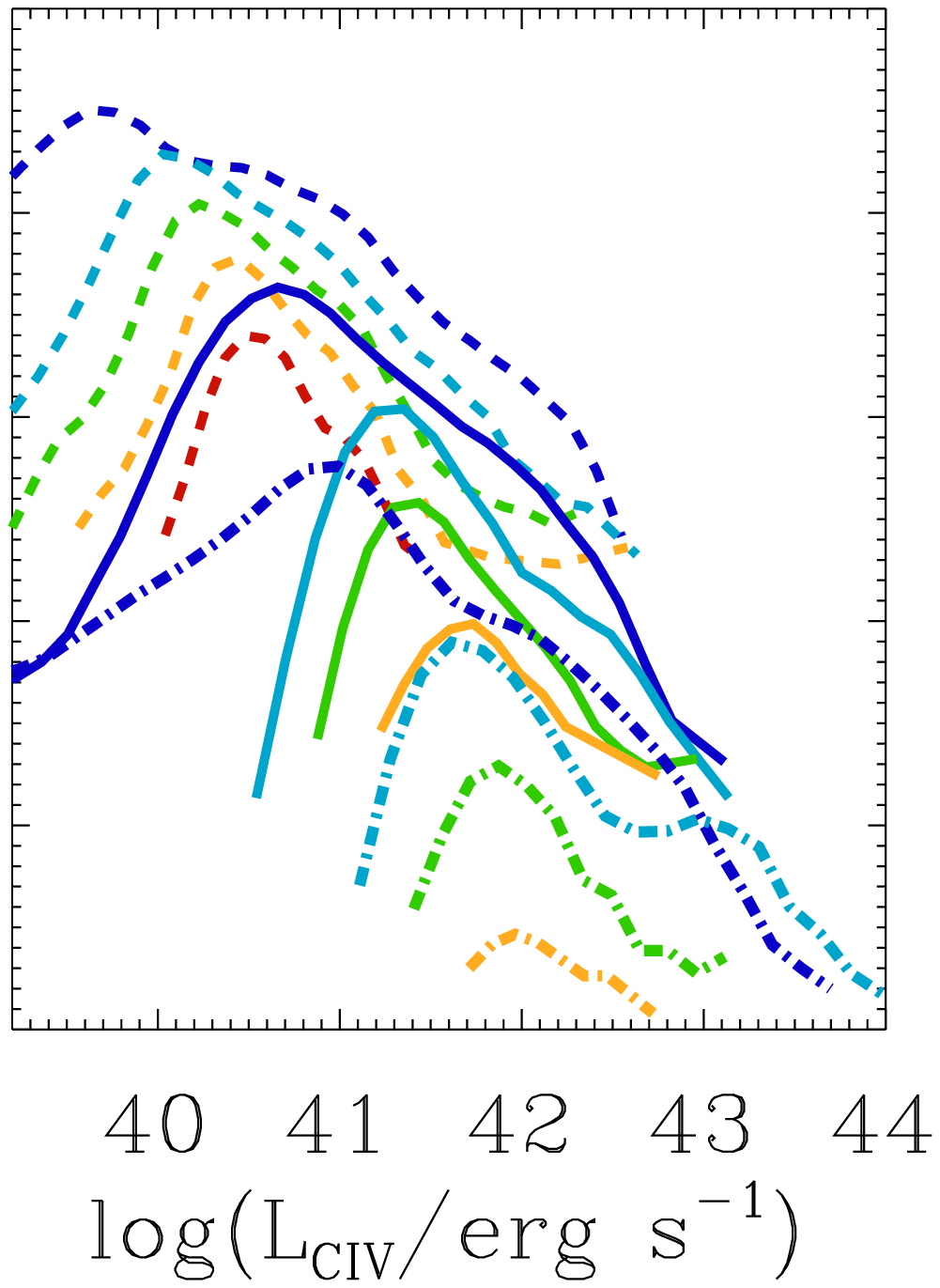,
  width=0.25\textwidth}
\epsfig{file=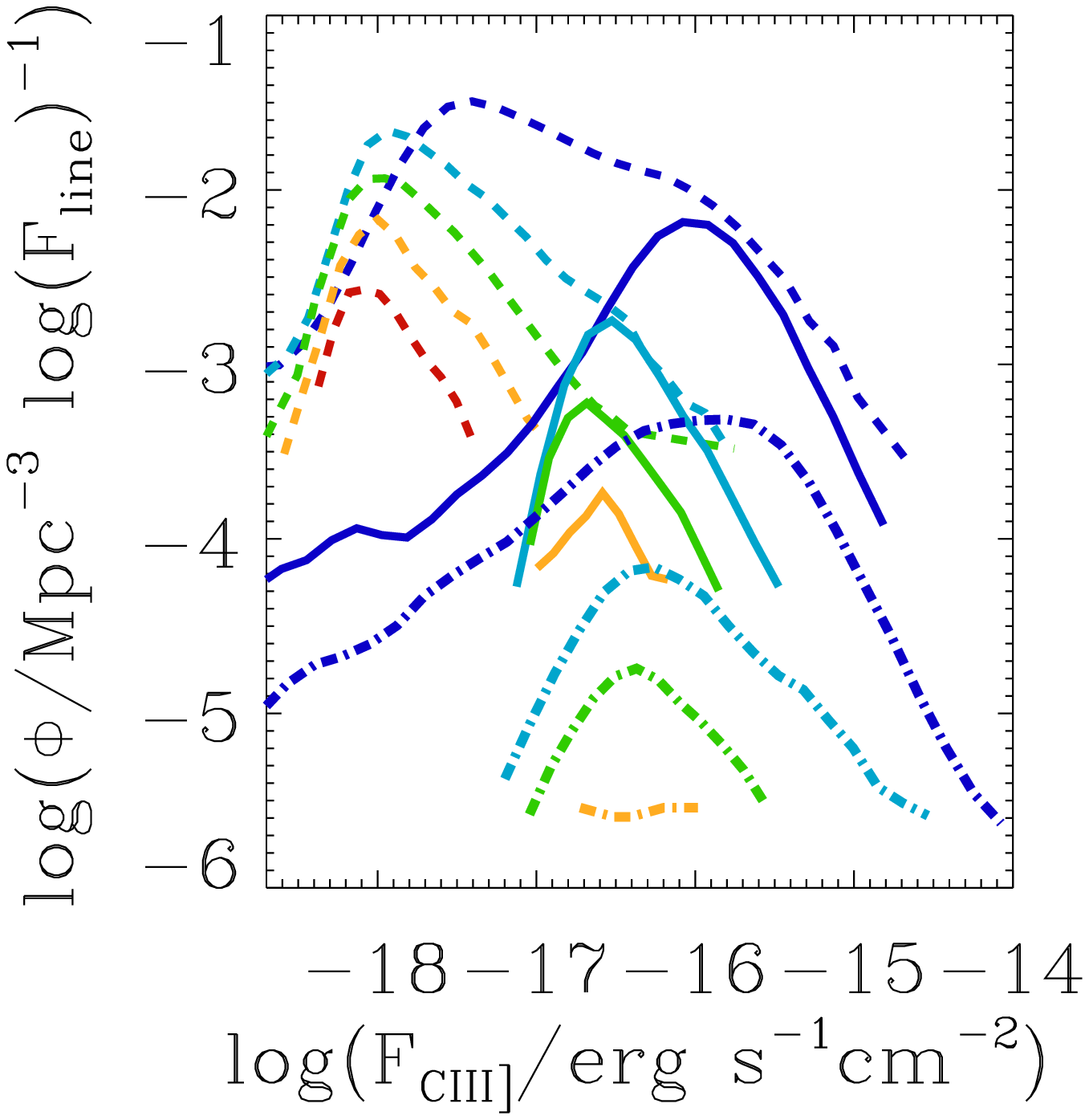,
  width=0.25\textwidth}\hspace{-1.3cm}
\epsfig{file=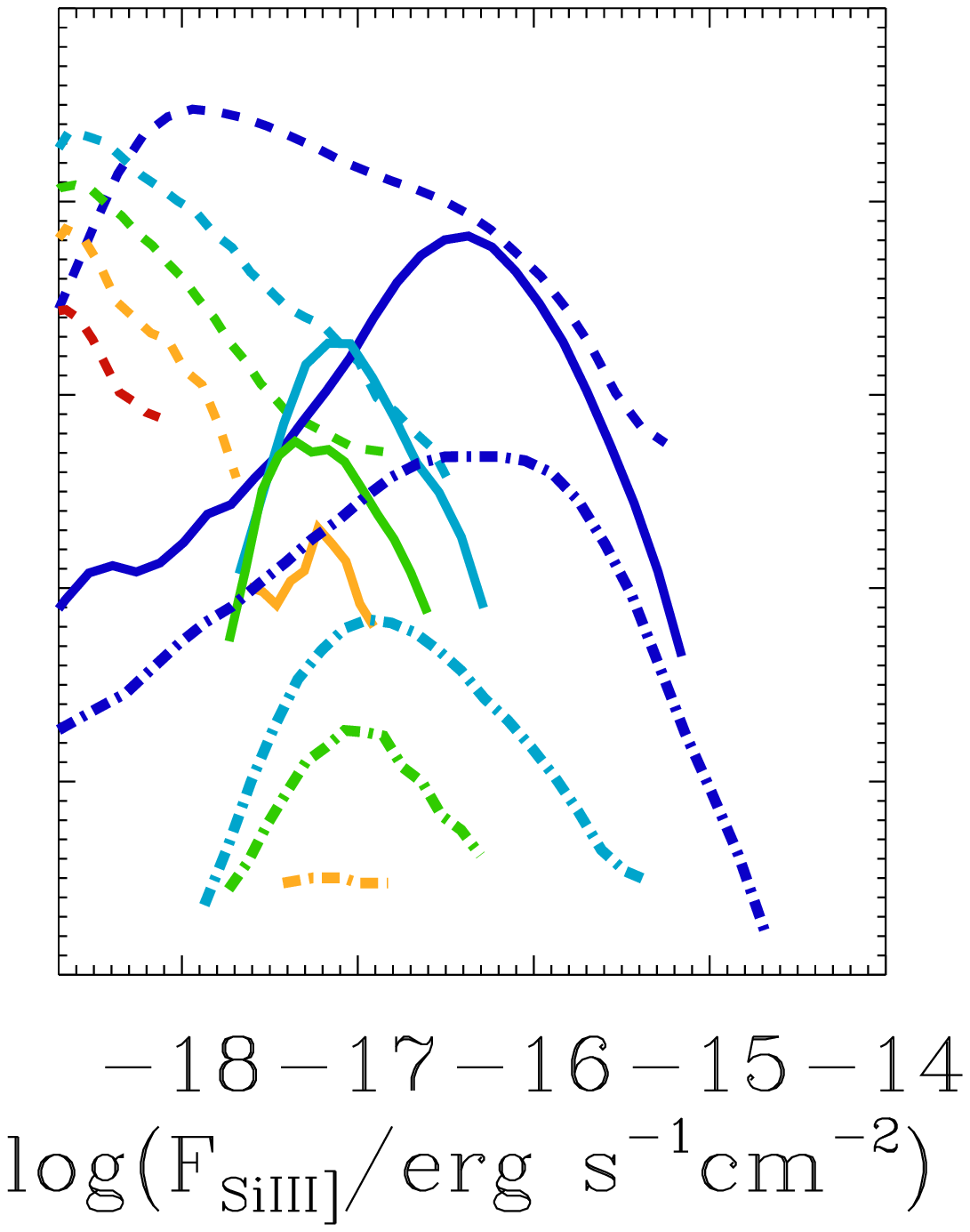,
  width=0.25\textwidth}\hspace{-1.3cm}
\epsfig{file=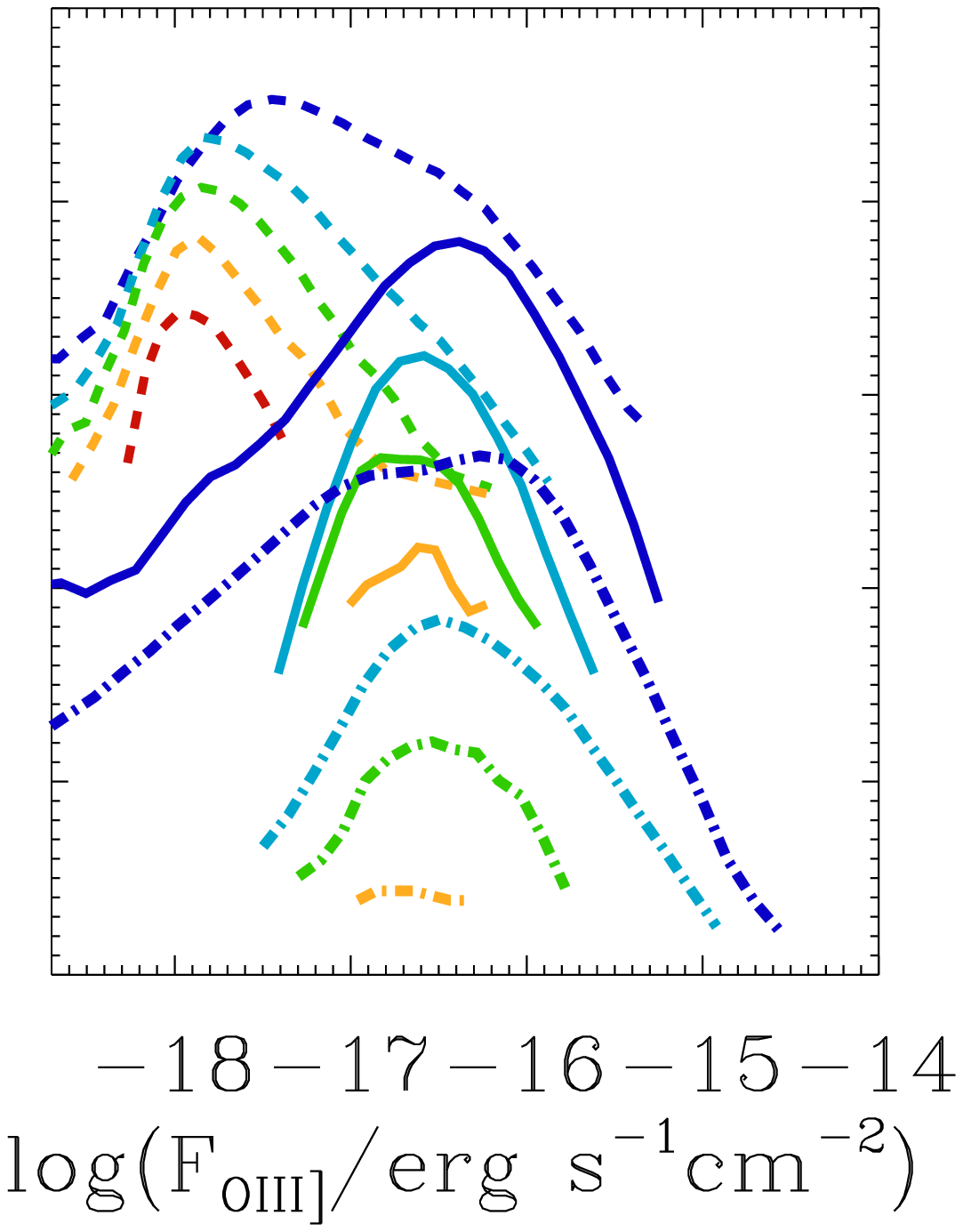,
  width=0.25\textwidth}\hspace{-1.3cm}
\epsfig{file=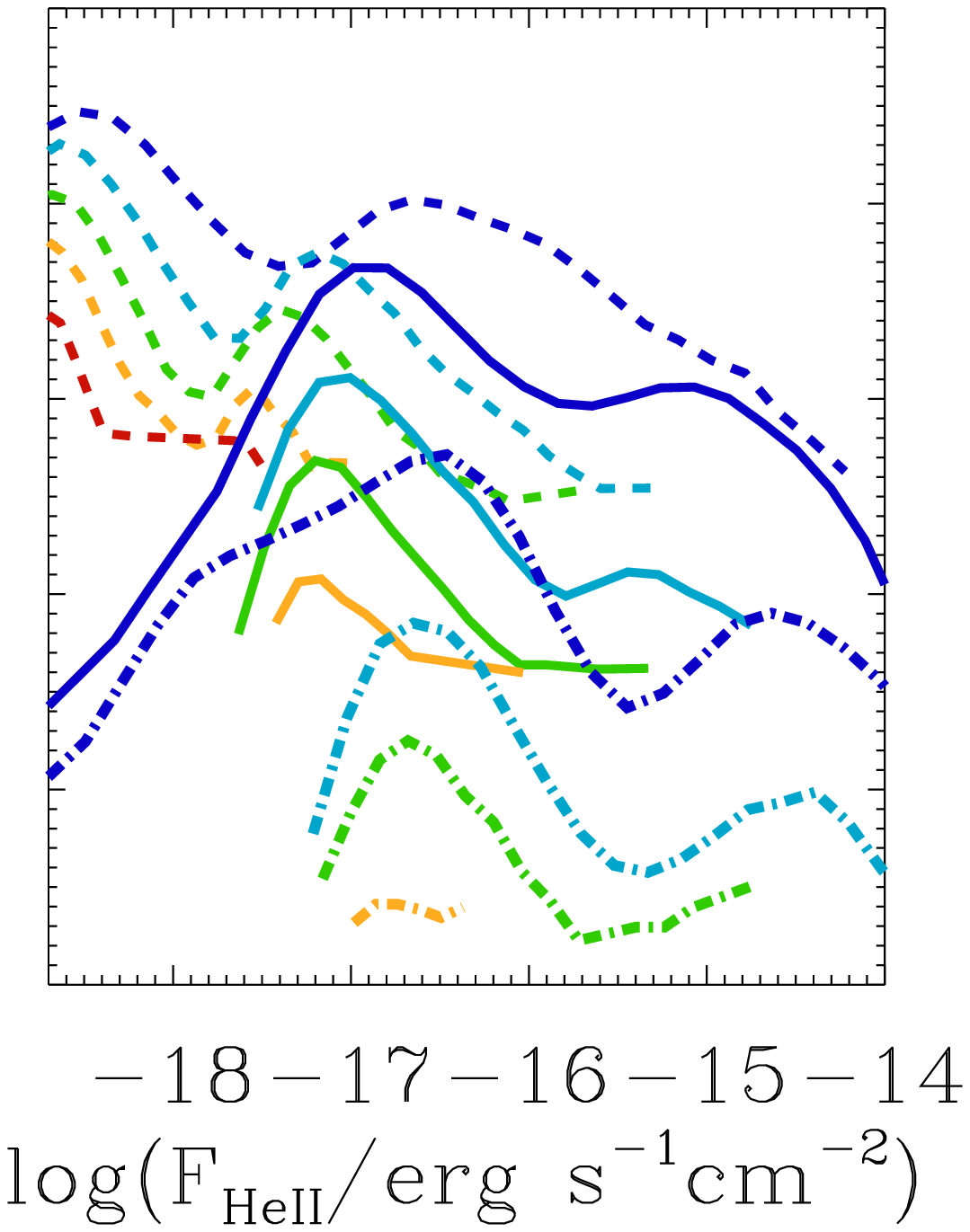,
  width=0.25\textwidth}\hspace{-1.3cm}
\epsfig{file=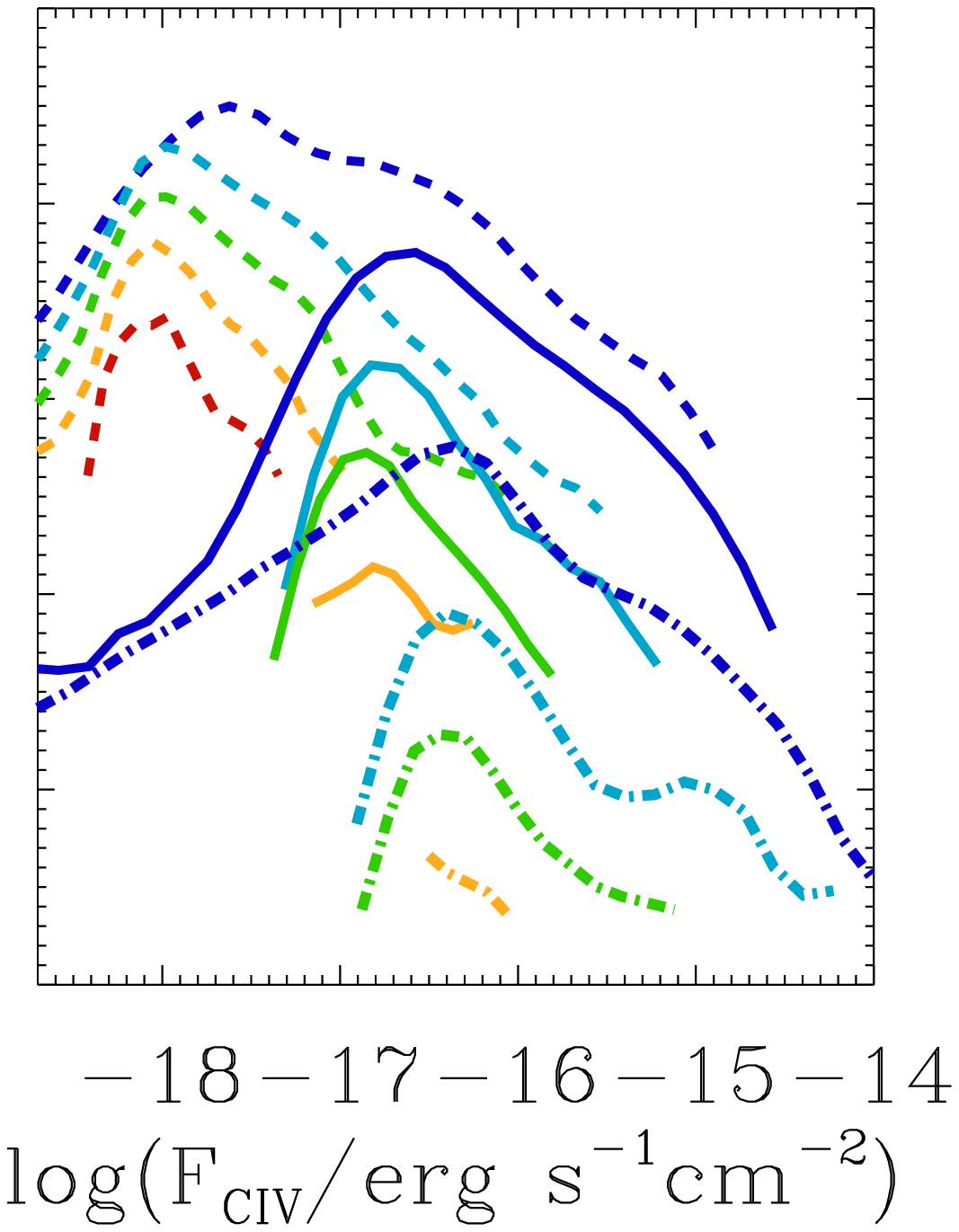,
  width=0.25\textwidth}
\caption{\ciii, \siliii, \oiiiuv, \heii\ and \civ\  luminosity functions (panels from 
left to right, top row) and flux functions (bottom row) of TNG50 
(dashed lines), TNG100 (solid lines) and TNG300 
(dot-dashed lines) galaxies more massive than $10^8\, \Msun$, 
$3 \times 10^9 \, \Msun$ and $10^{10} \,  \Msun$, respectively, at 
$z=2$, 4, 5, 6 and 7 (different colours, as indicated).}\label{UVLumfct}    
\end{figure*}
\begin{figure*}
  \centering{\bf Two-hour exposure time}
  
\epsfig{file=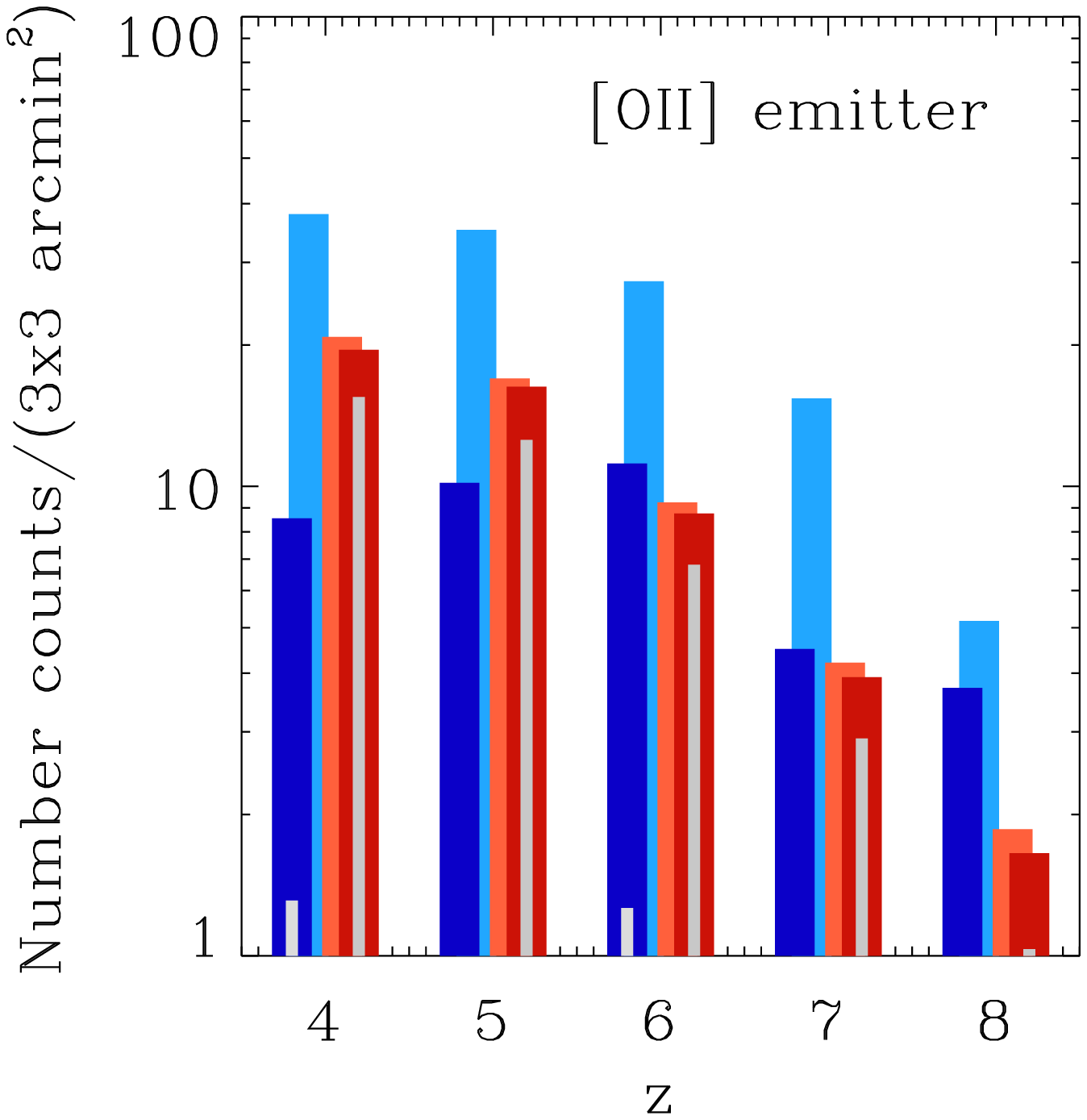,
  width=0.25\textwidth}\hspace{-1.4cm}
\epsfig{file=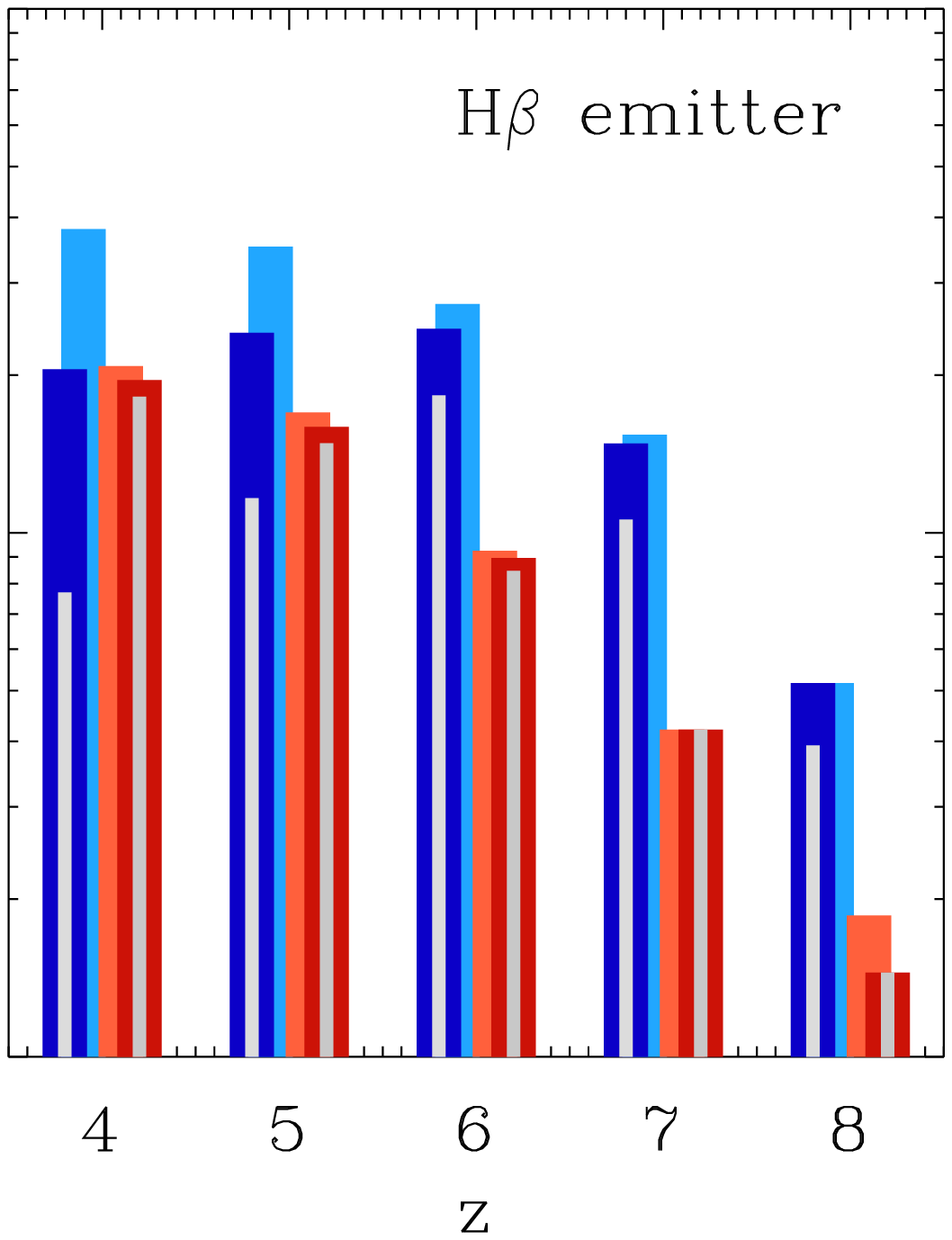,
  width=0.25\textwidth}\hspace{-1.4cm}
\epsfig{file=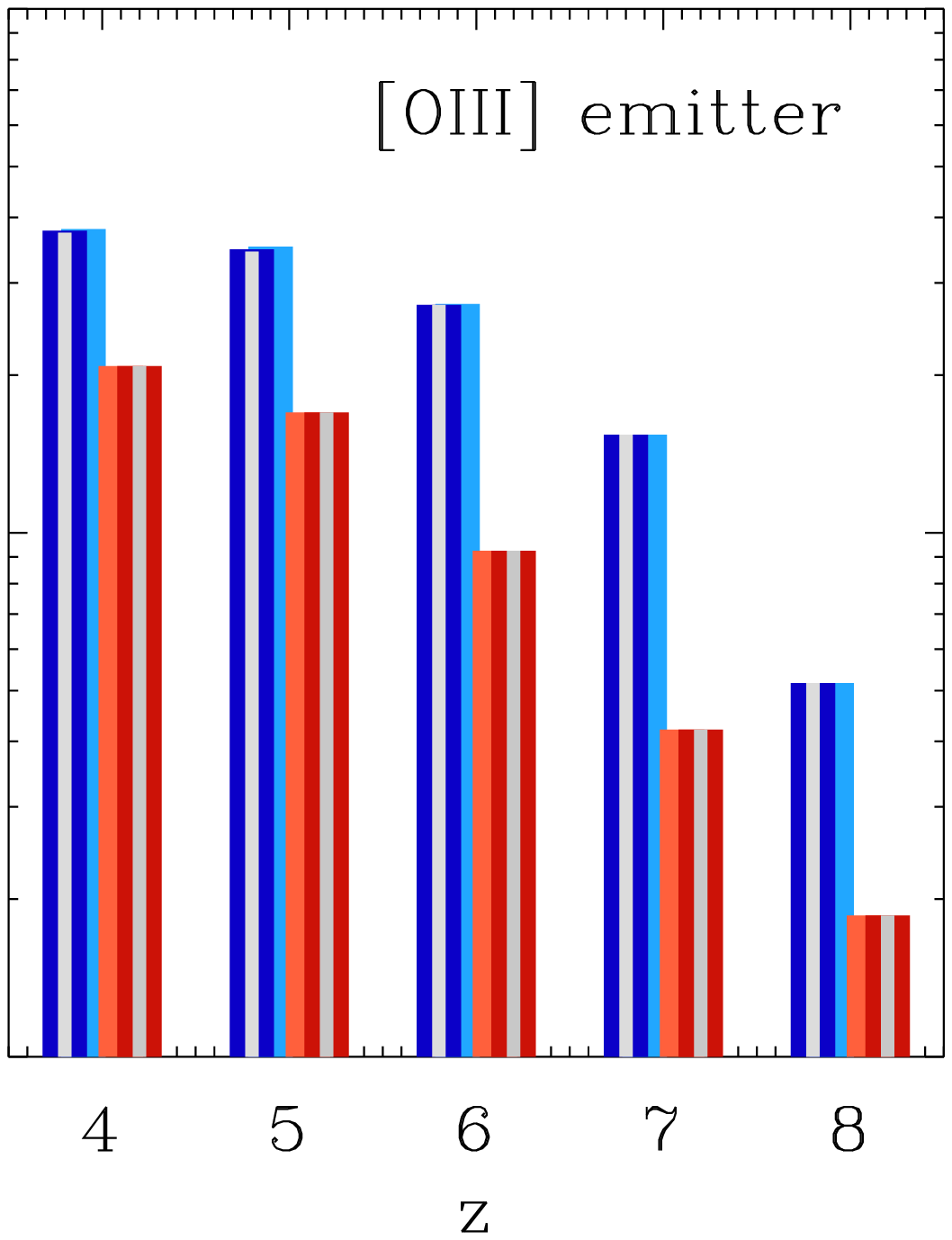,
  width=0.25\textwidth}\hspace{-1.4cm}
\epsfig{file=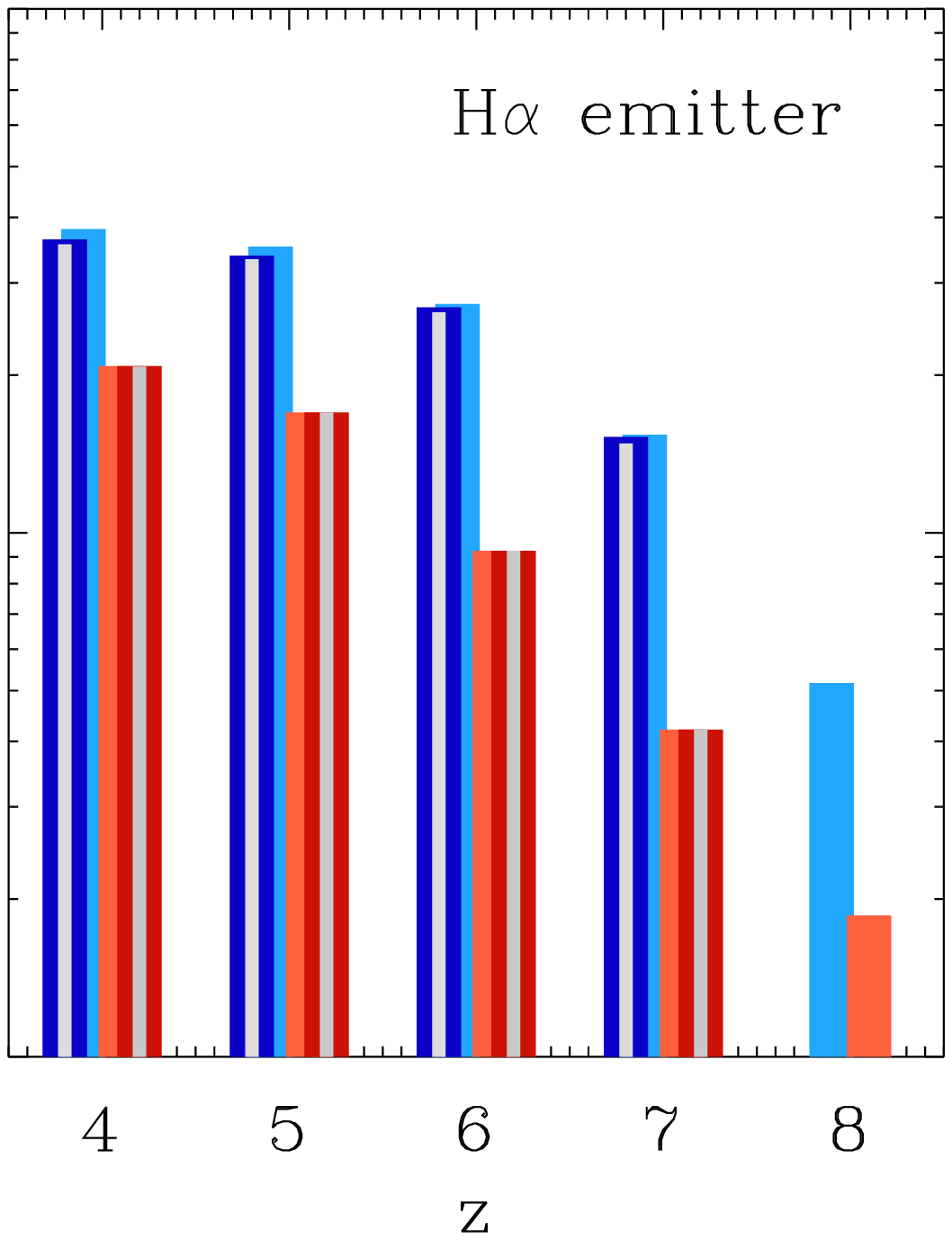,
  width=0.25\textwidth}\hspace{-1.4cm}
\epsfig{file=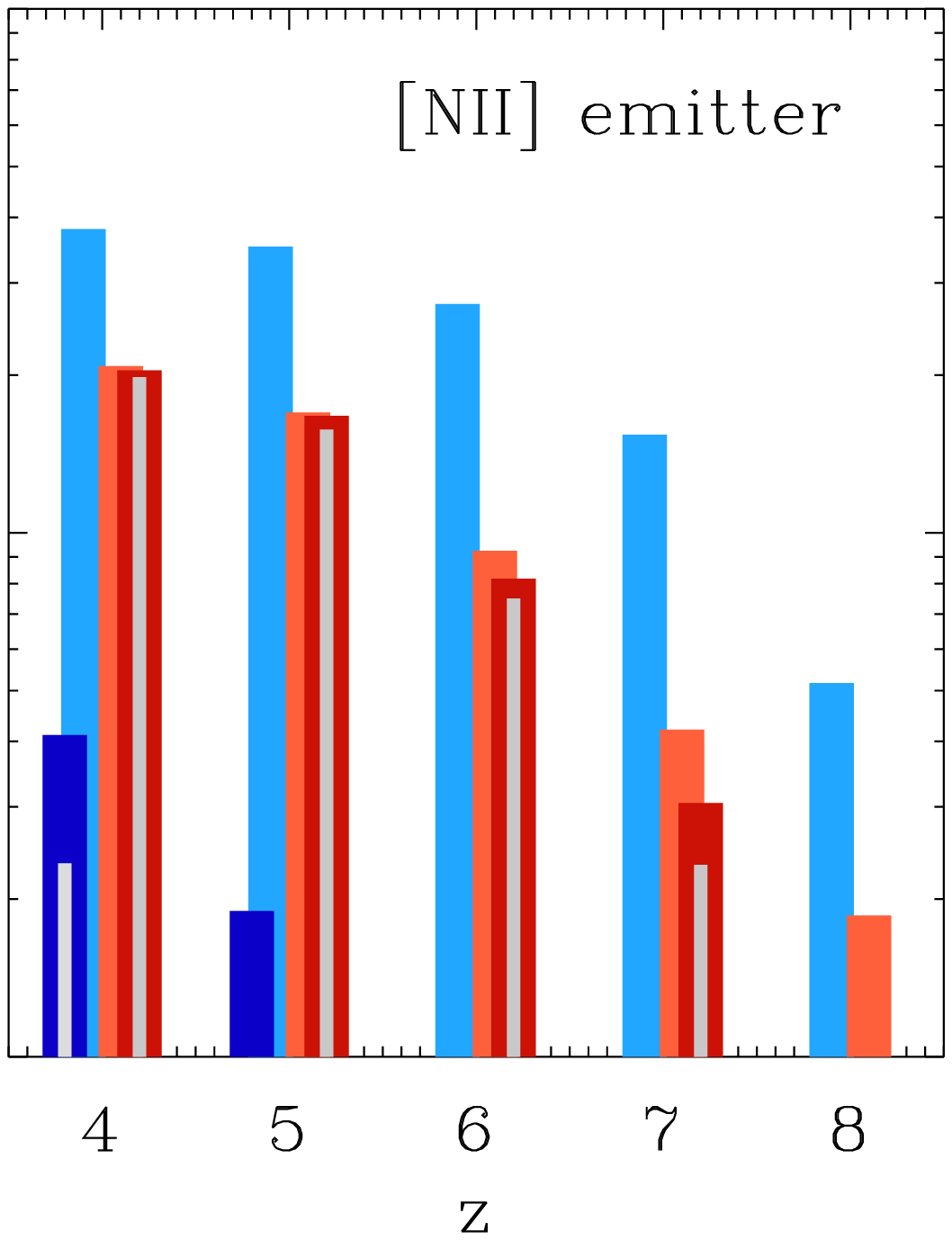,
  width=0.25\textwidth}

\centering{\bf Five-hour exposure time}

\epsfig{file=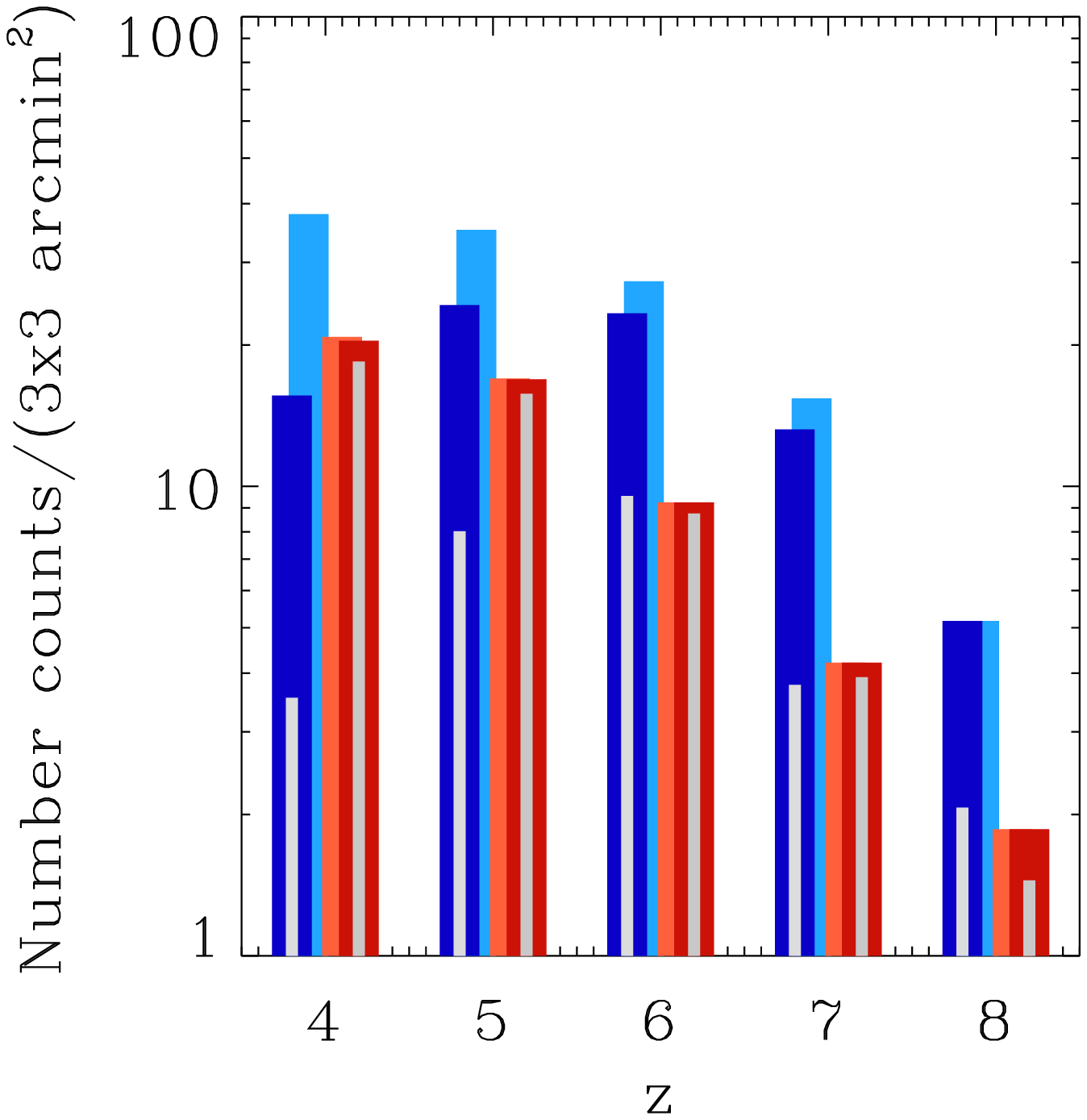,
  width=0.25\textwidth}\hspace{-1.4cm}
\epsfig{file=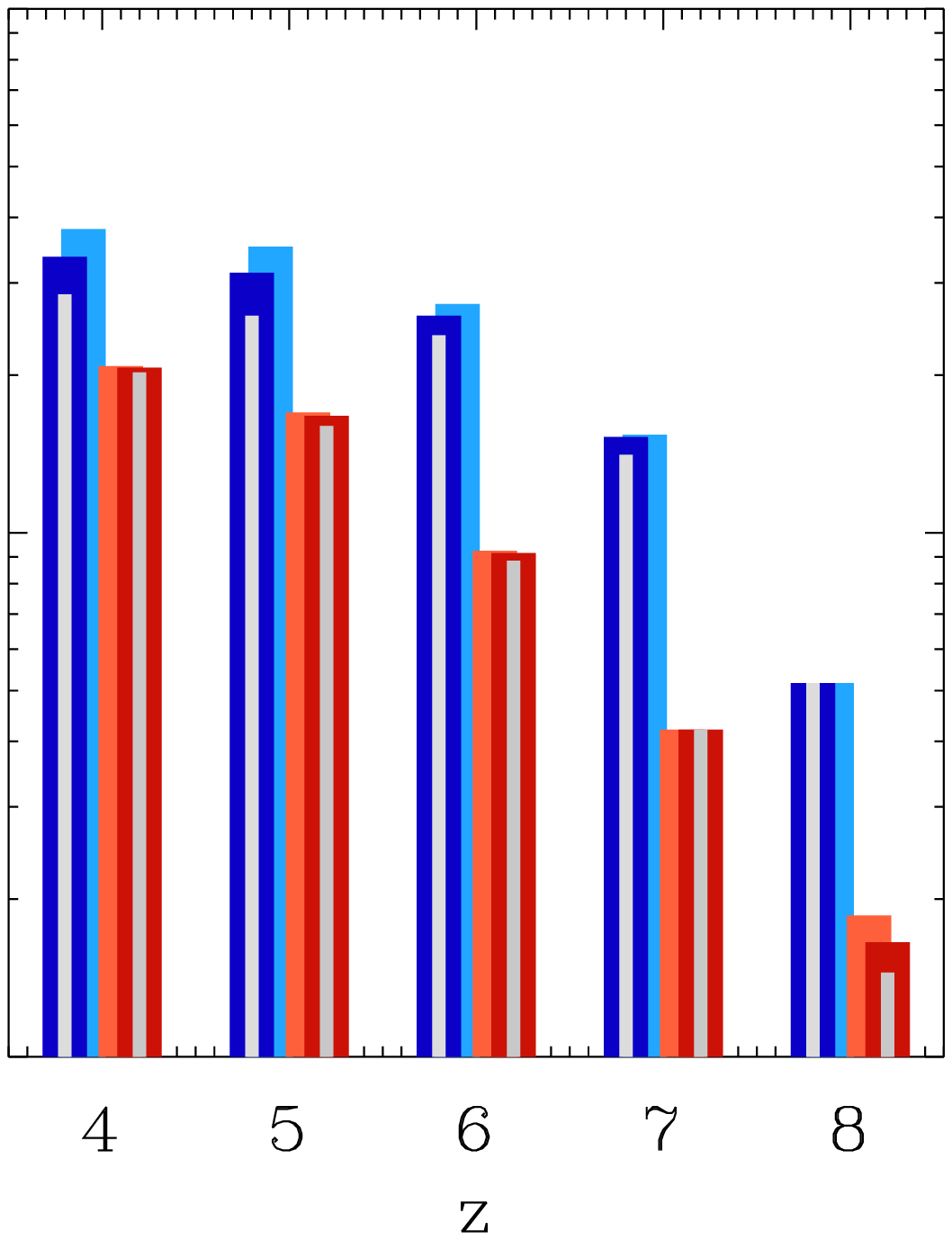,
  width=0.25\textwidth}\hspace{-1.4cm}
\epsfig{file=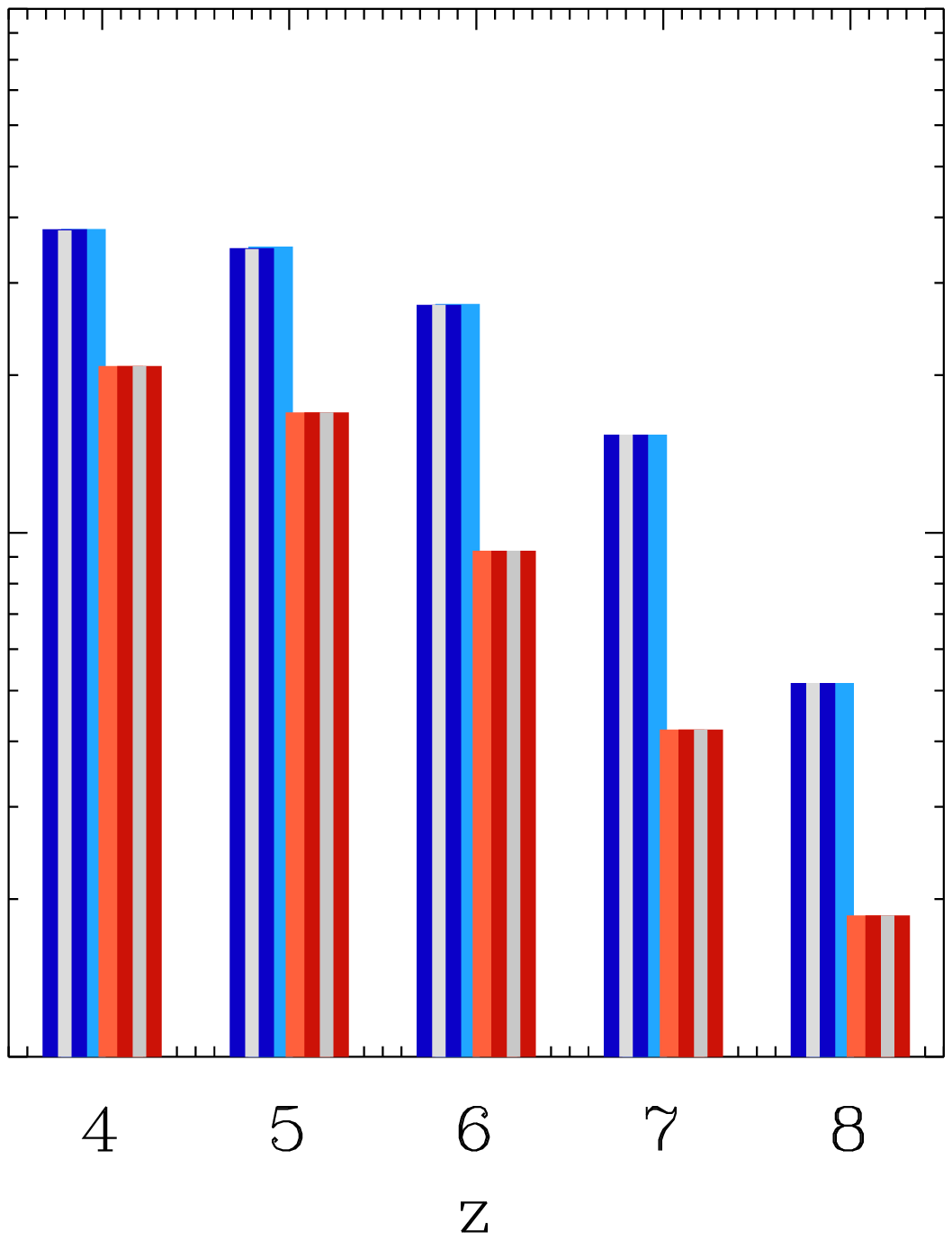,
  width=0.25\textwidth}\hspace{-1.4cm}
\epsfig{file=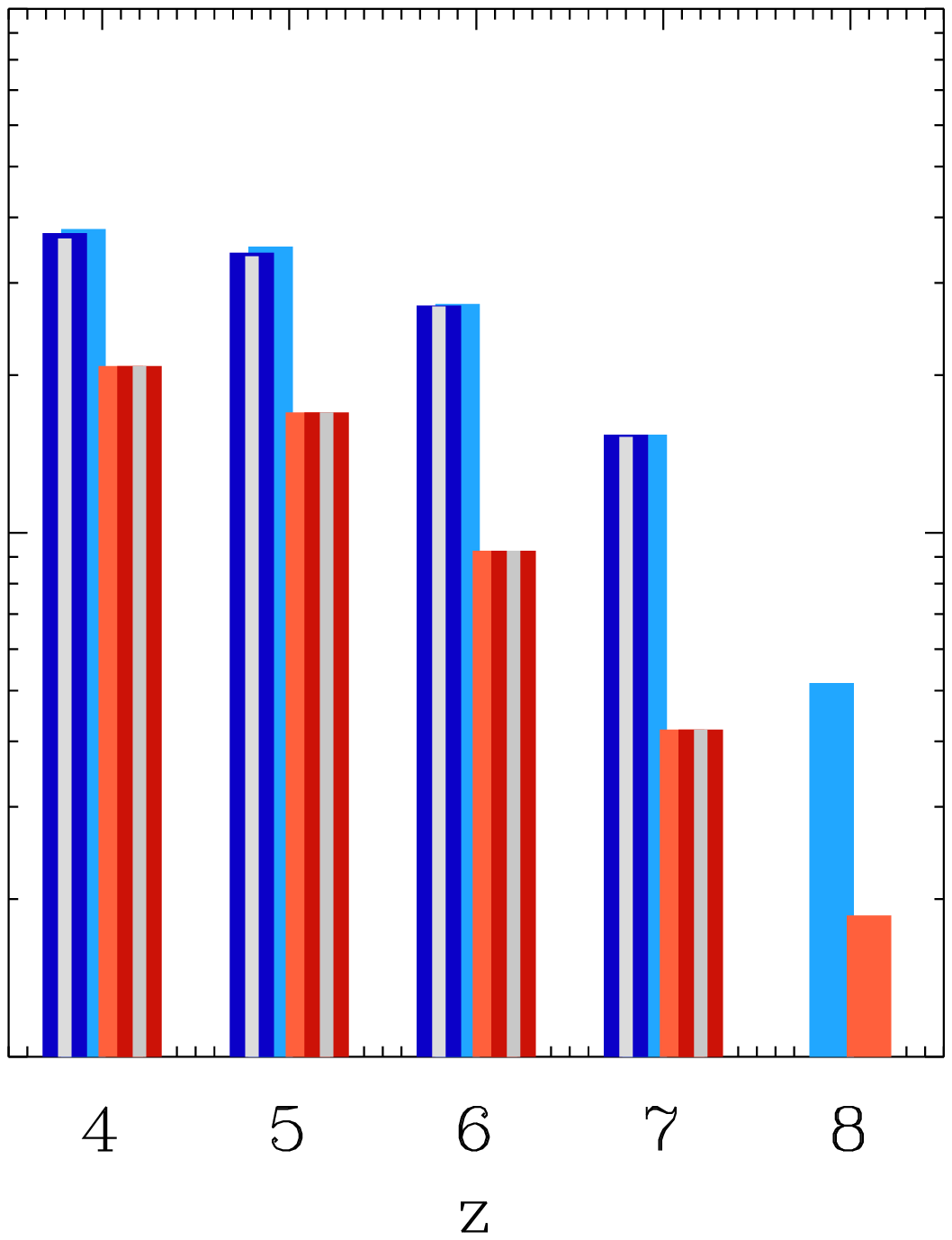,
  width=0.25\textwidth}\hspace{-1.4cm}
\epsfig{file=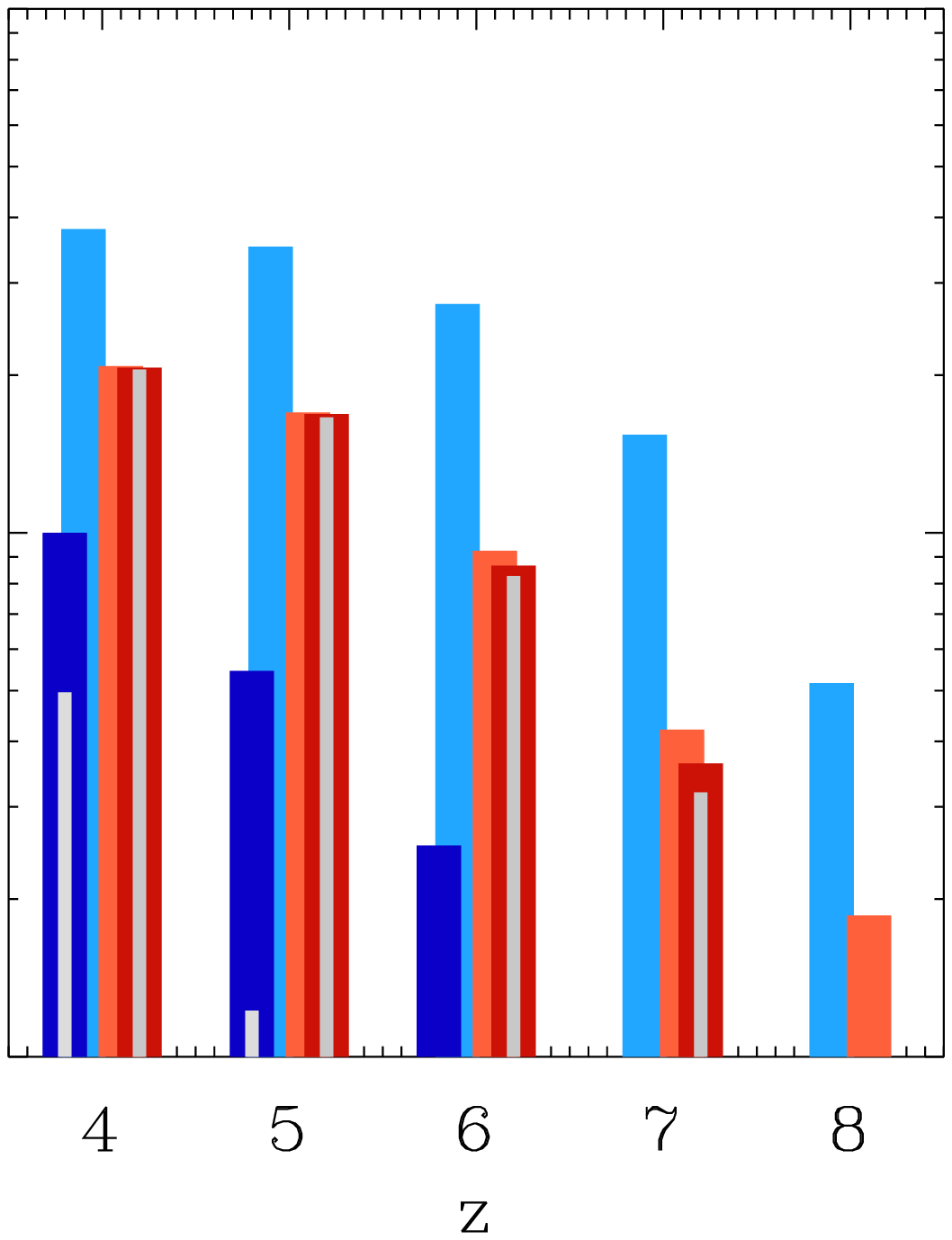,
  width=0.25\textwidth}

\centering{\bf Ten-hour exposure time}

\epsfig{file=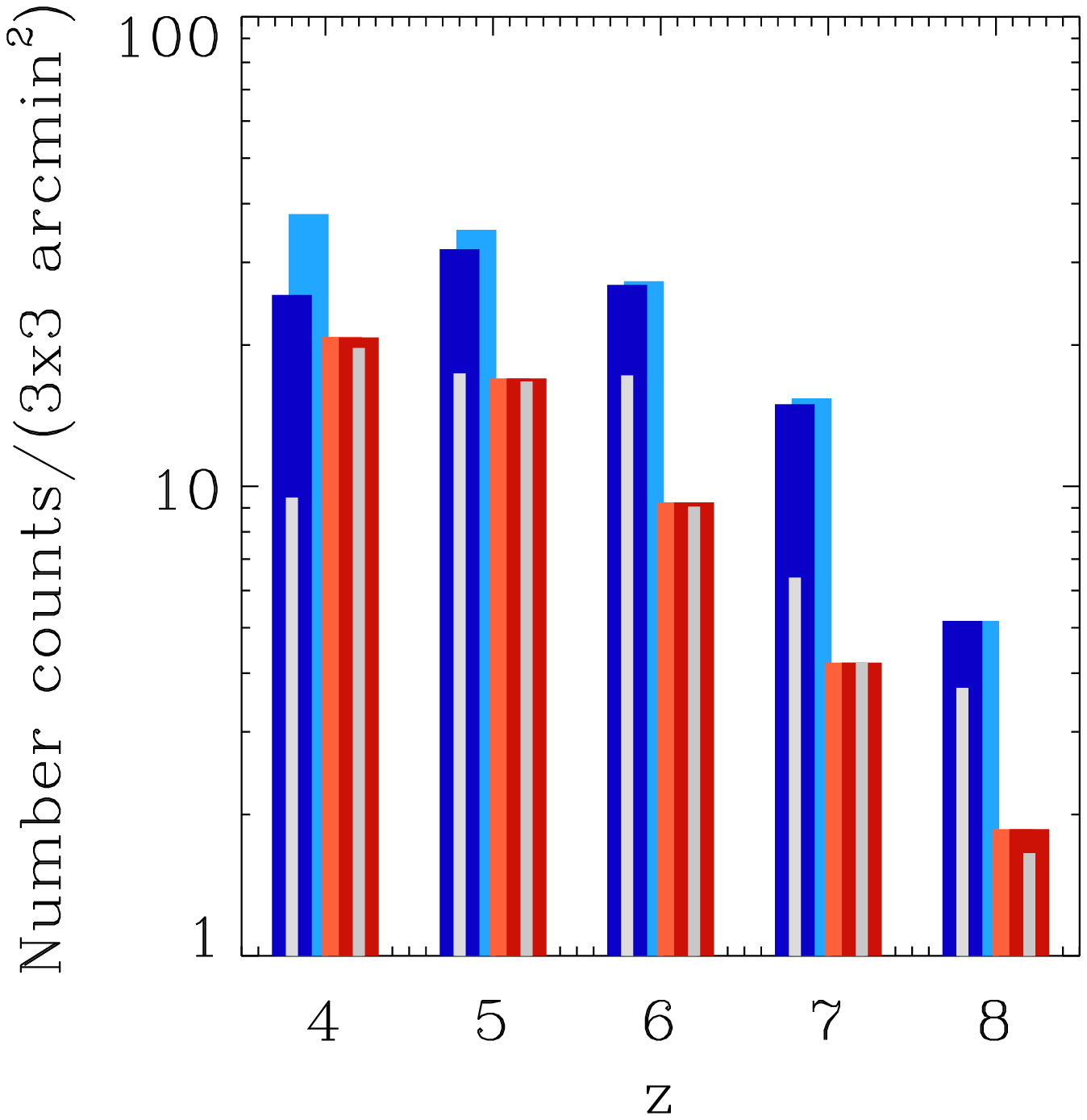,
  width=0.25\textwidth}\hspace{-1.4cm}
\epsfig{file=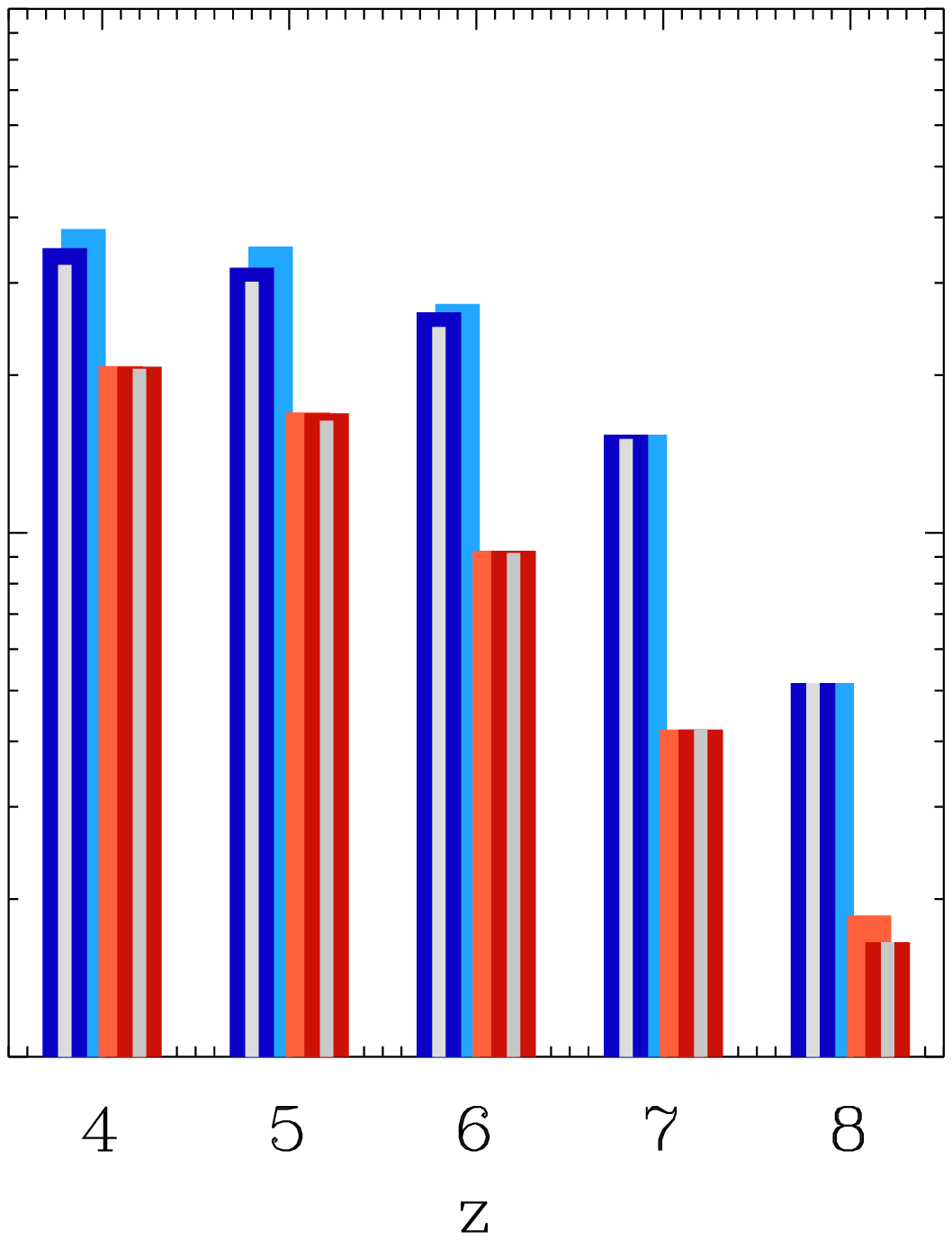,
  width=0.25\textwidth}\hspace{-1.4cm}
\epsfig{file=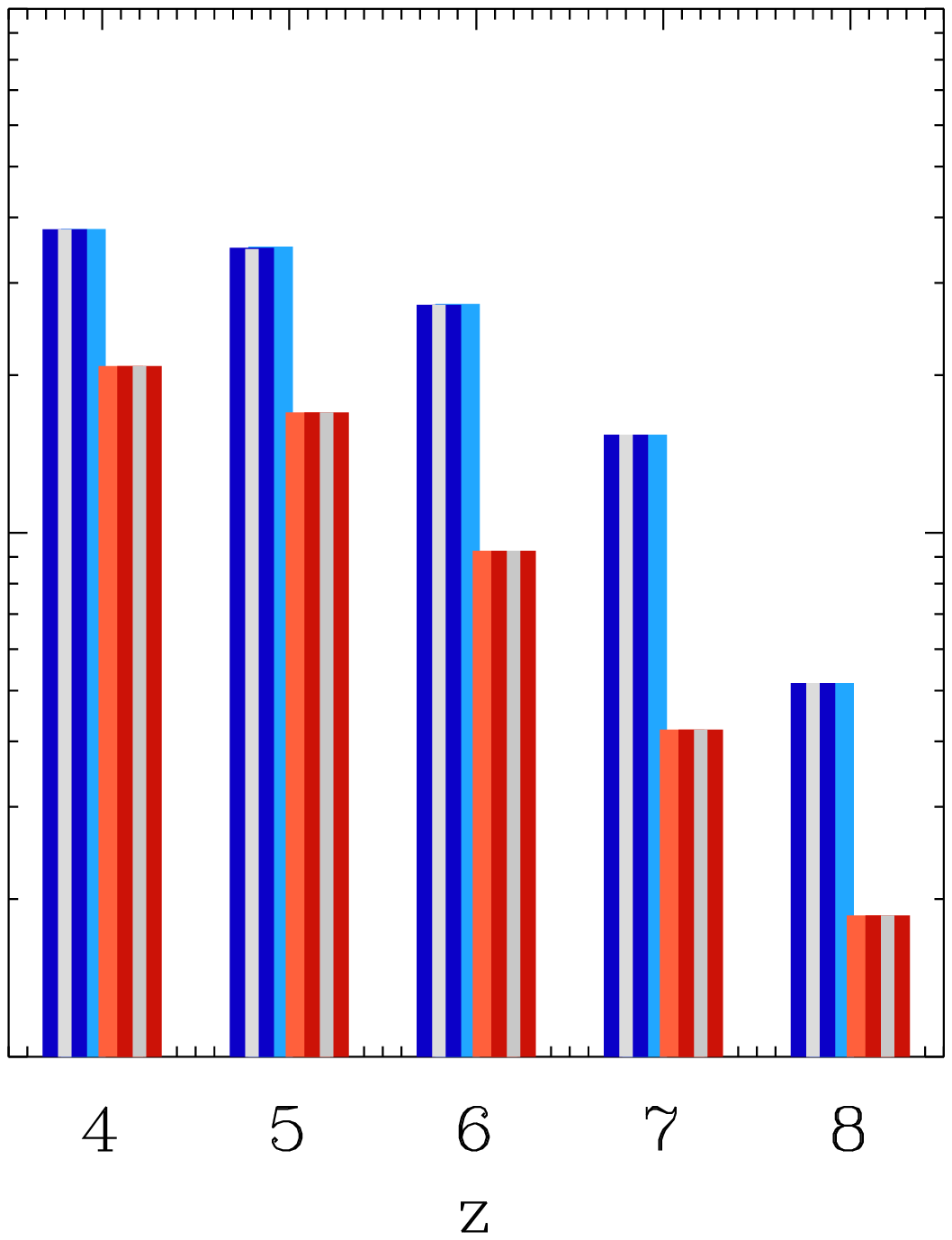,
  width=0.25\textwidth}\hspace{-1.4cm}
\epsfig{file=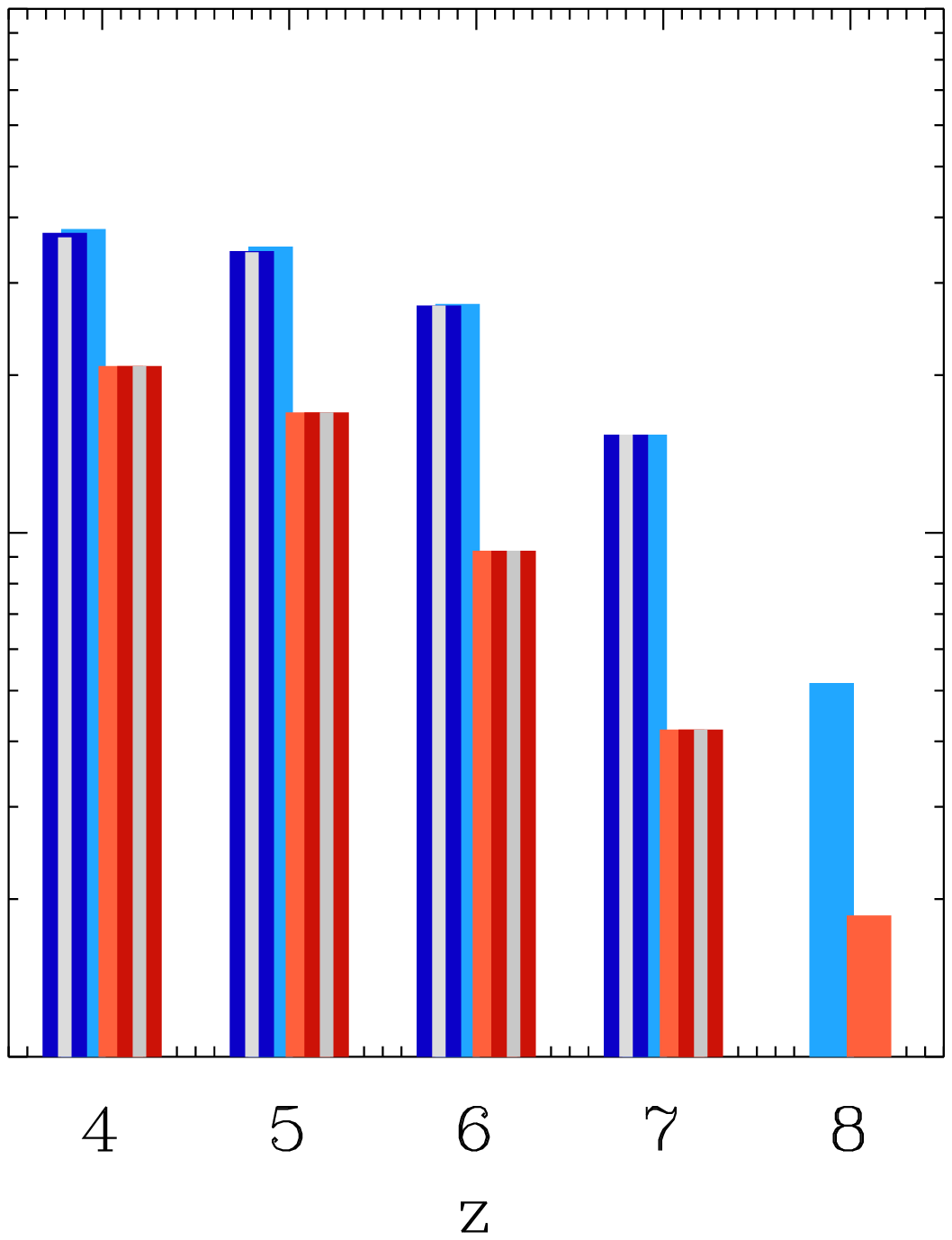,
  width=0.25\textwidth}\hspace{-1.4cm}
\epsfig{file=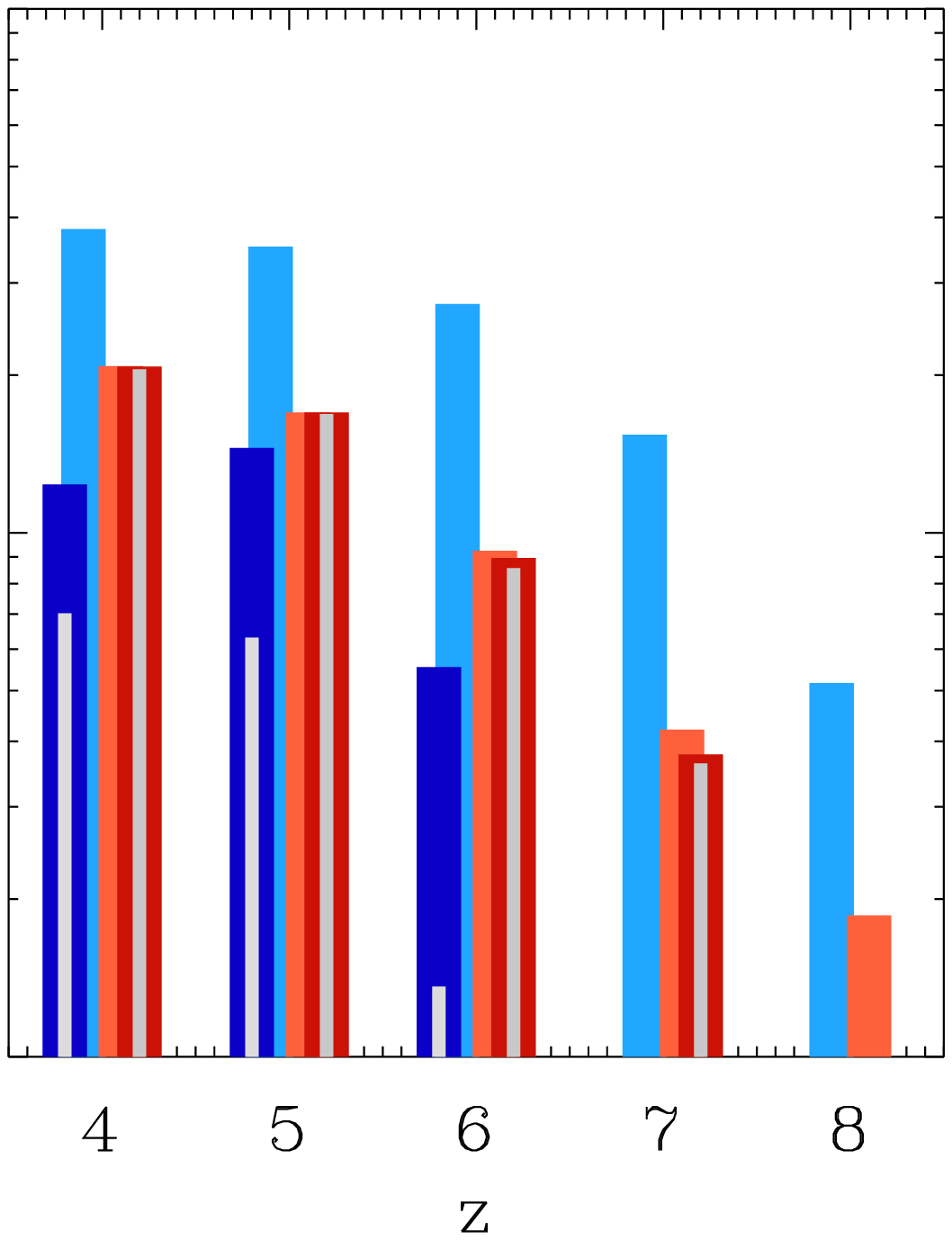,
  width=0.25\textwidth}
\caption{Number counts of \oii, \hb, \oiii, \ha\ and \nii\ emitters
(from left to right) as a function of redshift in a {\it JWST}/NIRSpec 
field of view ($3 \times 3\ \mathrm{arcmin}^2$). The light blue and red 
bars, which are the same in all panels, show the number of, respectively, 
SF-dominated and active (AGN-dominated and composite) galaxies 
more massive than $10^8\, \Msun$ predicted in redshift bins centred 
on $z=4$, 5, 6, 7 and 8 by the TNG50 simulation (the fixed 
boxlength of the simulation translates into redshift-bin widths shrinking 
from $\Delta z =10^{-3}$ to $5 \times 10^{-4}$ over this range). The 
dark blue and red bars show the subsets of these galaxies whose 
line emission can be detected with NIRSpec with 5$\sigma$ significance,
assuming exposure times of 2, 5 and 10\,hr (from top to bottom). The light-grey
thin bars carved in the dark-colour bars show the effect of including 
dust attenuation with $A_{\rm V}=0.5$ and a \citet{Calzetti00} curve.}
\label{NumberCountsOpt}    
\end{figure*}

%*****************************************************************************************************
%*****************************************************************************************************
\subsection{Evolution of UV-line luminosity functions}\label{uvlf} 
%*****************************************************************************************************
%*****************************************************************************************************

Since UV emission lines become prominent in metal-poor 
galaxies and can be detected with {\it JWST} out to high redshifts
\citep[e.g.,][]{Stark14}, it is of interest to explore the predictions
of our models for the evolution of UV-line luminosity and flux
functions at $z \geq 2$. 
We can also investigate which UV lines potentially offer the best SFR
estimators in high-redshift galaxies, at least when corrected for 
attenuation by dust, when standard (in particular, Balmer) 
optical lines fall out of the observational window.

Fig.~\ref{sfr_luv} shows the relations between the average 
luminosities of different UV lines (lilac: \heii; blue: \civ; green:
\oiii; yellow: \siliii; red: \ciii) and SFR in the TNG100
and TNG50 galaxy populations (solid and dashed lines,
respectively) over the redshift range $z=4$-7 (shaded, coloured
areas show the 1$\sigma$ scatter about the TNG50 relations).
All UV line luminosities appear to correlate with SFR over several 
orders of magnitude, the tightest relations being obtained for \ciii, 
\oiiiuv, and \siliii\  (1$\sigma$ scatter below 0.5~dex), compared to
\heii\ and \civ. This is because the latter two lines are more sensitive
to contamination by an AGN component.

In the top row of Fig.~\ref{UVLumfct}, we show the luminosity
functions of the \ciii, \siliii, \oiiiuv, \heii\ and \civ\ emission lines 
for TNG50 (solid lines), TNG100 (dashed lines) 
and TNG300 (dot-dashed lines) galaxies at different redshifts 
from $z=2$ to $z=7$ (different colours, as indicated). As in the
case of optical-line luminosity functions (Section~\ref{opticallf}), 
the TNG50,  TNG100 and TNG300 simulations
predominantly sample, respectively, the faint, intermediate and bright 
ranges of UV luminosity functions. In the luminosity ranges where
predictions from different simulations overlap, these are in good 
general agreement, except for galaxies at $z\ga 6$, where the
limitations arising from resolution effects become worse. 

Regardless of the UV line considered, we find that at a given 
luminosity, the number density drops by up to an order of magnitude
from $z=2$ to $z=6$, mainly for galaxies with luminosities below 
the exponential cutoff. Instead, the luminous end is largely in 
place as early as $z=6$, and thus evolves less strongly toward
low reshift. The fact that the area under the luminosity function (i.e., 
the integrated luminosity function) decreases from $z=2$ to $z=7$ 
is a consequence of the decline in cosmic SFR density over this 
redshift interval, as shown by, e.g., \citet{Pillepich18a}.

It is interesting to note that the \heii\ and \civ\ luminosity functions 
exhibit bi-modal distributions. This is because, while the
lower-luminosity peak is dominated by line emission from young star
clusters, a  second peak arises from the high \heii\ and \civ\
luminosities in  AGN-dominated galaxies. This bi-modality is not
observed in the  other UV luminosity functions in Fig.~\ref{UVLumfct},
because \ciii,  \siliii\ and \oiiiuv\ require lower ionization
energies, and hence, are less  sensitive than \heii\ and \civ\ to the
hard ionizing radiation from accreting BH.

As a complement to the rest-frame line-luminosity functions, we 
show in the bottom row of Fig.~\ref{UVLumfct} the evolution 
of the associated line-flux functions, which are directly observable
quantities. The observed-flux functions differ from the rest-frame 
luminosity functions because of the redshift-dependent luminosity 
distance. Although the main trends are similar to those described 
above for the luminosity functions, the main difference between flux 
and luminosity functions is the shift of the high-luminosity end
toward lower relative fluxes at increasing redshift (since at fixed
luminosity, the observed flux decreases with increasing
redshift).

Overall, it will be instructive to explore the extent to which future 
observational studies at high redshift, for example with {\it JWST}/NIRSpec, 
will be consistent with the predicted optical/UV
luminosity\footnote{We note that the optical and UV luminosity
functions in Figs~\ref{Halum} and \ref{UVLumfct} pertain to a mass-limited 
sample and might differ for flux-limited samples.} and flux functions
in Figs~\ref{Halum} and \ref{UVLumfct},
%and whether they will confirm the strong
%decline of the cosmic SFR above $z=7$ as suggested by \citet{Oesch18},
and whether potential inconsistencies between models and observations 
might require fundamental changes in models of galaxy evolution.

%*****************************************************************************************************
%*****************************************************************************************************
\subsection{Predicted number counts in {\it JWST}/NIRSpec observations}\label{numbercounts} 
%*****************************************************************************************************
%*****************************************************************************************************

The models presented in the previous sections allow us to predict 
the number of high-redshift galaxies that can be detected down to a given
line-flux limit with the near-infrared spectrograph NIRSpec on {\it
  JWST} \citep{Jakobsen22}. An interesting case study is that of the
{\it JWST} Advanced Deep Extragalactic  
Survey
(JADES),\footnote{\url{https://www.cosmos.esa.int/web/jwst-nirspec-gto/jades}} 
which combines NIRCam imaging and NIRSpec spectroscopy distributed in a 
`DEEP' and `MEDIUM'  surveys (a guaranteed-time-observation program led 
by N.~L\"utzgendorf and D.~Eisenstein). Among the different components of 
JADES, we focus on the medium-resolution spectroscopy (resolving power 
$R\sim1000$) that will be achieved on a few hundred galaxies in the DEEP survey 
(two NIRSpec pointings with exposure times of 8.4--25.2\,ksec,
i.e. from 2.3 to 7\,hr, in each of several 
grating/filter configurations) and a few thousand galaxies in the MEDIUM survey 
(24 NIRSpec pointings with exposure times ranging from 2.7 to
9.3\,ksec, i.e. 0.75 to  2.6\,hr, per grating/filter configuration). We adopt
the flux limits listed in Table \ref{tableflux}, computed with the pre-flight {\it JWST}
exposure-time calculator, for different lines at different redshifts and for
different exposure times.

\begin{figure*}
  \centering{\bf Two-hour exposure time}
  
\epsfig{file=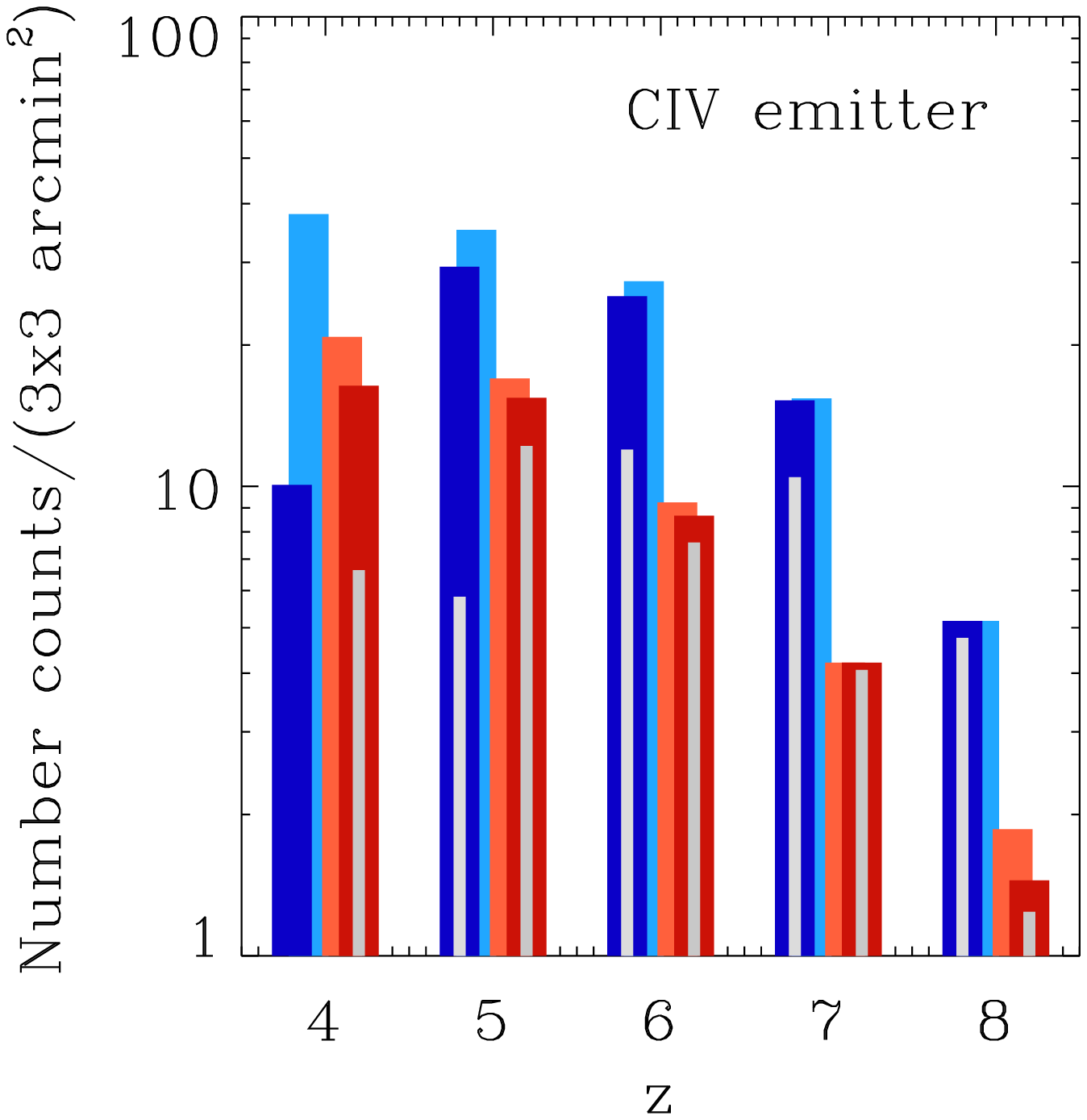,
  width=0.25\textwidth}\hspace{-1.4cm}
\epsfig{file=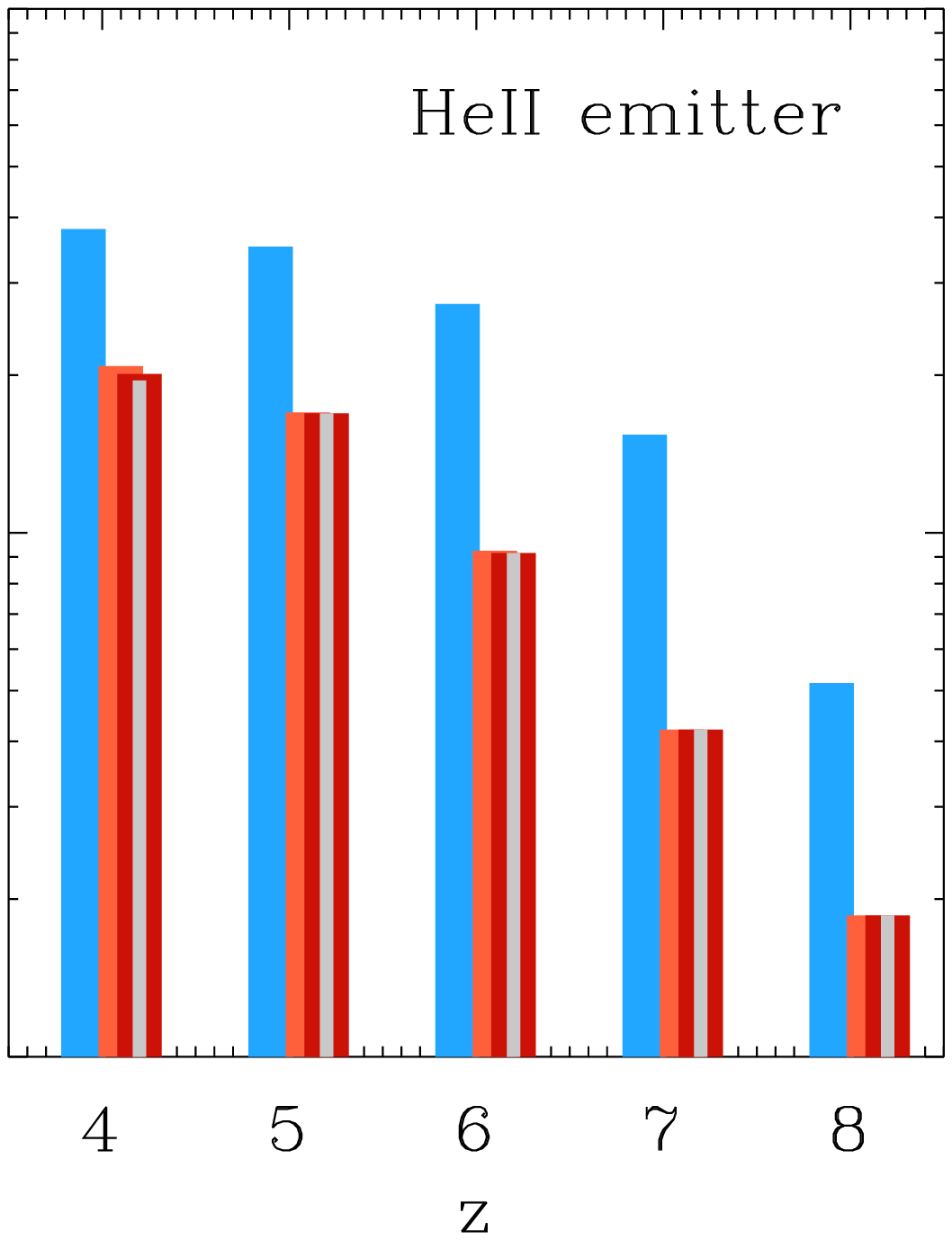,
  width=0.25\textwidth}\hspace{-1.4cm}
\epsfig{file=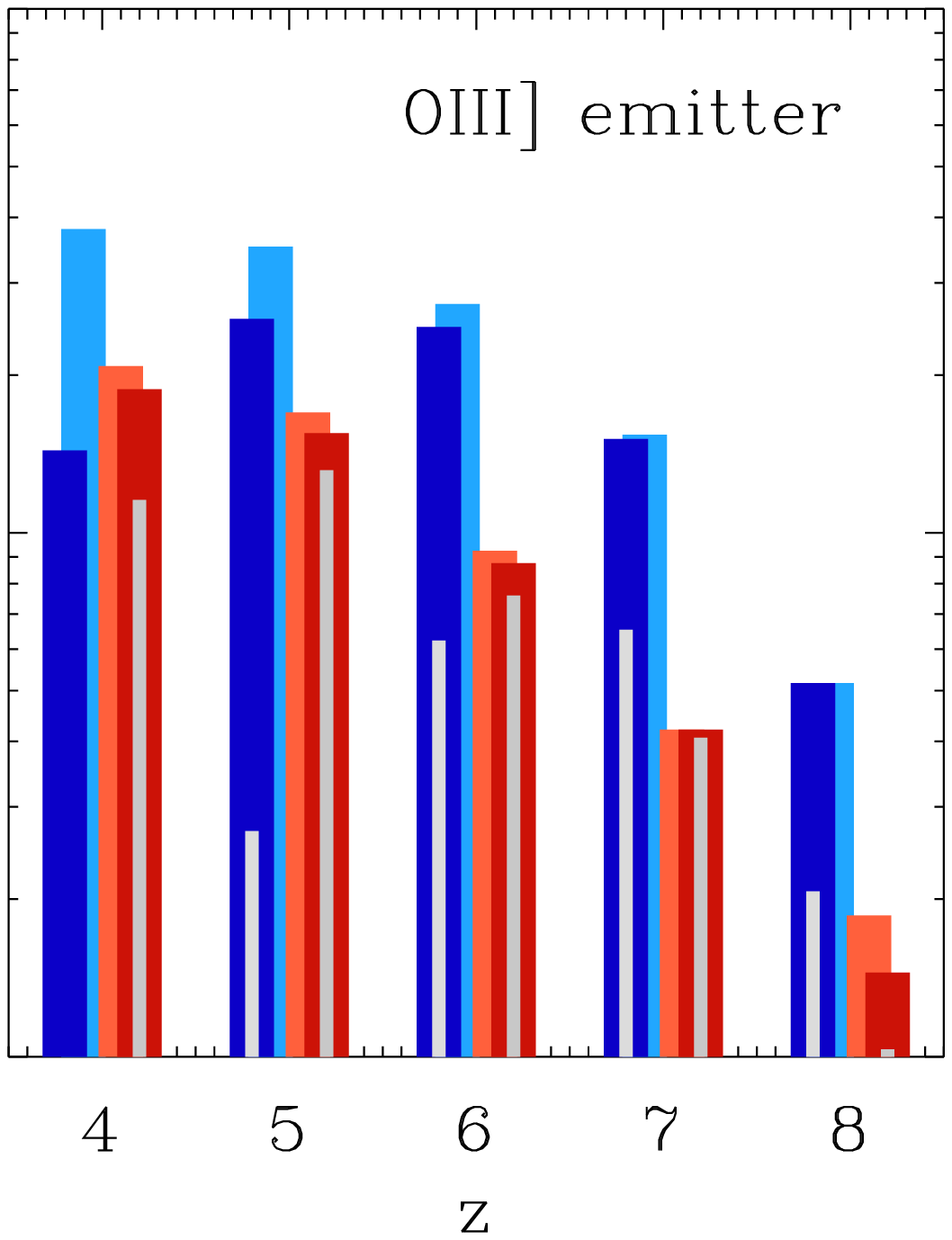,
  width=0.25\textwidth}\hspace{-1.4cm}
\epsfig{file=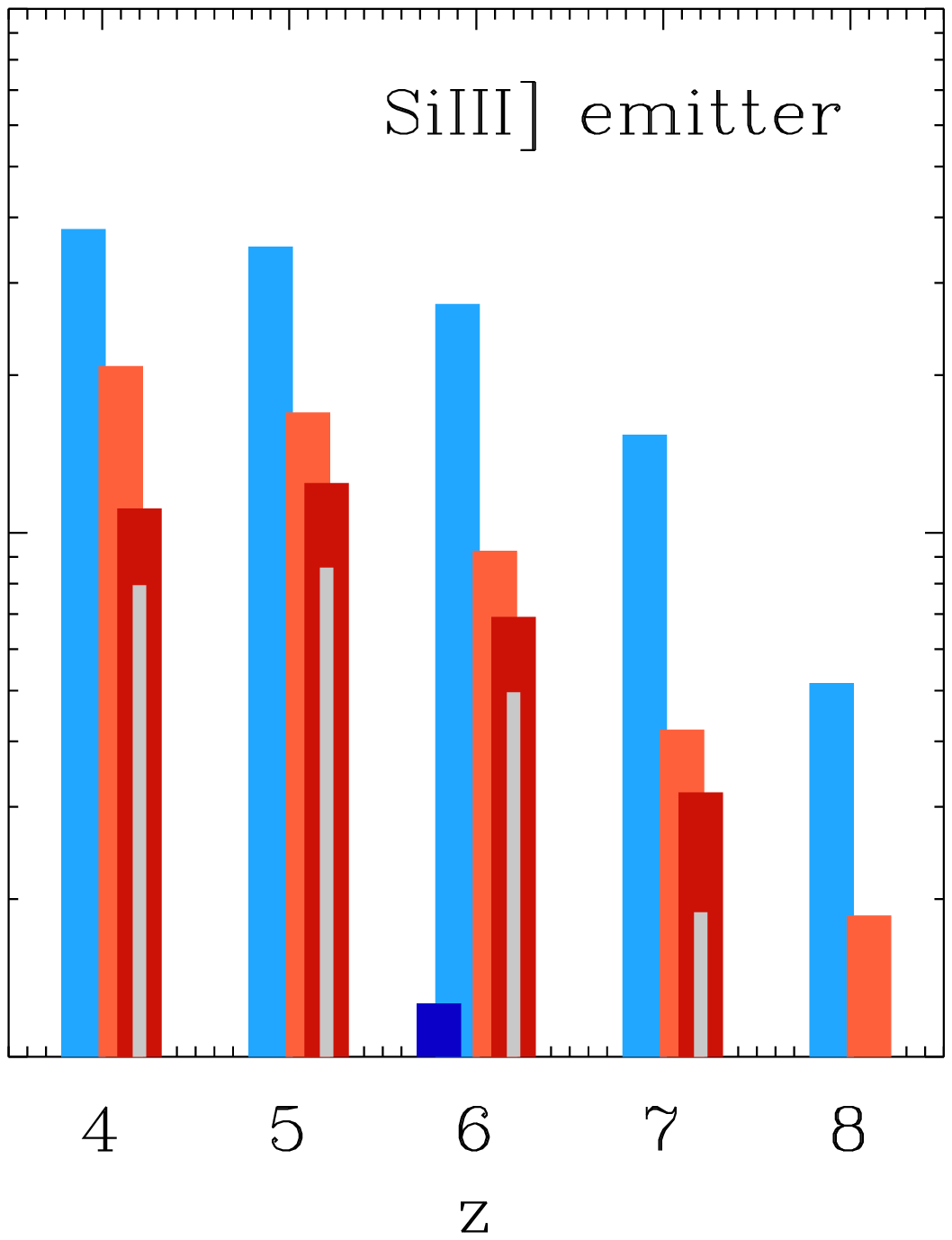,
  width=0.25\textwidth}\hspace{-1.4cm}
\epsfig{file=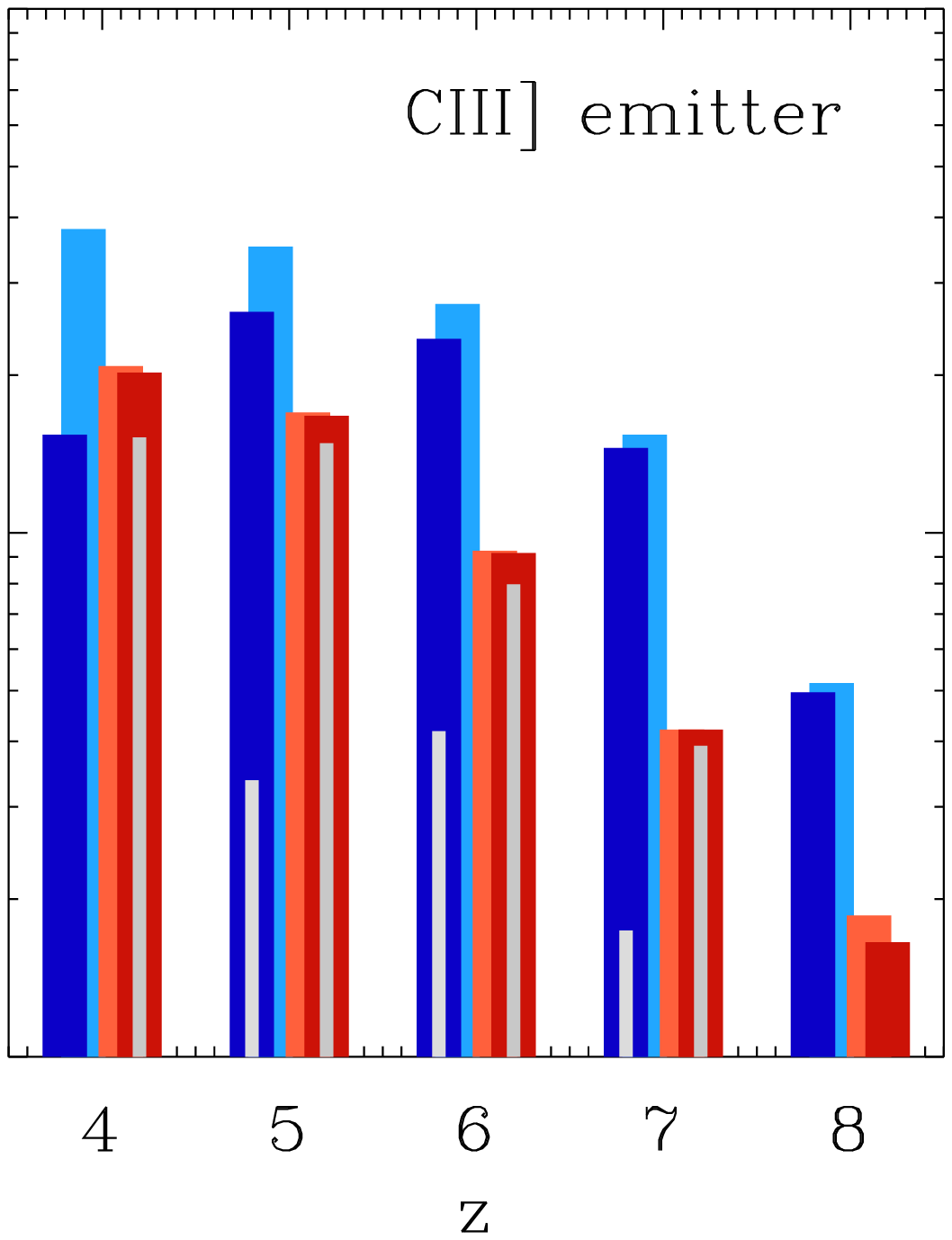,
  width=0.25\textwidth}

\centering{\bf Five-hour exposure time}

\epsfig{file=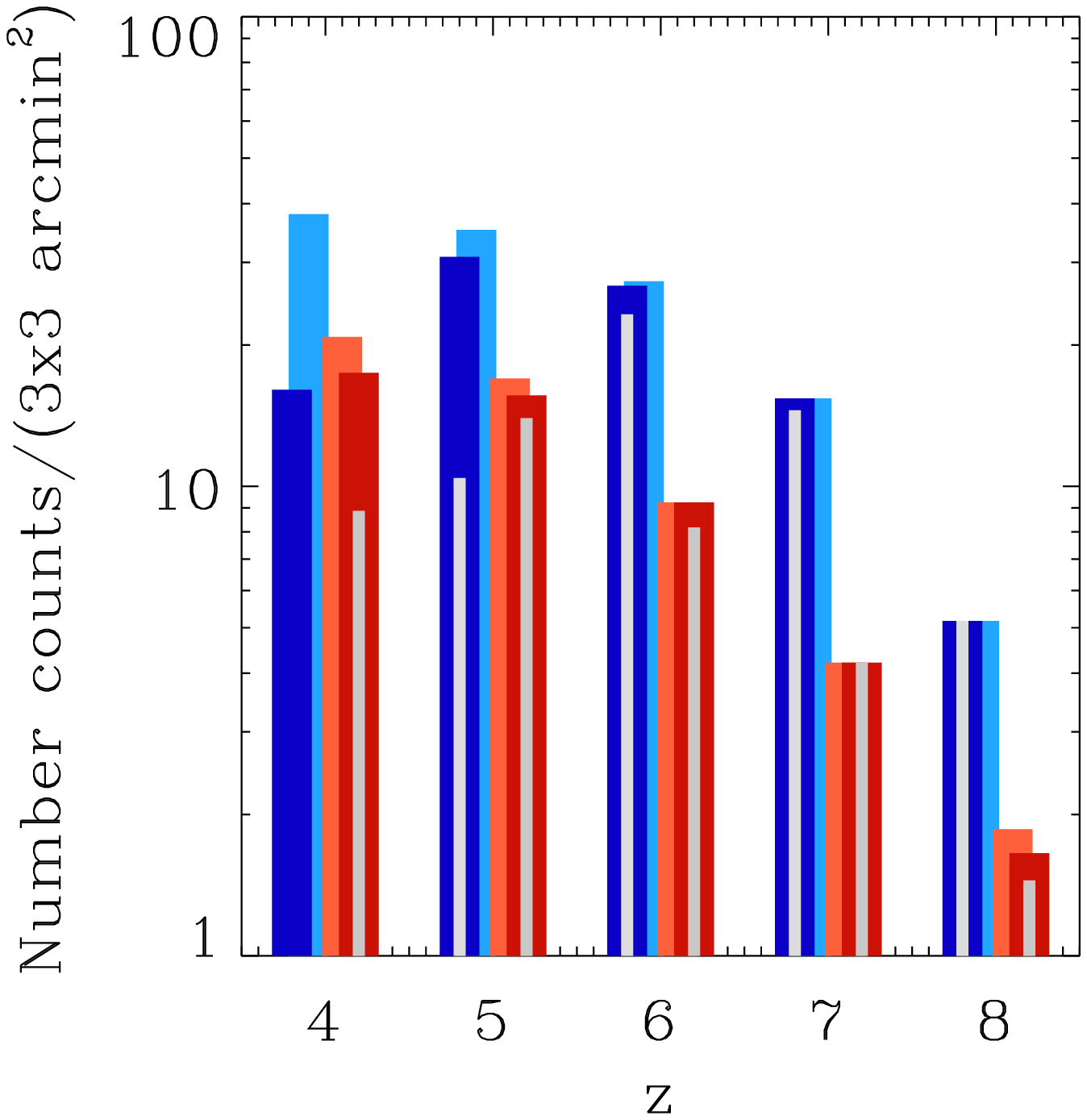,
  width=0.25\textwidth}\hspace{-1.4cm}
\epsfig{file=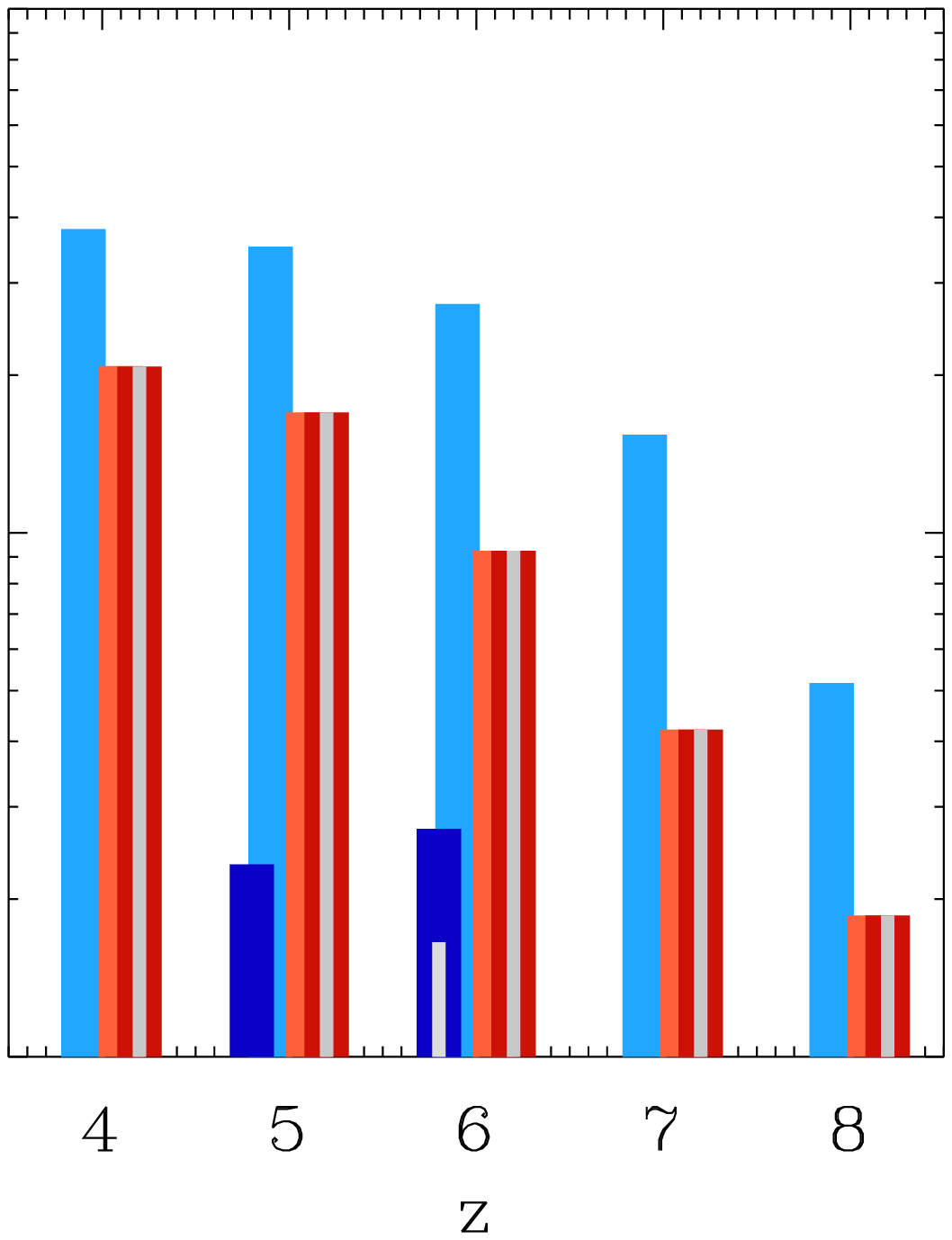,
  width=0.25\textwidth}\hspace{-1.4cm}
\epsfig{file=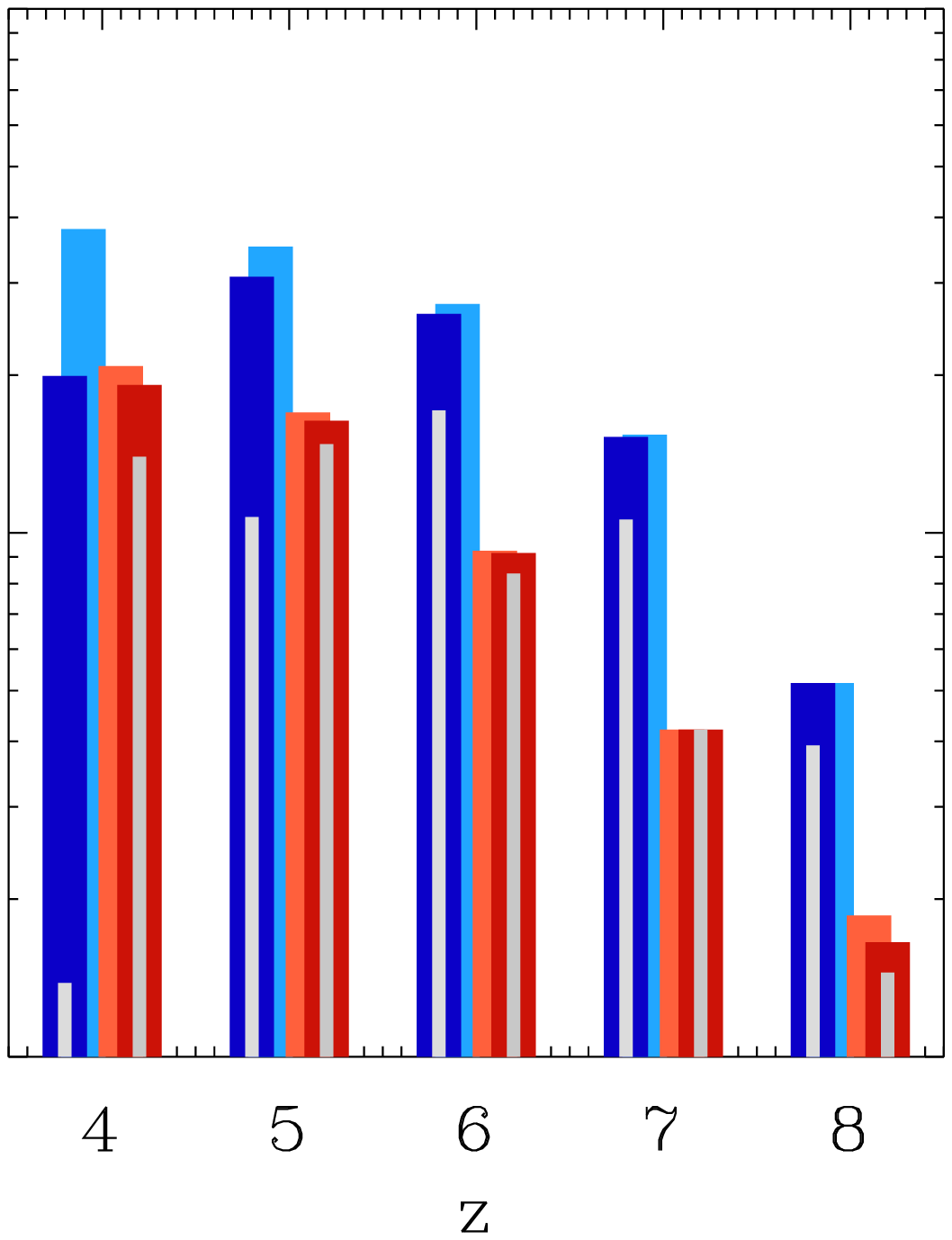,
  width=0.25\textwidth}\hspace{-1.4cm}
\epsfig{file=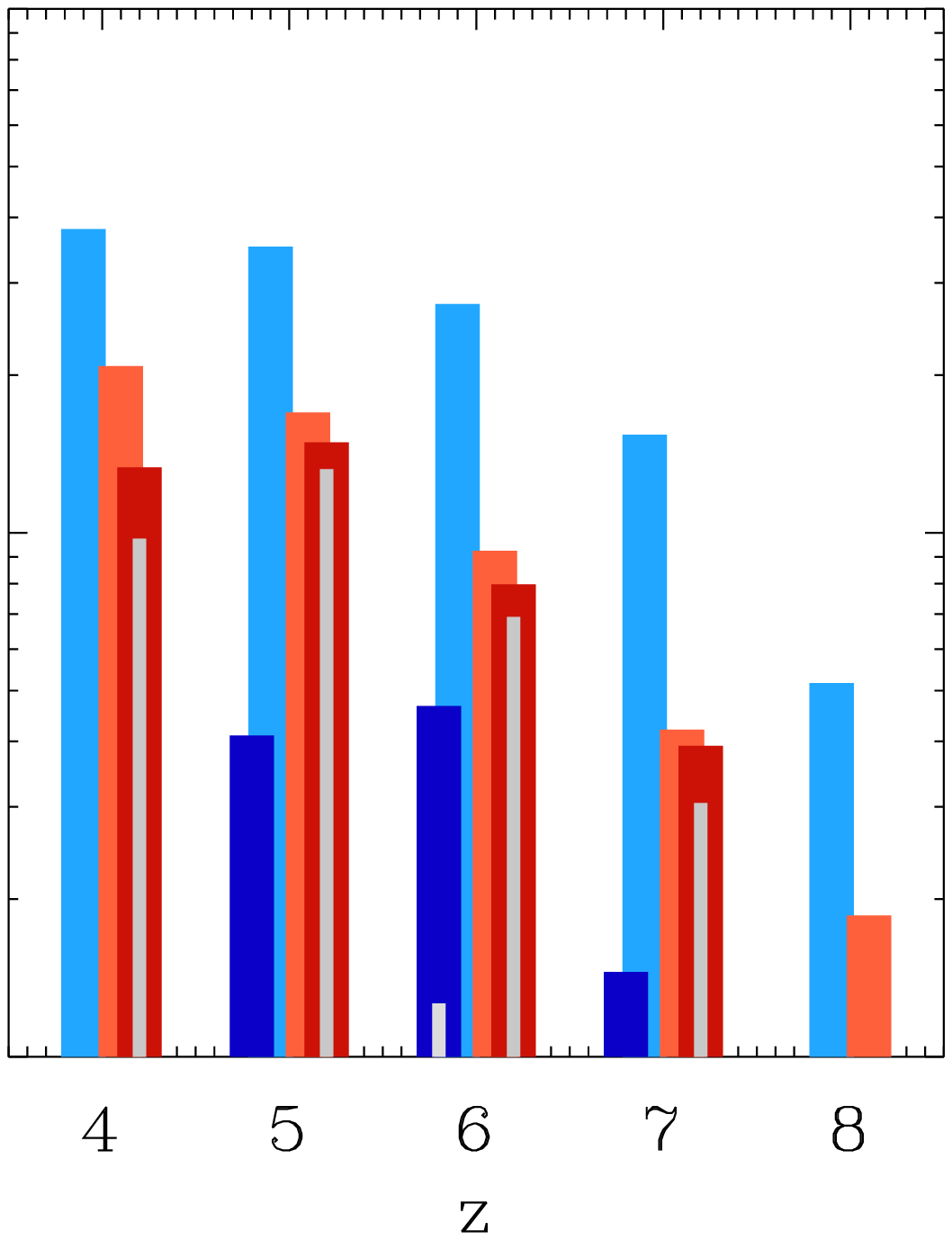,
  width=0.25\textwidth}\hspace{-1.4cm}
\epsfig{file=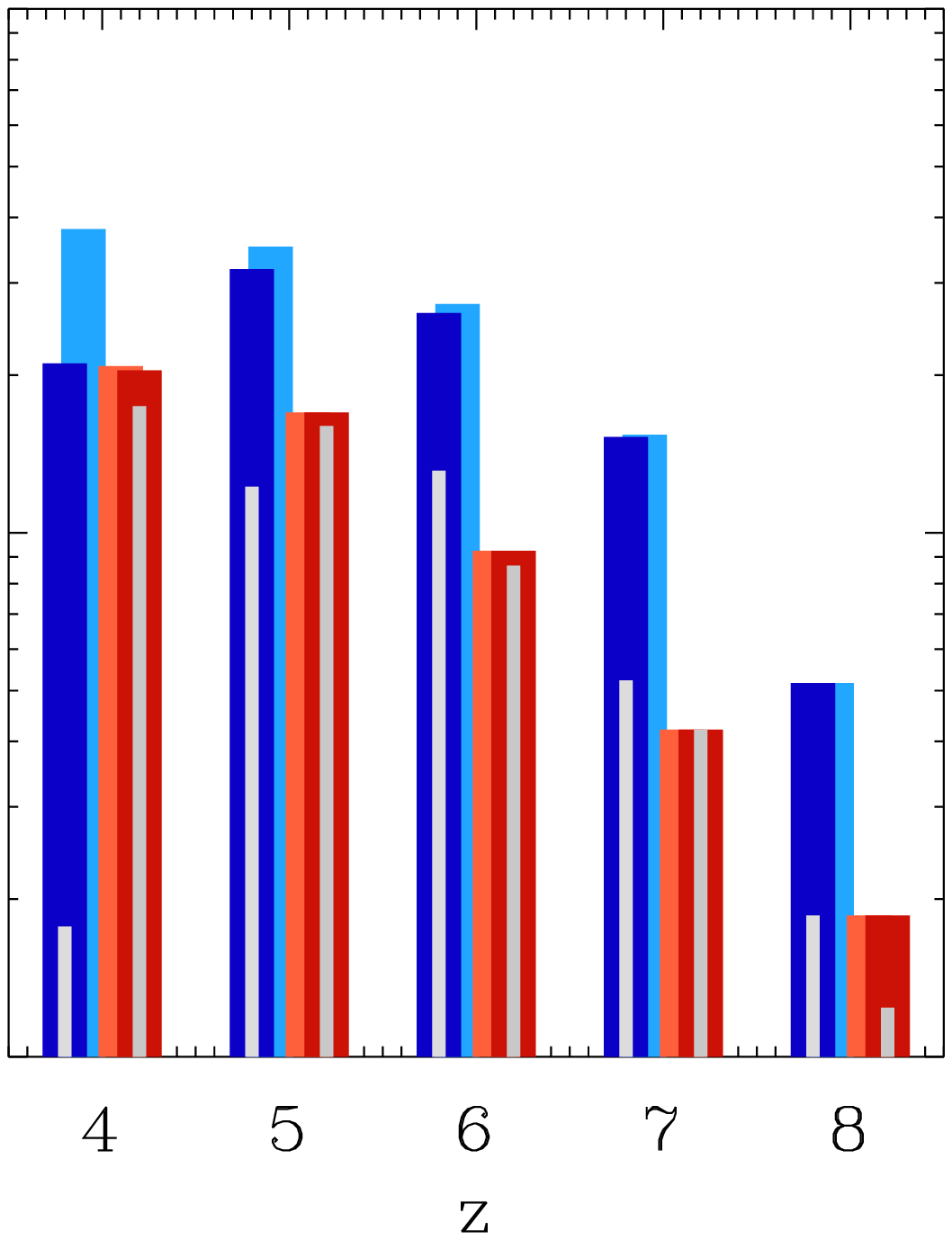,
  width=0.25\textwidth}

\centering{\bf Ten-hour exposure time}

\epsfig{file=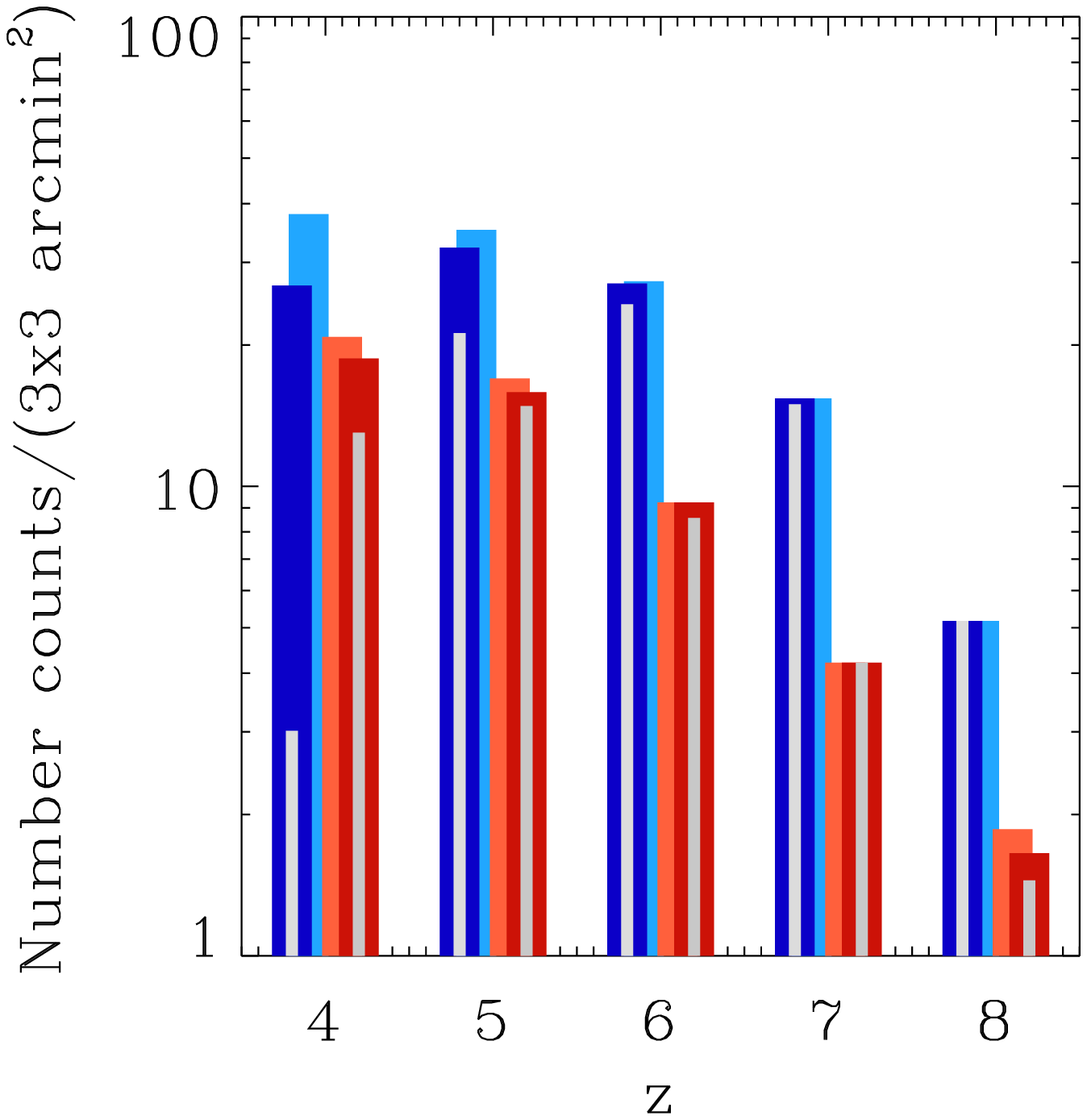,
  width=0.25\textwidth}\hspace{-1.4cm}
\epsfig{file=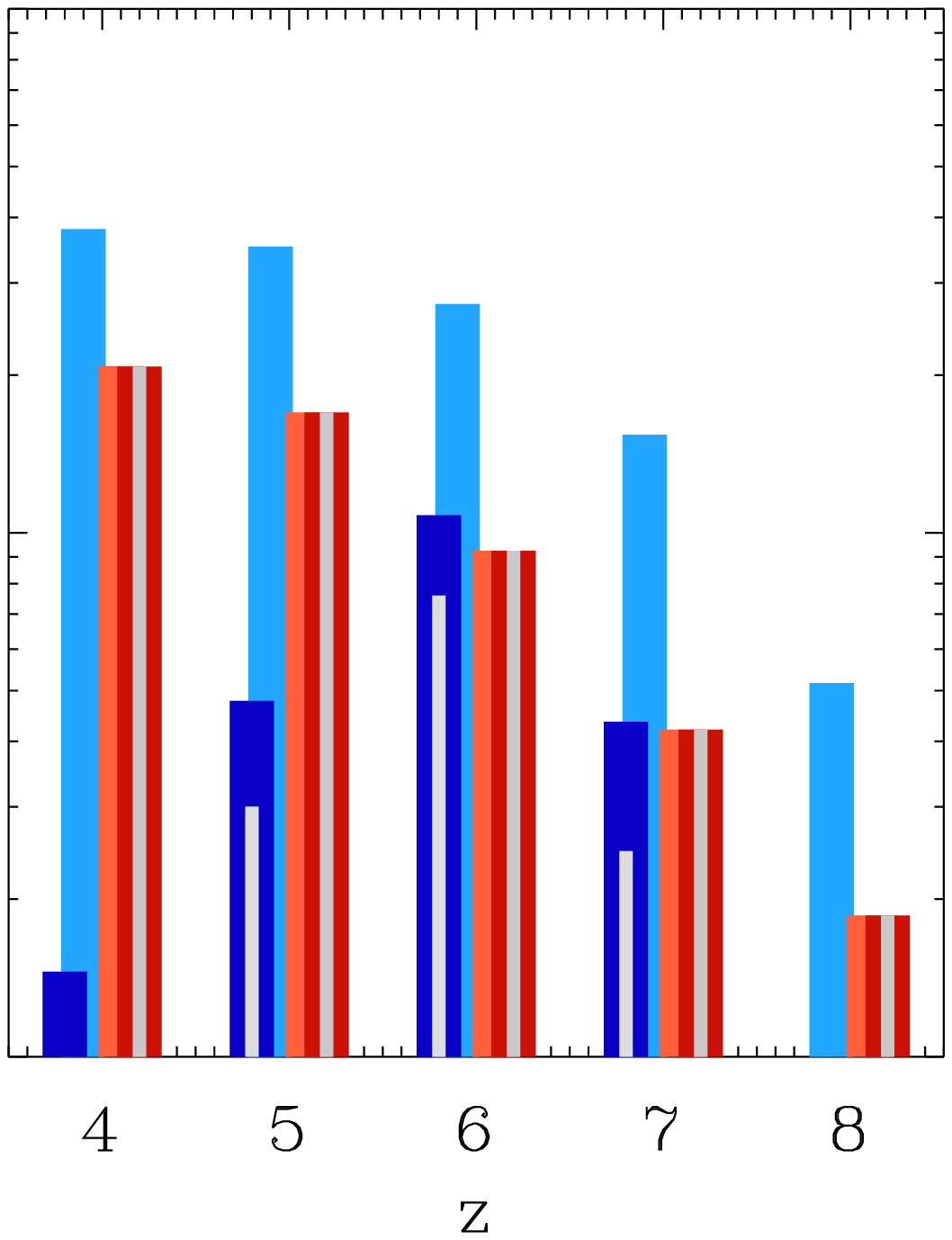,
  width=0.25\textwidth}\hspace{-1.4cm}
\epsfig{file=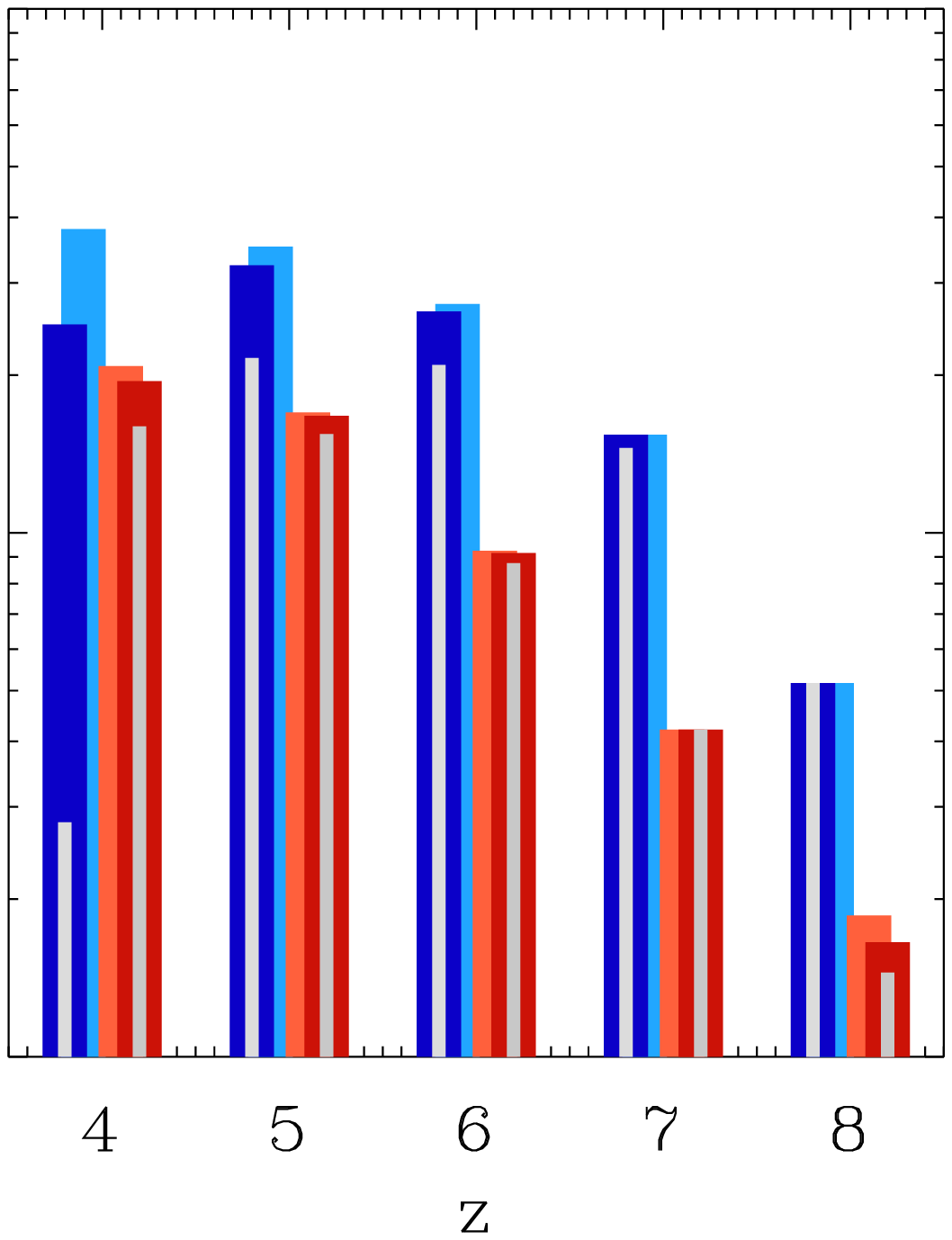,
  width=0.25\textwidth}\hspace{-1.4cm}
\epsfig{file=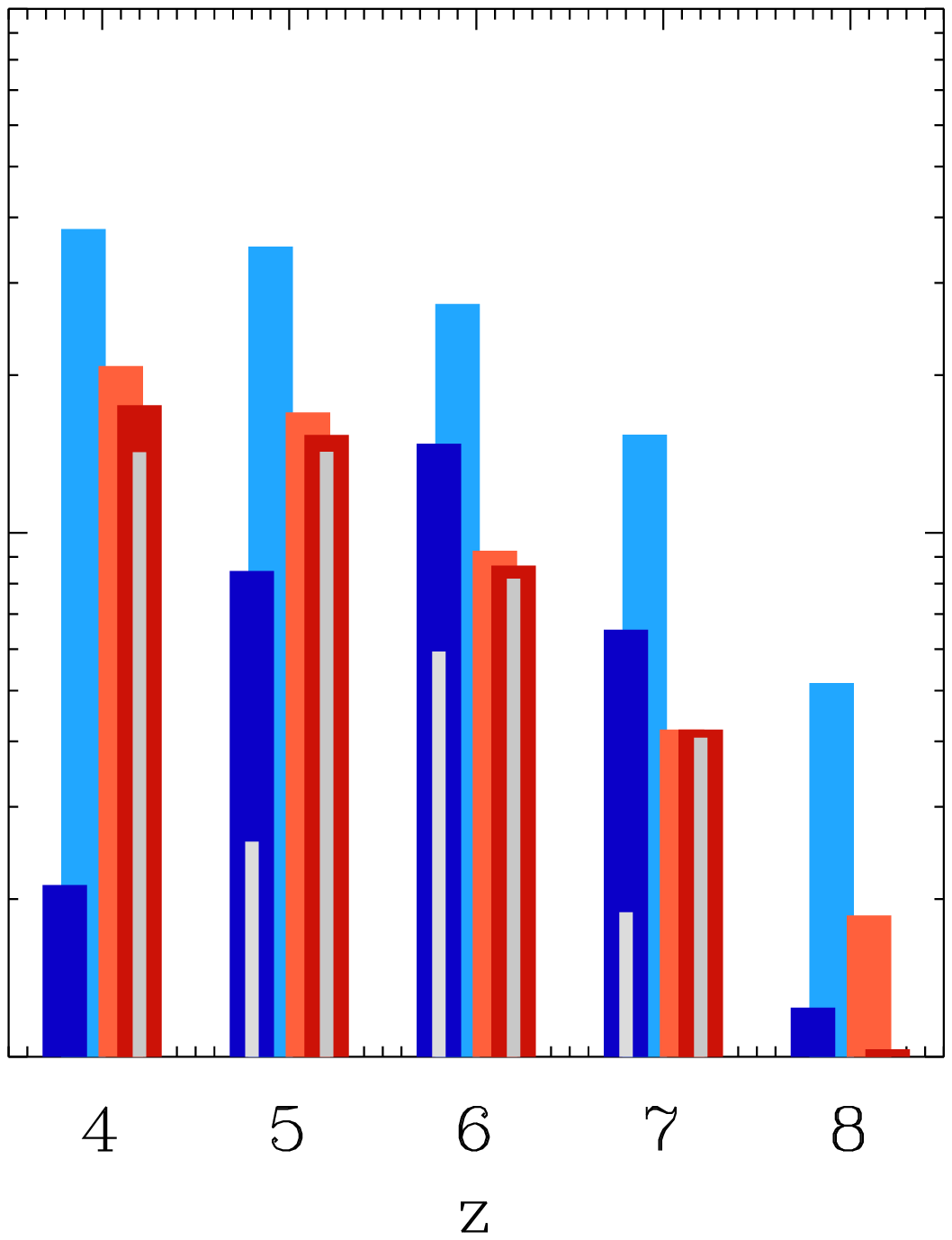,
  width=0.25\textwidth}\hspace{-1.4cm}
\epsfig{file=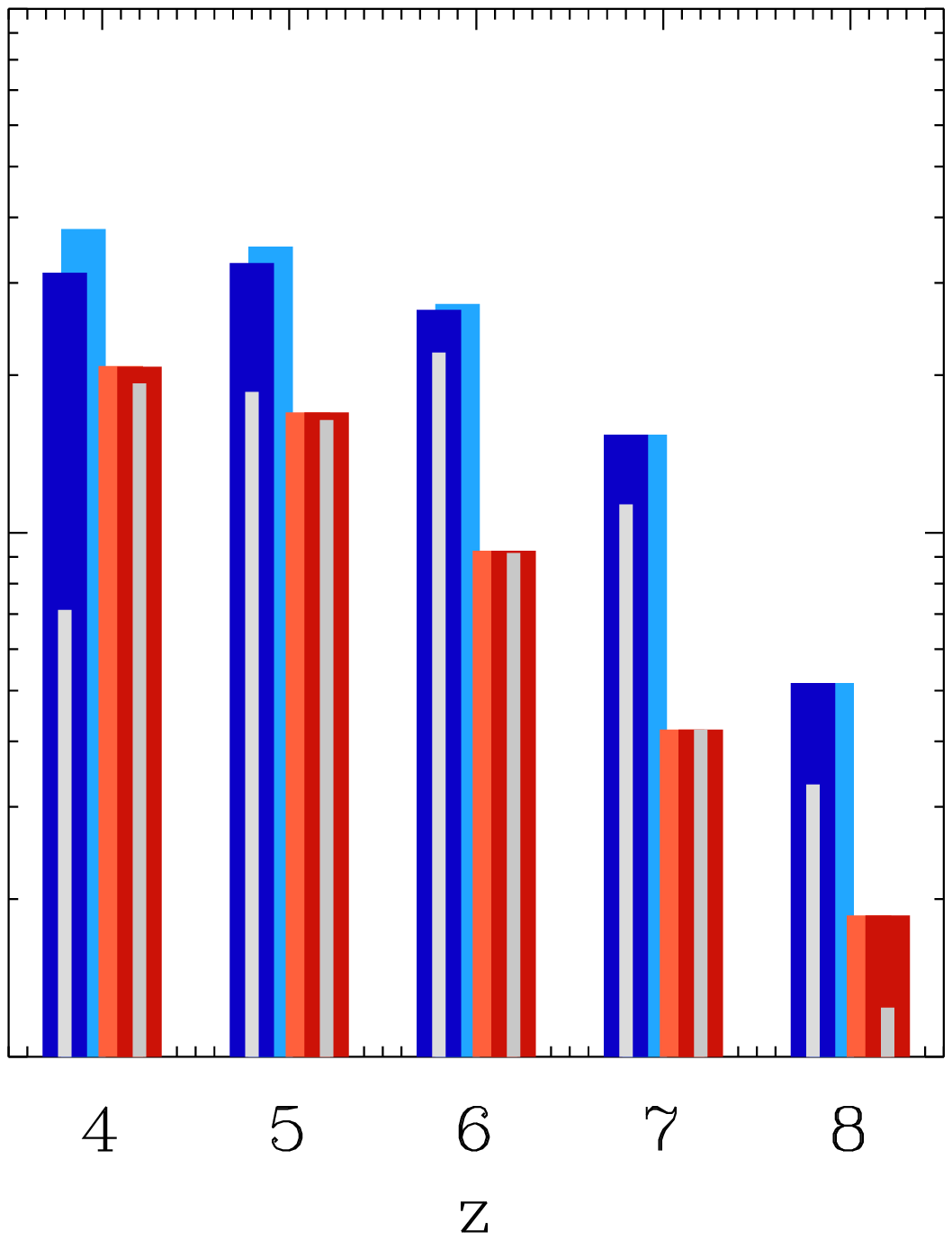,
  width=0.25\textwidth}
\caption{Same as Fig.~\ref{NumberCountsOpt}, but for the
\civ, \heii, \oiiiuv, \siliii\ and \ciii\ emission lines. The predicted number counts
of \civ\ emitters in the presence of dust are likely overestimated, as they  
ignore the effect of resonant scattering.}\label{NumberCountsUV}    
\end{figure*}

\begin{table}
  \centering
\begin{tabular}{ p{1.5cm} | p{0.9cm} | p{0.9cm} | p{0.9cm} | p{0.9cm} | p{0.9cm}}
  Line  &    $z=4$	&	$z=5$	&	$z=6$  &  $z=7$ &  $z=8$
  \\\hline
\end{tabular}
Two-hour exposure time\\\vspace{0.1cm}%\hline \vspace{0.2cm}
\begin{tabular}{ p{1.5cm} | p{0.9cm} | p{0.9cm} | p{0.9cm} | p{0.9cm} | p{0.9cm}}
  \oii \vspace{-0.3cm}  & $-18.3$  & $-18.5$ & $-18.6$ & $-18.6$ & $-18.7$ \\
  \hb \vspace{-0.3cm} & $-18.4$ & $-18.6$ & $-18.8$ & $-18.8$ & $-18.8$ \\
  \oiii \vspace{-0.3cm}  & $-18.5$ & $-18.6$ & $-18.8$ & $-18.8$ & $-18.8$ \\
  \ha  \vspace{-0.3cm} & $-18.7$ & $-18.8$ & $-18.8$ & $-18.6$ & --  \\
  \nii  \vspace{-0.3cm} & $-18.7$ & $-18.8$ & $-18.8$ & $-18.6$ & -- \\
  \civ \vspace{-0.3cm}  &  $-17.4$   & $-17.9$  & $-18.1$  & $-18.3$  & $-18.4$ \\%\vspace{-0.3cm}
  \heii \vspace{-0.3cm} & $-17.5$ & $-18.0$  & $-18.2$ &  $-18.3$ & $-18.4$ \\
   \oiiiuv \vspace{-0.3cm} & $-17.7$  & $-18.0$  & $-18.2$ & $-18.3$ & $-18.4$  \\
  \siliii \vspace{-0.3cm} &	$-17.9$  & $-18.2$ & $-18.3$ & $-18.4$  & $-18.3$ \\
  \ciii \vspace{-0.3cm} & $-17.9$ & $-18.2$  & $-18.3$ & $-18.35$ & $-18.3$ \\\hline
\end{tabular}
 Five-hour exposure time\\\vspace{0.1cm}%\hline \vspace{0.2cm}
\begin{tabular}{ p{1.5cm} | p{0.9cm} | p{0.9cm} | p{0.9cm} | p{0.9cm} | p{0.9cm}}
  % Line  &    z=4	&	z=5	&	z=6  &  z=7 &  z=8 \\\hline
  \oii \vspace{-0.3cm} & $-18.5$ & $-18.7$ & $-18.8$ & $-18.8$ & $-19.0$\\
  \hb \vspace{-0.3cm} & $-18.8$ & $-18.8$ & $-19.0$ & $-19.0$ & $-18.9$ \\
  \oiii \vspace{-0.3cm} & $-18.8$ & $-18.8$ & $-19.0$ & $-19.0$ & $-19.0$ \\
  \ha  \vspace{-0.3cm} & $-18.9$ & $-19.0$ & $-19.0$ & $-18.8$ & -- \\
  \nii  \vspace{-0.3cm} & $-18.9$ & $-19.0$ & $-19.0$ & $-18.8$ & -- \\
  \civ  \vspace{-0.3cm} &  $-17.5$  & $-18.0$   & $-18.4$  & $-18.5$  & $-18.5$ \\
  \heii \vspace{-0.3cm}  & $-18.0$ & $-18.4$ & $-18.5$ & $-18.5$  & $-18.5$ \\
   \oiiiuv \vspace{-0.3cm}  & $-17.8$ & $-18.2$ & $-18.4$ & $-18.5$ & $-18.5$ \\
   \siliii  \vspace{-0.3cm} &	$-18.0$ & $-18.4$ & $-18.5$ & $-18.5$  & $-18.5$ \\
  \ciii \vspace{-0.3cm} & $-18.0$ & $-18.4$  & $-18.5$ & $-18.5$ & $-18.5$ \\\hline
\end{tabular}
 Ten-hour exposure time\\\vspace{0.1cm}%\hline \vspace{0.2cm}
\begin{tabular}{ p{1.5cm} | p{0.9cm} | p{0.9cm} | p{0.9cm} | p{0.9cm} | p{0.9cm}}
  % Line  &    z=4	&	z=5	&	z=6  &  z=7 &  z=8 \\\hline
  \oii  \vspace{-0.3cm} & $-18.6$ & $-18.9$ & $-19.0$ & $-19.0$ & $-18.9$ \\
  \hb  \vspace{-0.3cm} & $-18.9$ & $-19.0$ & $-19.1$ & $-19.2$ & $-19.1$ \\
  \oiii  \vspace{-0.3cm} & $-19.0$ & $-19.0$ & $-19.1$ & $-19.2$ & $-19.1$ \\
  \ha  \vspace{-0.3cm} & $-19.0$ & $-19.2$ & $-19.1$ & $-18.8$ & -- \\
  \nii  \vspace{-0.3cm} & $-19.0$ & $-19.2$ & $-19.1$ & $-19.0$ & --  \\
  \civ  \vspace{-0.3cm} & $-17.7$  & $-18.2$  & $-18.5$ & $-18.6$ & $-18.7$ \\
  \heii \vspace{-0.3cm} & $-18.2$ & $-18.5$  & $-18.7$  & $-18.7$  & $-18.6$ \\
   \oiiiuv \vspace{-0.3cm} & $-17.9$  & $-18.4$ & $-18.5$ & $-18.5$ & $-18.7$ \\
  \siliii  \vspace{-0.3cm} &	$-18.2$ & $-18.5$ & $-18.7$ & $-18.7$  & $-18.6$ \\
  \ciii  \vspace{-0.3cm} &	$-18.2$  & $-18.5$  & $-18.7$ & $-18.7$ & $-18.6$ \\
\end{tabular}
\caption{NIRSpec-specific flux-detection limits (in units of
  log[erg s$^{-1}$cm$^{-2}$]) for different optical/UV emission lines at
  redshifts z=4--8 for three different exposure times, as
  indicated. The 5$\sigma$ flux limits were computed with
  the pre-flight (version 1.3) of the {\it JWST}
  exposure-time calculator.}\label{tableflux}
\end{table}

In this context, it is of interest to compute the number of galaxies expected 
to be detected in a given emission line, given an exposure time, in a 
NIRSpec field of view ($3 \times 3\ \mathrm{arcmin}^2$). This is shown 
as a function of redshift in Fig.~\ref{NumberCountsOpt} for the \oii, \hb, 
\oiii, \ha\ and \nii\ lines (from left to right), assuming exposure times of 
2, 5 and 10\,hr (from top to bottom).
In each panel of Fig.~\ref{NumberCountsOpt}, the light blue and red 
bars show the number of, respectively, SF-dominated and active 
(AGN-dominated and composite) galaxies more massive than $10^8\, \Msun$
predicted in redshift bins centred on $z=4$, 5, 6, 7 and 8 by the TNG50 
simulation. The fixed boxlength of the simulation translates into 
redshift-bin widths shrinking from $\Delta z =10^{-3}$ to $5 \times 10^{-4}$ 
over this range. These predicted numbers are the same in all panels of 
Fig.~\ref{NumberCountsOpt}, independent of emission line and exposure time. 
In contrast, the dark blue and red bars show the subsets of these galaxies 
whose line emission can be detected with NIRSpec with 5$\sigma$ significance. 
These numbers change with emission line and exposure time.

So far, we have not considered attenuation of emission lines by dust,
while galaxies may contain significant amounts of dust, even
at early cosmic epochs \citep[see e.g.,][]{Schneider04, Zavala22}.
We estimate the impact of attenuation by dust on the number counts 
of SF and active galaxies in Fig. \ref{NumberCountsOpt} by adopting
the dust attenuation curve of \citet{Calzetti00} with a $V$-band attenuation
of $A_{\rm V}=0.5\,$mag. The resulting drop in expected number counts is
shown by the light-grey thin bars carved in the dark-colour bars.

Fig.~\ref{NumberCountsOpt} shows that the TNG50 simulation
predicts from $\sim40$ (20) SF-dominated (active) galaxies per NIRSpec 
field in the $z=4$ bin, to $\sim5$ (2) at $z=8$. Virtually all of these can 
be detected in a two-hour exposure with NIRSpec in \ha\ and \oiii, typically 
the brightest optical emission lines, out to $z=7$ (when \ha\ drops out of
the observational window) and $z=8$, irrespective
of the presence of dust. Instead, exposures of at least 5 hours are
required to achieve a nearly unbiased census of \hb. Even 10-hour
exposures will not be sufficient to achieve an unbiased census of 
\oii\ SF-dominated emitters, and even less so of \nii\ emitters, especially 
in the presence of attenuation by dust. At any of
the survey depths in Fig.~\ref{NumberCountsOpt}, the observed
populations in these lines would be biased toward active galaxies.

All this suggests that
the JADES/MEDIUM survey will be able to detect
\ha\ and \oiii\ in the targeted galaxies out to $z=7$ and $z=8$,
respectively, and the 
DEEP survey extend this to \hb, while a
substantial fraction of SF-dominated, \oii\ and \nii\ emitters might remain
below the detectability limit of these surveys. It is also worth
noting that some galaxies less massive than $10^8 M_\odot$
may have line fluxes above the survey limits and therefore
be detectable. As stated before, we do not consider such
low-mass galaxies here as they are unresolved in TNG50.

In Fig.~\ref{NumberCountsUV}, we show the analogue of 
Fig.~\ref{NumberCountsOpt} for the \civ, \heii, \oiiiuv, \siliii\ and
\ciii\ UV emission lines (from left to right).
If galaxies were dust-free, the most complete censuses in 
this wavelength range would be achievable for \ciii, \oiiiuv\ and \civ\
emitters at $z\ga6$, especially for long exposure
times, although significant fractions of
SF-dominated galaxies would still be missed at lower redshifts. 
Instead, the weaker \heii\ and \siliii\ lines can be properly sampled only in 
active galaxies, because \heii\ requires highly energetic photons, while 
\siliii, hindered by the low abundance of Si relative to  C and O, is boosted 
by AGN radiation. Exposure times of at least 5\,hours would be required to start
detecting populations of dust-free, SF-dominated emitters in
these lines with NIRSpec, and even 10 hours would not suffice to
achieve proper censuses of these populations.

In the presence of dust, the expected number counts in 
Fig.~\ref{NumberCountsUV} drop significantly, particularly for 
SF-dominated galaxies, whose weaker lines compared to 
AGN-dominated galaxies fall more easily below the detectability
limits of surveys. Yet, significant fractions of \ciii, \oiiiuv\ and \civ\ emitters 
still appear detectable with the adopted $A_{\rm V}=0.5$, at least for 
exposure times exceeding 10 hours. We caution however 
that the predicted counts of \civ\ emitters are likely to be
overestimated, as we have ignored the enhanced absorption 
of \civ\ photons by dust due to resonant scattering
\citep[e.g.,][]{Senchyna22}.

The results of Fig.~\ref{NumberCountsUV} therefore
suggest that the UV lines most accessible to the JADES/MEDIUM and 
DEEP surveys, modulo dust attenuation, are likely to be \ciii\ and \oiiiuv,
and, in cases where enhanced attenuation by resonant scattering
is particularly modest, \civ. SF-dominated \heii\ and \siliii\ emitters are
likely to remain more elusive. While this
prediction relies on SF models having difficulty to reproduce the 
observed \heii\ emission of low-redshift analogues of distant SF galaxies
\citep[e.g.,][]{Plat19}, it is interesting to note that a non-detection of \heii\ 
combined with detections of \ciii\ and \oiiiuv\ (and perhaps \civ) could 
point to a population of SF-dominated galaxies (as expected from the 
\heii-based diagnostic diagrams in Fig.~\ref{uvdiagnostics}).

%*****************************************************************************************************
%*****************************************************************************************************
\section{Discussion}\label{discussion} 
%*****************************************************************************************************
%*****************************************************************************************************

In Sections~\ref{ionizingsources} and \ref{luminosityfunctions}, we
have shown that the predicted optical and UV emission-line 
properties of the IllustrisTNG galaxy populations are consistent with
numerous observational constraints out to redshifts $z \sim 2$--3. 
Based on this success, we could make predictions about the 
observability of these lines in more distant galaxies, out to $z\sim7$--8, 
and propose associated spectral diagnostics as guidance for the 
interpretation of new spectroscopic surveys at high redshift. These results
represent a major statistical extension of earlier 
work by \citet{Hirschmann17, Hirschmann19}.
%, based on a limited set 
%of 20 zoom-in simulations of massive galaxies not including the 
%emission from fast, radiative shocks.

Despite this success, the emission-line catalogues of IllustrisTNG 
galaxies used in our study may be affected by several caveats related
to current photoionization models, modern cosmological simulations and
our coupling methodology. We discuss these in 
Sections~\ref{cavphotoionization} and \ref{cavillustris} below and also
compare in Section~\ref{comparison} our emission-line catalogues to
those from previous studies of optical and UV emission lines of simulated 
galaxies.

\subsection{Caveats in the photoionization models}\label{cavphotoionization}

All photoionization models considered in this work were performed
with recent versions of the \textsc{Cloudy} and \textsc{Mappings}
photoionization codes, adopting a common set of element abundances
down to metallicities of a few per cent solar (Section~\ref{ELmodels}).
However, these models are computed via 1D calculations of a gas 
patch irradiated by ionizing photons, assuming a simplified geometry 
(e.g., spherical, plane-parallel) and constant gas density, which do not
reflect the complex 3D gas distributions found in nature. Also, 
while the \hii-region and AGN-NLR photoionization models
employed here are ionization-bounded, some actively star-forming 
galaxies show evidence of leakage of Lyman-continuum photons 
\citep[e.g.,][]{deBarros16, Shapley16, Bian17, Flury22}. This 
generally tends to reduce the intensities of low-ionization relative to 
high-ionization lines \citep[see, e.g., the density-bounded 
calculations of][]{Plat19}.  

The PAGB models we adopted to describe gas photoionized by 
evolved stellar populations also suffer from uncertainties. In 
particular, the most recent version of the \citet{Bruzual03} 
stellar-population-synthesis code we rely on incorporates PAGB 
evolutionary tracks from \citet[][]{MBertolami16}, which have significantly
shorter lifetimes -- although compensated in part by higher luminosities -- 
than previous prescriptions. Adopting older PAGB models could 
enhance the predicted EW(\ha) of PAGB-dominated galaxies
by a factor of up to 2.  We have checked that this would not 
qualitatively affect our results (in the sense that more galaxies
would be PAGB-dominated, but they would still separate
robustly in the WHAN diagram of Fig.~\ref{whan}).
%\SC{We should probably double-check  
%this number, and this last statement.}\MH{I checked this, and it is
%roughly a factor of two. You can see that in Fig. 3.: the old
%models were largely consistent with the EW limit proposed by Cid
%Fernandes; compared to the red line, which is the limit for our current models,
%there is difference of WHa by a factor of 2.} 
For the ionizing radiation from
an accreting black hole in the AGN-NLR models, for simplicity, we adopted 
the fixed broken power-law shape described by \citet{Feltre16}, with 
$S_\nu\propto\nu^\alpha$ and $\alpha = -1.7$ at wavelengths below 
2500\,\AA. As shown by \citet{Hirschmann17, Hirschmann19}, adopting 
a different value for $\alpha$ would not significantly affect the predicted 
luminosities of the optical and UV emission lines considered in the present work.

Another, perhaps more serious limitation of the \hii-region and PAGB 
models used in our study is that they do not include the hot ionizing radiation 
from binary-star products (envelope-stripped or spun-up stars), as they
are based on single-star population-synthesis models. Hard radiation from
binary stars will enhance the luminosities of lines from species requiring very 
high ionizing energies, such as \heiinl\ \citep[see, e.g.,][]{Xiao18}. This point
is of particular relevance, as no current stellar-population-synthesis model 
can account for the strong \heii\ and \heiiopt\ emission exhibited by some 
very metal-poor ($\la0.2\,\Zsun$) SF galaxies \citep[see, e.g.,][]{Plat19, 
Stanway19}. In fact, this is true even for the BPASS\,v2.2.1 models of 
\citet{Eldridge17}, which do include binary-star populations. Interestingly, 
the latest version of the \citet{Bruzual03} single-star models used here, 
which incorporate updated calculations of stellar evolution and atmospheres
\citep{Bressan12, Chen15}, produce more \heiinl-ionizing photons than
BPASS\,v2.2.1 \citep[see][]{Plat19}. Other potential sources of \heiinl\
emission in such metal-poor galaxies could be mini-AGN, radiative shocks
and massive X-ray binaries, although these are heavily debated 
\citep[e.g.,][]{Plat19, Senchyna20, Simmonds21, Umeda22}.

Despite the above caveats, the predictions of the \hii-region and AGN-NLR 
models used in this study have been repeatedly shown to perform remarkably 
well in comparisons with observations of common emission lines from other 
elements than \heiinl\ \citep[e.g.,][]{Chevallard18, Mignoli19, Tang19}. As 
shown in Section~\ref{uvdiagrams}, increasing the \heii-line luminosity 
predicted by our \hii-region models by a factor of 4 of would not reduce much the 
effectiveness of the suggested selection criteria in UV diagnostic diagrams in
Fig.~\ref{uvdiagnostics} \citep[in line with the result of][]{Hirschmann19}. We 
also find that this would attenuate the bi-modality of the \heii-luminosity function 
(Fig.~\ref{UVLumfct}) and increase the number counts of SF-dominated, \heii\
emitters by a factor of 2 to 3 (Fig.~\ref{NumberCountsUV}). 

Finally, we note that the radiative-shock models of \citet{Alarie19} used in
this study update the older calculations of \citet{Allen08}, which were based 
on a previous version of the \textsc{Mappings} code and spanned a smaller 
range of metallicities. Both model sets are limited to shocks 
fast enough  ($\ge 100\,$km\,s$^{-1}$) to generate sufficient radiation to 
ionize the pre-shock gas before it enters the shock front \citep[i.e., to allow the
pre-shock gas to reach ionization equilibrium before 
it is shocked; see also][]{Sutherland17}. The models also assume that shocks
are propagating in a neutral medium ionized by the hot post-shock gas,
which might not always be the case in practice (e.g., if a shock propagates
through an ionized medium). As we shall see in the next section, the resolution and
identification of shocks in cosmological simulations tend to include stronger 
assumptions and limitations than do the grids of radiative-shock models.

\subsection{Caveats in the IllustrisTNG simulations and their coupling
with emission-line models}\label{cavillustris}

A potential source of inaccuracy of all current large-scale cosmological
simulations, including IllustrisTNG, is that they do not resolve the 
ISM, which forces one to appeal to often simplified and ad-hoc 
sub-resolution models to describe baryonic processes \citep[see][for 
a discussion]{Naab16}. Specifically, the choice of stellar- and
AGN-feedback models has been shown to significantly affect
various galaxy properties \citep[e.g.][]{Pillepich18b}. In this context, it is
important to stress that IllustrisTNG is one of the best-tested 
and explored, large-scale cosmological simulations 
to date, providing fairly realistic galaxy populations up
to  $z=3$, for example in terms of chemical enrichment, SFR and 
black-hole accretion-rate histories \citep[although the observed AGN luminosity 
function is not always well reproduced, especially at low redshift, 
see][]{Habouzit19, Habouzit22}, stellar populations and the associated
scaling relations.
This success provides a solid foundation for our 
predictions of the emission-line luminosities controlled by these 
physical quantities.

Furthermore, because IllustrisTNG, like other modern large-scale 
cosmological simulations, does not resolve gas properties, such
as density, in individual ionized regions, nor does it track the
formation, evolution and destruction of dust grains in the ISM, we
must adopt a fixed  hydrogen gas density in ionized regions
($10^3\,$cm$^{-3}$ for NLR, $10^2\,$cm$^{-3}$  
for \hii\ regions and $10\,$cm$^{-3}$ for line emission from post-AGB  
stars), as well as a fixed dust-to-metal mass ratio ($\xi_d =
0.3$). As described in section 5.1.4 of \citet{Hirschmann17}, both
recent observations \citep{DeCia16, Wiseman17} and semi-analytic 
models \citep{Popping16} suggest that, at given gas metallicity, the
dust-to-metal mass ratio hardly changes with redshift out to $z = 6$.
The models further predict that this is the case also at fixed
stellar mass (see figs~5 and 6 of \citealp{Popping16}) and that galaxies
more massive than $M_{\mathrm{stellar}}>3 \times 10^{9} M_\odot$ are
expected to have dust-to-metal mass ratios around 
0.2, reasonably close to our adopted value. In
\citet{Hirschmann17, Hirschmann19}, we further studied the impact of these 
parameters on optical and UV line ratios, and found that they have a 
smaller effect on observed line ratios than the parameters of the 
photoionization models that can be determined from the simulations.

Another consequence of not resolving individual \hii\ regions in
IllustrisTNG is that we have to adopt galaxy-wide properties of
\hii-regions. We follow here the methodology of 
\citet[][see Section~\ref{theory} above]{Charlot01}, which
convolves the time evolution of an \hii\ region with the galaxy
star-formation history. While this approach goes beyond most 
current photoionization models (which typically treat a galaxy 
as a single region ionized by the entire stellar population), it 
suffers from shortcomings: even though we assume that a galaxy 
is composed of several \hii\ regions (with different star-cluster
ages), any variation in the gas properties of the regions across
the galaxy (such as density, filling factor and metallicity) is neglected.
Regardless, these galaxy-wide line-emission models may be 
justified by their long-term success in reproducing observed 
galaxy spectra (Section~\ref{cavphotoionization}). More detailed 
studies will be needed to fully validate this approach, but they 
lie beyond the scope of the present paper. 

Furthermore, the inability to resolve the ISM in large-scale
cosmological simulations such as IllustrisTNG implies that
shocks cannot be tracked and
identified in unresolved cool gas phases. In fact, the
on-the-fly shock finder described in Section~\ref{theory} is by 
default disabled for star-forming gas. We have therefore
intentionally neglected shock-related emission from the dense
star-forming ISM itself, which could arise for example from 
supernova explosions. Instead, in IllustrisTNG, we
are primarily capturing shocks caused by (AGN-driven) outflows,
gas accretion as well as merger events arising in the
warm/hot component of the ISM and in more distant regions, 
such as the disk-halo interface. Thus, in our analysis, we 
cannot draw any robust conclusion on line emission due to shocks in
higher-redshift galaxies, where 
AGN-driven outflows seem to be less prevalent, and instead shocks due
to stellar feedback, neglected in our approach, may become
important. Also, at the moment, IllustrisTNG includes 
radiation-driven feedback only as thermal energy input, which may
be insufficient to describe reality. We might therefore miss shocks linked 
to radiation-driven winds from radiatively efficient AGN, which may 
become particularly important in high-redshift galaxies.

While our modelling of synthetic emission lines of
IllustrisTNG galaxies accounts for the effects of dust in \hii\ regions
and AGN NLR (Section~\ref{coupling}), it neglects attenuation by 
dust in the diffuse ISM (with the exception of
  Figs~\ref{NumberCountsOpt} and  \ref{NumberCountsUV} in Section
  \ref{numbercounts}). To some extent, this neglect is  
justified by the fact that we compare our models with observed line
ratios and line-luminosity  
functions corrected for this effect. Moreover, as expected from our
results in section~5.1 of \citet{Hirschmann19}, we find that adopting a 
\citet{Calzetti00} attenuation law with a $V$-band attenuation of 
$A_V=0.5\,$mag would negligibly impact most UV diagnostic diagrams in 
Fig.~\ref{uvdiagnostics}. This however would reduce the predicted 
number counts of line emitters at high redshift by factors of up to several, 
as shown by Figs~\ref{NumberCountsOpt} and \ref{NumberCountsUV}. 
A more sophisticated modelling of dust (in the spirit of the on-the-fly 
dust modelling of \citealp{McKinnon18}), beyond the scope of the 
present paper, will be required for more robust predictions.

Finally, we note that our methodology does not account for the effect of
interstellar-line absorption in stellar birth clouds nor in the diffuse 
ISM. As shown by  \citet{Vidal17},  while some prominent UV emission 
lines, such as \oiiiuv, \siliii\ and \ciii, are little sensitive to interstellar 
absorption, the luminosities of other lines, such as \civ\ and \nv, can 
be significantly reduced through this effect. While quantifying interstellar-line 
absorption in simulated galaxies is important, we postpone such a complex 
analysis to a future study.

\subsection{Comparison with previous emission-line studies}\label{comparison}

In recent years, an increasing number of theoretical studies have emerged,
which provide models for the nebular emission from young star clusters in 
galaxies in cosmological simulations. Below, we briefly outline the different 
methodologies proposed to compute line emission, summarise the main 
associated goals and results, and discuss how they compare with the 
method and analysis presented in this work. 
%\SC{Please double-check that
%I have not made errors in interpreting your original
%text.}\MH{Everything seems to be consistent with what I wrote.}

%{\bf Kewley}: just considers SFR and chemical enrichment histories

{\bf The {\sc Sags} semi-analytic model \citep{Orsi14}}:
In this work, the gas metallicity and ionization
parameter (via a dependency on gas metallicity) of {\sc Sags} galaxies
are used to select \hii-region models from a \textsc{Mappings iii}
library \citep{Levesque10}. With this method, the authors reproduce the
local BPT diagram and the evolution of optical-line luminosity
functions (we note that the predicted \oii-luminosity function
underestimates the observed one locally, as for our models in 
Fig.~\ref{Halum}). However, their BPT diagrams, specifically the 
\oiiihb-ratio, do not evolve with redshift as observed. This model 
has been further applied to the {\sc Galacticus} and {\sc Sage} 
semi-analytic models (run over merger trees from the MultiDark
simulation) to explore the luminosity functions and clustering of 
\oii-emitting galaxies \citep{Favole20}.

{\bf The {\sc Sage} semi-analytic model applied to the UNIT project
simulations \citep{Knebe22}:} Emission lines for the {\sc Sage}
semi-analytic model are constructed using the same method as in
\citet{Orsi14}. Thanks to the large volume of the UNIT simulation
(effectively $\sim5\,$Gpc$^3$), this study presents galaxy catalogues for
{\it Euclid} forecasts. The authors start by validating their model 
against the observed \ha\ luminosity function, and then explore 
the abundance and clustering of model \ha-emitting galaxies.

{\bf The {\sc Galacticus} semi-analytic model \citep{Zhai19}:}
The emission-line luminosities of {\sc Galacticus} galaxies are
computed using the \textsc{Cloudy} photoionization code \citep{Ferland13}. 
The key step is to generate and interpolate tabulated libraries of
emission-line luminosities using \textsc{Cloudy} as a function of the number of
ionizing photons for various species, the ISM metallicity, the
hydrogen gas density and the volume filling factor of \hii\ regions,
which can be computed for {\sc Galacticus} galaxies. The
authors validate their emission-line results against observed \ha\
and \oiii\ luminosity functions and make predictions of number
counts of line-emitting galaxies for galaxy surveys with the {\it Roman
Space Telescope}. 

{\bf The {\sc{L-Galaxies}} semi-analytic model
  \citep{Izquierdo-Villalba19}:} This study adds the \citet{Orsi14}
model to {\sc L-Galaxies} lightcones (run over the Millennium
simulation) to create synthetic galaxy catalogues including emission
lines for narrow-band photometric surveys. The authors focus on optical lines
(including \neiii) and find fairly good agreement with the observed
evolution of the \ha, \oiii\ and \oii\ luminosity functions.

{\bf The {\sc Galform} semi-analytic model
  (\citealp{Gonzalez-Perez18, Gonzalez-Perez20, Baugh22})}:
\citet{Gonzalez-Perez18, Gonzalez-Perez20} assign \hii-region models from
\citet{Stasinska90} to {\sc Galform} galaxies and investigate line-luminosity
functions, halo-occupation distributions and the distribution of
these galaxies in the cosmic web. While the original {\sc Galform}
emission-line model provided a poor reproduction of local
BPT diagrams, \citet{Baugh22} propose an updated 
model based on the published version of the \citet{Gutkin16} 
photoionization grid and a relation between ionization parameter, 
gas metallicity, gas density and specific SFR from \citet{KashinoInoue19}. 
Their study suggests that the offset in \oiiihb\ of high-redshift relative to
local galaxies results primarily from a higher ionization parameter 
and higher gas density, largely in line with our results in \citet{Hirschmann17}
and in Section~\ref{opticaldiagrams_evol} above.

{\bf The $z>7$ cosmological simulation of \citet{Shimizu16}:} This
study explores the optical- and UV-line emission of $z>7$
galaxies in cosmological simulations and their detectability with
different current and future observational facilities ({\it JWST}, 
extremely large telescopes). The authors compute Ly$\alpha$, 
\ha\ and \hb\ line luminosities using \textsc{Pegase 2} \citep{Fioc97}. 
The luminosities of metal lines are scaled to the \hb\ luminosity 
based on \textsc{Cloudy} models and assuming a metallicity-dependent 
emission efficiency for each line. \citet{Shimizu16} find that \oiii, \hb, \civ\ 
and \ciii\ are the brightest lines in high-redshift galaxies and suggest
that these could be strong targets for future observational facilities out
to even for $z\sim15$. 

{\bf The {\sc Simba} cosmological simulation 
  \citep{Garg22}}: In this study, the nebular emission of a simulated galaxy
is computed by summing the contributions from all young-star particles.
The authors employ a sub-grid model to describe the star-cluster mass distribution
per star particle and perform \textsc{Cloudy} calculations, using ages 
and metallicities appropriate for each star particle, as well as an empirical 
dependence of N/O on O/H \citep[in the spirit of CloudyFSPS,][]{Byler17}.
Even though this methodology provides spatially-resolved information 
about line emission in a galaxy, it neglects all important dust processes
in \hii\ regions \citep{vanHoof04}. With this approach, \citet{Garg22}
reproduce the star-forming branch of the local BPT diagram and find
that the \oiiihb\ offset at high redshift arises primarily from selection 
effects. Yet, they find a mismatch between the predicted and
observed \oiiihb\ ratios of $z \sim 2$ galaxies, which can be alleviated 
by invoking higher N/O ratio or a lower ionization parameter at fixed
O/H ratio at high redshift. While \citet{Hirschmann17} had a similar 
conclusion with regard to N/O, the finding of \citet{Garg22} concerning
the ionization parameter disagrees with our result in 
Section~\ref{opticaldiagrams_evol} above, as well as with those of 
several previous studies \citep[e.g.,][]{Brinchmann08, Hirschmann17, 
Kashino17, Kewley19, Baugh22}. 

{\bf The {\sc BlueTides} cosmological simulation  at $z=8-13$ 
\citep{Wilkins20}}: As in the {\sc Simba} simulation of \citet{Garg22}, 
the nebular emission of a simulated galaxy is computed by summing 
the contributions from all young-star particles, connecting 
ionization-bounded \textsc{Cloudy} models to the predicted gas 
metallicity and ionization parameter of each star particle, and assuming 
a fixed gas density and dust-to-metal mass ratio. The authors focus on 
predictions of a few optical (\ha, \hb\ and \oiii) and UV (\cii, \ciii\ and \neiii)
emission lines for simulated galaxies at $z=8$--13. They find good 
agreement of the predicted equivalent-width distributions of \ciii\ and 
\oiii+\hb\ with the few available observational constraints, but poorer 
agreement between the predicted and observed \citep[from][]{deBarros16}
\oiii+\hb\ luminosity function. As we have seen, at slightly lower redshift
($z=5$--6), our predicted \oiiinl+\hb\ luminosity function based on the 
IllustrisTNG simulations roughly matches this observational constraint
(Fig.~\ref{Halum}).

{\bf The IllustrisTNG cosmological simulation at high redshifts
  \citep{Shen20}:} In this work, the \ha, \hb\ and \oiii\ luminosities
of IllustrisTNG galaxies at $z=2$--8 are computed through the
coupling with a \textsc{Mappings iii} library of young ionizing stars 
clusters \citep[][considering only stellar age and metallicity for the
coupling, and including dust physics]{Groves08}. The authors find
that the predicted \oiii+\hb\ luminosity function is consistent with 
observations at $z=2$--3, but that the \ha\ luminosity function at 
$z\sim2$ is $\sim0.3$\,dex dimmer than observed. 
%\SC{Are we 
%the only ones to include the \oiiiaur\ line in the \oiiinl+\hb\ LF? Should 
%we change that?} \MH{No, I don't think we should change that as
%narrow-band survey typically don't have the resolution to exlude the
%auroral oxygen line...} 
At higher redshift, the models show hardly any
evolution of the luminosity functions out to $z\sim5$, and are slightly
dimmer than the observed \oiii+\hb\ luminosity of \citet{DeBarros19}. 
In contrast, as illustrated by Fig.~\ref{Halum} above, our emission-line
treatment of the same cosmological simulations allows us to capture 
the observed \ha\ luminosity function of $z=2$ galaxies, and we find a 
marked decline of the \ha\ and \oiiinl+\hb\ luminosity functions from
$z=3$ to $z=6$. These contradictory emission-line predictions will be
testable with upcoming {\it JWST} surveys. 

To summarise, the vast majority of recent studies on emission-line
predictions of simulated galaxies focus on optical-line luminosity
functions, clustering of line emitters and number-count predictions
for high-redshift galaxy surveys expected to emerge from current and
future observational facilities. Despite the wide variety of modelling 
approaches, and the differences in photoionization models, basic 
predictions, such as optical-line luminosity functions are (surprisingly) 
widely consistent across different studies, including the work 
presented here. Only two such published studies \citep[][aside
from our own previous work]{Garg22, Baugh22} explored the location 
of simulated galaxies in BPT diagrams, and the evolution of the 
associated line ratios with redshift. The results of \citet{Baugh22} agree
with those presented here \citep[and in][]{Hirschmann17} by favouring
an evolution of the ionization parameter as the main driver of the rise 
in \oiiihb\ ratio at high redshift, in stark contrast to the theoretical results 
of \citet{Garg22}, which instead favour selection effects as
an explanation for the observed trends.

Compared to the above previous studies, our approach is unique 
in that we consider a large parameter space of physical quantities, 
as well as a more self-consistent modelling of the ionization parameter 
to select \hii-region photoionization models. Even though in some 
cosmological simulations, line emission is modelled on a 
spatially resolved basis, it is not clear how much more reliable 
such an approach is than the one presented here: modern, 
large-scale cosmological simulations can by no means resolve
ionized regions in the ISM and masses of young star clusters, so that
additional assumptions/sub-grid models become necessary (often just
considering star-particle age and metallicity to model line emission).

{\it Most importantly, we account not only for stellar nebular emission
of simulated galaxy populations, as done in literature so far, 
but also simultaneously for line emission due to AGN,
post-AGB stars and fast radiative shocks}, resulting in comprehensive
emission-line galaxy catalogues at different cosmic epochs.  

%*****************************************************************************************************
%*****************************************************************************************************
\section{Summary}\label{summary} 
%*****************************************************************************************************
%*****************************************************************************************************

In this paper, we have presented the {\it first
multi-component}, optical and UV emission-line catalogues of
large simulated galaxy populations at different cosmic epochs.
Following the methodology of \citet{Hirschmann17,   Hirschmann19},
the emission-line catalogues have been constructed by employing the
IllustrisTNG cosmological simulations and by self-consistently
connecting them to modern, state-of-the-art photoionization models
\citep{Gutkin16, Feltre16, Hirschmann17, Alarie19}. This allows us to 
compute in a self-consistent way the line emission from multiple components: young star clusters, 
AGN NLR, PAGB stellar populations and, for the first time, fast
radiative shocks. Specifically, we consider predictions from IllustrisTNG
galaxies for the redshift evolution of global and central interstellar
metallicity, C/O abundance ratio, SFR, BH  accretion rate, global and
central average gas densities, the age and metallicity of PAGB
stellar populations, and shock properties 
(velocity, magnetic-field strength, interstellar metallicity and
shock surface). Based on these, we select SF, AGN, PAGB and shock
nebular-emission models for each simulated galaxy (above the
mass-resolution limit) at any redshift. By default, we adopt a fixed
dust-to-metal mass ratio, ionized-gas hydrogen density, power-law
index of the AGN ionizing radiation and pre-shock gas density.

%Thus, in
%this work, we follow and extend the methodology of
%\citet{Hirschmann17,   Hirschmann19} by applying it to large
%cosmological simulations and by adding the contribution from shocks to
%line emission.
%This study demonstrates that our model predictions statistically
%confirm results presented in \citet{Hirschmann17,
%  Hirschmann19}.

Taking advantage of these emission-line catalogues of IllustrisTNG
galaxy populations, we have presented a variety of optical
and UV diagnostic diagrams to identify ionizing sources in galaxies
in the local Universe and at high redshift, together with the predicted
evolution of line-luminosity functions and number counts of distant
line-emitting galaxies expected to be detectable in deep spectroscopic 
observations with the NIRSpec instrument on board {\it JWST}. We can 
summarise our main results as follows: 

\begin{itemize}
 \item The synthetic \oiiihb, \niiha, \siiha, \oiha\ and \oiiioii\
   emission-line ratios and \ha\ equivalent widths of
   IllustrisTNG galaxies are in excellent statistical agreement with
   observations of star-forming, active, shock-dominated and
   retired (PAGB) SDSS galaxies in the local Universe. This
   confirms, for large galaxy populations (and with the different
   physical models adopted), the result originally 
   obtained by \citet{Hirschmann17, Hirschmann19} for only 20 
   re-simulated, massive galaxies. 

 \item Our results are consistent with the classical selection criteria
   in standard BPT diagnostic diagrams to distinguish SF and active
   galaxies in the local Universe. Furthermore, we confirm that the WHAN 
   diagram can robustly identify retired galaxies, and the \oiiioii-versus-\oiha\
   diagram shock-dominated galaxies.

 \item Present-day galaxies classified as `SF-dominated' and 
  `composites' (i.e., SF+AGN) 
   are found to be primarily star-forming, main-sequence galaxies with 
   large cold-gas fractions. Then, SFR and cold-gas content successively 
   decrease from `AGN-dominated' to `shock-dominated' and 
   `PAGB-dominated' galaxies: shock-dominated galaxies are
   predominantly located in the green valley of the SFR-stellar mass
   plane, and PAGB-dominated galaxies on the red sequence. This is
   primarily due to AGN feedback causing shocks in outflowing gas and
   suppressing star formation and black-hole accretion. This allows the 
   (intrinsically weak) line emission from shocks and PAGB stars 
   to become dominant.
   
 \item Shock- and PAGB-dominated galaxies make up between 3 and 
   10~per cent of the present-day galaxy population, depending on box size
   and resolution of the simulation. At redshifts $z>1$, these fractions drop
   below 1~per cent, with SF-dominated galaxies being
   the most abundant population (between 40 and 70~per cent) at all redshifts.

 \item At fixed \niiha\ ratio, the \oiiihb\ ratio is predicted to
   increase from low to high redshift. For SF-dominated galaxies,
   the increase is similar to that reported in several 
   observational studies \citep[e.g.,][]{Steidel14}. We confirm our previous
   finding in \citet{Hirschmann17} that at fixed galaxy stellar mass, the
   increase in \oiiihb\ from low to high redshift
   is primarily a consequence of a higher 
   ionization parameter in high-redshift galaxies, regulated by a
   larger specific SFR and a higher global gas density (consistent with
   the results of \citealp{Baugh22}).
   \item At high redshifts, classical optical-selection criteria to 
     identify the dominant ionizing sources in galaxies
     are no longer applicable: a drop in \niiha\ causes
     composite and AGN-dominated galaxies to shift toward the SF-dominated
     region of the classical \oiiihb-versus-\niiha\ BPT diagram. Instead, nine
     UV diagnostic diagrams and related selection criteria, originally presented
     in \citet[][for 20 re-simulated massive galaxies and
     their progenitors]{Hirschmann19}, provide a powerful means to distinguish SF
     from active galaxy populations out to early cosmic times (e.g., \ciii/\heii\
     versus \oiiiuv/\heii).
 \item The synthetic emission-line properties of IllustrisTNG galaxies
   can largely reproduce the observed 
   evolution of the \ha, \oiii\ and \oii\ luminosity functions out to
   $z=3$ (albeit with some tension for \oii). We provide
   predictions of optical- and UV-line luminosity and flux functions
   at $z>3$, potentially useful for future surveys with, e.g., {\it
     JWST}, {\it Euclid} and the {\it Roman Space Telescope}. From
   $z=3$ to $z=7$, the predicted number densities of optical
   and UV emission-line galaxies can drop by up to three orders of
   magnitude (primarily at faint luminosities), reflecting the
   declining SFR density over that redshift range.
 \item Finally, we present number counts of SF and active galaxies 
   expected to be detectable at redshifts between $z=4$ to $z=8$  in
   deep, medium-resolution spectroscopic observations with the NIRSpec
   instrument on board {\it JWST}.  We find that 2-hour-long  exposures 
   are sufficient to obtain unbiased censuses of \ha\ and \oiii\ emitters, 
   while at least 5 hours are required  for \hb, irrespective of the presence 
   of dust. Instead, for \oii, \oiiiuv, \ciii, \civ, \nii, \heii\ and
   \siliii\  emitters, even though the
   active-galaxy population can still be well sampled, progressively smaller
   fractions of the SF-galaxy population are likely to be detectable, 
   especially in the presence of dust, even with exposure times 
   exceeding 10 hours.
 \end{itemize}

 Overall, the multi-component optical and UV emission-line galaxy catalogues
 presented in this paper {\it include for the first time not only
   stellar nebular emission of simulated galaxy populations, as done
   in literature so far, but also simultaneously line emission due to
   AGN, post-AGB stars and fast radiative shocks}.
 These catalogues provide  useful  insights into various
 diagnostic diagrams and statistical properties of optical 
 and UV emission-line  galaxies. These mock observables may be further
 helpful for the  interpretation of near- and far-future emission-line
 surveys with, e.g.,  {\it JWST}, {\it Euclid} and the {\it VLT/MOONs}. 
 This study is the first of a  series, in which we also plan to explore, for example,
 the clustering of emission-line galaxies and the use of strong-line ratios
 as tracers of interstellar  metallicity, the ionization parameter
 and other physical quantities of the ISM  in high-redshift galaxies.

\section*{Acknowledgements}
MH acknowledges funding from the Swiss National Science Foundation
(SNF) via a PRIMA Grant PR00P2\_193577
``From cosmic dawn to high noon: the role of black holes for young
galaxies''.
ECL acknowledges support of an STFC Webb Fellowship (ST/W001438/1).
RSS is supported by the Simons Foundation. 
JC acknowledges funding from the ERC Advanced Grant 789056
``FirstGalaxies'' (under the European Union’s Horizon 2020 research and
innovation programme).
CM aknowledges grant from UNAM/DGAPA/PAPIIT project IN101220.
DN acknowledges funding from the Deutsche
Forschungsgemeinschaft (DFG) through an Emmy Noether Research
Group(grant number NE 2441/1-1).
The TNG simulations were run with
compute time granted by the Gauss Centre for Supercomputing (GCS)
under Large-Scale Projects GCS-ILLU and GCS-DWAR on the GCS share of
the supercomputer Hazel Hen at the High Performance Computing Center
Stuttgart (HLRS). 

\bibliographystyle{mnras}
\bibliography{Literaturdatenbank}

\label{lastpage}

\end{document}